\documentclass[a4paper,12pt,twoside]{report}
\usepackage[left=2cm,right=2cm,top=2.5cm,bottom=2.75cm]{geometry}

\usepackage[english]{babel}     
\usepackage[utf8]{inputenc}
\usepackage{csquotes}
\usepackage{CJKutf8}            

\usepackage{tgpagella}                  

\usepackage{latexsym}
\usepackage{mathchars}
\usepackage{amsmath,amssymb,amsfonts}   
\usepackage{numprint}                   
\usepackage{nicefrac}

\usepackage{setspace}       
\usepackage{pdflscape}      
\usepackage{emptypage}      

\usepackage[hang]{footmisc} 
\usepackage{graphicx}       
\usepackage{adjustbox} 

\usepackage{multirow}       
\usepackage{longtable}

\usepackage{caption}        
\usepackage{subcaption}
\usepackage{enumitem}       
\setlist[enumerate]{leftmargin=*, topsep=0.25em, itemsep=0pt}
\setlist[itemize]{leftmargin=*, topsep=0.25em, itemsep=0pt}

\usepackage{verbatim}

\usepackage[backend=biber, backref=true, maxbibnames=99, style=numeric, sorting=nyt]{biblatex}             
\renewbibmacro*{urldate}{
    (accessed \printfield{urlday}/\printfield{urlmonth}/\printfield{urlyear})
}

\DefineBibliographyStrings{english}{   
  backrefpage = {cited on p.},  
  backrefpages = {cited on pp.},
}

\usepackage{tocloft}            

\usepackage{chngcntr}           

\usepackage{hyperref}       
\usepackage{url}            

\input{uhead.sty}
\input{boxit.sty}
\input{icthesis.sty}

\makeatletter
\newcommand{\vast}{\bBigg@{4}}
\newcommand{\Vast}{\bBigg@{5}}
\makeatother

\usepackage{pifont}                             
\newcommand{\cmark}{\text{\ding{51}}}%
\newcommand{\xmark}{\text{\ding{55}}}%

\newcommand{\indep}{\perp \kern-0.6em \perp}    

\begin{document}

\title{\LARGE {\bf Some Statistical and Data Challenges When Building Early-Stage Digital Experimentation and Measurement Capabilities}\\
}

\author{C. H. Bryan Liu}
\submitdate{3rd October 2023}

\maketitle

\preface
\begin{statementoforiginality}
I confirm that the thesis reports original research work done by myself during the programme of study. The thesis has not been submitted elsewhere for examination for a PhD degree. Part of the thesis is adapted from research papers that I first-authored and published during the programme of study. They are listed in Section~\ref{sec:introduction_publications} and stated at the beginning of relevant chapters where applicable.
In addition, I have appropriately referenced ideas from others in the thesis and sought the necessary permission to use copyrighted materials.

The thesis is written at a time when large language models (LLMs) are sufficiently advanced to produce "human-like" writing on an arbitrary topic and refine the text based on external prompts. I see much potential with such technology in research and industry practice. That said, the use of LLMs in academic writing remains highly controversial as of 2023, with many unresolved issues on correctness, academic integrity and plagiarism, and publishing ethics~\cite{alahdab2023potential}.

As such, I also confirm that I did not use any LLMs or services derived from them to write this thesis.
That said, I used (the non-LLM part of) Grammarly, an online contextual proofreading platform, to help correct the written text's spelling, punctuation, and grammar. The platform also flags redundant, overused, or informal phrases and sentences based on set rules to help improve the overall delivery of the thesis. However, the decision to incorporate the suggestions solely rests with me.

\clearpage

The copyright of this thesis rests with the author. Unless otherwise indicated, its contents are licensed under a Creative Commons Attribution 4.0 International Licence (CC BY 4.0).

Under this licence, you may copy and redistribute the material in any medium or format for both commercial and non-commercial purposes. You may also create and distribute modified versions of the work. This on the condition that you credit the author.

When reusing or sharing this work, ensure you make the licence terms clear to others by naming the licence and linking to the licence text. Where a work has been adapted, you should indicate that the work has been changed and describe those changes.

Please seek permission from the copyright holder for uses of this work that are not included in this licence or permitted under UK Copyright Law.

\end{statementoforiginality}

\cleardoublepage
\addcontentsline{toc}{chapter}{Abstract}

\begin{abstract}

Digital experimentation and measurement (DEM) capabilities -- the knowledge and tools necessary to run experiments with digital products, services, or experiences and measure their impact -- are fast becoming part of the standard toolkit of digital/data-driven organisations in guiding business decisions.  Many large technology companies report having mature DEM capabilities, and several businesses have been established purely to manage experiments for others. Given the growing evidence that data-driven organisations tend to outperform their non-data-driven counterparts, there has never been a greater need for organisations to build/acquire DEM capabilities to thrive in the current digital era.

This thesis presents several novel approaches to statistical and data challenges for organisations building DEM capabilities. We focus on the fundamentals associated with building DEM capabilities, which lead to a richer understanding of the underlying assumptions and thus enable us to develop more appropriate capabilities.
We address \emph{why} one should engage in DEM by quantifying the benefits and risks of acquiring DEM capabilities. This is done using a ranking under lower uncertainty model, enabling one to construct a business case. We also examine \emph{what} ingredients are necessary to run digital experiments. In addition to clarifying the existing literature around statistical tests, datasets, and methods in experimental design and causal inference, we construct an additional dataset and detailed case studies on applying state-of-the-art methods. Finally, we investigate \emph{when} a digital experiment design would outperform another, leading to an evaluation framework that compares competing designs' data efficiency.

These approaches aim to enable one to run experiments that produce less biased estimates more quickly and adapt to various business constraints. As we maintain the theoretical rigour, we also emphasise applied use cases, interpretable processes/results, and practical tradeoffs, all to ensure that the contributions are accessible to researchers and practitioners from diverse scientific backgrounds.

\end{abstract}
\cleardoublepage

\addcontentsline{toc}{chapter}{Acknowledgements}

\begin{acknowledgements}

Academic research is never a solo effort, and I would like to acknowledge the many people who have helped, in various ways, for me to deliver this thesis.

First and foremost, I would like to thank my supervisor, Prof Emma McCoy, for her consistent support and guidance throughout the years. She has suggested many worthy research directions, challenged many potential ideas, and helped clarify my exposition on many topics. I would also like to thank Dr Seth Flaxman and Prof Robin Evans for the initial project idea discussions, as well as Prof Niall Adams, Dr Ciara Pike-Burke, Prof Nick Heard, and the anonymous reviewers for their feedback during the initial stages of the PhD.

I am fortunate to have met many collaborators whom I enjoy the many intellectually stimulating discussions and opportunities to publicise the latest results in high-impact venues with: Dr Elaine Bettaney, Dr \^{A}ngelo Cardoso, Dr Benjamin Chamberlain, Paul Couturier, Prof Marc Deisenroth, Georg Grob, Dr Duncan Little, and Dr Roberto Pagliari. I have also met many bright minds at Imperial College London and the University of Oxford, many affiliated with the EPSRC CDT in Modern Statistics and Statistical Machine Learning (StatML CDT), whose conversations provided me with new perspectives on the challenges at hand and enabled me to discover new opportunities.

The thesis is one of the many outputs from a fruitful academic-industry collaboration between the StatML CDT at Imperial and Oxford and the AI and Data Science Platform at ASOS.com, with myself also employed at the latter as a data/machine learning scientist. I am blessed to have also worked with and learnt from many talented colleagues at ASOS, including my managers Dr Benjamin Chamberlain, Papinder Dosanjh, and Dawn Rollocks; my teammates in the Customer Understanding (2016--18), Experimentation (2018--19), Prophecy (2019--20), Cassandra (2020), Kernel (2021--22), Experimentation (2022), and the Alchemists (2022--23) teams; and many more unnamed. Among other things, they have ensured that my research remains relevant to industry applications. The research is also generously funded by the two organisations mentioned above (EPSRC grant no. EP/S023151/1), without which I probably would not have taken on the PhD programme at all.

\newpage

Finally, I would like to thank my friends and family, especially my mum Yvonne, my dad Haston, my sister Nicole, and my grandmother Cecilia, for their love and warm encouragement while I took on the ambitious task of studying a PhD while holding an industry job. My PhD years were dominated by the COVID-19 pandemic, and travel restrictions have made visiting them virtually impossible for a few years. Fortunately, the availability of economical video calls has made the isolation much more bearable.

\end{acknowledgements}
\cleardoublepage


\vspace*{\baselineskip}

\begin{quote}
    \textit{``The true method of knowledge is experiment.''}
    
    \hfill --  William Blake\\

    \textit{``Negative results are just what I want. They’re just as valuable to me as positive results. I can never find the thing that does the job best until I find the ones that don’t.''}
    
    \hfill --  Thomas A. Edison\\

    \textit{``An experiment is a question which science poses to Nature and a measurement is the recording of Nature's answer.''}
    
    \hfill -- Max Planck\\

    \textit{``Measurement is the first step that leads to control and eventually to improvement. If you can’t measure something, you can’t understand it. If you can’t understand it, you can’t control it. If you can’t control it, you can’t improve it.''}

    \hfill -- H. James Harrington\\
\end{quote}


\body
\counterwithin{equation}{chapter}
\chapter{Introduction}
\label{chap:introduction}

    

\section{Motivation}

The value of making data-driven or data-informed decisions has become increasingly clear in recent years~\cite{mcafee2012bigdata,stobierski2019advantages}.
Key to making data-driven decisions is the ability to accurately measure a given choice's impact and experiment with possible alternatives.
We define digital experimentation and measurement (DEM) capabilities as the knowledge and tools necessary to run experiments (controlled or otherwise) with different digital products, services, or experiences and measure their impact. The capabilities may be an online controlled experiment framework, a team of analysts, or a system capable of performing machine learning-aided causal inference.

\begin{figure}
  \begin{center}
  \includegraphics[width=0.8\textwidth, trim =15mm 12mm 15mm 12mm, clip]{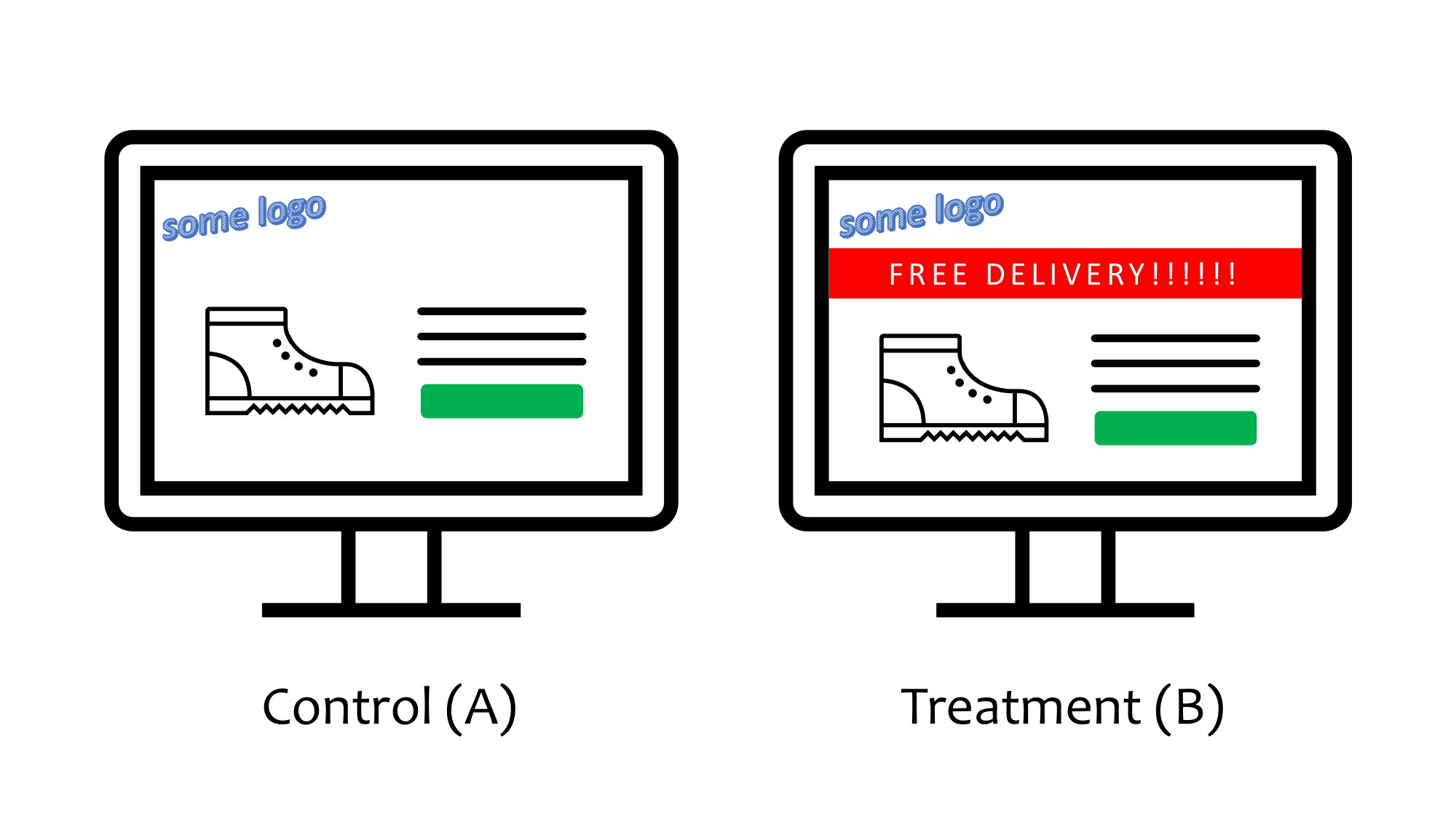}
  \end{center}
  \vspace*{-0.5em}
  \caption[Illustration of an A/B test.]{Illustration of an A/B test (a randomised controlled trial with two parallel groups) designed to test the claim that ``offering free delivery to users of an e-commerce website will lead to a larger proportion of the said users making a purchase.'' Incoming users are split randomly into two groups, where one group (A) acts as the control, and the other group (B) is shown a "free delivery" banner on the website as the treatment.}
  \label{fig:intro_exp_abtestillustration_freedelivery}
  \vspace*{\baselineskip}
\end{figure}

The simplest example of a digital experiment is what is commonly known as an A/B test~\cite{kohavi20trustworthy}. Suppose we are interested in whether offering free delivery to users of an e-commerce website will lead to more of them making a purchase (i.e., an increased \textit{conversion rate} in business speak). We set up an experiment where incoming users are randomly split into two groups, where one group is shown a "free delivery" banner on the website (the \emph{treatment}), while the other acts as the \emph{control} -- being shown the original website without any mention of free delivery (see Figure~\ref{fig:intro_exp_abtestillustration_freedelivery}). We calculate a \emph{decision metric} (here, the conversion rate) for both groups based on responses from their members and compare the decision metrics using a statistical test to draw causal statements about the treatment. The approach is popular in digital organisations, with the largest technology companies having reported running hundreds or thousands of experiments at any given time~\cite{kohavi13online,tang10overlapping,xu15frominfrastructure}.

From a statistical point of view, digital experimentation and measurement is essentially the application of experimental design and causal inference methods in a digital setting. Likewise, the free delivery experiment mentioned above is an online randomised controlled trial. Readers will agree that the underlying experimental design and causal inference methods have been established and frequently applied in agriculture, medicine, and economics for decades. 

That said, the arrival of the Web and, subsequently, the ability to run an experiment end-to-end online have brought different opportunities and challenges. The differences are well documented in~\cite{kohavi2020ocemedicine}, which necessitate the development of a different set of tools and processes exclusively for digital experiments. In some cases, they also require a complete rethink of how we approach experiment design and analysis. Under such premises, we examine the statistical and data challenges one faces when building DEM capabilities.

\paragraph{Guiding Principles}
Before we state the research questions in detail (see section below), we outline the three guiding principles that lead us to the questions. Firstly, similar to many contemporaries in digital experimentation, we seek methods that enable one to make better decisions more quickly and thus drive business/organisational growth. The methods may be those that produce less biased estimates, generate measurements more quickly, or give alternatives that can measure the impact of an intervention under various constraints in practice (e.g., inability to perform user-level random assignment in the ``free delivery'' example above).

Secondly, we seek interpretable processes and results alongside performance to maintain engagement from those beyond our field. We, humans, value the understanding of why a decision is made as such, and the use of black-box models or overly complex procedures in decision-making often faces additional barriers in gaining trust from stakeholders despite its potential to improve performance~\cite{rudin2019why}.


Lastly, and perhaps more importantly, we seek to make the knowledge and tools involved in digital experimentation accessible, thus ensuring widened participation.\footnote{Clearly, a PhD thesis is a poor medium to engage with a non-technical audience. We thus limit the scope of ``widened participation'' to include researchers and practitioners with diverse academic backgrounds.} Many recent advances in digital experimentation are proposed by researchers and practitioners who operate in an environment with mature capabilities. Focusing on those works may give the impression that mature capabilities are the norm and that we only need a little work to explain the basics on top of what is already available. Both are not true -- the DEM capabilities of many individuals and organisations are still in their infancy. These individuals and organisations will benefit from us continuing to consolidate and clarify the building blocks in digital experiments, e.g., statistical tests, datasets, and decision metrics.

\section{Key Research Questions}
\label{sec:introduction_research_q}

The thesis aims to address digital experimenters' and organisations' challenges as they build DEM capabilities. As the thesis title suggests, most of the work concerns statistical and data challenges one encounters at the early stage of the process. It also touches upon topics encountered by those with established capabilities, enabling readers to take away something wherever they are on the experimentation maturity spectrum~\cite{fabijan2018online}. We ask the following questions:

\paragraph{Why should one engage in digital experimentation and measurement?}
While running experiment(s) to evaluate a hypothesis is second nature to scientists, it is unnatural in the business world. Business decision-makers prefer having business cases, ideally complete with a cost-benefit analysis. The ability to quantify the benefits of acquiring DEM capabilities (the knowledge and tools necessary to experiment and measure) will strengthen our business case. It enables us to speak in the language of the business and thus stand a better chance of garnering their support.

\paragraph{What ingredients do we require to run experiments successfully in a digital setting?}
Committing to investing in DEM capabilities is the first of many steps to reap the benefits. To run their first experiment successfully, organisations require, at a minimum, a clear hypothesis and evaluation criteria, detailed procedures to deliver different treatments to and collect responses from experiment participants, plus basic knowledge in data processing and statistical testing. They would also require knowledge of advanced mathematical and statistical methods, powerful computer systems with sufficient computing power and data storage capacity, and a growth mindset organisational culture that embraces scientific, data-driven discoveries to scale their experiment operations. Each topic, whilst important, warrants extensive consideration. The thesis will focus on statistical testing, data, and advanced methods. We will provide a brief overview and pointers to relevant work for other topics for those interested.

\paragraph{When would an experiment design outperform another?}
In the parable of two woodcutters, the more experienced woodcutter emerged victorious in a woodchopping contest by spending most of their time sharpening their axe. In the same spirit, it is vital to continuously improve our tools, one of which is experiment designs, as we engage in experimentation efforts.
While there are many qualities that we can use to compare two experiment designs, in digital experimentation, we usually care about \emph{data efficiency}. This is represented by the minimum number of experiment participants required to reach a statistically sound conclusion.\footnote{This is closely related to optimal design~\cite{atkinson2011optimum} and the asymptotic efficiency of a statistical test~\cite{ems2011efficiency} in the statistical theory literature. We will approach the challenge from a more applied point of view.}  Increasing the data efficiency (i.e., requiring fewer experiment participants) enables experimenters to make decisions sooner, leading to better organisational performance as one can deploy improvements and roll back harmful changes more quickly.


\section{Contributions and Chapter Organisation}

The thesis is divided into seven chapters, each detailing contributions to a different statistical and data topic necessitated by developing early-stage DEM capabilities. The broad coverage of topics means the thesis embeds the background knowledge required for each topic within the individual chapters.

The remainder of \emph{Chapter~\ref{chap:introduction}} covers housekeeping items: a list of publications from the research programme and a list of mathematical symbols and the quantities/concepts they represent.

\emph{Chapter~\ref{chap:vem}} introduces the ranking under lower uncertainty problem as a novel model to value DEM capabilities. This arises from the observation that DEM capabilities reduce measurement uncertainty on the value of individual business propositions, improving the inherent value of business propositions chosen during prioritisation. We derive the expected gain and the variance of our gain estimate and provide case studies based on large-scale meta-analyses. This addresses the first research question and enables one to build a quantitative business case to justify investment in the capabilities.

\emph{Chapter~\ref{chap:statstest}} provides an introduction to statistical testing in digital experimentation that is both mathematically rigorous and accessible to researchers and practitioners with different backgrounds. 
It starts from the very beginning of a statistical test (i.e., specifying the statistical hypotheses), provides a detailed treatment on the popular null hypothesis significance testing (NHST) framework, and involves other more advanced test paradigms (e.g., sequential and Bayesian testing) and alternatives. It aims to provide a sufficient theoretical grounding to appreciate what is happening behind the scenes while highlighting the pitfalls in test design and interpretation that can trap even the most experienced.

\emph{Chapter~\ref{chap:oced}} describes a systemic investigation of datasets in digital experiments. We create the first ever taxonomy for digital experiment datasets and compile the first ever survey for publicly available online controlled experiment datasets. We also map the relationship between the dataset taxonomy and statistical tests discussed in Chapter~\ref{chap:statstest}. The taxonomy and survey also identify gaps in dataset availability, leading to the development of the ASOS Digital Experiments Dataset. This first real, multi-experiment time series dataset enables the design and running of experiments with adaptive stopping.

\emph{Chapter~\ref{chap:rade}} reviews existing challenges in digital experimentation and the established and emerging methods that address such challenges. 
We also complement the review by viewing some of the reported methodological advances through a practical lens. This is done by interleaving the review with case studies on randomised controlled trials with dependent responses and quasi-experiments with geographical regions as treatment units. They provide evidence on the extent of reported challenges and practical considerations when implementing relevant methods. Together with Chapters~\ref{chap:statstest} and~\ref{chap:oced}, these chapters address the second research question.

\emph{Chapter~\ref{chap:pse}} develops an evaluation framework for personalisation strategy experiment designs. These experiment designs face unique challenges, such as low test power and partial non-compliance from users, issues that we will cover in Chapters~\ref{chap:statstest} and~\ref{chap:rade}. The evaluation framework enables one to compare two experiment setups given the circumstances they face with their target audience, thus addressing the third research question. We also derive interpretable rules of thumb from the framework to enable experimenters to compare typical experiment setups quickly.

We conclude in \emph{Chapter~\ref{chap:conclusion}} with an outline of promising investigation pathways arising from each workstream.
In addition, we make all the code used for experiments and simulations described in the thesis, as well as the ASOS Digital Experiment Dataset, publicly available. Readers can find the links to the code repositories and the dataset at the bottom of this page\footnote{The link to code repositories and the dataset are as follows:
\begin{itemize}[leftmargin=2.8em]
    \item Code repository associated with the ranking under lower uncertainty framework (as described in Chapter~\ref{chap:vem}): \url{https://github.com/liuchbryan/ranking\_under\_lower\_uncertainty};
    \item ASOS Digital Experiments Dataset and accompanying datasheet (as described in Chapter~\ref{chap:oced}): \url{https://osf.io/64jsb/}. The OSF project also embeds a code repository with experiments on the dataset (\url{https://github.com/liuchbryan/oce-dataset/});
    \item Code repository and results on the two publicly available datasets associated with e-commerce experiments with dependent responses (as described in Section~\ref{sec:rade_abv}): \url{https://github.com/liuchbryan/oce-ecomm-abv-calculation}; and 
    \item Code repository associated with the evaluation framework for personalisation strategy experiment designs (as described in Chapter~\ref{chap:pse}):  \url{https://github.com/liuchbryan/experiment\_design\_evaluation}.
\end{itemize}
}
and again in the footnotes of the relevant chapters.

\section{Publications}
\label{sec:introduction_publications}

This thesis contains research that has been published during PhD study. We list where readers can find them in the thesis, together with details of the associated conference and journal papers, below:

\begin{itemize}
    \item Chapter~\ref{chap:vem} and Appendix \ref{sec:miscmath_prob_Ir_Js}: C. H. B. Liu and B. P. Chamberlain (2019). What is the Value of Experimentation \& Measurement? In: \textit{2019 IEEE International Conference on Data Mining (\textbf{ICDM '19})}, 1222--1227. DOI:~\href{https://doi.org/10.1109/ICDM.2019.00151}{\texttt{10.1109/ICDM.2019.00151}}.
    \item Chapter~\ref{chap:vem} and Appendix \ref{sec:miscmath_prob_Ir_Js}: C. H. B. Liu, B. P. Chamberlain and E. J. McCoy (2020). What is the Value of Experimentation and Measurement? \textit{\textbf{Data Science and Engineering}~5}, 152--167. DOI:~\href{https://doi.org/10.1007/s41019-020-00121-5}{\texttt{10.1007/s41019-020-00121-5}}. Part of Special Issue: Highly-Rated Short Papers of ICDM 2019.
    \item Chapter~\ref{chap:oced}:  C. H. B. Liu, \^{A}. Cardoso, P. Couturier and E. J. McCoy (2021). Datasets for Online Controlled Experiments. In: \textit{Proceedings of the Neural Information Processing Systems Track on Datasets and Benchmarks (\textbf{NeurIPS Datasets and Benchmarks '21})}. URL:\linebreak\href{https://datasets-benchmarks-proceedings.neurips.cc/paper/2021/file/274ad4786c3abca69fa097b85867d9a4-Paper-round2.pdf}{\texttt{https://datasets-benchmarks-proceedings.neurips.cc/paper/2021/file/\linebreak274ad4786c3abca69fa097b85867d9a4-Paper-round2.pdf}}.
    \item Chapter~\ref{chap:rade} (Section~\ref{sec:rade_abv}): C. H. B. Liu and E. J. McCoy (2023). Measuring e-Commerce Metric Changes in Online Experiments. In: \textit{Companion Proceedings of the ACM Web Conference 2023 (\textbf{WWW '23 Companion})}. DOI:~\href{https://doi.org/10.1145/3543873.3584654}{\texttt{10.1145/3543873.3584654}}.
    \item Chapter~\ref{chap:pse}, Appendix~\ref{sec:miscmath_metric_dilution}, and Appendix~\ref{sec:miscmath_dual_control}: C. H. B. Liu and E. J. McCoy (2020). An Evaluation Framework for Personalization Strategy Experiment Designs. \href{https://arxiv.org/abs/2007.11638}{\textit{arXiv:} \texttt{2007.11638}} \texttt{[stat.ME]}. Presented in \textit{\textbf{AdKDD '20 Workshop} (in conjunction with KDD '20)},\linebreak awarded Best Student Paper.
\end{itemize}


\section{List of Acronyms and Mathematical Symbols}
\label{sec:intro_mathsymbols}

Readers may encounter the following acronyms recurring in multiple thesis sections. We will define the acronyms in the main text at least once before referring to the acronyms.
\begin{itemize}
  \item ATE: Average treatment effect
  \item CATE: Conditional average treatment effect
  \item CI: Confidence interval
  \item CDF: Cumulative density function
  \item CLT: Central limit Theorem
  \item CVR: Con\textbf{v}ersion rate
  \item DEM: Digital experimentation and measurement (usually followed by the word ``capability/capabilities'')
  \item MDE: Minimum detectable effect
  \item NHST: Null hypothesis significance test(ing)
  \item OCE: Online controlled experiment
  \item PDF: Probability density function
  \item RCT: Randomised controlled trial
\end{itemize}

\hspace*{0pt}

We also list the mathematical symbols and the quantities/concepts they represent below. Each entry is indexed by the leading symbol and follows the following format:

(Applicable chapter(s)) \emph{Symbol(s)} -- (Type) Quantity/concept the symbol represents.

\paragraph{Applicable chapter(s)} 
This states the thesis chapter(s) that a notation applies to. The notations used in Chapter~\ref{chap:vem} also apply to Appendix~\ref{sec:miscmath_prob_Ir_Js}, and those used in Chapter~\ref{chap:pse} also apply to Appendices~\ref{sec:miscmath_metric_dilution} and~\ref{sec:miscmath_dual_control}.

\paragraph{Symbol(s)}
The thesis covers topics traditionally belonging to multiple statistical disciplines. Thus, it is inevitable to see some notation clash. We follow the order set out below when deciding which set of notations takes precedence:
\begin{enumerate}
\item Elementary statistical constructs found in introductory statistical texts (e.g., $X$ for a random variable, $f_X(\cdot)$ for the probability density function of $X$, and $\mathbb{E}(X)=\mu_X$ for the expected value of $X$); 
\item Notation specific to statistical testing (e.g., $H_0$ and $H_1$ for null and alternate hypotheses, $\alpha$ for the significance level, $\beta$ for the power);
\item Common notation in other sub-fields, including causal inference, order statistics, and econometrics; and
\item Specific notation introduced by individual research articles.
\end{enumerate}
Notations with lower precedence are generally assigned a new symbol to minimise confusion. That said, in a couple of cases, the same symbol may represent two different quantities/concepts in the same chapter -- their meaning should be apparent given the context surrounding its use.

\paragraph{Type}
We also abbreviate the types of quantities and concepts in the tables as follows:
\begin{itemize}
    \item calc.: calculated quantity,
    \item const.: constant,
    \item dist.: distribution,
    \item func.: function,
    \item id.: identifier, and
    \item r.v.: random variable.
\end{itemize}

\begin{table}
  \onehalfspacing
  \setlength\extrarowheight{0.42em}
  \begin{center}
  \caption[List of mathematical symbols and the quantity/concept represented by them.]{List of mathematical symbols and the quantity/concept represented by them (seven pages). See Section~\ref{sec:intro_mathsymbols} for further information.}
  \begin{tabular}[t]{c|l}
    \begin{tabular}[t]{@{}c@{}}
      Leading \\[-0.5em] Symbol
    \end{tabular} & 
    Quantity/Concept Represented \\\hline
    $A/a$ & 
    \begin{tabular}[t]{@{}p{0.85\textwidth}@{}}
      (Chapter~\ref{chap:rade}) $\mathcal{A}$ -- (r.v.) Treatment \textbf{A}ssignment variable in the potential outcomes framework, with realised values denoted $a$.\\
      (Chapter~\ref{chap:pse}) $A$ (and $A1$, $A2$) -- (id.) An \textbf{A}nalysis group in a personalisation strategy experiment, often referenced together with $B$.\\
    \end{tabular}\\
    $B/b$ & 
    \begin{tabular}[t]{@{}p{0.85\textwidth}@{}}
      (Chapter~\ref{chap:statstest}) $B$ -- (calc. / r.v.) Binomial test of proportion statistic.\\
      (Chapters~\ref{chap:statstest}--\ref{chap:oced}) $BF_{n, m}$ -- (calc.) Bayes Factor (in a Bayesian test, upon observing first $n$ and $m$ items in the first and second sample, respectively).\\
      (Chapter~\ref{chap:pse}) $B$ (and $B1$, $B2$) -- (id.) An analysis group in a  personalisation strategy experiment, often referenced together with $A$.\\
    \end{tabular}\\
    $C/c$ & 
    \begin{tabular}[t]{@{}p{0.85\textwidth}@{}}
      (All chapters) $\textrm{Cov}(\cdot \,, \cdot)$ -- (func.) Covariance.\\
      (Chapter~\ref{chap:vem}) $c$ -- (const.) Adjustment to the quantile of an order statistic.\\
      (Chapter~\ref{chap:vem}) $c$ -- (const.) Risk-free value in the Sharpe ratio.\\
      (Chapter~\ref{chap:statstest}) $C^2$ -- (calc. / r.v.)  $\chi^2$ goodness-of-fit test statistic.\\
      (Chapter~\ref{chap:rade}) $\mathcal{C}$ -- (r.v.) Covariate(s) under the potential outcomes framework.\\
      (Chapter~\ref{chap:pse}) $C0, C1, C2, C3$ -- (id.) User groups in a personalisation strategy experiment that acts as \textbf{C}ontrol.\\
      (Appendix \ref{sec:miscmath_prob_Ir_Js}) $C$ -- (r.v.) \textbf{C}ount.\\
    \end{tabular}\\
    $D/d$ & 
    \begin{tabular}[t]{@{}p{0.85\textwidth}@{}}
      (Chapter~\ref{chap:vem}) $\mathcal{D}$ -- (r.v.) \textbf{D}ifference in value of items selected in the ranking under lower uncertainty problem.\\
      (Chapters~\ref{chap:statstest}--\ref{chap:oced}) $d$ -- (calc.) Cohen's $d$.\\
    \end{tabular}\\
    $E/e$ & 
    \begin{tabular}[t]{@{}p{0.85\textwidth}@{}}
      (All chapters) $\mathbb{E}(\cdot)$ -- (func.) Expected value.\\
      (Chapter~\ref{chap:vem}) $\mathcal{E}_i$ / $\mathcal{E}_{(r)}$ -- (r.v.) \textbf{E}stimated value of item $i$ / $r$th-ranked item under estimation uncertainty in the ranking under lower uncertainty problem, often referenced together with $\mathcal{V}_i$ / $\mathcal{V}_{\mathcal{I}(r)}$.\\
      (Chapters~\ref{chap:statstest}) $E_i$ -- (r.v.) \textbf{E}xpected frequency for category $i$ in a $\chi^2$ goodness-of-fit test.\\
      (Chapters~\ref{chap:statstest}--\ref{chap:oced}) $E_{n, m}$ -- (calc.) \textbf{E}ffective sample size (in a Bayesian test, upon observing first $n$ and $m$ items in the first and second sample, respectively).\\
    \end{tabular}\\
  \end{tabular}
  \pagebreak
  \end{center}
\end{table}

\begin{table}
  \onehalfspacing
  \setlength\extrarowheight{0.5em}
  \begin{center}
  \begin{tabular}[t]{c|l}
    \begin{tabular}[t]{@{}c@{}}
      Leading \\[-0.5em] Symbol
    \end{tabular} & 
    Quantity/Concept Represented \\\hline
    $F/f$ & 
    \begin{tabular}[t]{@{}p{0.85\textwidth}@{}}
      (All chapters) $F_X(\cdot)$ -- (func.) Cumulative density function (CDF) of an r.v. $X$.\\
      (All chapters) $f_X(\cdot)$ -- (func.) Probability density function (PDF) of an r.v. $X$.\\
      (All chapters) $f'_X(\cdot)$ -- (func.) Derivative of the PDF of an r.v. $X$.\\
      (All chapters) $f(\cdot \,|\, \cdot)$ -- (func.) Conditional PDF.\\
    \end{tabular}\\
    $G/g$ & 
    \begin{tabular}[t]{@{}p{0.85\textwidth}@{}}
      (Chapter~\ref{chap:rade}) $g(\cdot)$ -- (func.) A generic function that maps the treatment and covariate(s) to the observed/potential response under the potential outcomes framework.\\
      (Chapter~\ref{chap:pse}) $G$ -- (id.) A generic user \textbf{G}roup-scenario combination in a personalisation strategy experiment. See Table~\ref{tab:pse_group_id} for possible values.\\
    \end{tabular}\\
    $H/h$ & 
    \begin{tabular}[t]{@{}p{0.85\textwidth}@{}}
      (Chapter~\ref{chap:vem}) $\mathcal{H}_i$ / $\mathcal{H}_{(r)}$ -- (r.v.) Estimated value of item $i$ / $r$th-ranked item under \textbf{h}igh estimation noise in the ranking under lower uncertainty problem, often referenced together with $\mathcal{V}_i$ / $\mathcal{V}_{\mathcal{I}(r)}$ and $\mathcal{L}_j$ / $\mathcal{L}_{(s)}$.\\
      (Chapters~\ref{chap:statstest}--\ref{chap:pse}) $H_0$, $H_1$ -- (id.) Null and alternate hypotheses.\\
      (Chapters~\ref{chap:statstest}--\ref{chap:oced}) $H$ -- (id.) The mixing distribution in a mixed Sequential\\ Probability Ratio Test (mSPRT).\\
      (Chapters~\ref{chap:statstest}--\ref{chap:oced}) $H$ -- (id.) A generic statistical hypothesis in a Bayesian test.\\
    \end{tabular}\\
    $I/i$ & 
    \begin{tabular}[t]{@{}p{0.85\textwidth}@{}}
      (All chapters) $i$ -- (id.) \textbf{I}ndex identifying an r.v. within a set of i.i.d. r.v.'s, often referenced together with $j$.\\
      (Chapter~\ref{chap:vem}) $\mathcal{I}(\cdot)$ -- (func.) \textbf{I}ndex function that maps the rank of an r.v. within a set of independently identically distributed (i.i.d.) r.v.'s $\mathcal{E}$ / $\mathcal{H}$ to its index, often referenced together with $\mathcal{J}(\cdot)$.\\
      (Chapter~\ref{chap:pse}) $I1, I2, I\phi, I\psi$ -- (id.) User groups in a personalisation strategy experiment under \textbf{I}ntervention.\\
      (Chapter~\ref{sec:miscmath_prob_Ir_Js}) $\mathbb{I}$ -- (r.v.) \textbf{I}ndicator variable, often with the condition expressed as a subscript.\\
    \end{tabular}\\
    \\[0.09\textheight]
  \end{tabular}
  \end{center}
\end{table}

\begin{table}
  \onehalfspacing
  \setlength\extrarowheight{0.5em}
  \begin{center}
  \begin{tabular}[t]{c|l}
    \begin{tabular}[t]{@{}c@{}}
      Leading \\[-0.5em] Symbol
    \end{tabular} & 
    Quantity/Concept Represented \\\hline
    $J/j$ & 
    \begin{tabular}[t]{@{}p{0.85\textwidth}@{}}
      (All chapters) $j$ -- (id.) Index identifying an r.v. within a set of i.i.d. r.v.'s, often referenced together with $i$.\\
      (Chapter~\ref{chap:vem}) $\mathcal{J}(\cdot)$ -- (func.) Index function that maps the rank of an r.v. within a set of independently identically distributed (i.i.d.) r.v.'s. $\mathcal{L}$ to its index, often referenced together with $\mathcal{I}(\cdot)$.\\
    \end{tabular}\\
    $L/l$ & 
    \begin{tabular}[t]{@{}p{0.85\textwidth}@{}}
        (Chapter~\ref{chap:vem}) $\mathcal{L}_i$ / $\mathcal{L}_{(s)}$ -- (r.v.) Estimated value of item $i$ / $s$th-ranked item under \textbf{l}ow estimation noise in the ranking under lower uncertainty problem, often referenced together with $\mathcal{V}_i$ / $\mathcal{V}_{\mathcal{J}(s)}$ and $\mathcal{H}_i$ / $\mathcal{H}_{(r)}$.\\
        (Appendix~\ref{sec:miscmath_prob_Ir_Js}) $\mathcal{L}^*$ -- (r.v.) $\mathcal{L}_{\mathcal{I}(r)}$ normalised by the mean and variance of $\mathcal{L}_j$ under normal assumptions.\\
    \end{tabular}\\
    $M/m$ & 
    \begin{tabular}[t]{@{}p{0.85\textwidth}@{}}
      (Chapter~\ref{chap:vem}) $M$ -- (const.) Number of items one can select (i.e., capacity) in the ranking under lower uncertainty problem.\\
      (Chapters~\ref{chap:statstest}--\ref{chap:pse}) $m$ -- (const.) Number of items in (or size of) a statistical sample~$Y$, either total or that observed so far within a sequential/Bayesian test setting; often referenced together with $n$.\\
    \end{tabular}\\
    $N/n$ & 
    \begin{tabular}[t]{@{}p{0.85\textwidth}@{}}
    (All chapters) $\mathcal{N}(\cdot \,, \cdot)$ -- (dist.) A normal distribution, with its mean and variance as parameters.\\
    (Chapter~\ref{chap:vem}) $N$ -- (const.) \textbf{N}umber of items available for selection in the ranking under lower uncertainty problem.\\
    (Chapters~\ref{chap:statstest}--\ref{chap:pse}) $n$ -- (const.) Number of items in (or size of) a statistical sample~$X$, either total or that observed so far within a sequential/Bayesian test setting; often referenced together with $m$.\\
    (Chapter~\ref{chap:pse}) $n_\mathcal{G}$ -- (const.) \textbf{N}umber of samples in a user/analysis group~$\mathcal{G}$, where $\mathcal{G}\in \{0, 1, 2, 3, A, B, A1, A2, B1, B2\}$, in a personalisation strategy experiment.\\
    \end{tabular}\\
    $O/o$ & 
    \begin{tabular}[t]{@{}p{0.85\textwidth}@{}}
    (Chapters~\ref{chap:statstest}) $O_i$ -- (r.v.) \textbf{O}bserved frequency for category $i$ in a $\chi^2$ goodness-of-fit test.\\
    (Chapter~\ref{chap:pse}) $O(\cdot)$ -- (func.) Big-O notation describing a function's limiting behaviour.\\
    \end{tabular}\\
    $P/p$ & 
    \begin{tabular}[t]{@{}p{0.85\textwidth}@{}}
    (Chapter~\ref{chap:vem}) $\mathbb{P}(\cdot)$ -- (func.) Probability.\\
    (Chapters~\ref{chap:statstest}--\ref{chap:pse}) $p$ -- (calc.) $p$-value of a null hypothesis significance test.\\
    \end{tabular}\\
  \end{tabular}
  \end{center}
\end{table}

\begin{table}
  \onehalfspacing
  \setlength\extrarowheight{0.5em}
  \begin{center}
  \begin{tabular}[t]{c|l}
    \begin{tabular}[t]{@{}c@{}}
      Leading \\[-0.5em] Symbol
    \end{tabular} & 
    Quantity/Concept Represented \\\hline
    $R/r$ & 
    \begin{tabular}[t]{@{}p{0.85\textwidth}@{}}
    (Chapter~\ref{chap:vem}) $r$ -- (id.) \textbf{R}ank of an r.v. (in order statistics terms) within a set of r.v.'s $\mathcal{E}$ / $\mathcal{H}$, often referenced together with $s$.\\
    (Chapter~\ref{chap:pse}) $R$ -- (id.) A personalisation strategy experiment setup, often referenced together with $S$.\\
    \end{tabular}\\
    $S/s$ & 
    \begin{tabular}[t]{@{}p{0.85\textwidth}@{}}
    (Chapter~\ref{chap:vem}) $s$ -- (id.) Rank of an r.v. (in order statistics terms) within a set of r.v.'s $\mathcal{L}$, often referenced together with $r$.\\
    (Chapters~\ref{chap:statstest}) $S_n$ -- (r.v.) Sequential probability ratio test (SPRT) statistic.\\
    (Chapters~\ref{chap:statstest}--\ref{chap:oced}) $S(\cdot \,, \cdot)$ -- (func.) Sign function (in a Mann-Whitney $U$ test).\\
    (Chapters~\ref{chap:statstest}--\ref{chap:pse}) $s^2_X$ -- (calc.) Sample variance of a sample $X$.\\
    (Chapter~\ref{chap:pse}) $S$ -- (id.) A personalisation strategy experiment \textbf{S}etup, often referenced together with $R$.\\
    \end{tabular}\\
    $T/t$ & 
    \begin{tabular}[t]{@{}p{0.85\textwidth}@{}}
    (All chapters) $t_{\nu}$ -- (dist.) A Student's $t$ distribution with $\nu$ degrees of freedom.\\
    (All chapters) $T_{\nu}(\cdot)$ -- (func.) CDF of a Student's $t$-distribution with $\nu$ d.f.\\
    (All chapters) $t_{\nu, q}$ -- (calc.) $q$th quantile of a Student's $t$-distribution with $\nu$ d.f.\\
    (Chapter~\ref{chap:vem}) $T_i$ / $T_i'$ -- (r.v.) Student's $t$-distributed r.v.'s.\\
    (Chapters~\ref{chap:statstest}--\ref{chap:pse}) $T$ -- (calc. / r.v.) Test statistic (generic, or that of a $t$-test).\\
    (Chapter~\ref{chap:oced}) $t, t+1, ...$ -- (id.) Points in time (in statistical tests with adaptive stopping).\\
    (Appendix~\ref{sec:miscmath_prob_Ir_Js}) $T(\cdot\,,\cdot)$ -- (func.) Owen's $T$-function.\\
    \end{tabular}\\
    $U/u$ & 
    \begin{tabular}[t]{@{}p{0.85\textwidth}@{}}
    (Chapters~\ref{chap:vem} \& \ref{chap:pse}) $U(a,b)$ -- (dist.) A uniform distribution with bounds $a$ and $b$.\\
    (Chapters~\ref{chap:statstest}--\ref{chap:oced}) $U$ -- (calc. / r.v.) Mann-Whitney $U$-test statistic.\\
    \end{tabular}\\
    $V/v$ & 
    \begin{tabular}[t]{@{}p{0.85\textwidth}@{}}
    (All chapters) $\textrm{Var}(\cdot)$ -- (func.) Variance.\\
    (Chapter~\ref{chap:vem}) $\mathcal{V}_i$, $\mathcal{V}_{\mathcal{I}(r)}$, $\mathcal{V}_{\mathcal{J}(s)}$ -- (r.v.) True \textbf{V}alue of items $i$, $\mathcal{I}(r)$, and $\mathcal{J}(s)$ in the ranking under lower uncertainty problem, often referenced together with $\mathcal{E}_i$ / $\mathcal{E}_{(r)}$, $\mathcal{H}_i$ / $\mathcal{H}_{(r)}$, and $\mathcal{L}_j$ / $\mathcal{L}_{(s)}$.\\
    (Chapters~\ref{chap:statstest}--\ref{chap:oced}) $V^2$ -- (const.) Hyperparameter, in a Bayesian test, which represents the variance of the effect size prior.\\
    \end{tabular}\\
    \\[0.023\textheight]
  \end{tabular}
  \end{center}
\end{table}

\begin{table}
  \onehalfspacing
  \setlength\extrarowheight{0.5em}
  \begin{center}
  \begin{tabular}[t]{c|l}
    \begin{tabular}[t]{@{}c@{}}
      Leading \\[-0.5em] Symbol
    \end{tabular} & 
    Quantity/Concept Represented \\\hline
    $W/w$ & 
    \begin{tabular}[t]{@{}p{0.85\textwidth}@{}}
    (Chapter~\ref{chap:vem}) $\mathcal{W}$, $\mathcal{W}_1$, $\mathcal{W}_2$ -- (r.v.) Average true value of items selected under some estimation uncertainty, high estimation noise, and low estimation noise, respectively, in the ranking under lower uncertainty problem.\\
    (Chapters~\ref{chap:statstest}--\ref{chap:oced}) $W_{n, m}$ -- (calc. / r.v.) \textbf{W}ald test statistic (in a sequential/Bayesian test, upon observing first $n$ and $m$ items in the first and second sample, respectively.
    \end{tabular}\\
    $X/x$ & 
    \begin{tabular}[t]{@{}p{0.85\textwidth}@{}}
    (Chapters~\ref{chap:statstest}--\ref{chap:pse}) $X_i$ -- (r.v.) Item $i$ in a sample $X$, often referenced together with~$Y_j$ in two-sample testing.\\
    (Chapters~\ref{chap:statstest}--\ref{chap:pse}) $\bar{X}$ -- (r.v.) Sample mean of a sample $X$.\\
    (Chapters~\ref{chap:statstest}--\ref{chap:pse}) $\bar{X}_n$ -- (r.v.) Sample mean of a sample $X$ (in a sequential/Bayesian test, upon observing the first $n$ items in the sample).\\
    (Chapters~\ref{chap:statstest}--\ref{chap:pse}) $\bar{x}$ -- (calc.) Realised sample mean of a sample $X$.\\
    (Chapters~\ref{chap:rade}) $X(a)$ / $X_i(a)$ -- (r.v.) Potential response/outcome of $X$/$X_i$ under the treatment assignment $a \in \{0, 1, ...\}$ in the potential outcomes framework, often referenced together with $Y(a)$ / $Y_j(a)$.\\
    \end{tabular}\\
    $Y/y$ & 
    \begin{tabular}[t]{@{}p{0.85\textwidth}@{}}
      (Chapters~\ref{chap:statstest}--\ref{chap:pse})  $Y_j$ -- (r.v.) Item $j$ in a sample $Y$, often referenced together with $X_i$ in two-sample testing.\\
      (Chapters~\ref{chap:statstest}--\ref{chap:pse})  $\bar{Y}$ -- (r.v.) Sample mean of a sample $Y$.\\
      (Chapters~\ref{chap:statstest}--\ref{chap:pse})  $\bar{Y}_n$ -- (r.v.) Sample mean of a sample $Y$ (in a sequential/\\Bayesian test, upon observing the first $n$ items in the sample).\\
      (Chapters~\ref{chap:rade}) $Y(a)$ / $Y_j(a)$ -- (r.v.) Potential response/outcome of $Y$/$Y_j$ under the treatment assignment $a \in \{0, 1, ...\}$ in the potential outcomes framework, often referenced together with $X(a)$ / $X_i(a)$.\\
    \end{tabular}\\
    $Z/z$ & 
    \begin{tabular}[t]{@{}p{0.85\textwidth}@{}}
     (Chapters~\ref{chap:statstest}--\ref{chap:pse})  $z_q$ -- (calc.) $q$th quantile of a standard normal r.v.\\
     (Chapter~\ref{chap:statstest}) $Z$ -- (calc. / r.v.) Standard score / $z$-test statistic.\\
     (Chapter~\ref{chap:pse}) $z$ -- (id.) Mathematical shorthand -- the difference between two normal quantiles (see Equation~\eqref{eq:pse_setup23_maths_shorthand}).\\
     (Appendix~\ref{sec:miscmath_prob_Ir_Js}) $Z$ -- (r.v.) A standard normal r.v.\\
    \end{tabular}\\
    \\[0.02\textheight]
  \end{tabular}
  \end{center}
\end{table}

\begin{table}
  \onehalfspacing
  \setlength\extrarowheight{0.5em}
  \begin{center}
  \begin{tabular}[h]{c|l}
    \begin{tabular}[t]{@{}c@{}}
      Leading \\[-0.5em] Symbol
    \end{tabular} & 
    Quantity/Concept Represented \\\hline
    $\alpha$ & 
    \begin{tabular}[t]{@{}p{0.85\textwidth}@{}}
    (Chapters~\ref{chap:statstest}--\ref{chap:pse}) $\alpha$ -- (const.) Significance level of a statistical test.\\
    (Appendix~\ref{sec:miscmath_prob_Ir_Js}) $\alpha_{\mathcal{P}}$ -- (const.) Beta distribution parameter, often referenced together with $\beta_{\mathcal{P}}$.\\
    \end{tabular}\\
    $\beta$ & 
    \begin{tabular}[t]{@{}p{0.85\textwidth}@{}}
      (Chapters~\ref{chap:statstest}--\ref{chap:pse}) $\beta_{\theta}$ -- (const.) Power of a statistical test under a specific hypothesis with parameter $\theta$.\\
      (Appendix~\ref{sec:miscmath_prob_Ir_Js}) $\beta_{\mathcal{P}}$ -- (const.) Beta distribution parameter, often referenced together with $\alpha_{\mathcal{P}}$.\\
    \end{tabular}\\
    $\Delta / \delta$ & 
    \begin{tabular}[t]{@{}p{0.85\textwidth}@{}}
      (All chapters) $\Delta$ -- (calc. / r.v.) Effect size.\\
      (Chapters~\ref{chap:statstest}--\ref{chap:oced}) $\Delta^{\!\circ}$ -- (calc. / r.v.) Effect size standardised by the pooled variance in a Bayesian test.\\
      (Chapters~\ref{chap:statstest}--\ref{chap:oced}) $\delta_{n, m}$ -- (calc. / r.v.) ``Test statistic'' of a Bayesian test, upon observing the first $n$ and $m$ items in the first and second sample, respectively.\\
      (Chapter~\ref{chap:pse}) $\Delta_{S}$ -- (calc.) Actual effect size of a personalisation strategy setup~$S$.\\
    \end{tabular}\\
    $\epsilon$ & 
    \begin{tabular}[t]{@{}p{0.85\textwidth}@{}}
      (Chapter~\ref{chap:vem}) -- $\epsilon, \epsilon_1, \epsilon_2$ -- (r.v.) Estimation noise.\\
    \end{tabular}\\
    $\eta$ & 
    \begin{tabular}[t]{@{}p{0.85\textwidth}@{}}
      (Chapter~\ref{chap:pse}) $\eta$ -- (id.) Mathematical shorthand (see Equation~\eqref{eq:pse_setup23_maths_shorthand}).\\
    \end{tabular}\\
    $\Theta / \theta$ & 
    \begin{tabular}[t]{@{}p{0.85\textwidth}@{}}
      (All chapters) $\theta$ -- (const.) Parameter of interest in a hypothesis test.\\
      (All chapters) $\theta_0$ -- (const.) Value of the parameter of interest under the null hypothesis.\\
      (All chapters) $\theta^*$ -- (const.) A specific parameter of interest.\\
      (Chapter~\ref{chap:pse}) $\theta^*_S$ -- (const.) Minimum detectable effect size of a personalisation strategy experiment setup $S$.\\
    \end{tabular}\\
    $\Lambda / \lambda$ & 
    \begin{tabular}[t]{@{}p{0.85\textwidth}@{}}
      (Chapters~\ref{chap:statstest}--\ref{chap:oced}) $\Lambda^{H, \theta_0}_{n}$ / $\tilde{\Lambda}^{H, \theta_0}_{n}$ -- (calc. / r.v.) One-sample / two-sample mixed sequential probability ratio test (mSPRT) statistic with mixing distribution $H$ and null hypothesis parameter $\theta_0$.\\
    \end{tabular}\\
    $\mu$ & 
    \begin{tabular}[c]{@{}p{0.85\textwidth}@{}}
      (All chapters) $\mu_X$ -- (const.) Mean of an r.v., a population, or a group $X$.\\
    \end{tabular}\\
    $\nu$ & 
    \begin{tabular}[t]{@{}p{0.85\textwidth}@{}}
      (All chapters) $\nu$ -- (const.) Degrees of freedom (d.f.).\\
    \end{tabular}\\
    $\Xi / \xi$ & 
    \begin{tabular}[t]{@{}p{0.85\textwidth}@{}}
      (Chapter~\ref{chap:pse}) $\xi$ -- (id.) Mathematical shorthand (see Equation~\eqref{eq:pse_setup23_maths_shorthand}).\\
    \end{tabular}\\
    \\[0.01\textheight]
  \end{tabular}
  \end{center}
\end{table}

\begin{table}
  \onehalfspacing
  \setlength\extrarowheight{0.5em}
  \begin{center}
  \begin{tabular}[t]{c|l}
    \begin{tabular}[t]{@{}c@{}}
      Leading \\[-0.5em] Symbol
    \end{tabular} & 
    Quantity/Concept Represented \\\hline
    $\Pi / \pi$ & 
    \begin{tabular}[t]{@{}p{0.85\textwidth}@{}}
      (Chapter~\ref{chap:statstest}) $\pi_X$ -- (const.) Success rate of a Bernoulli distributed r.v. $X$.\\
      (Chapter~\ref{chap:pse}) $\pi_{\textrm{min}}$ -- (const.) Minimum test power.\\
    \end{tabular}\\
    $\rho$ & 
    \begin{tabular}[c]{@{}p{0.85\textwidth}@{}}
      (All chapters) $\rho_{XY}$ -- (func.) Correlation between two r.v.'s $X$ and $Y$.\\
    \end{tabular}\\
    $\Sigma / \sigma$ & 
    \begin{tabular}[t]{@{}p{0.85\textwidth}@{}}
      (All chapters) $\sigma_X$ -- (const.) Standard deviation of an r.v., a population, or a group $X$.\\
      (All chapters) $\sigma^2_X$ -- (const.) Variance of an r.v., a population, or a group $X$.\\
    \end{tabular}\\
    $\tau$ & 
    \begin{tabular}[t]{@{}p{0.85\textwidth}@{}}
      (Chapters~\ref{chap:statstest}--\ref{chap:oced}) $\tau^2$ -- (const.) Hyperparameter in a mixed sequential probability ratio test (mSPRT), representing the variance of the mixture distribution.\\
    \end{tabular}\\
    $\Phi / \phi$ & 
    \begin{tabular}[t]{@{}p{0.85\textwidth}@{}}
      (All chapters) $\Phi(\cdot)$ -- (func.) Cumulative density function (CDF) of a standard normal r.v.\\
      (All chapters) $\phi(\cdot)$ / $\phi(\,\cdot \,; \mu, \sigma^2)$ -- (func.) Probability density function (PDF) of a standard normal r.v. / a normal r.v. with mean $\mu$ and variance $\sigma^2$.\\
    \end{tabular}\\
    $\chi$ & 
    \begin{tabular}[t]{@{}p{0.85\textwidth}@{}}
      (Chapter~\ref{chap:statstest}) $\chi^2_\nu$ -- (dist.) A $\chi^2$-distribution with $\nu$ d.f.\\
    \end{tabular}\\
    \\[0.45\textheight]
  \end{tabular}
  \end{center}
\end{table}
\cleardoublepage

\chapter{What is the Value of Digital Experimentation and Measurement Capabilities?}
\label{chap:vem}

From the thesis title, it is reasonable for readers to expect content covering methodological advances in statistical tests, experiment design, and causal inference. While we will address such mainstream topics in later chapters of the thesis, we first present a ranking under lower uncertainty problem as an application of order statistics to motivate our topics. As the chapter title suggests, addressing the problem enables us to value digital experimentation and measurement (DEM) capabilities quantitatively, which, in turn, helps us to answer \textit{why} we require such capabilities in the first place.

This chapter is adapted from the research paper ``\textit{What is the Value of Experimentation \& Measurement?}'' It was first presented at the \textit{2019 IEEE International Conference on Data Mining (ICDM)} \cite{liu19whatisthevalue} and later extended and published in \textit{Data Science and Engineering} as part of a special issue on Highly-Rated Short Papers of ICDM 2019~\cite{liu20whatisthevalue}.
In Sections~\ref{sec:vem_introduction} and~\ref{sec:vem_related_work}, as well as introducing related work, we will motivate how the requirement to value DEM capabilities arises and leads to our upcoming ranking under lower uncertainty problem. Readers who are more interested in the mathematical formulation of the ranking under lower uncertainty problem can skip straight to Section~\ref{sec:vem_mathematical_formulation}, though they may find the chapter organisation at the end of Section~\ref{sec:vem_introduction} useful.

\section{Motivation}
\label{sec:vem_introduction}

The value of DEM capabilities is currently best reflected in the success of organisations that have adopted and advocated for them in the past decade. 
Many major technology companies report having mature infrastructure for online controlled experiments (OCEs, e.g., Google~\cite{tang10overlapping}, Linkedin~\cite{xu15frominfrastructure}, and Microsoft~\cite{kohavi13online}) or are heavily investing in state-of-the-art techniques (e.g., Airbnb~\cite{lee2018winner},
 Netflix~\cite{xie16improving}, and Yandex~\cite{poyarkov16boosted}), or both.
Amazon~\cite{hill15measuring}, Facebook~\cite{gordon2019comparison}, and Uber~\cite{barajas20advertising} have also reported the use of various causal inference techniques to measure the incrementality of advertising campaigns. Several start-ups (e.g. Optimizely~\cite{johari17peeking} and Qubit~\cite{browne17whatworks}) have also recently been
established purely to manage OCEs for
businesses.

While mature DEM capabilities can quantify the value of a business proposition, it remains a substantial challenge to ``measure the measurer'' -- to quantify the value of the DEM capabilities themselves. To the best of our knowledge, no work addresses the question, ``Should we invest in DEM capabilities?'' or how to value these capabilities when we first present our ideas. The lack of  prior work makes it hard to build a compelling business case to justify investment in the related personnel and infrastructure. We address this problem by calculating both the expected value and the risk, allowing the Sharpe ratio~\cite{sharpe1966mutual} for a DEM capability to be calculated and compared to other potential investments.

\begin{figure*}
\begin{center}
\begin{subfigure}[t]{.9\textwidth}
  \includegraphics[width=\textwidth, trim=0 0 0 0, clip]{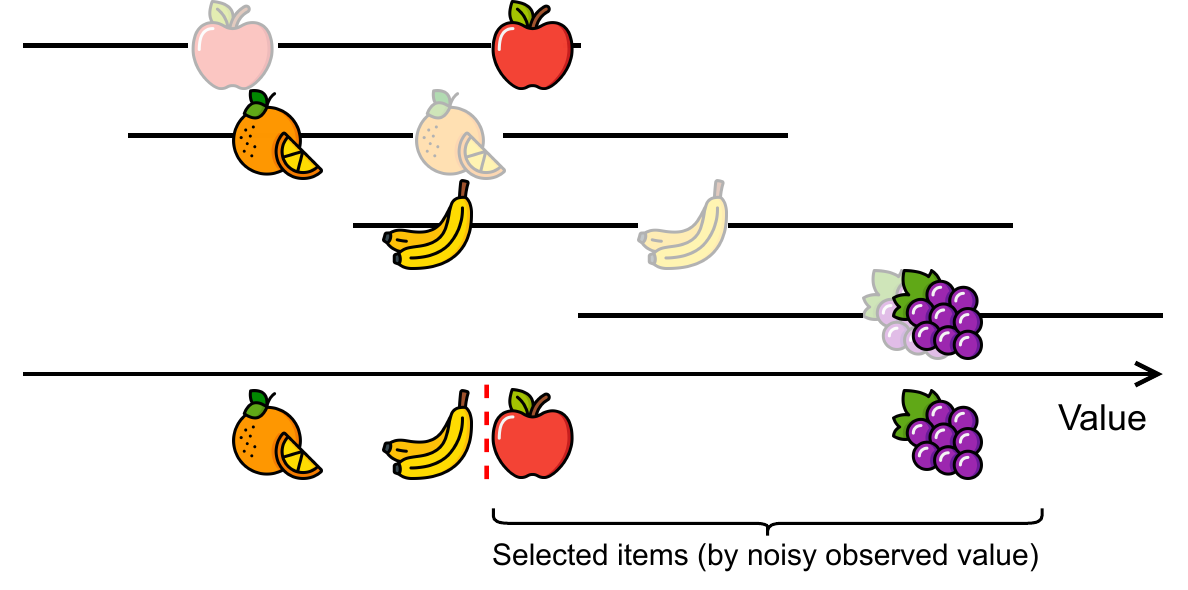}
  \caption{High estimation noise}
  \label{fig:vem_prioritisation_high_noise}
\end{subfigure}\\
\vspace*{2em}
\begin{subfigure}[t]{.9\textwidth}
  \includegraphics[width=\textwidth, trim=0 0 0 0, clip]{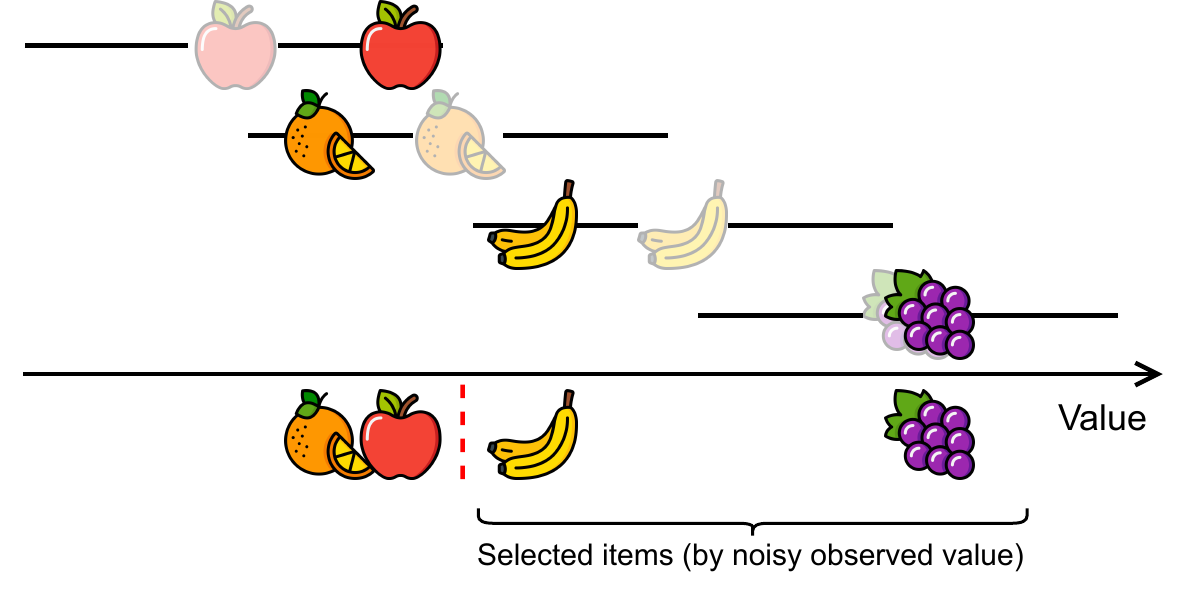}
  \caption{Low estimation noise}
  \label{fig:vem_prioritisation_low_noise}
\end{subfigure}
\end{center}
\caption[Prioritising projects according to their value under different noise levels in value estimation.]{Prioritising four projects (the fruits) according to their value (x-axes). The semi-opaque icons represent the projects' true value, and the solid icons represent possible project value estimates under some level of uncertainty (horizontal lines) in the estimation process. (Top) Under a noisy estimation process, projects with a low true value (e.g., project apple) may appear to have a high value and be prioritised erroneously. (Bottom) DEM reduces the estimation noise, enabling a better prioritisation with value estimates closer to the truth. The figures include icons designed by Smashicons from Flaticon.com.}
\label{fig:vem_prioritisation_noise}
\end{figure*}

The value created by DEM capabilities comes from the following three sources: 
\begin{enumerate}
\item \textbf{Recognising value}: DEM capabilities enable one to attribute value to a digital product, business proposition or service. They also prevent damage from business propositions that have a negative value. This is important for dynamic organisations with large numbers of business propositions, as the damage caused by individual rollouts can be compartmentalised and contained in a similar fashion, similar to unit and integration testing in software development.
\item \textbf{Prioritisation}: Without DEM capabilities, one prioritises based on back-of-envelope estimates or gut feel, which has high uncertainty. DEM reduces the magnitude of the noise arising from estimation, enabling prioritisation based on estimates closer to the true values and improving long-term decision-making (see Figure~\ref{fig:vem_prioritisation_noise}).
\item \textbf{Optimisation}: DEM capabilities allow one to evaluate large numbers of variants against each other and the best to be selected efficiently. Without such capabilities, we can still experiment with different business propositions sequentially, though it is slow and introduces noise from the changing environment.
\end{enumerate}

Quantifying the value of a DEM capability is relatively straightforward once it is in place.
For example, the value of Item 1 comes from rolling back negative business propositions: we can calculate it by summing the negative contributions of unsuccessful business propositions. Likewise, we can calculate the value of Item 3 by summing the difference between the maximum and the mean value for each variant over the business propositions. Indeed, these are the approaches taken by~\cite{mao2021quantifying} when valuing LinkedIn's experimentation platform using past experiment data.

The approaches above are not feasible for organisations building DEM capabilities from scratch. One can attempt to estimate the values from Items 1 and 3 using generic value distribution across business propositions, which is given across industries in~\cite{johnson2017online} and~\cite{browne17whatworks}, though these estimates are generally no better than back-of-envelope calculations. Without an accurate valuation, organisations remain in a chicken and egg situation -- no valuation, no investment; no investment, no capability; no capability, no valuation.

Given the above, we see quantifying the value of Item 2 as the less explored yet more compelling approach and the subject of the remainder of this chapter. DEM capabilities improve prioritisation by reducing uncertainty in the value estimates of each business proposition. This is a form of ranking under uncertainty, a well-studied problem in statistics and operations research. However, in all previous work, the variance is assumed to be a fixed constant or changed without measuring the corresponding change in the value of the ranked items. Here, we wish to understand the value of ranking under \emph{lower} uncertainty through DEM capabilities.

Our contribution is as follows. We 
\begin{enumerate}
\item Specify the first model that values the contribution of a DEM capability in terms of better prioritisation due to reduced estimation noise for business propositions (Sections~\ref{sec:vem_mathematical_formulation} and~\ref{sec:vem_E_D}); 
\item Derive the variance of our estimator, allowing one to calculate a Sharpe ratio to guide organisations considering an investment in DEM (Section~\ref{sec:vem_var}); and finally 
\item Provide two case studies based on large-scale meta-analyses that reflect how one can apply our model to real-world practice (Section~\ref{sec:vem_case_study}) and two extensions that open the door to future work in this area (Section~\ref{sec:vem_extensions}).
\end{enumerate}

\section{Related work}
\label{sec:vem_related_work}

There is a vast literature on using controlled or natural experiments in a digital technology context. In addition to that mentioned in Section~\ref{sec:vem_introduction}, there are works dedicated to running trustworthy online controlled experiments~\cite{dmitriev2017adirtydozen}, choosing good metrics~\cite{hohnhold15focusing} and designing experiments where samples are dependent due to external confounders~\cite{backstrom11network,bakshy13uncertainty}.\footnote{We will describe the works in greater detail in Chapter~\ref{chap:rade}.} However, these works all assume the existence of DEM capabilities, and to the best of our knowledge, no literature that helps organisations justify the acquisition of DEM capabilities exists. While~\cite{mao2021quantifying} asked a similar question as this chapter does, they aimed to value an experimentation platform, i.e., an existing DEM capability. We believe that filling this gap is necessary for widespread adoption and that increased participation will accelerate the development of the field.

This chapter is related to existing work in statistics and operations research, particularly on decision-making under uncertainty, which has been extensively studied since the 1980s. Notable work includes proposals for additional components in a decision maker's utility function~\cite{bell82regret}, alternate risk measures~\cite{xu09alternativerisk}, and a general framework for decision-making with incomplete information (i.e., uncertainty)~\cite{weber87decision}. These works assume the inability to change the noise associated with estimation or measurement (or both).

The sub-problem of ranking under uncertainty has also attracted considerable attention, partially due to the advent of large databases and the requirement to rank results with a certain ambiguity in relevance~\cite{soliman09ranking}. While~\cite{zuk07ranking} measured the influence of noise levels in their work, they focused on the quality of the ranks themselves but not the value associated with the ranks.

The project selection problem is a related topic in optimisation, where the goal is to find the optimal set of business propositions using mixed integer linear programming, possibly under uncertainty. Work in this domain generally seeks methods that cope with existing risk/noise~\cite{mavrotas2013trichotomic}, and to the best of our knowledge, no work considers the value of reducing risk. While~\cite{shakhsiniaei11comprehensive} have discussed lowering the uncertainty level during the selection process, they refer to the uncertainty of decision parameters instead of the general noise level.

\section{Mathematical Formulation}
\label{sec:vem_mathematical_formulation}

We formulate the ranking under lower uncertainty problem, which has a wide variety of important applications in its own right~\cite{singh2021fairness,zuk07ranking}, as follows. Consider the scenario where we select $M$ business propositions from $N$ candidates, where $M < N$.
The \textbf{E}stimated value of each business proposition is given by ${\mathcal{E}_i = \mathcal{V}_i + \epsilon_i}$, where~$\mathcal{V}_i$ are the \textit{true yet unobserved} \textbf{V}alues  estimated with error~$\epsilon_i$. 
The business propositions are labelled in ascending order of estimated value $\mathcal{E}_i$ to get the order statistics ${\mathcal{E}_{(1)}, \mathcal{E}_{(2)}, ..., \mathcal{E}_{(N)}}$,
and we select the $M$ business proposition with the highest estimated values: ${\mathcal{E}_{(N-M+1)}, \mathcal{E}_{(N-M+2)}, ..., \mathcal{E}_{(N)}}$.
We are interested in the true value of the selected business propositions, given by
\begin{align}
    \mathcal{V}_{\mathcal{I}(N-M+1)}, \mathcal{V}_{\mathcal{I}(N-M+2)}, ..., \mathcal{V}_{\mathcal{I}(N)} \,,
    \label{eq:vem_corresponding_true_val}
\end{align}
where $\mathcal{I}(\cdot)$ denotes the index function that maps the ranking to the index of the business proposition. Note one should not confuse the set in Expression~\eqref{eq:vem_corresponding_true_val} with the set $\{\mathcal{V}_{(N-M+1)},$ $\mathcal{V}_{(N-M+2)}, ..., \mathcal{V}_{(N)}\}$  --  the latter denotes the top $M$ business propositions ranked by their true value and are likely to be different~\cite{zuk07ranking}.

We define the mean true value of the $M$ selected business propositions as
\begin{align}
    \mathcal{W} \triangleq 
    \frac{1}{M}\left(
    \mathcal{V}_{\mathcal{I}(N-M+1)} + 
    \mathcal{V}_{\mathcal{I}(N-M+2)} + ... + 
    \mathcal{V}_{\mathcal{I}(N)}
    \right) \,,
    \label{eq:vem_mean_true_value}
\end{align}
where the best prioritisation maximises $\mathcal{W}$.\footnote{Readers can internalise the $\mathcal{W}$ notation by drawing parallels between the name of the letter in French (\textit{double v\'{e}}) and Spanish (\textit{doble ve}) -- both literally meaning ``double v'' -- and the fact that we use $\mathcal{W}$ to denote an aggregation of two or more $\mathcal{V}$.}

Part of the value of ranking under lower uncertainty, which DEM capabilities help bring, arises from the observation that $\mathcal{W}$ increases when the magnitude of the uncertainties arising from estimation ($\epsilon_i$) decreases.  We are interested in the value gained by reducing estimation uncertainty \textbf{without changing the set of business propositions} (i.e., retaining all $\mathcal{V}_i$), as the true value of the business propositions does not depend on the measurement method used:
\begin{align}
    \mathcal{D} \triangleq \mathcal{W}\, |_{\textrm{ lower noise}} - \mathcal{W}\, |_{\textrm{ higher noise}} \,.
    \label{eq:vem_D_pseudo}
\end{align}

Obtaining the value gained ($\mathcal{D}$) enables us to compare DEM capabilities with other potential investments. The Sharpe ratio is a standard quantity used in finance to compare different investments. This standard score-like ratio represents the risk-adjusted expected excess return on investment compared to a risk-free investment~\cite{sharpe1966mutual}. In our case, $\mathbb{E}(\mathcal{D})$ represents the expected return on investment of DEM capabilities, $\textrm{Var}(\mathcal{D})$ represents the risk, and we can specify the return on investment of the risk-free investment as a constant $c$. This enables us to calculate the Sharpe ratio as
\begin{align}
    \frac{\mathbb{E}(\mathcal{D})-c}{\sqrt{\textrm{Var}(\mathcal{D})}} \,.
    \label{eq:vem_sharpe_ratio}
\end{align}


\subsection{Modelling values with statistical distributions}
\label{sec:vem_mathematical_formulation_distributions}

\begin{figure}
\begin{center}
    \includegraphics[width=0.25\textwidth, trim = 0 0 0 0, clip]{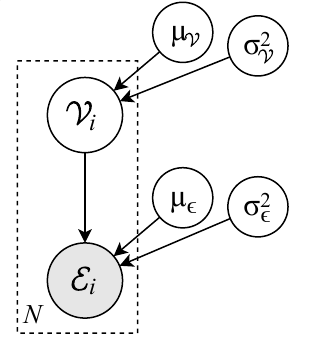}
\end{center}
\vspace*{-\baselineskip}
\caption[The generative model in the ranking under lower uncertainty problem in plate notation.]{The generative model in the ranking under lower uncertainty problem in plate notation. $\mathcal{V}_i$ represents the true, unobserved values of the items to be ranked. $\mathcal{E}_i$ represents the observed values under some estimation noise level~$\sigma^2_\epsilon$.}
\label{fig:vem_rulu_generative_model_one_sample}
\vspace*{\baselineskip}
\end{figure}

To value a DEM capability using such a generic framework that one can apply in many different ways across diverse organisations, it is first necessary to make some simplifying assumptions about the statistical properties of the business propositions under consideration. We assume the value of the business propositions~($\mathcal{V}_i$) and the estimation noises ($\epsilon_i$) are randomly distributed:
\begin{align}
    \mathcal{V}_i \overset{\textrm{i.i.d.}}{\sim} F_\mathcal{V}(\cdot) \,, \textrm{ where } & \mathbb{E}(\mathcal{V}_i) = \mu_\mathcal{V}, \textrm{Var}(\mathcal{V}_i) = \sigma^2_\mathcal{V} \,, \nonumber\\
    \epsilon_i \overset{\textrm{i.i.d.}}{\sim} F_\epsilon(\cdot) \,, \textrm{ where } & \mathbb{E}(\epsilon_i) = \mu_\epsilon, \textrm{Var}(\epsilon_i) = \sigma^2_\epsilon \,, \label{eq:vem_X_n_eps_n_modelling_general}
\end{align}
where 
$\mathcal{V}_i \perp \epsilon_j\, \forall\, i, j$ (see Figure~\ref{fig:vem_rulu_generative_model_one_sample}).

We note two special cases, one when both the value and the noises are assumed to be normally distributed:
\begin{align}
    \mathcal{V}_i \overset{\textrm{i.i.d.}}{\sim} \mathcal{N}(\mu_\mathcal{V}, \sigma^2_\mathcal{V})\,,\;
    \epsilon_i \overset{\textrm{i.i.d.}}{\sim} \mathcal{N}(\mu_\epsilon, \sigma^2_\epsilon) \,;
    \label{eq:vem_X_n_eps_n_modelling_normal}
\end{align}
and the other when both the value and the noises are assumed to follow some Generalized Student's $t$-distributions:
\begin{align}
    \mathcal{V}_i = \mu_\mathcal{V} + \sigma_\mathcal{V}\sqrt{\frac{ (\nu - 2)}{\nu}}  T_i \,, \quad T_i \overset{\textrm{i.i.d.}}{\sim} t_\nu \,, \nonumber\\
    \epsilon_i = \mu_\epsilon + \sigma_\epsilon \sqrt{\frac{(\nu - 2)}{\nu}}  T^{'}_i \,, \quad T^{'}_i \overset{\textrm{i.i.d.}}{\sim} t_\nu \,,
    \label{eq:vem_X_n_eps_n_modelling_t}
\end{align}
where~$t_{\nu}$ is a standard Student's $t$-distribution with~$\nu$ degrees of freedom. The location and scaling parameters ensure $\mathcal{V}_i$ and $\epsilon_i$ have the mean and variance specified in Expression~\eqref{eq:vem_X_n_eps_n_modelling_general}.

These two cases are particularly relevant as meta-analyses compiled on the results of~6,700 e-commerce~\cite{browne17whatworks} and~432 marketing experiments~\cite{johnson2017online}, respectively, indicate the uplifts measured by the experiments, and hence the value of the business propositions under some estimation noise, exhibit the following properties:
\begin{enumerate}
    \item They can be positive or negative,
    \item They are usually clustered around an average instead of uniformly spreading across a specific range, and
    \item The distributions are heavy-tailed. 
\end{enumerate}
The normal assumptions cover the first two properties only, yet enable one to draw on the wealth of results in order statistics and Bayesian inference related to normal distributions to get started. The $t$-distributed assumptions also cover property~3, though valuation under such assumptions is more complicated as $t$-distributions do not have conjugate priors. 

We will include the valuation under $t$-distributed assumptions under the general case for brevity. We will, however, present empirical results in Section~\ref{sec:vem_extension_t_distributed}, showing that the value gained under $t$-distributed assumptions has a higher mean and variance, thus demonstrating that the model can capture the ``higher risk, higher reward'' concept.

\subsection{Key results}

In the following two sections, we will derive the expected value and variance for~$\mathcal{W}$, the mean true value of the top~$M$ business propositions selected after being ranked by their estimated value (as defined in Equation~\eqref{eq:vem_mean_true_value}), as well as the expected value and the variance of~$\mathcal{D}$, the value gained when the estimation noise is reduced. This enables us to calculate the Sharpe ratio defined in Expression~\eqref{eq:vem_sharpe_ratio}.

We will also provide two key insights. Firstly,
the expected mean true value of the selected business propositions~($\mathcal{W}$) increases when the estimation noise~($\sigma^2_\epsilon$) decreases, and the relative increase in value depends on how much noise we can reduce. Secondly, when $M$ is small, reducing the estimation noise may not lead to a statistically significant improvement in the true value of the business propositions selected. As a result, improvements in prioritisation driven by DEM may only be justified for larger organisations.

\section{Calculating the Expectation}
\label{sec:vem_E_D}

We first derive the expected value for $\mathcal{D}$. This requires the expected values of, in order:
\begin{enumerate}
    \item $\mathcal{E}_{(r)}$ -- the \emph{estimated} value of the $r^{\textrm{th}}$ ranked business proposition in estimated value;
    \item $\mathcal{V}_{\mathcal{I}(r)}$ -- the \emph{true} value of the $r^{\textrm{th}}$ ranked business proposition in estimated value; and
    \item $\mathcal{W}$ -- the mean of the \emph{true} value for the $M$ most valuable business propositions ranked by their estimated values.
\end{enumerate}


To obtain the expected value for $\mathcal{E}_{(r)}$, we apply a result in Section 4.6 of~\cite{david2003order}, which approximates the expected value of the order statistics $\mathcal{E}_{(r)}$ using the quantile function of $\mathcal{E}_i$:
\begin{align}
    \mathbb{E}(\mathcal{E}_{(r)}) \approx  
    F_\mathcal{E}^{-1}\Big(\frac{r - c}{N - 2c + 1}\Big),
    \label{eq:vem_E_Yr_general}
\end{align}
where $F_\mathcal{E}^{-1}(\cdot)$ denotes the quantile function for $\mathcal{E}_i$ and $c$ is a constant correcting the quantile of the ranks.\footnote{Many values for $c$ were proposed. Early works advocate $c=0$ or $c=0.5$ depending on how one approaches continuity correction between ranks and quantiles~\cite{harter61expected}. \cite{blom1958statistical} proposes a compromise value of $c=\frac{3}{8}$ based on a tabulation of $c$ required to yield the correct expected value for different $r$ and $N$. \cite{harter61expected} expands the tabulation and proposes using a different $c$ for each $N$, with the values of $c$ hovering around 0.4. For simplicity, we take $c=0.4$ for all $N$ in all our calculations.} The formula results from applying the first-order delta method on the underlying compound beta-$F^{-1}_{\mathcal{E}}$ distribution~\cite{probabilityislogic2015approximate}. Such estimation will inevitably incur a small bias, particularly for extreme order statistics (i.e., $r$ close to one or $N$) and when $N$ is small. However, as verified empirically in Section~\ref{sec:vem_experiments}, we consider such an estimation accurate enough
for our application.\footnote{\cite{harter61expected} also noted for normal order statistics, the maximum error is 0.018 (i.e., less than one percent relative to the value of extreme order statistics) for some choice of $c$.}
One can also include higher-order terms in the corresponding Taylor expansion to improve estimation accuracy, though we believe the additional complexity and, thus, reduced interpretability in the resultant formula is not worth the marginal gain in accuracy.

We then obtain the expected value of $\mathcal{V}_{\mathcal{I}(r)}$ by applying Equation 6.8.3a of~\cite{david2003order}:
\begin{align}
    \mathbb{E}(\mathcal{V}_{\mathcal{I}(r)}) = \,& \mu_\mathcal{V} + \rho_{\mathcal{V}\mathcal{E}}\,\sigma_\mathcal{V} \,\mathbb{E}\left(\frac{\mathcal{E}_{(r)} - (\mu_\mathcal{V} + \mu_\epsilon)}{\sqrt{\sigma^2_\mathcal{V} + \sigma^2_\epsilon}}\right) \nonumber\\
    = \,&  \mu_\mathcal{V} + \frac{\sigma^2_\mathcal{V}}{\sigma^2_\mathcal{V} + \sigma^2_\epsilon}\left(\mathbb{E}(\mathcal{E}_{(r)}) - (\mu_\mathcal{V} + \mu_\epsilon)\right) ,
    \label{eq:vem_E_XIr_general}
\end{align}
where $\rho_{\mathcal{V}\mathcal{E}} = \sqrt{\sigma^2_\mathcal{V}} / \sqrt{\sigma^2_\mathcal{V} + \sigma^2_\epsilon}$ is the correlation between $\mathcal{V}_i$ and $\mathcal{E}_i$. 

Equation~\eqref{eq:vem_E_XIr_general} shows that decreasing the estimation noise $\sigma^2_\epsilon$ will increase $\mathbb{E}(\mathcal{V}_{\mathcal{I}(r)})$ for any
${r > (N+1)\cdot F_\mathcal{E}(\mu_\mathcal{V} + \mu_\epsilon)}$. It follows that the mean true value of the top $M$ business propositions, selected according to their estimated value, will generally increase with a lower estimation noise. We show this by applying the expectation function to~$\mathcal{W}$ defined in Equation~\eqref{eq:vem_mean_true_value} to obtain
\begin{align}
    \mathbb{E}(\mathcal{W}) = \,& \frac{1}{M} \sum_{r=N-M+1}^{N} \mathbb{E}(\mathcal{V}_{\mathcal{I}(r)}) \nonumber\\
    = \,& \mu_\mathcal{V} + \frac{\sigma^2_\mathcal{V}}{\sigma^2_\mathcal{V} + \sigma^2_\epsilon} \left[\left(\frac{1}{M} \sum_{r=N-M+1}^{N} \mathbb{E}(\mathcal{E}_{(r)}) \right) - \left(\mu_\mathcal{V} + \mu_\epsilon \right)\right] \,. \label{eq:vem_E_V_general}
\end{align}

We finally consider the improvement when we reduce the estimation noise from $\sigma^2_\epsilon = \sigma^2_1$ to~$\sigma^2_\epsilon = \sigma^2_2$. This will be the expected value gained by having better DEM capabilities:
\begin{align}
    \mathbb{E}(\mathcal{D}) = \, &  \mathbb{E}\left(\mathcal{W} \,|\,{\sigma^2_\epsilon = \sigma^2_2}\right) - \mathbb{E}\left(\mathcal{W} \,|\, {\sigma^2_\epsilon = \sigma^2_1}\right) \nonumber\\
    = \, &  \frac{\sigma^2_\mathcal{V}}{\sigma^2_\mathcal{V} + \sigma^2_2}\left[\left(\frac{1}{M}\sum_{r=N-M+1}^{N} \mathbb{E}\left(\mathcal{E}_{(r)} \,|\, {\sigma^2_\epsilon = \sigma^2_2}\right)\right) - (\mu_\mathcal{V} + \mu_\epsilon) \right] - \nonumber \\
    & \quad \frac{\sigma^2_\mathcal{V}}{\sigma^2_\mathcal{V} + \sigma^2_1}\left[\left(\frac{1}{M}\sum_{r=N-M+1}^{N} \mathbb{E}\left(\mathcal{E}_{(r)} \,|\, {\sigma^2_\epsilon = \sigma^2_1}\right)\right) - (\mu_\mathcal{V} + \mu_\epsilon) \right] \,. \label{eq:vem_E_D_general}
\end{align}

\subsection{Expectation under normal assumptions}

In the special case where $\mathcal{E}_i$ are normally distributed (with mean $\mu_\mathcal{V} + \mu_\epsilon$ and variance $\sigma^2_\mathcal{V} + \sigma^2_\epsilon$), the expected value for the normal order statistics $\mathcal{E}_{(r)}$ is approximately
\begin{align}
    \mathbb{E}(\mathcal{E}_{(r)}) \approx 
    \mu_\mathcal{V} + \mu_\epsilon + 
    \sqrt{\sigma^2_\mathcal{V} + \sigma^2_\epsilon}\;\Phi^{-1}\left(\frac{r - c}{N - 2c + 1}\right),
    \label{eq:vem_E_Yr_normal}
\end{align}
where $\Phi^{-1}(\cdot)$ denotes the quantile function of a standard normal distribution~\cite{blom1958statistical}. It is worth noting that decreasing the estimation noise $\sigma^2_\epsilon$ will decrease $\mathbb{E}(\mathcal{E}_{(r)})$ for any $r > \frac{N+1}{2}$, appearing to lower the average value of the top~$M$ business propositions. This is a common pitfall -- we are not optimising the combined estimated value of the selected business propositions. What matters is the true yet unobserved value of that proposition, $\mathcal{V}_{\mathcal{I}(r)}$, as shown below.

For $\mathcal{V}_{\mathcal{I}(r)}$, we can simplify Equation~\eqref{eq:vem_E_XIr_general} by substituting Equation~\eqref{eq:vem_E_Yr_normal}. We can also evaluate from first principles by noting a standard result in Bayesian inference, which states that the posterior distribution of~$\mathcal{V}_i$ once $\mathcal{E}_i$ is observed is also normally distributed with mean
\begin{align}
    \mu_{\mathcal{V}_i \,|\, \left (\mathcal{E}_i=e \right)} & = \frac{\sigma^2_\mathcal{V}}{\sigma^2_\mathcal{V} + \sigma^2_\epsilon} \,(e - \mu_\epsilon) + \frac{\sigma^2_\epsilon}{\sigma^2_\mathcal{V} + \sigma^2_\epsilon} \,\mu_\mathcal{V} \,,
    \label{eq:vem_true_val_posterior_mean}
\end{align}
and applying the law of iterated expectations to obtain\footnote{$\mathcal{E}_{(r)} \equiv \mathcal{E}_{\mathcal{I}(r)}$, as we rank the business propositions by their estimated values.\label{footnote:vem_noisy_rank_index_equiv}}
\begin{align}
    \mathbb{E}(\mathcal{V}_{\mathcal{I}(r)}) = \mathbb{E}\left(\mathbb{E}(\mathcal{V}_{\mathcal{I}(r)} \,|\, \mathcal{E}_{(r)})\right) 
    \approx \mu_\mathcal{V} + 
    \frac{\sigma^2_\mathcal{V}}{\sqrt{\sigma^2_\mathcal{V} + \sigma^2_\epsilon}}
   \;\Phi^{-1}\left(\frac{r - c}{N - 2c + 1}\right) \,.
    \label{eq:vem_E_XIr_normal}
\end{align}
Here, decreasing the estimation noise $\sigma^2_\epsilon$ will lead to an increase in $\mathbb{E}(\mathcal{V}_{\mathcal{I}(r)})$ for any
${r > \frac{N+1}{2}}$.\footnote{While $\mathcal{V}_{\mathcal{I}(r)}$ are no longer normally distributed with the ranking information, it remains valid to use the law of iterated expectations to obtain the expected value for all $r$.\label{footnote:vem_loie_validity}}

The value of business propositions chosen ($\mathcal{W}$) under normal assumptions then evaluates to 
\begin{align}
    \mathbb{E}(\mathcal{W}) \approx
    \mu_\mathcal{V} + \frac{\sigma^2_\mathcal{V}}{\sqrt{\sigma^2_\mathcal{V} + \sigma^2_\epsilon}} \frac{1}{M} \sum\limits_{r=N-M+1}^{N} \Phi^{-1}\left(\frac{r-c}{N - 2c + 1}\right) \,. 
    \label{eq:vem_E_V_normal}
\end{align}
This is done by substituting Equation~\eqref{eq:vem_E_XIr_normal} into Equation~\eqref{eq:vem_E_V_general}. Note the absence of $\mu_\epsilon$ in Equation~\eqref{eq:vem_E_V_normal}, which suggests that systematic bias in estimation will not affect the true value of the chosen business propositions in the normal case.

Finally, the expression for the expected value of $\mathcal{D}$ when we reduce the estimation noise from $\sigma^2_\epsilon = \sigma^2_1$ to $\sigma^2_2$ is much neater under normal assumptions, as many terms cancel out in Equation~\eqref{eq:vem_E_D_general}, leading to
\begin{align}
    \mathbb{E}(\mathcal{D})
    \approx \left(\frac{\sigma^2_\mathcal{V}}{\sqrt{\sigma^2_\mathcal{V} + \sigma^2_2}} - \frac{\sigma^2_\mathcal{V}}{\sqrt{\sigma^2_\mathcal{V} + \sigma^2_1}} \right) \frac{1}{M} \sum\limits_{r=N-M+1}^{N} \Phi^{-1}\left(\frac{r-c}{N - 2c + 1}\right).
    \label{eq:vem_E_D_normal}
\end{align}
If we further assume that $\mu_\mathcal{V} = 0$ (i.e., the true value of the business propositions centres around zero), then the relative gain is entirely dependent on $\sigma^2_\mathcal{V}$, $\sigma^2_1$, and $\sigma^2_2$:
\begin{align}
    & \frac{\mathbb{E}(\mathcal{D}\,|\,\mu_\mathcal{V} = 0)}{\mathbb{E}(\mathcal{W}\,|\,\sigma^2_\epsilon = \sigma^2_1,\, \mu_\mathcal{V} = 0)} 
    = \, \frac{\sqrt{\sigma^2_\mathcal{V} + \sigma^2_1}}{\sqrt{\sigma^2_\mathcal{V} + \sigma^2_2}} - 1 \,.
\label{eq:vem_gauss_res}
\end{align}

To calculate the relative improvement in prioritisation delivered by DEM under these assumptions, we plug the following into Equation~\eqref{eq:vem_gauss_res} to obtain an estimate of how much one will gain from acquiring such capabilities:
\begin{enumerate}
    \item The estimated spread of the values ($\sigma^2_\mathcal{V}$),
    \item The estimated deviation of the current estimation process ($\sigma^2_1$), and
    \item The estimated deviation to the actual value upon acquisition of DEM capabilities ($\sigma^2_2$).
\end{enumerate}
For example, if Example Company Ltd's project values are spread with a standard deviation of 1 unit and their current estimation has a standard error of 0.5 units, then by acquiring an A/B test framework that is capable of measuring with an error of 0.4 units, the company gains~3.8\% of extra value simply due to the ability to prioritise with more accurate measurements under normal assumptions. 

We conclude this section by suggesting how one may estimate  $\sigma^2_{\mathcal{V}}$, $\sigma^2_{1}$, and $\sigma^2_{2}$, especially when they have yet to build any DEM capabilities. Clearly, it is impossible to recommend specific values for the three parameters as organisations seeking to build DEM capabilities come in all shapes and sizes. That said, one may consider using industry averages published in meta-analyses~\cite{browne17whatworks,johnson2017online} when estimating $\sigma^2_{\mathcal{V}}$. For $\sigma^2_{1}$, they may solicit value estimates for several business propositions from current decision makers and take the variance over such estimates.\footnote{It does not matter if the decision makers provide value estimates far from the actual value of the business propositions. In fact, one need not know the business propositions' actual value. What matters here is the spread of the estimates.} Finally, they may leverage the user numbers and the variance of user responses that form the business metric to estimate $\sigma^2_{2}$ under an A/B test.\footnote{For example, in an A/B test measuring conversion rate (CVR) uplift using a practical $t$-test, the estimation noise ($\sigma^2_{\epsilon}$ or~$\sigma^2_{2}$) is the variance of the difference in CVR between two variants under a nil null hypothesis. This equals $2\cdot\frac{p(1-p)}{n/2}=\frac{4p(1-p)}{n}$, where~$p$ is the base CVR and~$n$ is the total number of users across both groups. See Chapter~\ref{chap:statstest} for further discussions on statistical tests, particularly  Sections~\ref{sec:statstest_nhstexamples_ttest} and~\ref{sec:statstest_nhstexamples_binomial} on practical $t$-tests and how they are applied to business metrics based on binary responses such as CVR. \label{footnote:vem_abtest_estimation_noise}} We also encourage one to explore parameter values around their estimates for robustness, as we will show in Section~\ref{sec:vem_case_study} when we provide two case studies.

\section{Calculating the Variance}
\label{sec:vem_var}
It is crucial to understand both the expected gain and the risk or uncertainty of ranking under lower uncertainty to make effective investment decisions.
Therefore, having derived the expected value in Equation~\eqref{eq:vem_E_D_general}, we address the investment risk given by the variance of~$\mathcal{D}$ in this section. 

The variance calculation features new challenges in addition to that identified in the section above, the most prominent of which concerns the interactions between quantities generated under different estimation noise levels. While these interactions do not affect the expected value, they influence the variance via the covariance terms. Failure to account for the covariance terms may lead to a large error in the variance estimate.

\begin{figure}
    \begin{center}
        \includegraphics[width=0.4\textwidth, trim = 0 0 0 0, clip]{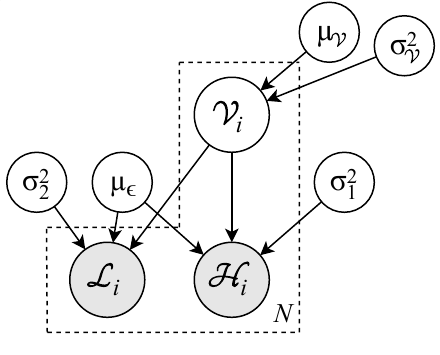}
    \end{center}
    \vspace*{-\baselineskip}
    \caption[The generative model in the ranking under lower uncertainty problem when two distinct noise levels are involved in plate notation.]{The generative model in the ranking under lower uncertainty problem when two distinct noise levels are involved in plate notation. When we change the noise level of our ranking under uncertainty setup from~$\sigma^2_1$ to~$\sigma^2_2$ (see Figure~\ref{fig:vem_rulu_generative_model_one_sample}), we obtain two sets of observed values, $\mathcal{H}_i$ and $\mathcal{L}_i$, for each noise level.}
    \label{fig:vem_rulu_generative_model_two_sample}
    \vspace*{\baselineskip}
\end{figure}

To address the challenges, we first extend the notation to clarify the interactions. Instead of a single set of noise shown in Section~\ref{sec:vem_mathematical_formulation}, we define two sets of random noise with different noise levels:
\begin{align}
    \epsilon_{1i} \overset{\textrm{i.i.d.}}{\sim} F_{\epsilon_1}(\cdot)\,,
    & \textrm{ where } \mathbb{E}(\epsilon_{1i}) = \mu_\epsilon \,, \textrm{Var}(\epsilon_{1i}) = \sigma^2_1 \,, \textrm{ and}\nonumber\\
    \epsilon_{2i} \overset{\textrm{i.i.d.}}{\sim} F_{\epsilon_2}(\cdot)\,,
    & \textrm{ where } \mathbb{E}(\epsilon_{2i}) = \mu_\epsilon \,, \textrm{Var}(\epsilon_{2i}) = \sigma^2_2 \,.
    \label{eq:vem_eps_1_eps_2_def}
\end{align}

Similar to how the problem is modelled in Section~\ref{sec:vem_mathematical_formulation}, we assume all true item values ($\mathcal{V}$) and noises ($\epsilon_1$ and $\epsilon_2$) are independent, i.e., $\mathcal{V}_i \perp \epsilon_{1j}$, $\mathcal{V}_i \perp \epsilon_{2j}$, and $\epsilon_{1i} \perp \epsilon_{2j}$ $\forall\, i, j$.
The estimated value of each item is then given by
\begin{align}
    \mathcal{H}_i = \mathcal{V}_i + \epsilon_{1i} & \quad\textrm{ under \textbf{H}igher noise level } \sigma^2_1 \textrm{ and }\nonumber\\ 
    \mathcal{L}_i = \mathcal{V}_i + \epsilon_{2i} & \quad\textrm{ under \textbf{L}ower noise level } \sigma^2_2 .
    \label{eq:vem_Yn_Zn_def}
\end{align}
The setup is illustrated in Figure~\ref{fig:vem_rulu_generative_model_two_sample}.

Having obtained two sets of estimated values, we rank and trace the corresponding indices for each set separately. For $\mathcal{H}$, we denote~$\mathcal{H}_{(r)}$ as the~$r^\textrm{th}$ order statistic of $\mathcal{H}_i$, the estimated value of the $r^\textrm{th}$ ranked item under noise level $\sigma^2_1$, followed by $\mathcal{V}_{\mathcal{I}(r)}$ as the concomitant~\cite{david2003order} of~$\mathcal{H}_{(r)}$, i.e., the true value of the $r^\textrm{th}$ item ranked by its estimated value. We repeat the process for $\mathcal{L}$: we denote~$\mathcal{L}_{(s)}$ as the~$s^\textrm{th}$ order statistic of~$\mathcal{L}_i$ and~$\mathcal{V}_{\mathcal{J}(s)}$ as the concomitant of~$\mathcal{L}_{(s)}$.\footnote{$\mathcal{I}(\cdot)$ and $\mathcal{J}(\cdot)$ are the index functions that map the ranking to the index for $\mathcal{H}$ and $\mathcal{L}$, respectively.}

We also define the mean true value of the top $M$ items, ranked by their estimated value, under both noise levels as
\begin{align}
    \mathcal{W}_1 = \frac{1}{M}\sum_{r=N-M+1}^{N} \mathcal{V}_{\mathcal{I}(r)} \,, \quad
    \mathcal{W}_2 = \frac{1}{M}\sum_{s=N-M+1}^{N} \mathcal{V}_{\mathcal{J}(s)} \,,
    \label{eq:vem_V1_V2_def}
\end{align}
where $\mathcal{W}_1$ is the mean true value under $\sigma^2_1$, and $\mathcal{W}_2$ is the mean true value under $\sigma^2_2$.
Finally, we denote the difference between the mean true values as $\mathcal{D} \triangleq \mathcal{W}_2 - \mathcal{W}_1$.

\begin{figure}
\begin{center}
    \includegraphics[width=0.9\textwidth, trim = 0 0 0 0, clip]{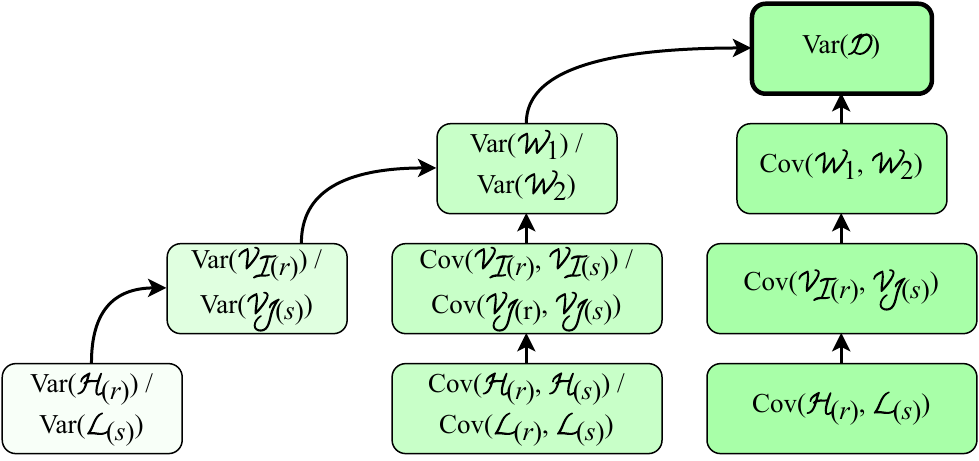}
\end{center}
\caption[Relationship between different variances/covariances used to calculate the variance of the value gained when the estimation noise is reduced.]{Relationship between different variances/covariances used to calculate the variance of $\mathcal{D}$, the value gained when the estimation noise is reduced. An arrow from quantity A to B means the value of B depends on the value of A.}
\label{fig:vem_var_cov_quantity_relation}
\vspace*{\baselineskip}
\end{figure}

Deriving the variance is similar to deriving the expectation -- one has to obtain the variances for (in order) $\mathcal{H}_{(r)}$/$\mathcal{L}_{(s)}$, $\mathcal{V}_{\mathcal{I}(r)}$/$\mathcal{V}_{\mathcal{J}(s)}$, $\mathcal{W}_1$/$\mathcal{W}_2$, and $\mathcal{D}$. The relationship between these quantities is shown in Figure~\ref{fig:vem_var_cov_quantity_relation}.

\subsection{\texorpdfstring{$\textrm{Var}(\mathcal{H}_{(r)})$/$\textrm{Var}(\mathcal{L}_{(s)})$}{Var(H\_(r))/Var(L\_(s))}}

We apply a result from~\cite{david54statitical}, which states that the variance of $\mathcal{H}_{(r)}$ and~$\mathcal{L}_{(s)}$ can be approximated as
\begin{align}
    \textrm{Var}(\mathcal{H}_{(r)}) & \approx
    \frac{r(N-r+1)}{(N+1)^2 (N+2)}
    \frac{1}{\left(f_\mathcal{H}\big(F_\mathcal{H}^{-1}\big(\frac{r}{N+1}\big)\big)\right)^2} \,, 
    \\[0.2em]
    \textrm{Var}(\mathcal{L}_{(s)}) & \approx
    \frac{s(N-s+1)}{(N+1)^2 (N+2)}
    \frac{1}{\left(f_\mathcal{L}\big(F_\mathcal{L}^{-1}\big(\frac{s}{N+1}\big)\big)\right)^2} \,,
    \label{eq:vem_var_Yr}
\end{align}
where $f(\cdot)$ and $F^{-1}(\cdot)$ denote the probability density function and quantile function of the corresponding r.v., respectively. As is the case with calculating the expected value (see Section~\ref{sec:vem_E_D}), the formula results from applying the first-order delta method~\cite{liu2022variance}, with the same strength and weakness considerations. In the special case where~$\mathcal{V}_i$, $\epsilon_{1i}$, and~$\epsilon_{2i}$ are all normally distributed, the variances are
\begin{align}
    \textrm{Var}(\mathcal{H}_{(r)}) & \approx
    \frac{r(N-r+1)}{(N+1)^2 (N+2)}
    \frac{\sigma^2_X + \sigma^2_1}{\big(\phi\big(\Phi^{-1}\big(\frac{r}{N+1}\big)\big)\big)^2} \,, 
    \\[0.2em]
    \textrm{Var}(\mathcal{L}_{(s)}) & \approx
    \frac{s(N-s+1)}{(N+1)^2 (N+2)}
    \frac{\sigma^2_X + \sigma^2_2}{\big(\phi\big(\Phi^{-1}\big(\frac{s}{N+1}\big)\big)\big)^2} \,,
    \label{eq:vem_observed_val_normal_order_stat_variance_approx}
\end{align}
where $\phi(\cdot)$ is the probability density function, and $\Phi^{-1}(\cdot)$ is the quantile function of a standard normal distribution.

\subsection{\texorpdfstring{$\textrm{Var}(\mathcal{V}_{\mathcal{I}(r)})$/$\textrm{Var}(\mathcal{V}_{\mathcal{J}(s)})$}{Var(V\_I(r))/Var(V\_J(s))}}

The variance for $\mathcal{V}_{\mathcal{I}(r)}$ is obtained using properties of the concomitants of order statistics~\cite{DAVID1998487}:\footnote{$\mathcal{H}/\mathcal{L}$ and $\mathcal{V}$ in this thesis correspond to $X$ and $Y$ in~\cite{DAVID1998487}.}
\begin{align}
    \textrm{Var}(\mathcal{V}_{\mathcal{I}(r)})
    = \; & \sigma^2_\mathcal{V} \left(\rho_{\mathcal{V}\mathcal{H}}^2 \frac{\textrm{Var}(\mathcal{H}_{(r)})}{\sigma^2_\mathcal{V} + \sigma^2_1} + 1 - \rho_{\mathcal{V}\mathcal{H}}^2 \right)
    = 
    \frac{\sigma^2_1 \sigma^2_\mathcal{V}}{\sigma^2_\mathcal{V} + \sigma^2_1} + 
    \frac{\sigma^4_\mathcal{V}}{\left(\sigma^2_\mathcal{V} + \sigma^2_1\right)^2}  \,\textrm{Var}(\mathcal{H}_{(r)}) \,, \label{eq:vem_var_xir}
    \\[0.5em]
    \textrm{Var}(\mathcal{V}_{\mathcal{J}(s)})
    = \; & \sigma^2_\mathcal{V} \left(\rho_{\mathcal{V}\mathcal{L}}^2 \frac{\textrm{Var}(\mathcal{L}_{(s)})}{\sigma^2_\mathcal{V} + \sigma^2_2} + 1 - \rho_{\mathcal{V}\mathcal{L}}^2 \right)
    =   
    \frac{\sigma^2_2 \sigma^2_\mathcal{V}}{\sigma^2_\mathcal{V} + \sigma^2_2} + 
    \frac{\sigma^4_\mathcal{V}}{\left(\sigma^2_\mathcal{V} + \sigma^2_2\right)^2}  \,\textrm{Var}(\mathcal{L}_{(s)}) \,,
    \label{eq:vem_var_xjs}
\end{align}
where $\rho_{\mathcal{V}\mathcal{H}} = {\sigma_\mathcal{V}}/{\sqrt{\sigma^2_\mathcal{V} + \sigma^2_1}}$ denotes the correlation between $\mathcal{V}_i$ and $\mathcal{H}_i$, and $\rho_{\mathcal{V}\mathcal{L}} = {\sigma_\mathcal{V}}/{\sqrt{\sigma^2_\mathcal{V} + \sigma^2_2}}$ denotes the correlation between $\mathcal{V}_i$ and $\mathcal{L}_i$.

In the multivariate normal case, the same result can also be obtained from first principles. 
We first recall Equation~\eqref{eq:vem_true_val_posterior_mean}, which describes the mean of $\mathcal{V}_i$ once we observe $\mathcal{E}_i$ (now $\mathcal{H}_i$ or $\mathcal{L}_i$), and note its variance counterpart is
\begin{align}
    \sigma^2_{\mathcal{V}_i \,|\, \mathcal{H}_i}
    = \frac{\sigma^2_1 \sigma^2_\mathcal{V}}{\sigma^2_\mathcal{V} + \sigma^2_1} \,, \quad
    \sigma^2_{\mathcal{V}_i \,|\, \mathcal{L}_i}
    = \frac{\sigma^2_2 \sigma^2_\mathcal{V}}{\sigma^2_\mathcal{V} + \sigma^2_2} \,.
    \label{eq:vem_true_val_posterior_variance}
\end{align}
Using the results in Equations~\eqref{eq:vem_true_val_posterior_mean} and~\eqref{eq:vem_true_val_posterior_variance}, we apply the law of total variance to obtain\footnote{Similar to what has been mentioned in Footnotes~\ref{footnote:vem_noisy_rank_index_equiv} and~\ref{footnote:vem_loie_validity}, $\mathcal{H}_{(r)} \equiv \mathcal{H}_{\mathcal{I}(r)}$ and $\mathcal{L}_{(s)} \equiv \mathcal{L}_{\mathcal{J}(s)}$, since we rank the business propositions by their estimated values. Moreover, while $\mathcal{V}_{\mathcal{I}(r)}$ and $\mathcal{V}_{\mathcal{J}(s)}$ are no longer normally distributed with the ranking information, using the law of total variance to obtain the variance for all~$r$ and~$s$ remains valid.} 
\begin{align}
    \textrm{Var}(\mathcal{V}_{\mathcal{I}(r)}) 
    = \,& 
    \mathbb{E}\big(\textrm{Var}(\mathcal{V}_{\mathcal{I}(r)} \,|\, \mathcal{H}_{(r)})\big) +
    \textrm{Var}\big(\mathbb{E}(\mathcal{V}_{\mathcal{I}(r)} \,|\, \mathcal{H}_{(r)})\big) \nonumber\\
    \approx\, &  
    \frac{\sigma^2_1 \sigma^2_\mathcal{V}}{\sigma^2_\mathcal{V} + \sigma^2_1} + 
    \frac{\sigma^4_\mathcal{V}}{\sigma^2_\mathcal{V} + \sigma^2_1} \frac{r(N-r+1)}{(N+1)^2 (N+2)}
    \frac{1}{\big(\phi\big(\Phi^{-1}(\frac{r}{N+1})\big)\big)^2} \,,
    \label{eq:vem_var_vir_normal} \\[0.5em]
    \textrm{Var}(\mathcal{V}_{\mathcal{J}(s)}) 
    = \,& 
    \mathbb{E}\big(\textrm{Var}(\mathcal{V}_{\mathcal{J}(s)} \,|\, \mathcal{L}_{(s)})\big) +
    \textrm{Var}\big(\mathbb{E}(\mathcal{V}_{\mathcal{J}(s)} \,|\, \mathcal{L}_{(s)})\big) \nonumber\\
    \approx\, &  
    \frac{\sigma^2_2 \sigma^2_\mathcal{V}}{\sigma^2_\mathcal{V} + \sigma^2_2} + 
    \frac{\sigma^4_\mathcal{V}}{\sigma^2_\mathcal{V} + \sigma^2_2} \frac{s(N-s+1)}{(N+1)^2 (N+2)}
    \frac{1}{\big(\phi\big(\Phi^{-1}(\frac{s}{N+1})\big)\big)^2} \,.
    \label{eq:vem_var_vjs_normal}
\end{align}

\subsection{\texorpdfstring{$\textrm{Var}(\mathcal{W}_1)$/$\textrm{Var}(\mathcal{W}_2)$}{Var(W\_1)/Var(W\_2)}}

To derive the variance of $\mathcal{W}_1$, we require the covariance between $\mathcal{H}_{(r)}$ and $\mathcal{H}_{(s)}$, as well as that between $\mathcal{V}_{\mathcal{I}(r)}$ and $\mathcal{V}_{\mathcal{I}(s)}$ $\forall r, s$. The same goes for $\mathcal{W}_2$, where we require the covariance between~$\mathcal{L}_{(r)}$ and $\mathcal{L}_{(s)}$, as well as that between $\mathcal{V}_{\mathcal{J}(r)}$ and $\mathcal{V}_{\mathcal{J}(s)}$ $\forall r, s$. This is necessary as the terms of~$\mathcal{W}_1$ (see Equation~\eqref{eq:vem_mean_true_value}), which result from removing noise from successive order statistics, are highly correlated.

Equation 4.6.5 of~\cite{david2003order} provided a formula to estimate the covariance between~$\mathcal{H}_{(r)}$ and~$\mathcal{H}_{(s)}$ and between~$\mathcal{L}_{(r)}$ and~$\mathcal{L}_{(s)}$ for any ${r < s \leq N}$:\footnote{For $r > s$, simply swap $r$ and $s$ as covariance functions are symmetrical.}
\begin{align}
    \textrm{Cov}(\mathcal{H}_{(r)}, \mathcal{H}_{(s)})
    & \, \approx 
    \frac{r(N-s+1)}{(N+1)^2(N+2)}
    \frac{1}{f_\mathcal{H}\big(F_\mathcal{H}^{-1}(\frac{r}{N+1})\big)\,f_\mathcal{H}\big(F_\mathcal{H}^{-1}(\frac{s}{N+1})\big)} \,, \label{eq:vem_cov_yr_ys}
    \\[0.2em]
    \textrm{Cov}(\mathcal{L}_{(r)}, \mathcal{L}_{(s)}) & \, \approx 
    \frac{r(N-s+1)}{(N+1)^2(N+2)}
    \frac{1}{f_\mathcal{L}\big(F_\mathcal{L}^{-1}(\frac{r}{N+1})\big)\,f_\mathcal{L}\big(F_\mathcal{L}^{-1}(\frac{s}{N+1})\big)} \,. 
    \label{eq:vem_cov_zr_zs}
\end{align}

To obtain the covariance between $\mathcal{V}_{\mathcal{I}(r)}$ and $\mathcal{V}_{\mathcal{I}(s)}$ for any $r, s \leq N$, we again refer to~\cite{DAVID1998487} (Equation 2.3d):
\begin{align}
    \textrm{Cov}(\mathcal{V}_{\mathcal{I}(r)}, \mathcal{V}_{\mathcal{I}(s)})
    & = \rho^2_{\mathcal{V}\mathcal{H}}\sigma^2_\mathcal{V}\,\frac{\textrm{Cov}(\mathcal{H}_{(r)}, \mathcal{H}_{(s)})}{\sigma^2_\mathcal{V} + \sigma^2_1}\, = \frac{\sigma^4_\mathcal{V}}{\left(\sigma^2_\mathcal{V} + \sigma^2_1\right)^2} \,\textrm{Cov}(\mathcal{H}_{(r)}, \mathcal{H}_{(s)}) \,, \label{eq:vem_cov_xir_xis_final}
    \\[0.2em]
    \textrm{Cov}(\mathcal{V}_{\mathcal{J}(r)}, \mathcal{V}_{\mathcal{J}(s)})
    & = \rho^2_{\mathcal{V}\mathcal{L}}\sigma^2_\mathcal{V}\,\frac{\textrm{Cov}(\mathcal{L}_{(r)}, \mathcal{L}_{(s)})}{\sigma^2_\mathcal{V} + \sigma^2_2}\, = \frac{\sigma^4_\mathcal{V}}{\left(\sigma^2_\mathcal{V} + \sigma^2_2\right)^2} \,\textrm{Cov}(\mathcal{L}_{(r)}, \mathcal{L}_{(s)}) \,.
    \label{eq:vem_cov_xjr_xjs_final}
\end{align}

Equation~\eqref{eq:vem_cov_xir_xis_final} and~\eqref{eq:vem_cov_xjr_xjs_final} affirms the claim that $\mathcal{V}_{\mathcal{I}(r)}$ are positively correlated $\forall r$. Unlike $\mathcal{V}_i$, which are independent by definition, they become correlated under the presence of ranking information.

We can now state the variance of $\mathcal{W}_1$ and $\mathcal{W}_2$. Applying the variance function to Equation~\eqref{eq:vem_V1_V2_def}, we get
\begin{align}
    \textrm{Var}(\mathcal{W}_1)
    = \, & \frac{1}{M^2}\left(\sum_{r=N-M+1}^{N} \textrm{Var}\left(\mathcal{V}_{\mathcal{I}(r)}\right) \, + 
    \sum_{r=N-M+1}^{N} \, \sum_{s=r+1}^{N} 2 \cdot \textrm{Cov}\left(\mathcal{V}_{\mathcal{I}(r)}, \mathcal{V}_{\mathcal{I}(s)}\right)\right) ,
    \label{eq:vem_var_V1}
    \\[0.2em]
    \textrm{Var}(\mathcal{W}_2)
    = \, & \frac{1}{M^2}\left(\sum_{s=N-M+1}^{N} \textrm{Var}\left(\mathcal{V}_{\mathcal{J}(s)}\right) \, + \sum_{r=N-M+1}^{N} \, \sum_{s=r+1}^{N} 2 \cdot \textrm{Cov}\left(\mathcal{V}_{\mathcal{J}(r)}, \mathcal{V}_{\mathcal{J}(s)}\right)\right) ,
    \label{eq:vem_var_V2}
\end{align}
where the constituent variances and covariances are derived in Equations~\eqref{eq:vem_var_xir}/\eqref{eq:vem_var_xjs} and \eqref{eq:vem_cov_xir_xis_final}/\eqref{eq:vem_cov_xjr_xjs_final}, respectively.

\subsection{\texorpdfstring{$\textrm{Var}(\mathcal{D})$}{Var(D)}}
\label{sec:vem_var_D}

Finally, we derive the variance of $D$. In addition to the variance of $\mathcal{W}_1$ and $\mathcal{W}_2$ derived, respectively, in Equations~\eqref{eq:vem_var_V1} and~\eqref{eq:vem_var_V2}, we require the covariance between these two terms. This, in turn, requires the covariance between $\mathcal{H}_{(r)}$ and $\mathcal{L}_{(s)}$ and that between $X_{\mathcal{I}(r)}$ and $X_{\mathcal{J}(s)}$.

The covariance between $\mathcal{H}_{(r)}$ and $\mathcal{L}_{(s)}$ can be derived using results in~\cite{david2003order}:
\begin{align}
    \textrm{Cov}(\mathcal{H}_{(r)},  \mathcal{L}_{(s)}) = \, & \rho_{\mathcal{V}\mathcal{H}} \rho_{\mathcal{V}\mathcal{L}} \,\textrm{Cov}(\mathcal{V}_{(r)},  \mathcal{V}_{(s)}) \nonumber\\
    = \, & \frac{\sigma^2_\mathcal{V}}{\sqrt{\sigma^2_\mathcal{V} + \sigma^2_1}\sqrt{\sigma^2_\mathcal{V} + \sigma^2_2}} 
    \frac{r(N-s+1)}{(N+1)^2(N+2)}
    \frac{1}{f_\mathcal{V}(F^{-1}_\mathcal{V}(\frac{r}{N+1})) \, f_\mathcal{V}(F^{-1}_\mathcal{V}(\frac{s}{N+1}))} \,, 
    \label{eq:vem_cov_Yr_Zs}
\end{align}
where $f_\mathcal{V}(\cdot)$ and $F^{-1}_\mathcal{V}(\cdot)$ are the probability density function and quantile function for $\mathcal{V}_i$, respectively.
 
Deriving the covariance between $\mathcal{V}_{\mathcal{I}(r)}$ and $\mathcal{V}_{\mathcal{J}(s)}$ is perhaps the most challenging problem within the work, as they take two forms depending on the indices:
\begin{align}
    \textrm{Cov}(\mathcal{V}_{\mathcal{I}(r)}, \mathcal{V}_{\mathcal{J}(s)})
    = \begin{cases}
    \begin{array}{ll}
       \textrm{Var}(\mathcal{V}_{\mathcal{I}(r)}) = \textrm{Var}(\mathcal{V}_{\mathcal{J}(s)}) & \textrm{if } \mathcal{I}(r) = \mathcal{J}(s)  \\
       \frac{\sigma^2_\mathcal{V}}{\sigma^2_\mathcal{V} + \sigma^2_1} \frac{\sigma^2_\mathcal{V}}{\sigma^2_\mathcal{V} + \sigma^2_2} \,\textrm{Cov}(\mathcal{H}_{(r)}, \mathcal{L}_{(s)}) & \textrm{if } \mathcal{I}(r) \neq \mathcal{J}(s)
    \end{array},
    \end{cases}
    \label{eq:vem_cov_XIr_XJs}
\end{align}
where the second case is a standard Bayesian inference result.

\begin{figure}
\begin{center}
    \includegraphics[width=0.8\textwidth, trim = 0 0 0 0, clip]{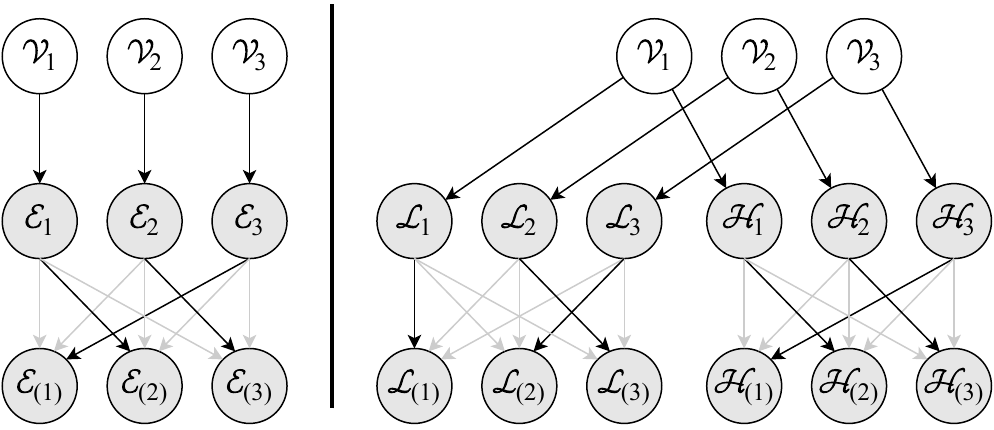}
\end{center}
\caption[Relationship between different quantities in a three-item generative model under the ranking under lower uncertainty problem.]{Relationship between different quantities in a three-item generative model under the ranking under lower uncertainty problem. $\mathcal{V}_i$, $\mathcal{H}_i$/$\mathcal{L}_i$, and $\mathcal{H}_{(r)}$/$\mathcal{L}_{(s)}$ represent the true value, the unranked noisy estimates, and the ranked noisy estimates of the items, respectively, for ${i, r, s \in \{1, 2, 3\}}$. (L)~Under one estimation noise level, $\exists$ a bijection between $\mathcal{V}_i$ and $\mathcal{H}_{(r)}$ within a set of generated samples. (R)~With two noise levels, $\mathcal{H}_{(r)}$ and~$\mathcal{L}_{(s)}$ may be generated by the same~$\mathcal{V}_i$ for some~$r$ and~$s$.}
\label{fig:vem_rulu_ranking_three_items}
\vspace*{\baselineskip}
\end{figure}

The problem arises as the~$r^\textrm{th}$ ranked~$\mathcal{H}$ and the~$s^\textrm{th}$ ranked~$\mathcal{L}$ can be generated by the same~$\mathcal{V}_i$ for some $i$. It will not arise if we have only~$\mathcal{H}$ or~$\mathcal{L}$ (see Figure~\ref{fig:vem_rulu_ranking_three_items} for an example). In this case, when we consider the covariance of the concomitants $\mathcal{V}_{\mathcal{I}(r)}$/$\mathcal{V}_{\mathcal{J}(s)}$, we have to take into account both the existing variance of $\mathcal{V}_i$ and the ranking information provided by~$\mathcal{H}_{(r)}$ and~$\mathcal{L}_{(s)}$. If the order statistics are generated by different~$\mathcal{V}$, we only need to consider the latter as we assumed~$\mathcal{V}_i$ to be independent and thus uncorrelated.

As we are interested in the overall behaviour, we only need to derive the two cases on the RHS of Equation~\eqref{eq:vem_cov_XIr_XJs} and weigh them using the probability that $\mathcal{I}(r) = \mathcal{J}(s)$ without worrying about which case applies to each $(r, s)$ pair.
The first case (when $\mathcal{I}(r) = \mathcal{J}(s)$) can be evaluated using the law of total variance with multiple conditioning random variables: 
\begin{align}
    & \textrm{Var}(\mathcal{V}_{\mathcal{I}(r)}) = \textrm{Var}(\mathcal{V}_{\mathcal{J}(s)}) \nonumber\\
    = \,&
    \mathbb{E}(\textrm{Var}(\mathcal{V}_{\mathcal{I}(r)} | \mathcal{H}_{(r)}, \mathcal{L}_{(s)})) \,+
    \mathbb{E}(\textrm{Var}(\mathbb{E}(\mathcal{V}_{\mathcal{I}(r)} | \mathcal{H}_{(r)}, \mathcal{L}_{(s)})| \mathcal{H}_{(r)})) +
    \textrm{Var}(\mathbb{E}(\mathcal{V}_{\mathcal{I}(r)} | \mathcal{H}_{(r)}))
     \nonumber\\ 
    = \, &
    \frac{\sigma^2_\mathcal{V}\sigma^2_1\sigma^2_2}{\sigma^2_\mathcal{V}\sigma^2_1 + \sigma^2_\mathcal{V}\sigma^2_2 + \sigma^2_1\sigma^2_2} \, + 
    \left(\frac{\sigma^2_\mathcal{V}\sigma^2_1}{\sigma^2_\mathcal{V}\sigma^2_1 + \sigma^2_\mathcal{V}\sigma^2_2 + \sigma^2_1\sigma^2_2}\right)^2 \textrm{Var}(\mathcal{L}_{(s)}) \,+ 
    \left(\frac{\sigma^2_\mathcal{V}}{\sigma^2_\mathcal{V} + \sigma^2_1}\right)^2 \textrm{Var}(\mathcal{H}_{(r)}) .
    \label{eq:vem_var_XIr_XJs_general}
\end{align}
The second case can be derived by substituting Equation~\eqref{eq:vem_cov_Yr_Zs} into Equation~\eqref{eq:vem_cov_XIr_XJs}.


For the weighting probability $\mathbb{P}(\mathcal{I}(r) = \mathcal{J}(s))$, we see its derivation as an interesting and potentially important problem in its own right, yet to the best of our knowledge, no proper treatment was given to the problem. In this work, we approximate the probability using beta-binomial distributions, with parameters derived from quantities calculated above.
Without distracting readers from the main question of quantifying the value and risk of DEM capabilities, we relegate the detailed discussion on approximating the quantity to Appendix~\ref{sec:miscmath_prob_Ir_Js}.


With the three components for the covariance between $\mathcal{V}_{\mathcal{I}(r)}$ and $\mathcal{V}_{\mathcal{J}(s)}$ in place, we can finally derive $\textrm{Cov}(\mathcal{W}_1, \mathcal{W}_2)$ and $\textrm{Var}(\mathcal{D})$ by applying the covariance and variance functions to the definitions of, respectively, $\mathcal{W}_1$/$\mathcal{W}_2$ and $\mathcal{D}$ (see~\eqref{eq:vem_V1_V2_def}) to obtain
\begin{align}
    \textrm{Cov}(\mathcal{W}_1, \mathcal{W}_2) = \, & \frac{1}{M^2}\sum_{r=N-M+1}^{N} \sum_{s=N-M+1}^{N} \textrm{Cov}(\mathcal{V}_{\mathcal{I}(r)}, \mathcal{V}_{\mathcal{J}(s)}), \label{eq:vem_cov_v1_v2} \\
    \textrm{Var}(\mathcal{W}) = \, & \textrm{Var}(\mathcal{W}_2 - \mathcal{W}_1) 
    = \textrm{Var}(\mathcal{W}_1)+ \textrm{Var}(\mathcal{W}_2) - 2 \,  \textrm{Cov}(\mathcal{W}_1 , \mathcal{W}_2) , \label{eq:vem_var_D}
\end{align}
where the first two terms on the RHS of Equation~\eqref{eq:vem_var_D} are that derived in Equations~\eqref{eq:vem_var_V1} and \eqref{eq:vem_var_V2}.

We conclude this section by observing that $M$ and $N$ influence $\textrm{Var}(\mathcal{D})$ considerably. In particular, $\textrm{Var}(\mathcal{D})$ is generally large when $M$ and $N$ are small with other parameters fixed. This is crucial as even in cases where $\mathbb{E}(\mathcal{D})$ is positive, the limited capacity of an organisation to introduce new business propositions may mean that the Sharpe ratio defined in Section~\ref{sec:vem_mathematical_formulation} (see Expression~\eqref{eq:vem_sharpe_ratio}) may not be high enough to justify investment in a DEM capability.

The exact threshold where an organisation should consider acquiring such capabilities depends on multiple factors, including their size (which affects $M$), the size of their backlog~($N$), the nature of their work ($\mu_\mathcal{V}$ and $\sigma^2_\mathcal{V}$), and how good they were at estimation ($\sigma^2_1$). Thus, we refrain from providing a one-size-fits-all recommendation but give examples in Section~\ref{sec:vem_case_study}.

\section{Empirical Verification}
\label{sec:vem_experiments}

Having performed theoretical calculations for the expectation and variance of the value DEM capabilities deliver through enhanced prioritisation, we verify those calculations using simulation results.\footnote{All code used in the experiments, case studies and extensions are available on GitHub: \url{https://github.com/liuchbryan/ranking\_under\_lower\_uncertainty}.}

We verify the result derived in Sections~\ref{sec:vem_E_D} and~\ref{sec:vem_var} empirically, in particular under the normal assumptions. We run multiple \emph{statistical tests} for each quantity of interest -- the mean and variance of $\mathcal{W}_1$/$\mathcal{W}_2$ and~$\mathcal{D}$, as well as the covariance between different pairs of order statistics and their concomitants. 

In each \emph{statistical test}, we randomly select and fix the value of the parameters, i.e., $N$, $M$, $\mu_\mathcal{V}$, $\mu_\epsilon$, $\sigma^2_\mathcal{V}$, $\sigma^2_1$, $\sigma^2_2$, $r$, and~$s$ (the latter two for the covariance of the order statistics only). We select~$N$ uniformly on a log scale, with the log scale favoured due to its better resemblance to the organisation size distribution in real life and bounds chosen to represent the usual number of items/propositions an organisation will consider during prioritisation. We also select $\mu_{\mathcal{V}}$, $\mu_{\epsilon}$, $\sigma^2_{\mathcal{V}}$, and $\sigma^2_1$ randomly from a uniform distribution (on a linear scale), with bounds chosen that are reflective of the usual business metrics an organisation uses to prioritise their items/propositions and their unit of measurement (e.g., percentage points change, revenue impact in hundred thousand pounds).\footnote{Denoting $U(a, b)$ as a uniform distribution with bounds $a$ and $b$, we draw $N \sim 10^{U(1, 3.5)}$, $\mu_{\mathcal{V}}, \mu_{\epsilon} \sim U(-10, 10)$ and $\sigma^2_{\mathcal{V}}, \sigma^2_1 \sim U(0.3, 10)^2$.}

Each parameter is drawn independently from other parameters where possible, with the only exceptions being when the range of a parameter is restricted (i.e., ${M < N}$, ${\sigma^2_2 < \sigma^2_1}$, and~${N-M+1} \leq r, s, \leq N$ for selected items/propositions). To draw $M$, we draw a uniformly random multiplier between 0.01 and 0.8, then apply the multiplier to $N$.\footnote{In other words, we draw $M \sim N \times U(0.01, 0.8)$, subjected to rounding with the floor function and a minimum value of one. Multipliers above 0.8 are ignored, as it is unrealistic for organisations to have such a high relative capacity.} We apply a similar process to draw~$\sigma^2_2$, though we use a different multiplier range and apply the multiplier on~$\sigma_1$ (the standard deviation) before squaring.\footnote{We draw $\sigma^2_2 \sim (\sigma_1 \times U(0.1, 0.99))^2$, subjected to a minimum value of $0.2^2$. Multipliers below 0.1 and above 0.99 are ignored, as we consider such an estimation noise reduction unrealistic and of negligible value, respectively.} Finally, we draw~$r$ and~$s$ independently and uniformly amongst integers between~$N-M+1$ and~$N$ (both inclusive).

Upon selecting and fixing the value of the parameters, we compare the theoretical value of the quantity of interest to the centred 95\% confidence interval (CI) generated from multiple \emph{empirical samples}. If the derivations above are exact, the 95\% CI should contain the theoretical value in around 95\% of the statistical tests. The histogram of the percentile rank of the theoretical quantity among the empirical samples should also follow a uniform distribution~\cite{talts2018validating}.

\begin{table}
    \caption[The number of statistical tests with the centred 95\% confidence interval containing the derived theoretical value for each quantity of interest.]{The number of statistical tests with the centred 95\% confidence interval containing the derived theoretical value for each quantity of interest. If a theoretical value derived in Sections~\ref{sec:vem_E_D} or~\ref{sec:vem_var} is exact, its associated 95\% CI should contain the theoretical value in 95\% of the statistical tests.}
    \label{tab:experiments_num_within_CI}
    \onehalfspacing
    \centering
    \begin{tabular}{c|c|c|c}
         Quantity & \# within CI & \# statistical tests & \% within CI \\\hline
         $\mathbb{E}(\mathcal{W})$ & 3,991 & 4,428 & 90.13\% \\
         $\mathbb{E}(\mathcal{D})$ & 4,162 & 4,428 & 93.99\% \\
         $\textrm{Var}(\mathcal{W})$ & 3,390 & 4,555 & 74.42\%\\
         $\textrm{Cov}(\mathcal{H}_{(r)}, \mathcal{L}_{(s)})$ & 4,663 & 4,940 & 94.39\% \\
         $\textrm{Cov}(\mathcal{V}_{\mathcal{I}(r)}, \mathcal{V}_{\mathcal{J}(s)})$ & 4,730 & 4,940 & 95.75\% \\
    \end{tabular}
    \vspace*{\baselineskip}
\end{table}

Each \emph{empirical sample} is generated using one of the following two methods depending on the quantity we are evaluating:

a) Bootstrap resampling -- We use the method to generate a sample for the mean/variance of $\mathcal{W}_1$/$\mathcal{W}_2$ and $\mathcal{D}$. We first generate the initial samples for $\mathcal{W}_1$, $\mathcal{W}_2$, and $\mathcal{D}$ by performing 10,000 \emph{simulation runs} (see below). We then resample the initial samples and calculate the mean/variance of the resample to obtain an empirical mean/variance sample. Finally, we repeat the resampling 2,000 times to obtain a representative empirical distribution for the mean/variance. 

b) Sampling for order statistics --
The bootstrapping approach is unlikely to work on the covariance between the order statistics (e.g., $\mathcal{H}_{(r)}$ and $\mathcal{L}_{(s)}$) and their concomitants (e.g., $\mathcal{V}_{\mathcal{I}(r)}$ and $\mathcal{V}_{\mathcal{J}(s)}$), as the ranking information may not be preserved during resampling. Hence, for these quantities, we opt for a more na\"ive sampling approach. We generate 200 samples for $\mathcal{H}_{(r)}$, $\mathcal{L}_{(s)}$, $\mathcal{V}_{\mathcal{I}(r)}$, and $\mathcal{V}_{\mathcal{J}(s)}$ via the same number of \emph{simulation runs}, and calculate the covariance between these quantities to obtain an empirical sample. The process is repeated 1,000 times to yield a representative empirical distribution for the covariance.

\begin{figure}
\begin{center}
\begin{subfigure}[t]{.4\textwidth}
  \includegraphics[width=.97\textwidth, trim=3mm 0 2.5mm 0, clip]{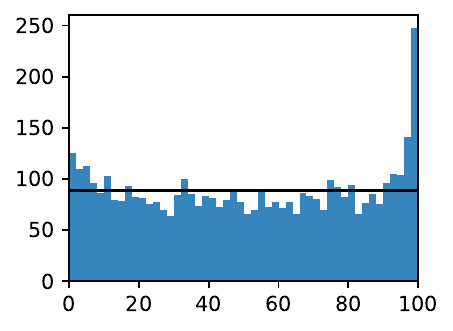}
  \caption{$\mathbb{E}(\mathcal{W})$}
  \label{fig:vem_E_V_validation}
\end{subfigure}
\begin{subfigure}[t]{.4\textwidth}
  \includegraphics[width=.97\textwidth, trim=3mm 0 2.5mm 0, clip]{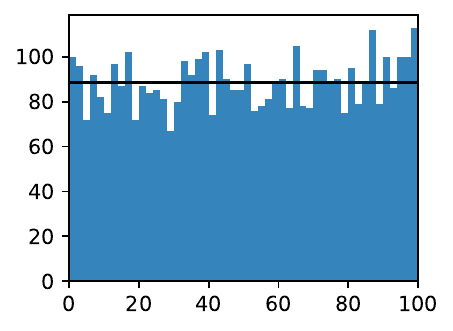}
  \caption{$\mathbb{E}(\mathcal{D})$}
  \label{fig:vem_E_D_validation}
\end{subfigure}
\end{center}
\vspace*{-0.5em}
\caption[Histogram of the theoretical quantity's percentile rank compared to the empirical samples across multiple statistical tests.]{Histogram of the theoretical quantity's percentile rank compared to the empirical samples across multiple statistical tests. If the theoretical value derived in Section~\ref{sec:vem_E_D} is exact, the histogram should show a uniform distribution with probability mass around the black line~\cite{talts2018validating}.}
\label{fig:vem_validation}
\vspace*{\baselineskip}
\end{figure}

Finally, in each \emph{simulation run}, we obtain one sample for each of $\mathcal{H}_{(r)}$/$\mathcal{L}_{(s)}$, $\mathcal{V}_{\mathcal{I}(r)}$/$\mathcal{V}_{\mathcal{J}(s)}$, $\mathcal{W}_1/\mathcal{W}_2$, and $D$ w.r.t. the parameters via the following four-step process:\footnote{Identifiers in \texttt{monospace} refer to variables used in software packages, which correspond to the random variables used in Sections \ref{sec:vem_mathematical_formulation}--\ref{sec:vem_var}.}
\begin{enumerate}
    \item Take $N$ samples from $\mathcal{N}(\mu_\mathcal{V}, \sigma^2_\mathcal{V})$ as \texttt{V\textsubscript{i}}; 
    \item Take $N$ samples from $\mathcal{N}(\mu_\epsilon, \sigma^2_1)$, and sum the \texttt{i}\textsuperscript{th}-indexed sample with \texttt{V\textsubscript{i}} $\forall$\texttt{i} to obtain \texttt{H\textsubscript{i}}:
    \begin{enumerate}
        \item Rank all \texttt{H\textsubscript{i}}, take the value of the \texttt{r}\textsuperscript{th}-ranked \texttt{H\textsubscript{i}} as \texttt{H\textsubscript{(r)}} and its index as~\texttt{I(r)},
        \item Take the value of the \texttt{I(r)}\textsuperscript{th}-indexed sample of \texttt{V\textsubscript{i}} as \texttt{V\textsubscript{I(r)}},
        \item Obtain the indices of the $M$ largest samples of the ranked \texttt{H\textsubscript{i}},
        \item Calculate the mean of \texttt{V\textsubscript{i}}, where \texttt{i} is in the set of indices in Step 2c, as \texttt{W\textsubscript{1}};
    \end{enumerate}
    \item Take $N$ samples from $\mathcal{N}(\mu_\epsilon, \sigma^2_2)$, and sum the \texttt{i}\textsuperscript{th}-indexed sample with \texttt{V\textsubscript{i}} $\forall$\texttt{i} to obtain \texttt{L\textsubscript{i}}:
    \begin{enumerate}
        \item Rank all \texttt{L\textsubscript{i}}, take the value of the \texttt{s}\textsuperscript{th}-ranked \texttt{L\textsubscript{i}} as \texttt{L\textsubscript{(s)}} and its index as~\texttt{J(s)},
        \item Take the value of the \texttt{J(s)}\textsuperscript{th}-indexed sample of \texttt{V\textsubscript{i}} as \texttt{V\textsubscript{J(s)}},
        \item Obtain the indices of the $M$ largest samples of the ranked \texttt{L\textsubscript{i}},
        \item Calculate the mean of \texttt{V\textsubscript{i}}, where \texttt{i} is in the set of indices in Step 3c, as \texttt{W\textsubscript{2}};
    \end{enumerate}
    \item Take the difference between \texttt{W\textsubscript{2}} obtained in Step~3d and \texttt{W\textsubscript{1}} from Step~2d to get~\texttt{D}.
\end{enumerate}

We show the results in Table~\ref{tab:experiments_num_within_CI} and Figure~\ref{fig:vem_validation}. We observed that the 95\% CI of the quantities $\mathbb{E}(\mathcal{W})$, $\mathbb{E}(\mathcal{D})$, $\textrm{Var}(\mathcal{W})$, $\textrm{Cov}(\mathcal{H}_{(r)}, \mathcal{L}_{(s)})$, and $\textrm{Cov}(\mathcal{V}_{\mathcal{I}(r)}, \mathcal{V}_{\mathcal{J}(s)})$ contain the derived theoretical value for roughly 90\%, 94\%, 74\%, 94\%, and 96\% of the times respectively. While these numbers are expected for the expectations and covariances, considering they are approximations, they are on the low side for the variances. Upon further investigation, we realised that most out-of-CI cases have a theoretical variance below the CI, suggesting a slight underestimate in our variance derivation. We believe that this is due to the omission of higher-order terms when using the formulas in~\cite{david54statitical}, leading to a small (less than one percent relative) bias, which is more apparent for smaller $N$ and $M$. Otherwise, we are satisfied with the soundness of the derived quantities.

\section{Case Study}
\label{sec:vem_case_study}

\emph{``What do e-commerce/marketing companies gain by acquiring digital experimentation and measurement capabilities?''}

It is difficult to verify any model that seeks to ascertain the value of DEM capabilities with real-life data. This is because of the inability to observe the true value of a digital product / business proposition / service without any measurement error and the lack of published measurements from organisations. The closest proxies are meta-analyses, including those compiled by Browne and Johnson~\cite{browne17whatworks} and Johnson et al.~\cite{johnson2017online}, which contain statistics on the measured uplift (in relative \%) over a large number of e-commerce and marketing experiments for many organisations.

The information presented by the two groups of researchers, which we describe in more detail below, is sufficient for us to ask the following question: If the same organisation conducted all the experiments presented by Browne and Johnson / Johnson et al., how much value did the DEM capabilities add due to improved prioritisation? We present results under normal assumptions in this section and will revisit the question when we discuss the model under $t$-distributed assumptions in Section~\ref{sec:vem_extension_t_distributed}.

\subsection{e-Commerce companies}
\label{sec:vem_case_study_ecommerce}

In~\cite{browne17whatworks}, Browne and Johnson reported running 6,700 A/B tests in e-commerce companies, with an overall effect in relative conversion rate (CVR) uplift centred at around zero, and the 5\% and 95\% percentiles at around $\pm [1.2\%, 1.3\%]$. We then divide the range by ${z_{0.95} \approx 1.645}$, the 95\textsuperscript{th} percentile of a standard normal, to estimate that the distribution reported has a standard deviation of around 0.75\%.
Based on this information, we take $\mu_\mathcal{V} = 0$ and $\sigma^2_\mathcal{V} = (0.6\%)^2$, considering that the reported distribution incorporated some estimation noise, and hence, the spread of the true values should be slightly lower.

Given an A/B test on CVR uplift run by the  most prolific organisations (e.g. one with five million visitors and a 5\% CVR) carries an estimation noise of around~$(0.28\%)^2$,\footref{footnote:vem_abtest_estimation_noise} we explore the scenarios where we reduce the noise level from $\sigma^2_1 = \{(1\%)^2, (0.8\%)^2, (0.6\%)^2\}$ to $\sigma^2_2 = \{(0.8\%)^2, (0.6\%)^2, (0.4\%)^2\}$, representing different levels of estimation abilities before and after the acquisition of DEM capabilities for companies of various sizes. We also calculate the value gained under different~$M$ (from 10 to 2,000) to simulate organisations with different yet realistic capacities while fixing $N = 6700$ (\# experiments). Finally, we set $\mu_\epsilon = 0$ as we do not assume any systematic bias during estimation in this case.

\begin{figure*}
\begin{center}
\begin{subfigure}[t]{.325\textwidth}
  \includegraphics[height=0.8\textwidth, trim=3mm 0 2.5mm 0, clip]{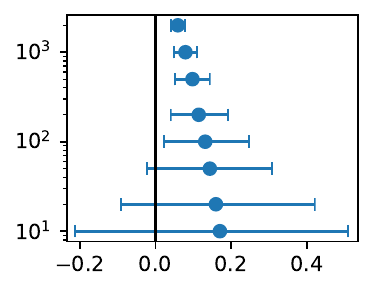}
  \vspace*{-6pt}
  \caption{$(1\%)^2, (0.8\%)^2$}
  \label{fig:vem_value_gained_browne_johnson_1_0-8}
\end{subfigure}
\begin{subfigure}[t]{.325\textwidth}
  \includegraphics[height=0.8\textwidth, trim=3mm 0 2.5mm 0, clip]{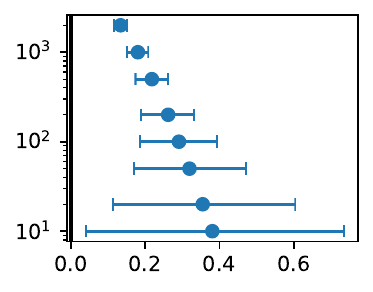}
  \vspace*{-6pt}
  \caption{$(1\%)^2, (0.6\%)^2$}
  \label{fig:vem_value_gained_browne_johnson_1_0-6}
\end{subfigure}
\begin{subfigure}[t]{.325\textwidth}
  \includegraphics[height=0.8\textwidth, trim=3mm 0 2.5mm 0, clip]{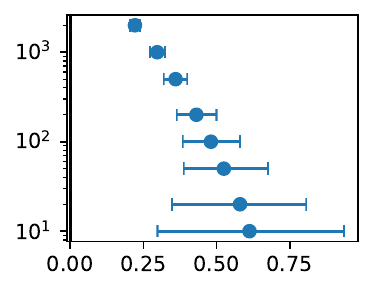}
  \vspace*{-6pt}
  \caption{$(1\%)^2, (0.4\%)^2$}
  \label{fig:vem_value_gained_browne_johnson_1_0-4}
\end{subfigure}
\\[0.5em]
\hspace*{.001\textwidth}
\begin{subfigure}[t]{.325\textwidth}
  \includegraphics[height=0.8\textwidth, trim=3mm 0 2.5mm 0, clip]{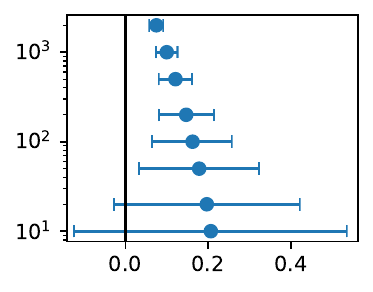}
  \vspace*{-6pt}
  \caption{$(0.8\%)^2, (0.6\%)^2$}
  \label{fig:vem_value_gained_browne_johnson_0-8_0-6}
\end{subfigure}
\begin{subfigure}[t]{.325\textwidth}
  \includegraphics[height=0.8\textwidth, trim=3mm 0 2.5mm 0, clip]{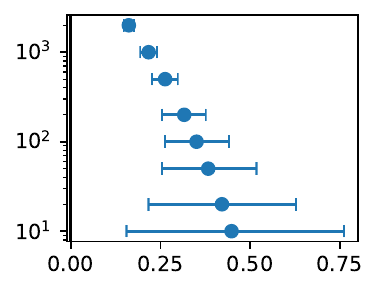}
  \vspace*{-6pt}
  \caption{$(0.8\%)^2, (0.4\%)^2$}
  \label{fig:vem_value_gained_browne_johnson_0-8_0-4}
\end{subfigure}
\begin{subfigure}[t]{.325\textwidth}
  \includegraphics[height=0.8\textwidth, trim=3mm 0 2.5mm 0, clip]{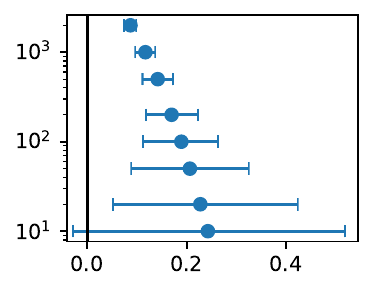}
  \vspace*{-5pt}
  \caption{$(0.6\%)^2, (0.4\%)^2$}
  \label{fig:vem_value_gained_browne_johnson_0-6_0-4}
\end{subfigure}
\end{center}
\caption[The value gained by having some digital experimentation and measurement capabilities~$\mathcal{D}$ under different capacity~$M$ in the case study on 6,700 e-commerce experiments reported by Browne and Johnson~\cite{browne17whatworks}.]{The value gained by having some digital experimentation and measurement (DEM) capabilities~$\mathcal{D}$ (x-axes, in per cent) under different capacity~$M$ ($y$-axes, in log scale) in the case study on 6,700 e-commerce experiments reported by Browne and Johnson~\cite{browne17whatworks} (see Section~\ref{sec:vem_case_study_ecommerce}). In each plot, the dot represents the mean, and the error bar represents the 5th--95th percentile of the empirical value distribution. Each sub-caption denotes the estimation noise before and after the acquisition of DEM capabilities (i.e., $\sigma^2_1, \sigma^2_2$). We fix $\mu_\mathcal{V}, \mu_\epsilon = 0$, $\sigma^2_\mathcal{V} = (0.6\%)^2$, and $N = 6700$.}
\label{fig:vem_value_gained_browne_johnson}
\end{figure*}

We show the results in Figure~\ref{fig:vem_value_gained_browne_johnson}, which shows the relationship between different $M$ and the value gained under different magnitudes of estimation noise reduction. One can observe that the expected gain in value decreases when~$M$ increases. This is expected: as one increases their capacity, they will run out of the most valuable work and have to settle for less valuable work that has many acceptable replacements with similar value, limiting the value DEM capabilities bring.

We can also see an inverse relationship between the size of~$M$ and the uncertainty of the value gained. As a result, while the expected value gain decreases with increasing $M$, the uncertainty drops quicker, such that at some $M$, we will see a statistically significant increase in value gained or an acceptable Sharpe ratio that justifies investment in DEM capabilities or both. The specific value that tips the balance is heavily dependent on individual circumstances.

\subsection{Marketing companies}
\label{sec:vem_case_study_marketing}

In the second case study, we repeat the process applied to e-commerce in Section~\ref{sec:vem_case_study_ecommerce} for the marketing experiments described in~\cite{johnson2017online}. In that work, Johnson et al. reported running~184 marketing experiments that measured relative CVR uplift, with a mean relative uplift of 19.9\% and standard error of 10.8\%. Thus, we take $\mu_\mathcal{V} = 19.9\%$ and $\sigma^2_\mathcal{V} = (10\%)^2$, which is slightly reduced to account for the estimation noise being included in the reported standard error.

Johnson et al. also noted that the average sample size in these experiments is over five million, which keeps the estimation noise low. However, the design of marketing experiments often comes with additional sources of noise compared to standard A/B tests~\cite{gordon2019comparison,liu2018designing}. Hence,  we assume the same estimation noise as in the e-commerce case study above (i.e., $\sigma^2_2 = \{(0.8\%)^2, (0.6\%)^2, (0.4\%)^2\}$). The larger variance in the uplifts allows us to assume a larger estimation error without DEM capabilities, and we explore the scenarios where $\sigma^2_1 = \{(5\%)^2, (2\%)^2, (1\%)^2, (0.8\%)^2, (0.6\%)^2\}$. We set~$N=184$ (\# experiments) and vary~$M$  between 10 and 100 for each combination of~$\sigma^2_1$ and~$\sigma^2_2$.

\begin{figure}
\begin{center}
\begin{subfigure}[t]{.32\textwidth}
  \includegraphics[height=0.8\textwidth, trim=3mm 0 2.5mm 0, clip]{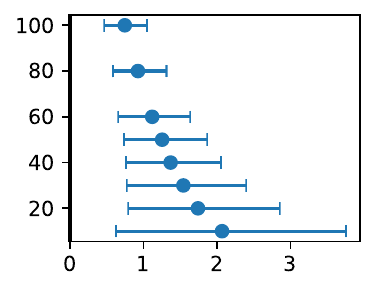}
  \vspace*{-8pt}
  \caption{$(5\%)^2, (0.8\%)^2$}
  \label{fig:vem_value_gained_johnson_etal_5_0-8}
\end{subfigure}
\begin{subfigure}[t]{.32\textwidth}
  \includegraphics[height=0.8\textwidth, trim=3mm 0 2.5mm 0, clip]{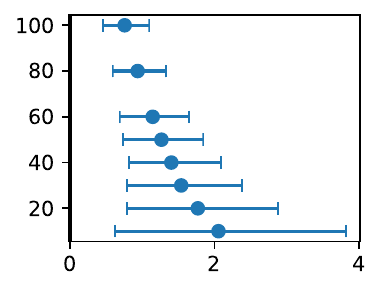}
  \vspace*{-8pt}
  \caption{$(5\%)^2, (0.6\%)^2$}
  \label{fig:vem_value_gained_johnson_etal_5_0-6}
\end{subfigure}
\begin{subfigure}[t]{.32\textwidth}
  \includegraphics[height=0.8\textwidth, trim=3mm 0 2.5mm 0, clip]{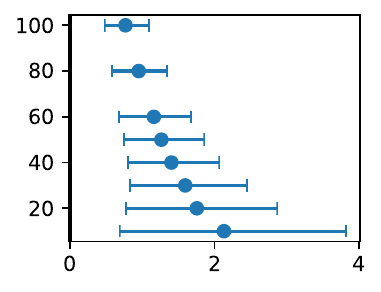}
  \vspace*{-8pt}
  \caption{$(5\%)^2, (0.4\%)^2$}
  \label{fig:vem_value_gained_johnson_etal_5_0-4}
\end{subfigure}
\\[0.5em]
\begin{subfigure}[t]{.32\textwidth}
  \includegraphics[height=0.8\textwidth, trim=3mm 0 2.5mm 0, clip]{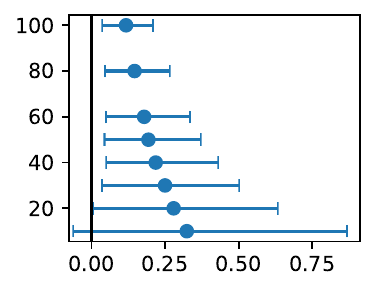}
  \vspace*{-8pt}
  \caption{$(2\%)^2, (0.8\%)^2$}
  \label{fig:vem_value_gained_johnson_etal_2_0-8}
\end{subfigure}
\begin{subfigure}[t]{.32\textwidth}
  \includegraphics[height=0.8\textwidth, trim=3mm 0 2.5mm 0, clip]{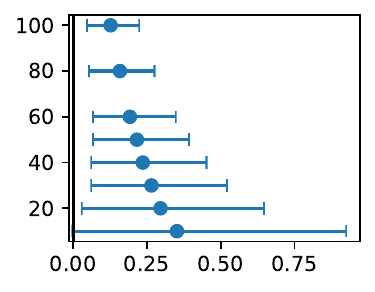}
  \vspace*{-8pt}
  \caption{$(2\%)^2, (0.6\%)^2$}
  \label{fig:vem_value_gained_johnson_etal_2_0-6}
\end{subfigure}
\begin{subfigure}[t]{.32\textwidth}
  \includegraphics[height=0.8\textwidth, trim=3mm 0 2.5mm 0, clip]{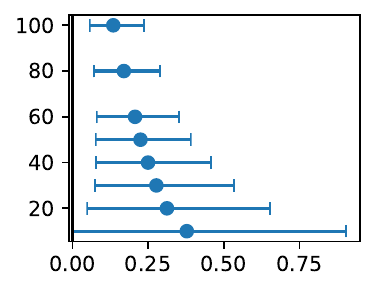}
  \vspace*{-8pt}
  \caption{$(2\%)^2, (0.4\%)^2$}
  \label{fig:vem_value_gained_johnson_etal_2_0-4}
\end{subfigure} 
\\[0.5em]
\begin{subfigure}[t]{.32\textwidth}
  \includegraphics[height=0.8\textwidth, trim=3mm 0 2.5mm 0, clip]{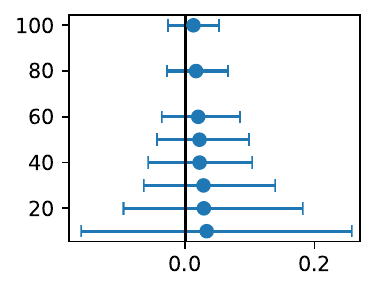}
  \vspace*{-8pt}
  \caption{$(1\%)^2, (0.8\%)^2$}
  \label{fig:vem_value_gained_johnson_etal_1_0-8}
\end{subfigure}
\begin{subfigure}[t]{.32\textwidth}
  \includegraphics[height=0.8\textwidth, trim=3mm 0 2.5mm 0, clip]{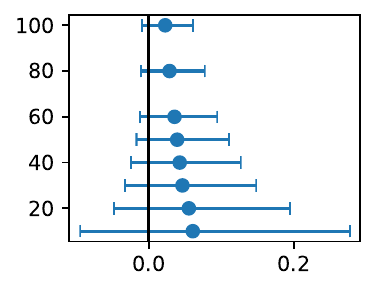}
  \vspace*{-8pt}
  \caption{$(1\%)^2, (0.6\%)^2$}
  \label{fig:vem_value_gained_johnson_etal_1_0-6}
\end{subfigure}
\begin{subfigure}[t]{.32\textwidth}
  \includegraphics[height=0.8\textwidth, trim=3mm 0 2.5mm 0, clip]{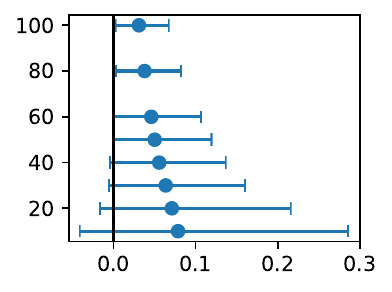}
  \vspace*{-8pt}
  \caption{$(1\%)^2, (0.4\%)^2$}
  \label{fig:vem_value_gained_johnson_etal_1_0-4}
\end{subfigure}
\\[0.5em]
\begin{subfigure}[t]{.32\textwidth}
  \includegraphics[height=0.8\textwidth, trim=3mm 0 2.5mm 0, clip]{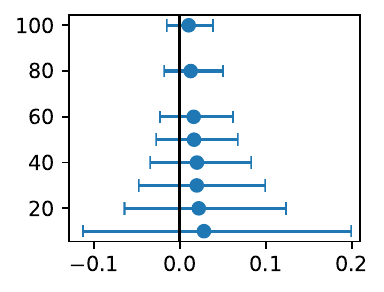}
  \vspace*{-8pt}
  \caption{$(0.8\%)^2, (0.6\%)^2$}
  \label{fig:vem_value_gained_johnson_etal_0-8_0-6}
\end{subfigure}
\begin{subfigure}[t]{.32\textwidth}
  \includegraphics[height=0.8\textwidth, trim=3mm 0 2.5mm 0, clip]{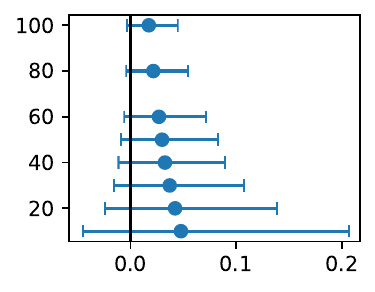}
  \vspace*{-8pt}
  \caption{$(0.8\%)^2, (0.4\%)^2$}
  \label{fig:vem_value_gained_johnson_etal_0-8_0-4}
\end{subfigure}
\begin{subfigure}[t]{.32\textwidth}
  \includegraphics[height=0.8\textwidth, trim=3mm 0 2.5mm 0, clip]{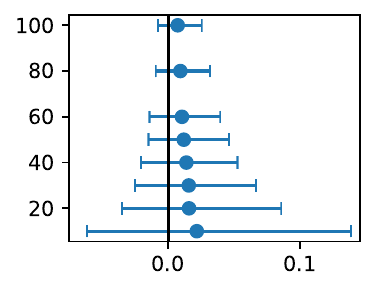}
  \vspace*{-8pt}
  \caption{$(0.6\%)^2, (0.4\%)^2$}
  \label{fig:vem_value_gained_johnson_etal_0-6_0-4}
\end{subfigure}
\end{center}
\vspace*{-0.5em}
\caption[The value gained by having some digital experimentation and capabilities~$\mathcal{D}$ under different capacity $M$ in the case study on 184 marketing experiments reported by Johnson et al.~\cite{johnson2017online}.]{The value gained by having some digital experimentation and measurement (DEM) capabilities~$\mathcal{D}$ (x-axes, in per cent) under different capacity $M$ ($y$-axes) in the case study on 184 marketing experiments reported by Johnson et al.~\cite{johnson2017online} (see Section~\ref{sec:vem_case_study_marketing}). In each plot, the dot represents the mean, and the error bar represents the 5th--95th percentile of the empirical value distribution. Each sub-caption denotes the estimation noise before and after the acquisition of DEM capabilities (i.e., $\sigma^2_1, \sigma^2_2$). Here, we fix $\mu_\mathcal{V} = 19.9\%$, $\mu_\epsilon = 0$, $\sigma^2_\mathcal{V} = (10\%)^2$, and $N = 184$.}
\label{fig:vem_value_gained_johnson_etal}
\end{figure}

Figure~\ref{fig:vem_value_gained_johnson_etal} shows the results. In the presence of a larger variability in the true uplift of the advertising campaigns ($\sigma^2_\mathcal{V}$) and lower capacity ($M$), the level of estimation noise reduction that gave a statistically significant value gained in the e-commerce example is no longer sufficient. Therefore, one needs a larger noise reduction or to increase their capacity to effectively control the risk in investing in DEM capabilities. They may also be better off increasing their limited number of existing business propositions.

\section{Empirical Extensions}
\label{sec:vem_extensions}

We also provide two extensions, evaluated empirically, that open the door for future work in this area.

\subsection{Valuation under independent \texorpdfstring{$t$}{t}-distributed assumptions}
\label{sec:vem_extension_t_distributed}

So far, we have spent much of the work assuming that the business propositions' true value and the estimation noise are normally distributed. While possessing decent mathematical properties, more is needed to explain the heavy tail in the distribution of uplifts shown in~\cite{browne17whatworks} or~\cite{johnson2017online}.

In this section, we model the true value of the business propositions and the estimation noise as Generalised Student's $t$-distributions (see Equation~\eqref{eq:vem_X_n_eps_n_modelling_t}).
It is difficult to derive the exact theoretical quantities under such model assumptions because Student's $t$-distributions do not have conjugate priors~\cite{robert2007bayesian}. We instead simulate the empirical distribution of the value gained under different parameter combinations to understand if this model is a better alternative to that under normal assumptions. The sampling procedure is similar to that described in Section~\ref{sec:vem_experiments}, with steps modified to generate samples using standard $t$-distributions, which are then scaled and located as specified by Equation~\eqref{eq:vem_X_n_eps_n_modelling_t}.

We compare the value gain estimates obtained under t-distributed and normal assumptions. For each comparison, we randomly sample values for $N$, $M$, $\mu_\mathcal{V}$, $\mu_\epsilon$, $\sigma^2_\mathcal{V}$, $\sigma^2_1$, $\sigma^2_2$, and perform 1,000 simulation runs of the four-step sampling procedure in Section~\ref{sec:vem_experiments} to obtain samples of \texttt{D} using both the~$t_3$ and normal distributions.\footnote{$t_3$ ($t$-distribution with three degrees of freedom (d.f.)) is used as it is the distribution with the longest tail under the $t$ family with a natural number of d.f. while retaining a finite variance.} We then compare the expected values, the 5th and 95th percentile of the value gained under the two distributions.

\begin{figure*}
\begin{center}
\begin{subfigure}[t]{.42\textwidth}
  \includegraphics[width=.97\textwidth, trim=3mm 0 2.5mm 0, clip]{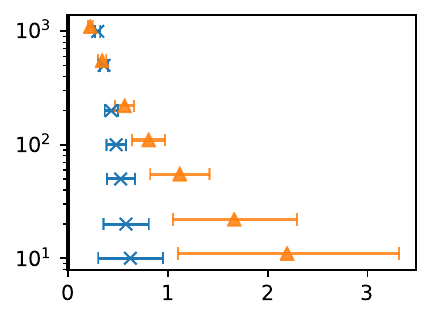}
  \vspace*{-5pt}
  \caption{e-Commmerce - $(1\%)^2, (0.4\%)^2$}
  \label{fig:vem_t_case_study_ecomm_1_0-4}
\end{subfigure}
\hspace*{.03\textwidth}
\begin{subfigure}[t]{.42\textwidth}
  \includegraphics[width=.97\textwidth, trim=3mm 0 2.5mm 0, clip]{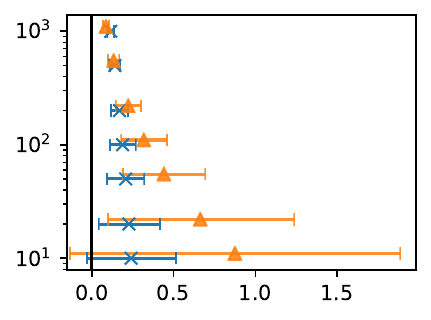}
  \vspace*{-5pt}
  \caption{e-Commerce - $(0.6\%)^2, (0.4\%)^2$}
  \label{fig:vem_t_case_study_ecomm_0-6_0-4}
\end{subfigure}
\\[1em]
\begin{subfigure}[t]{.42\textwidth}
  \includegraphics[width=.97\textwidth, trim=3mm 0 2.5mm 0, clip]{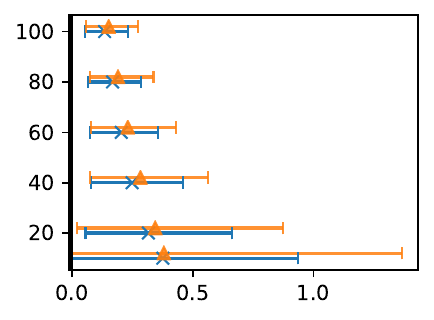}
  \vspace*{-5pt}
  \caption{Marketing - $(2\%)^2, (0.4\%)^2$}
  \label{fig:vem_t_case_study_marketing_2_0-4}
\end{subfigure}
\hspace*{.03\textwidth}
\begin{subfigure}[t]{.42\textwidth}
  \includegraphics[width=0.97\textwidth, trim=3mm 0 2.5mm 0, clip]{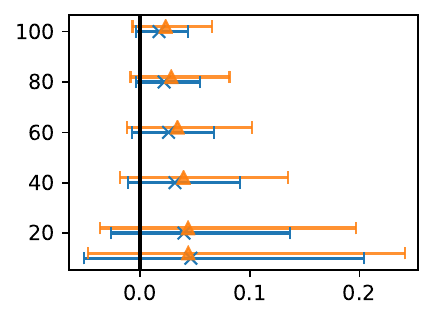}
  \vspace*{-5pt}
  \caption{Marketing - $(0.8\%)^2 , (0.4\%)^2$}
  \label{fig:vem_t_case_study_marketing_0.8_0-4}
\end{subfigure}
\end{center}
\caption[The value gained by having some some digital experimentation and measurement capabilities under normal assumptions and \texorpdfstring{$t$}{t}-distributed assumptions.]{The value gained by having some digital experimentation and measurement capabilities (x-axes, in per cent) under normal assumptions (blue error bars with crosses) and $t$-distributed assumptions (orange error bars with triangles). Both assumptions are evaluated under different but matching capacity $M$ ($y$-axes, with the error bars displaced for clarity). See Section~\ref{sec:vem_case_study_ecommerce}/Figure~\ref{fig:vem_value_gained_browne_johnson} and Section~\ref{sec:vem_case_study_marketing}/Figure~\ref{fig:vem_value_gained_johnson_etal} for details of the parameters for e-commerce and marketing experiments, respectively.}
\label{fig:vem_t_case_study}
\end{figure*}

We observed from 840 comparisons that overall, the value gained under the $t$-distributed assumptions has a higher mean (7\% higher mean) and variance (7\% higher in the 95\% percentile on average) than that under normal assumptions. The result arises despite us setting the mean/variance of the true value and estimation noise under the $t$-distributed assumptions to that under the normal assumptions. This suggests that the model under $t$-distributed assumptions can capture the ``higher risk, higher reward'' concept.

Individual comparisons paint a more nuanced picture, perhaps best illustrated by revisiting the case study in Section~\ref{sec:vem_case_study} under $t$-distributed assumptions. We select a few scenarios featured in the previous section and overlay the value gained by having DEM capabilities under $t$-distributed assumptions over that under normal assumptions in Figure~\ref{fig:vem_t_case_study}. One can see that while $t$-distributed assumptions generally yield a higher value gained, this is not always the case -- for the e-commerce case, as $M$ increases, the value gained decreases quicker under $t$-distributed assumptions than under normal assumptions. This shows that the valuation of DEM capabilities is sensitive to model assumptions.

\subsection{Partial estimation/measurement noise reduction}
\label{sec:vem_extension_partial_noise_reduction}

There are many situations when not all business propositions are immediately measurable upon acquiring DEM capabilities. This may be due to the extra work required to integrate additional capabilities in certain legacy systems or the limited ability to run experiments on online but not offline activities.
In the case where there is a single backlog, we ask the question, will an organisation still benefit from a partial noise reduction when some business propositions' values are obtained under reduced uncertainty while others are subject to the original noise level?

We address this by attempting to establish the relationship between the expected improvement in the mean true value of the selected business propositions and the proportion of business propositions that benefited from a reduced estimation noise (denoted ${p \in [0, 1]}$).\footnote{We can model the estimation noise using a two-component mixture distribution parameterised by $p$.}
The sampling procedure is similar to that described in Section~\ref{sec:vem_experiments}, with Step~3 modified: instead of generating all samples from $\mathcal{N}(\mu_\epsilon, \sigma^2_2)$, we generate~$p$ of the samples from $\mathcal{N}(\mu_\epsilon, \sigma^2_2)$ (the lowered estimation noise) and $1-p$ of the samples from $\mathcal{N}(\mu_\epsilon, \sigma^2_1)$ (the original estimation noise).

\begin{figure*}
\begin{center}
\begin{subfigure}[t]{.225\textwidth}
  \includegraphics[width=\textwidth, trim=3.5mm 0 2.8mm 0, clip]{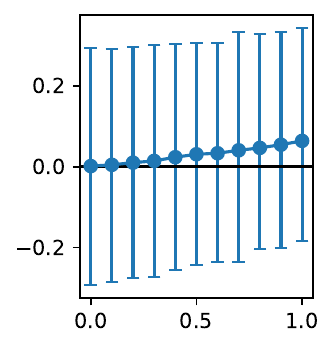}
  \caption{$N = 50$, $M = 5$}
  \label{fig:vem_partial_noise_50_5_1_0-5_0-4}
\end{subfigure}
\hspace*{.005\textwidth}
\begin{subfigure}[t]{.2355\textwidth}
  \includegraphics[width=\textwidth, trim=3.5mm 0 2.8mm 0, clip]{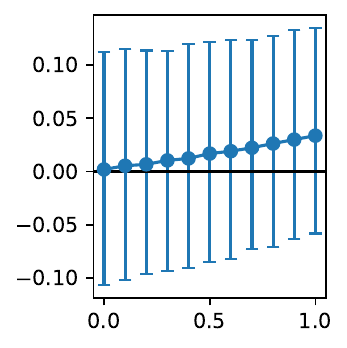}
  \caption{$N = 50$, $M = 20$}
  \label{fig:vem_partial_noise_50_20_1_0-5_0-4}
\end{subfigure}
\hspace*{.005\textwidth}
\begin{subfigure}[t]{.225\textwidth}
  \includegraphics[width=\textwidth, trim=2.8mm 0 2.8mm 0, clip]{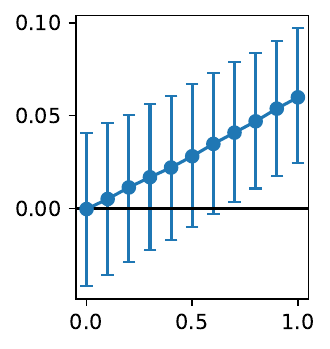}
  \caption{$N = 2500$,\\\hspace*{15pt} $M = 250$}
  \label{fig:vem_partial_noise_2500_250_1_0-5_0-4}
\end{subfigure}
\hspace*{.005\textwidth}
\begin{subfigure}[t]{.225\textwidth}
  \includegraphics[width=\textwidth, trim=2.8mm 0 2.8mm 0, clip]{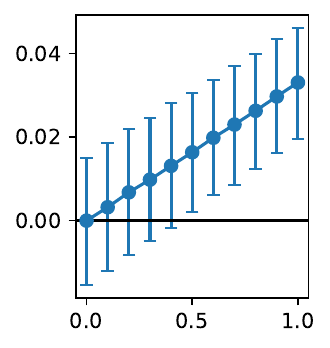}
  \caption{$N = 2500$,\\\hspace*{16pt} $M = 1000$}
  \label{fig:vem_partial_noise_2500_1000_1_0-5_0-4}
\end{subfigure}
\\[0.5em]
\begin{subfigure}[t]{.2355\textwidth}
  \includegraphics[width=\textwidth, trim=3.5mm 0 2.8mm 0, clip]{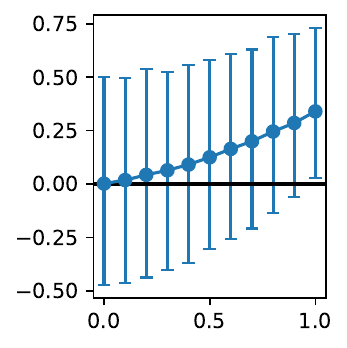}
  \caption{$N = 50$, $M = 5$}
  \label{fig:vem_partial_noise_50_5_1_0-8_0-2}
\end{subfigure}
\hspace*{.002\textwidth}
\begin{subfigure}[t]{.225\textwidth}
  \includegraphics[width=\textwidth, trim=3.5mm 0 2.8mm 0, clip]{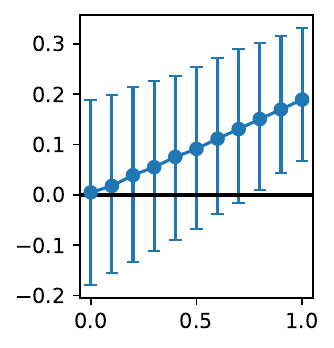}
  \caption{$N = 50$, $M = 20$}
  \label{fig:vem_partial_noise_50_20_1_0-8_0-2}
\end{subfigure}
\hspace*{.016\textwidth}
\begin{subfigure}[t]{.2145\textwidth}
  \includegraphics[width=\textwidth, trim=2.8mm 0 2.8mm 0, clip]{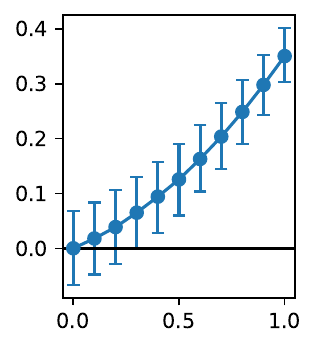}
  \caption{$N = 2500$,\\\hspace*{16pt} $M = 250$}
  \label{fig:vem_partial_noise_2500_250_1_0-8_0-2}
\end{subfigure}
\hspace*{.006\textwidth}
\begin{subfigure}[t]{.225\textwidth}
  \includegraphics[width=\textwidth, trim=2.8mm 0 2.8mm 0, clip]{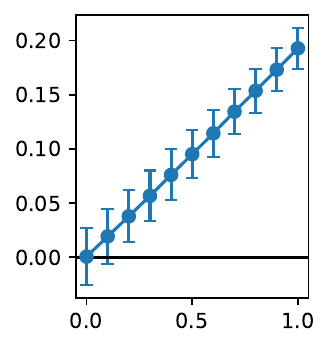}
  \caption{$N = 2500$,\\\hspace*{16pt} $M = 1000$}
  \label{fig:vem_partial_noise_2500_1000_1_0-8_0-2}
\end{subfigure}
\end{center}
\caption[The near-linear relationship between the proportion of business propositions with their value estimated under a lower estimation noise and the improvement in mean true value of the selected propositions under normal assumptions.]{The near-linear relationship between $p$ (proportion of business propositions with their value estimated under a lower estimation noise, $x$-axes) and the improvement in the mean true value of the selected business propositions ($y$-axes) under normal assumptions. In each plot, the dot represents the sample mean, and the error bar represents the 5\%--95\% percentile of the sample value gained. All figures assume $\sigma^2_\mathcal{V} = 1$, while the top four figures assume $\sigma^2_1 = 0.5^2$ and $\sigma^2_2 = 0.4^2$ (corresponding to a small reduction in estimation noise), and the bottom four figures assume $\sigma^2_1 = 0.8^2$ and $\sigma^2_2 = 0.2^2$ (corresponding to a large reduction in estimation noise).}
\label{fig:vem_partial_noise_experiment}
\end{figure*}

We run the procedure above under various scenarios, including a large/small $N$, a large/small ratio between an organisation's capacity and backlog (${M}/{N}$), and a large/small magnitude of noise reduction upon acquisition of DEM capabilities ($\sigma^2_1 - \sigma^2_2$).
Figure~\ref{fig:vem_partial_noise_experiment} shows the result. We can see that under most scenarios, the expected value gained increases with $p$ linearly, while there are a few scenarios where the expected improvement in mean true value of the selected business propositions curve upward for increasing~$p$. This shows that while organisations are incentivised to acquire DEM capabilities that cover
the majority of their work, in many scenarios, a partial acquisition yields proportional benefits. Potential experimenters need not consider the acquisition a zero-one decision or worry about any steep initial investment required to unlock returns.

\section{A Brief Recap}
\label{sec:vem_conclusion}

In this chapter, we have addressed the problem of valuing DEM capabilities, which enables one to justify acquiring the said capabilities. Such capabilities deliver three forms of value to organisations. These are
\begin{enumerate}
    \item Improved recognition of the value of business propositions,
    \item Enhanced capability to prioritise, and
    \item The ability to optimise individual business propositions.
\end{enumerate}
Of these, improving prioritisation is the most challenging to address while being the most applicable for organisations seeking to build DEM capabilities from scratch.

We have established a methodology to value better prioritisation through reduced estimation error using the framework of ranking under uncertainty. The key insight is that DEM capabilities reduce the estimation error in the value of individual business propositions, allowing prioritisation to follow the optimal order of projects more closely had the true values of business propositions been observable. In addition, we have provided simple formulas that give the value of prioritising under lower estimation error brought by DEM capabilities and the Sharpe ratio governing investment decisions. Finally, alongside two case studies that illustrate how we can apply the methodology in specific situations, we have provided general guidelines for conditions when such investments are inappropriate.

\cleardoublepage
\chapter{Statistical Testing}
\label{chap:statstest}

This chapter contains background material that appeared in multiple publications:
\begin{itemize}
  \item ``\textit{An Evaluation Framework for Personalization Strategy Experiment Designs}'', presented at and awarded Best Student Paper of \textit{AdKDD 2020 Workshop (in conjunction with SIGKDD '20)} \cite{liu2020evaluation};
  \item ``\textit{Datasets for Online Controlled Experiments}'', presented at the \textit{35th Conference on Neural Information Processing Systems (NeurIPS 2021)} \cite{liu2021datasets}; and
  \item ``\textit{Measuring e-Commerce Metric Changes in Online Experiments}'', presented at \textit{ACM Web Conference 2023 (WWW '23)} \cite{liu2023measuring}.
\end{itemize}

\section{Motivation}

After establishing the business case of engaging in digital experimentation and measurement, we turn our focus to two necessary ingredients for running digital experiments. They are data, which record what happens during an experiment, and statistical tests, which indicate whether groups in the experiment give different responses beyond randomness. 

We first discuss statistical testing in this chapter, which will complement our approach to datasets in Chapter~\ref{chap:oced}. We will also build upon the concepts introduced in this chapter in Chapter~\ref{chap:rade}, when we formally describe an experiment and its underlying causal reasoning, and in Chapter~\ref{chap:pse}, when we compare different experimental designs for digital experiments. Generally (and loosely) speaking, a statistical test is a statistical inference procedure that informs us whether the data collected is compatible with our hypothesis(es). In the context of digital experimentation, the data collected is often responses from two or more groups exposed to different treatments, and the hypothesis is usually whether or how much the responses from one group are different from another group.

Statistical testing may appear to be a simple and widely understood topic -- at the end of the day, it is taught in many upper secondary education curricula around the globe\footnote{A non-exhaustive list of examples includes (in alphabetical order):
\begin{itemize}[leftmargin=2.8em]
    \item China: (National College Entrance Examination) Mathematics - Extended Topic 3: Probability and Statistics (Item 3(3)2), Elective Group A: Probability and Statistics (Topic 4), and Mathematics Elective Group B: Applied Statistics (Topic 4)~\cite[pp. 48,57,62]{moechina2018maths}
    \item Germany (Berlin \& Brandenburg): (Abitur) Mathematics (Semester 4 / Q4, Advanced-level only, Theme 5 / L5)~\cite[p. 31]{mfbjs2022rahmenlehrplan}
    \item Japan: Mathematics B (Topic (2)A(D))~\cite[pp. 104, 108--109]{mext2018maths}
    \item United Kingdom (AQA): GCE A Level in Mathematics (Subject Content O)~\cite[p. 23]{aqa2018alevelmaths}
    \item United Kingdom (Edexcel): GCE A Level in Statistics (Topics 7--10, 15--17, and 19--20)~\cite[pp. 11--13, 15--19]{pearson2017alevelstats}
    \item USA: Advanced Placement Statistics (Units 6--9)~\cite[pp. 125--205]{collegeboard2020apstatistics}
    \item Worldwide - International Baccalaureate Mathematics: applications and interpretation (Topics SL 4.11 \& AHL 4.18)~\cite[pp. 57--58, 62]{ibo2019mathsaandi}
\end{itemize}
}
and forms part of the standard toolkit in scientific research. It is not. As we will show in this chapter, its development is riddled with assumptions that do not suit practical needs and its application is permeated with careless interpretations. Multiple generations of academics and practitioners have derided its misuse (and sometimes its use)~\cite{berkson1938somedifficulties,cohen1994earth,lew2020reckless,morrison1970thesignificance,wasserstein2019moving,ziliak2008cult}, with the latest round of controversy surrounding ``$p$-hacking'' necessitating an official statement from the American Statistical Association~\cite{asa2016pvalue}.


Despite its many associated problems, statistical testing remains useful for decision-making if applied correctly. 
This chapter aims to provide a comprehensive introduction to the area, with a good balance between the theoretical foundation and its practical use in digital experiments. We seek to enable practitioners in digital experimentation, having arrived from diverse backgrounds, to quickly acquire the knowledge required to perform a statistical test while being mindful of the nuances and pitfalls. 

\paragraph{Target audience}
The chapter is written with a mathematically-inclined audience in mind, i.e., those exposed to university-level introductory mathematics. As we place equal importance on the applications, we will not cover formal proofs and skip through many advanced theoretical concepts, relegating them to footnotes with pointers to relevant works. However, the chapter will involve a fair bit of mathematical notations and algebraic manipulations to help us properly navigate the many concepts in statistical testing. 
For brevity, we also assume readers are already familiar with concepts covered in introductory probability and statistics textbooks, including conditional probability, population vs sample, and probability distributions.\footnote{One can refer to, e.g., Chapters 4--7 of~\cite{mendenhall2018introduction} (more illustrative), Chapters 1--5 of~\cite{casella2002statistical} (more formal), Chapters 1--2 of~\cite{wasserman2004all} (more compact), or any relevant massive open online courses (MOOCs) delivered by a reputable higher education institution.}  
Readers may also find expositions aimed at a broader audience in digital experimentation, e.g., early chapters of~\cite{georgiev2019statistical}, Chapter 17 of~\cite{kohavi20trustworthy}, and \cite{miller10hownot} plus its sibling online articles, as well as works on various aspects of statistical testing written for researchers and practitioners in other fields~\cite{casella2002statistical,dong13powerup,ellis10essential,fleiss03determining,lew2020reckless,livingston05statistical,wasserman2004all} helpful.

\paragraph{A note of caution}
It is worth emphasising that everything in this chapter only helps us establish whether the groups being compared give different responses and, at best, whether the difference in responses is \emph{correlated} to the different treatments provided to the groups. As the old saying goes, ``Correlation does not imply causation''~\cite{aldrich1995correlations}. We require further theoretical foundations to properly establish the causal relationship between the treatment(s) and the change in responses. This will be explored in Chapter~\ref{chap:rade} when we discuss experimental design in digital experimentation.

\paragraph{Chapter organisation}
The rest of the section is organised as follows. We first define the basic terminologies and outline the considerations that drive the development of many different statistical tests in Section~\ref{sec:statstest_considerations}, followed by some non-examples but important concepts related to statistical testing in Section~\ref{sec:statstest_nonexamples}. We then discuss tests and quantities commonly used in digital experimentation from Section~\ref{sec:statstest_nhst} onward, categorised by the considerations outlined at the end of Section~\ref{sec:statstest_considerations}.

\vspace*{1.05\baselineskip}

\section{Pre-matter and Considerations}
\label{sec:statstest_considerations}

A statistical test starts with the experimenter stating one or more statistical hypotheses.\footnote{Not to be confused with a \emph{scientific/research hypothesis}. A scientific hypothesis describes what the experimenter intends to discover/validate about the relationship between variables in the real world. A statistical hypothesis is a much narrower statement on how a statistical model/procedure (usually a simplified view of the real-world phenomenon) behaves. 
See~\cite{bolles1962difference,lawler2021misalignment,lew2020reckless,mcpherson2001teaching} for further discussions.
}

Usually, one specifies a \emph{null hypothesis} $H_0$, the preferred/assumed state of matter,\footnote{Not to be confused with a \emph{nil hypothesis}, one that hypothesises the parameter of interest (e.g., $\Delta$ as defined in the same paragraph) is zero. Our comparison test example shows a nil hypothesis as the null, though other hypotheses, such as $H_0: \Delta = -3$, are also valid. See~\cite{cohen1994earth,good2012common} for further discussions.} and one or more \emph{alternative hypothesis(es)} $H_1, H_2, \cdots$. Together, the hypotheses usually cover all possible scenarios. 
In digital experiments, we may have two populations, $X$ and $Y$, and are interested in inferring the difference in their population mean, $\Delta = \mu_Y - \mu_X$, from their respective samples. Depending on the experimenter's claim, the hypotheses can be in the form of:
\begin{itemize}
    \item In a \emph{comparison} test, a.k.a. a \emph{two-sided} test in some cases  -- 
    \begin{align}
        H_0: \Delta = 0,\; H_1: \Delta \neq 0 \,. \nonumber
    \end{align}
    \item In a \emph{superiority} test, a variant of a \emph{one-sided} test~\cite{committee2001superioritytrial,walker2011understanding} -- 
    \begin{align}
        H_0: \Delta \leq 0,\; H_1: \Delta > 0 \,. \nonumber
    \end{align}
    \item In a \emph{non-inferiority} test, another variant of a one-sided test commonly used to show a population is not substantially worse than another -- 
    \begin{align}
        H_0: \Delta \leq  -\theta,\; H_1: \Delta > -\theta \,, \nonumber
    \end{align}
    where $\theta$ is a small margin that an experimenter is prepared to take to declare that $\mu_Y$ is not substantially lower than $\mu_X$.
    \item In an \emph{equivalence} test (commonly carried out using two one-sided tests, or TOSTs)  -- 
    \begin{align}
        & H_0: \Delta \leq -\theta_L,\; H_1: \Delta > -\theta_L & \textrm{in one and }\, \nonumber\\
        & H_0: \Delta \geq \theta_U, \; H_1:\Delta < \theta_U & \textrm{in the other.} \nonumber
    \end{align}
    Rejecting both null hypotheses would lead to the conclusion that ${-\theta_L < \Delta < \theta_U}$, where $\theta_L$ and $\theta_U$ are the negative and positive margins, respectively, that an experimenter is prepared to take to declare that $\mu_Y$ and $\mu_X$ are equal.
\end{itemize}

In practice, within and outside digital experimentation, it is common to treat the null hypothesis as one to be nullified, one set up to enable an experimenter to demonstrate the contrary despite its status as the assumed state of matter~\cite{cohen1994earth}.\footnote{One can draw a parallel between the reasoning behind nullifying the null hypothesis in a statistical test and that behind proof by contradiction in pure mathematics. That said, the introduction of probabilistic statements has complicated matters. See~\cite{cohen1994earth}, particularly the ``The Permanent Illusion'' section, for further discussions.} One major exception is A/A tests -- randomised controlled trials featuring two supposedly identical treatments. These tests contrast A/B tests that feature two different treatments (see Chapter~\ref{chap:introduction} / Figure~\ref{fig:intro_exp_abtestillustration_freedelivery}) and are often employed in digital experimentation to ensure one implements the systems and models correctly. Given identical treatments, one would expect results consistent with a nil null hypothesis (no difference in the population means) in a comparison test -- the alternative hypothesis is the unexpected artefact.\footnote{Instead of merely assuming the nil null hypothesis, some may instead put the hypothesis to test by running an equivalence test with~$-\theta_L$ and $\theta_U$ very close to zero. However, we are back to the ``nullifying the null hypothesis'' regime in that case.}

After specifying the statistical hypotheses, one then has to decide how to carry out the statistical test procedure. Many statistical tests have been proposed to date, each driven by a combination of different considerations:

\paragraph{Philosophical alignment}
The philosophical debates on the interpretation of probability and the role of models in statistical inference have led to different approaches to running a statistical test. The \emph{frequentist} versus \emph{Bayesian} debate dominates the former, and the latter manifests in whether a statistical test should solely inform one’s degree of belief in a particular hypothesis or act as a decision rule between competing hypotheses~\cite{lehmann1993fishernp,lenhard2006models}. To make things more confusing, statistical testing in modern times, known as \emph{null hypothesis significance testing (NHST)}, is a hybrid of the two procedures (and underlying schools of thought)~\cite{lew2020reckless}. We will introduce examples of NHST and Bayesian tests common in digital experiments in Sections~\ref{sec:statstest_nhst}/\ref{sec:statstest_nhstexamples} and~\ref{sec:statstest_bayesian}, respectively.

\paragraph{Model/distributional assumptions} 
As an inference procedure, statistical tests almost always utilise a statistical model -- a set of assumptions related to the sample data generation process. The model can be \emph{parametric} (i.e., assuming distributions with a finite number of parameters), \emph{non-parametric} (i.e., assuming distributions with an infinite number of parameters or even no distributions), or a mixture of the two~\cite{ghahramani2009brief}. All tests that we will introduce in Section~\ref{sec:statstest_nhst} are parametric. We will also discuss non-parametric tests in Section~\ref{sec:statstest_nonparametric}.

\paragraph{Number of samples} 
A test can involve \emph{one}, \emph{two}, or \emph{many} sample(s) (groups of responses in business speak). Some tests also utilise models that enable one to collapse a two-sample test into a one-sample test (which, one can argue, reduces the operational complexity). Examples of the latter include the $z$-test and $t$-test, which we will discuss in depth in Section~\ref{sec:statstest_nhst}.

\paragraph{Decision-making process}
Some tests are developed with a \emph{fixed-sample/fixed-horizon} experiment in mind. Other tests, which include those that arise from \emph{sequential} analysis and \emph{Bayesian} statistics, support adaptive stopping (see Section~\ref{sec:oced_introduction} for a further exposition). It is important to separate this consideration from that on philosophical alignment, e.g., there is both frequentist and Bayesian interpretation to sequential tests. We will introduce only fixed-horizon tests in Section~\ref{sec:statstest_nhst}, followed by sequential tests in Section~\ref{sec:statstest_sequential} and Bayesian tests in Section~\ref{sec:statstest_bayesian}.

\section{Alternative Approaches to Formal Tests}
\label{sec:statstest_nonexamples}

Before we proceed with the definitions of relevant statistical tests, we discuss some related concepts that are not formal statistical tests yet are often mentioned alongside statistical testing. They are the inter-ocular trauma test, effect size (of a treatment), and confidence intervals.

\subsection{The inter-ocular trauma test}

Given the number of considerations listed above, it is perhaps unsurprising that many works that collate and organise statistical tests already exist~\cite{marshall2016statistics}. Indeed, many guides for experimenters from different fields in need of statistical tests feature a flowchart of various sophistication -- see~\cite{queenborough2016choosing}, \cite{dias2010selecting,greene2006learning}, and \cite{deb2018under} for one intended for general use, clinical trials, and digital experimentation, respectively. However, many of these works may have neglected to ask an important question:
\begin{quote}
    \textit{``Do we require a statistical test at all?''}
\end{quote}

Many statistical tests are unnecessary from the outset. Some results are obvious and way beyond any possible random fluctuations. Other results come from such a noisy environment that it is statistically unfeasible to extract any signal. The inter-ocular trauma (IOT) test deals with the former:
\begin{quote}
    \textit{``You know what the data mean when the conclusion hits you between the eyes.''}~\cite{edwards1963bayesian}
\end{quote}
An IOT test is not a statistical test per se -- it is often performed visually via plots~\cite{buja2009statistical,munroe2020statistics} or tables without involving formal statistical inference. Nonetheless, it reminds experimenters that statistical testing is not just a rite of passage or a recipe to follow to obtain a ``significant result'', whatever ``significant'' might be.

We are not asserting that all statistical tests are unnecessary. Determining what constitutes inter-ocular trauma is more an art than science and requires the backing of substantial domain expertise. Indeed, Edwards et al.~\cite{edwards1963bayesian} added the following immediately after publicly coining the IOT test,
\begin{quote}
    \textit{``But the enthusiast's interocular trauma may be the skeptic's random error. A little arithmetic to verify the extent of the trauma can yield great peace of mind for little cost.''}~\cite{edwards1963bayesian}
\end{quote}

\subsection{Effect size}
\label{sec:statstest_nonexamples_effectsize}

We move on to effect size, which we touched upon without explicitly naming when we set up our running example in Section~\ref{sec:statstest_considerations}. Recall that we have two populations,~$X$ and $Y$, and are interested in the difference in their means
\begin{align}
    \Delta = \mu_Y - \mu_X \,.
    \label{eq:statstest_effectsize_def}
\end{align}
When $X$ and $Y$ represent responses from two groups receiving different treatments in a randomised controlled trial, $\Delta$ is known as the \emph{effect size} (of the mean difference) or the \emph{average treatment effect}. It enables us to tell whether a treatment is creating an impact that matters in practice (i.e., \emph{practically significant}). Effect size is just a measure,\footnote{In a layman's sense, i.e., ``a basis or standard of comparison.''~\cite{merriamwebstermeasure}\label{footnote:measure}} not a statistical test (as no inference is made). It is, though, a measure that many statistical tests are set up to infer and is one of the two quantities of interest when we compare digital experiment designs in Chapter~\ref{chap:pse}.

A standardised effect size enables us to compare the difference across many (related) experiments, thus useful in meta-analyses. One commonly used effect size is Cohen's~$d$, defined as the difference in sample means divided by the pooled sample standard deviation~\cite{cohen1988statistical}. More formally, let $X_1, \cdots, X_{n}$ and $Y_1, \cdots, Y_{m}$ be independent and identically distributed (i.i.d.) samples from the two populations in our running example, with sample means and variances $(\bar{X}, \bar{Y})$ and $(s^2_X, s^2_Y)$, respectively. Cohen's~$d$ is defined as
\begin{align}
    d = \frac{\bar{Y} - \bar{X}}{\sqrt{\frac{(n - 1) s^2_X + (m - 1) s^2_Y}{n + m - 2}}} \,.
    \label{eq:statstest_cohens_d_def}
\end{align}
Other commonly used effect sizes, including Glass' $\Delta$ and Hedge's $g$~\cite{hedges1985statistical}, also involve dividing the difference in sample means by a standalone or pooled standard deviation, with the latter also involving a correction factor dependent on the sample sizes $n$ and $m$.

\subsection{Confidence Intervals}

The final non-example is confidence intervals. Confidence intervals are range estimates for a parameter of interest $\theta$, which is unknown but fixed. Examples of such parameters include the population mean of a sample and the effect size introduced above.

We define confidence intervals as follows. Consider a sample $X$ generated from a statistical model with a parameter of interest $\theta$. A random interval constructed from $X$ is an interval delimited by two transformations of $X$, themselves r.v.'s: $(u(X), v(X))$.
It is said to be a $(1-\alpha)$ confidence interval of $\theta$ if the random interval covers $\theta$ in at least $(1-\alpha)$ of instances, i.e.,
\begin{align}
  \mathbb{P}(u(X) < \theta < v(X)) = 1-\alpha \,,
\end{align}
where $(1-\alpha)$ is known as the \emph{confidence level} and is usually chosen to be a value close to but less than one (e.g., 95\%).
The probability statement should hold for all possible values of $\theta$, as well as that for all other parameters in the model. This is achieved by choosing appropriate~$u(\cdot)$ and $v(\cdot)$. 

For example, consider an i.i.d. sample $X_1, \cdots, X_n$ generated from a normal distribution with mean $\mu_X$ and variance $\sigma^2_X$, estimated by the sample mean $\bar{X}$ and sample variance~$s^2_X$, respectively. A commonly used $(1-\alpha)$ confidence interval of $\mu_X$ is
\begin{align}
    \left(
    \bar{X} - t_{n-1,\, 1 - \alpha/2}\sqrt{\frac{s^2_X}{n}},\,
    \bar{X} + t_{n-1,\, 1 - \alpha/2}\sqrt{\frac{s^2_X}{n}}
    \right) \,,
    \label{eq:statstest_confidence_interval_normal}
\end{align}
where $t_{n-1, 1 - \alpha/2}$ is the~$(1 - \alpha/2)$ quantile of the Student's $t$-distribution with $n-1$ degrees of freedom.\footnote{For brevity, we will skip how such an interval is constructed and why such construction is valid. For a more extensive exposition, readers can refer to, e.g., Sections 7.2 \& 7.3 of~\cite{cox1974theoretical}.} Upon observing the empirical responses, they are substituted into the formula in Expression~\eqref{eq:statstest_confidence_interval_normal} to obtain the realised confidence interval specific to the experiment.

Confidence intervals are not statistical tests as they do not involve any statistical hypotheses. That said, a ``duality'' exists between a $(1-\alpha)$ confidence interval and the non-rejection region of a statistical test that utilises the same statistical model at significance level $\alpha$, which leads to confidence intervals frequently appearing in test results. The ``duality'' enables one to construct a confidence interval for $\theta$ by taking values that a corresponding statistical test on its point estimate (i.e., $H_0: \theta = \hat{\theta}$) fails to reject~\cite{cox1974theoretical}. It also enables one to reject the null hypothesis $H_0: \theta = \theta_0$ in a statistical test if the corresponding confidence interval does not cover $\theta_0$.



Some consider confidence intervals superior to a statistical test or $p$-values (see Section~\ref{sec:statstest_nhstconcepts_pvalue}). They see confidence intervals rebalancing the influence from the null and alternate hypotheses -- instead of giving the null a preferred status as one does in a (frequentist) test, confidence intervals make one look at the full range of effect sizes compatible with the data~\cite{greenland2016statistical}. Others consider confidence intervals a more effective heuristic on the interaction between effect size and sampling uncertainty
than any two numbers from a statistical test.

On the other hand, confidence intervals suffer from the same problem as $p$-values -- there is no guarantee that they are always useful in practice. It is possible for confidence interval procedures, even the ``good'' ones, to produce intervals that are overly wide or narrow (or even empty) while maintaining a $(1-\alpha)$ confidence level overall~\cite{morey2016fallacy}. Moreover, its root in frequentist, parametric statistics also prevents meaningful comparison with non-parametric or Bayesian tests on a like-for-like basis. Any attempt at such will likely revert to debates on the underlying philosophy.

In addition, many interpretation pitfalls prevent one from fully reaping the benefits outlined above. Some stem from losing sight of $\theta$, the parameter of interest, during the estimation procedure. Recall that a confidence interval is a random interval that estimates~$\theta$, a fixed quantity. It does not infer the samples themselves and thus does not provide any information on the proportion of the samples it may cover.\footnote{A random interval that covers $(1-\alpha)$ of the samples is known as a $(1-\alpha)$ \emph{prediction interval}. Generally, these two intervals are not comparable as they estimate different quantities. When one produces confidence intervals for the population mean of a normal distribution, a prediction interval usually shares the same mid-point with, yet is much wider than, a confidence interval with the same ``rating''.
} Readers may have also encountered texts claiming ``there is a $(1-\alpha)$ chance for $\theta$ to fall within the $(1-\alpha)$ confidence interval''. This is misleading. One may consider the phrases ``the interval covering $\theta$'' and ``$\theta$ falling within the interval'' to describe the same event from a different perspective. However, things get complicated once they form part of a probability statement. The description above, which has $\theta$ as the subject, incorrectly implies $\theta$ is random.

Other misconceptions arise from confusing the procedure, which produces random intervals from random samples, with a particular interval constructed from realised response values. The confidence level is a probability statement on the estimation procedure itself -- its value gives the proportion of confidence intervals that cover $\theta$ in the long run.\footnote{The notion of long-run proportion underpins the frequentist interpretation of probability.} It does not matter what value $\theta$ takes or what the parameter is -- all $(1-\alpha)$ confidence intervals that were and will be constructed should cover the corresponding~$\theta$ in~$(1-\alpha)$ of the instances.\footnote{This assumes all intervals involved are correctly constructed and all other model assumptions are met.}
On the other hand, once an interval is realised from the response values, it is no longer random. While~$\theta$ remains unknown, this particular interval either covers the fixed~$\theta$ or does not -- there is no probability statement to be made. Thus, the notion of confidence level is meaningless for realised intervals~\cite{neyman1937outline}.

\section{Null Hypothesis Significance Tests}
\label{sec:statstest_nhst}

We continue to discuss statistical tests by introducing null hypothesis significance tests (NHSTs), the most popular class of statistical tests featured in classrooms and most application areas, including digital experimentation. In this section, we describe the general procedure for running an NHST. Three examples follow in Section~\ref{sec:statstest_nhstexamples}, and a practical guide to concepts and quantities featured in NHSTs in Section~\ref{sec:statstest_nhstconcepts}.

The general procedure for running an NHST is as follows:

\begin{enumerate}
    \item Determine the statistical hypotheses, namely the null and alternative(s), as outlined in Section~\ref{sec:statstest_considerations}.
    \item Determine an appropriate test based on underlying considerations (see Section~\ref{sec:statstest_considerations}) and state the associated \emph{test statistic} $T$ -- see below for typical examples in digital experimentation.
    \item Determine the appropriate \emph{significance level} ($\alpha$), and as best practice, minimum \emph{power}~($\pi_{\min}$) and \emph{sample size} ($n$) via an \textit{a priori} power analysis -- see Section~\ref{sec:statstest_nhstconcepts} for the definition of the mentioned concepts and description of a power analysis. 
    \item Derive/obtain the distribution of $T$ under the null hypothesis. We call a hypothesis \emph{simple} if it completely determines the distribution of $T$ and \emph{composite} otherwise.
    \item Calculate the test statistic based on the observed data, $t_{\textrm{obs}}$.
    \item Determine the \emph{critical region} $\mathbb{T}_{\textrm{crit}}$ -- values of $T$ that lead to us rejecting the null hypothesis. The critical region should be chosen so that the probability of $T$ falling within the region is at most $\alpha$, as determined in Step 2 (see Section~\ref{sec:statstest_nhstconcepts}). For named tests, both the distribution and the optimal critical region\footnote{For one-sided parametric tests, this is given by the uniformly most powerful (UMP) test under the same statistical assumptions. See Section 8.3.2 of~\cite{casella2002statistical} or Chapter 3 of~\cite{lehmann2006testing} for further discussions.} are standard results.
    \item Reject the null hypothesis if the test statistic falls within the critical region, i.e., $t_{\textrm{obs}} \in \mathbb{T}_{\textrm{crit}}$. This indicates it is implausible for the null to generate the observed data, and one of the alternate hypotheses is now favoured.\footnote{It might sound innocuous to say that rejecting the null hypothesis indicates the null hypothesis is implausible given the observed data (or, in mathematical terms, a low $\mathbb{P}(H_0 | \textrm{Data})$). This is not true -- it indicates the observed data is implausible given the null hypothesis (or a low $\mathbb{P}(\textrm{Data} | H_0)$). We will explore this subtle difference in greater detail in Section~\ref{sec:statstest_nhstconcepts_pvalue}.} Otherwise, do not reject the null hypothesis.\footnote{Strictly speaking, there is no such notion of accepting the null hypothesis -- failing to reject the null simply means it remains the preferred hypothesis as determined at the beginning of a statistical test.}
\end{enumerate}

Another approach to the decision steps (Steps 6 and 7) is as follows:

\begin{enumerate}[leftmargin=24pt]
    \item[6A.] Calculate the \emph{$p$-value}, the highest probability that, given the null hypothesis, a sampled value from $T$ (itself an r.v.) will be at least or more extreme than observed test statistic~$t_{\textrm{obs}}$ (see Section~\ref{sec:statstest_nhstconcepts}).
    \item[7A.] Reject the null hypothesis if the $p$-value is less than the significance level, i.e. $p < \alpha$. Otherwise, do not reject the null hypothesis.
\end{enumerate}

The existence of the two approaches highlights the difference in philosophical alignment discussed in Section~\ref{sec:statstest_considerations}. The critical region was developed to provide an objective decision criterion. In contrast, the $p$-value was developed to measure the strength of evidence against the null hypothesis without making a formal decision. The latter approach is a result of mashing the two mathematical concepts together, perhaps driven by the desire to ``have your cake and eat it'', i.e., to have a measure of the strength of evidence while making a formal decision, and further aided by the happy coincidence that there is a one-to-one mapping between these two concepts. In practice, both approaches are commonly used. Some experimenters also engage in both approaches when running the same NHST.

\section{Examples of NHSTs}
\label{sec:statstest_nhstexamples}
We now focus on individual examples, i.e., some named tests commonly used in digital experimentation. These include the $z$-test, the (Welch's) $t$-test, and the binomial test of proportion. The examples are all parametric, fixed-horizon NHSTs.\footnote{Some regard these tests as \emph{classical} NHSTs, though there is no formal definition or consensus on what constitutes a classical NHST.} Needless to say, there are many other examples of parametric, fixed-horizon NHSTs (e.g., $F$-test on population variances, one-way analysis of variance test on population means, Pearson's correlation coefficient test on the correlation between two sets of data) that we have to omit for brevity.

We will follow a similar pattern when introducing each test. In the order laid out by the general procedure in Section~\ref{sec:statstest_nhst}, we first state the statistical assumptions, definition of the associated test statistic~$T$, distribution of~$T$, the optimal critical region~$\mathbb{T}_{\textrm{crit}}$, and~$p$-value. We then provide examples of their use in practice.

\subsection{\texorpdfstring{$z$}{z}-test}
\label{sec:statstest_nhstexamples_ztest}

The $z$-test (or test for normal population mean(s)) is one of the most recognised statistical tests, perhaps because of its early coverage in introductory statistical textbooks.

\paragraph{Assumptions \& Definitions} We first consider a one-sample test: let $X_1, \cdots, X_n$ be i.i.d. samples from a population with unknown mean~$\mu_X$ and known variance~$\sigma^2_X$. We are interested in inferring~$\mu_X$, estimated by the sample mean~$\bar{X}$. The $z$-test assumes all other parameters (in this case $\sigma^2_X$) are already known and defines the test statistic as
\begin{align}
    Z \triangleq \frac{\bar{X} - \mu_X}{\sqrt{\sigma^2_X / n}} \,.
    \label{eq:statstest_ztest_onesample}
\end{align}

\paragraph{Test statistic distribution}
The test also assumes the test statistic, also known as the standard score, follows the standard normal distribution, denoted $\mathcal{N}(0, 1)$.
Theoretically, having normally distributed data ensures that the test statistic is normally distributed. This can easily be shown by observing 
\begin{align}
    X_1, \cdots, X_n \overset{\textrm{i.i.d.}}{\sim} \mathcal{N}(\mu_X, \sigma^2_X) \implies \bar{X} \sim \mathcal{N}(\mu_X, \sigma^2_X / n) \implies Z \sim \mathcal{N}(0, 1) \,.
    \label{eq:statstest_ztest_normalassumption}
\end{align}
It leads to some textbooks (incorrectly) mandating normality assumptions in the \emph{data} for a~$z$-test.
In practice, one would be hard-pressed to find data that is strictly normally distributed. Instead, many appeal to the central limit theorem (CLT), which implies that the sample mean of any underlying distribution approximately follows the normal distribution given sufficient samples,\footnote{Formally, the Lindeberg-L\'{e}vy CLT states that $\sqrt{n}(\bar{X} - \mu_X)$ \emph{converges in distribution} to $\mathcal{N}(0, \sigma^2_X)$. The practical but na\"{i}ve interpretation of the statement above is that the distribution of $\bar{X}$ is getting closer to a normal distribution with mean $\mu_X$ and variance $\sigma^2_X / n$ as $n$ increases. It makes no claim on what constitutes a sufficient number of samples.} i.e., $\bar{X} \overset{\textrm{approx.}}{\sim} \mathcal{N}(\mu_X, \sigma^2_X / n)$. This enables one to run an approximate $z$-test using the same test statistic.

The implication above also means we can run an approximate $z$-test on two or more samples to infer a linear combination of the means, e.g., the effect size between two groups. We further let $Y_1, \cdots, Y_m$ be i.i.d. samples from a population with unknown mean~$\mu_Y$ (estimated by the sample mean $\bar{Y}$) and known variance~$\sigma^2_Y$. Using the same implication above, we can assume $\bar{Y} \overset{\textrm{approx.}}{\sim} \mathcal{N}(\mu_Y, \sigma^2_Y / m)$ and thus
\begin{align}
    \bar{Y} - \bar{X} \overset{\textrm{approx.}}{\sim} \mathcal{N}\left(\mu_Y - \mu_X, \frac{\sigma^2_X}{n} + \frac{\sigma^2_Y}{m}\right) \,.
\end{align}
Noting the effect size as $\Delta = \mu_Y - \mu_X$ (see Equation~\eqref{eq:statstest_effectsize_def}), the test statistic in this two-sample setting is
\begin{align}
    Z \triangleq \frac{(\bar{Y} - \bar{X}) - \Delta}{\sqrt{\frac{\sigma^2_X}{n} + \frac{\sigma^2_Y}{m}}} \,\overset{\textrm{approx.}}{\sim}\, \mathcal{N}(0, 1) \,.
    \label{eq:statstest_ztest_twosample_approx}
\end{align}

One may notice that Equations~\eqref{eq:statstest_ztest_onesample} and~\eqref{eq:statstest_ztest_twosample_approx} place the parameters of interest ($\mu_X$ and $\Delta$) in the definition of $Z$ rather than the distribution.\footnote{With $\sigma^2_X$ and $\sigma^2_Y$ known, this makes $Z$ a \emph{pivotal quantity}.} Such formulation enables one to use the same distribution for different statistical hypotheses on $\mu_X$ or $\Delta$ -- parameters' values only affect the value of the test statistic calculated using the observed data. This is appealing for old clinical trials and pedagogical purposes as one only needs to refer to a single $z$-table. Such appeal is less pronounced in digital experiments with modern computers.

\paragraph{Critical region \& $p$-value}
While the distribution of a $z$-test statistic does not depend on the statistical hypotheses, it is not the case for the critical region $\mathbb{T}_{\textrm{crit}}$ and $p$-value -- the direction of the hypotheses matters. Consider the one-sample setting, where we use $Z$ defined in Equation~\eqref{eq:statstest_ztest_onesample} to infer on $\mu_X$:
\begin{align}
    \mathbb{T}_{\textrm{crit}} = &
    \begin{cases}
        (z_{1 - \alpha}, \infty) & \mkern-36mu \textrm{under a \emph{greater/right-tailed} test } (H_0 \!: \mu_X \leq \theta_0, H_1 \!: \mu_X > \theta_0) \\
        (-\infty, z_{\alpha}) & \textrm{under a \emph{less/left-tailed} test } (H_0 \!: \mu_X \geq \theta_0, H_1 \!: \mu_X < \theta_0) \\
        (-\infty, z_{\alpha / 2}) \cup (z_{1 - \alpha / 2}, \infty) & \mkern34mu \textrm{under a \emph{two-sided} test } (H_0 \!: \mu_X = \theta_0, H_1 \!: \mu_X \neq \theta_0) \,,
    \end{cases}
    \label{eq:statstest_ztest_critregion_def}
    \\
    p = &
    \begin{cases}
        1 - \Phi(z_{\textrm{obs}}) & \textrm{under a \emph{greater/right-tailed} test \hspace*{115pt}} \\
        \Phi(z_{\textrm{obs}}) & \textrm{under a \emph{less/left-tailed} test} \\
        2 \cdot (1 - \Phi\left(|z_{\textrm{obs}}|\right)) & \textrm{under a \emph{two-sided} test} \,,
    \end{cases}
    \label{eq:statstest_ztest_pvalue_def}
\end{align}
where $z_{\textrm{obs}}$ denotes the observed test statistic, $z_q$ is the $q^{\textrm{th}}$ quantile of a standard normal, and~$\Phi(\cdot)$ is the CDF of a standard normal.
The exposition for a two-sample test is identical, with $\Delta$ and the definition of~$Z$ in Equation~\eqref{eq:statstest_ztest_twosample_approx} replacing $\mu_X$ and the definition in Equation~\eqref{eq:statstest_ztest_onesample}, respectively.

\paragraph{Practical usage} Despite its popularity in introductory statistical textbooks, a pure $z$-test is uncommon in digital experiments as one needs to know/assume the population variances ($\sigma^2_X$ and $\sigma^2_Y$) in advance. Nonetheless, experimenters have combined elements of the approximate $z$-test and the Student's/Welch's $t$-test to test whether the mean response of a group differs from that of another group. We will describe how such a combination works in practice below.

\subsection{\texorpdfstring{$t$}{t}-tests}
\label{sec:statstest_nhstexamples_ttest}

We now introduce $t$-tests, the most popular class of statistical tests in digital experimentation~\cite{kohavi20trustworthy}. It is a collection of closely-related variants -- in addition to the one- or two-sample distinction (similar to the $z$-test discussed above), there are also slightly different $t$-tests for independent or paired samples~\cite{hsu2005pairedt}, equal or unequal response variance~\cite{student08probable,welch1947generalization}, and full or partial range of data points~\cite{yuen74twosample}. In the rest of the section, we focus on Welch's $t$-test~\cite{welch1947generalization}, which deals with independent samples with potentially unequal variance across groups and takes all available data.

\paragraph{Assumptions \& Definitions} Let $X_1, \cdots, X_n$ and $Y_1, \cdots, Y_m$ be i.i.d. samples from two populations with unknown means $\mu_X$ and $\mu_Y$, respectively, and \emph{unknown} variances~$\sigma^2_X$ and~$\sigma^2_Y$, respectively. The population means and variances are estimated by the sample means ($\bar{X}, \bar{Y}$) and variances ({$s^2_X, s^2_Y$}), respectively. Similar to the $z$-test, we are interested in inferring $\mu_X$ and $\Delta = \mu_Y - \mu_X$ in a one-sample and a two-sample $t$-test, respectively. However, unlike the $z$-test, which only assumes the test statistic is normally distributed, a $t$-test assumes the \emph{data} (more precisely, the underlying r.v.'s generating the data) are normally distributed.\footnote{Strictly speaking, neither the original $t$-test from Gosset/Student nor Welch's $t$-test explicitly states the r.v.'s generating the data has to be normally distributed. Instead, they assume (1) $\bar{X}$ (or its two-sample counterpart) follows the normal distribution, (2) $s^2 (n-1)/ \sigma^2$ follows the $\chi^2$ distribution, and (3) the quantities in (1) and (2) are independent. However, the assumptions only hold when the underlying r.v.'s are normally distributed.} 

The test statistics are 
\begin{align}
    & T \triangleq \frac{\bar{X} - \mu_X}{\sqrt{s^2_X / n}} & \textrm{for a one-sample test on } \mu_X \textrm{ and}
    \label{eq:statstest_tstatistic_def_onesample}
\end{align}
\begin{align}
    & T \triangleq \frac{\bar{Y} - \bar{X} - \Delta}{ \sqrt{\frac{s^2_X}{n} + \frac{s^2_Y}{m}}} & \textrm{for a two-sample test on }\Delta \,.
    \label{eq:statstest_tstatistic_def_twosample}
\end{align}
They are similar to that in the $z$-test (cf. Equations~\eqref{eq:statstest_ztest_onesample} and~\eqref{eq:statstest_ztest_twosample_approx}), with the only difference being that the $t$-test statistics use the sample variances ($s^2_X$ and $s^2_Y$) instead of the population variance. The difference is critical -- as the sample variances estimate the population variances  (and hence can be formulated as estimators/r.v.'s), they introduce another source of uncertainty to the parameters we are inferring. Other variants of $t$-tests may pool the variance, resulting in slightly different denominators in the fraction within the square root.

\paragraph{Test statistic distribution} 
Both test statistics follow the Student's $t$-distribution with degrees of freedom $\nu$ determined by the Welch-Satterthwaite equation~\cite{satterthwaite46approximate}:
\begin{align}
    T \sim t_{\nu}, \textrm{ where }
    \nu = \frac{\left(\frac{s^2_X}{n} + \frac{s^2_Y}{m}\right)^2}{\frac{\left(s^2_X / n\right)^2}{n - 1} + \frac{\left(s^2_Y / m\right)^2}{m - 1}} \;\textrm{ for a two-sample test.}
\end{align}
In cases where $s^2_X \approx s^2_Y$ and $n \approx m$, the degrees of freedom are around $n + m - 2$. In a one-sample test on $\mu_X$, the degrees of freedom are simply $n - 1$.

\paragraph{Critical region \& $p$-value}
Similar to the $z$-test, the distribution of a $t$-test statistic does not depend on the statistical hypotheses. In contrast, the critical region and $p$-value do. For a one-sample test inferring on $\mu_X$, they are
\begin{align}
    \mathbb{T}_{\textrm{crit}} = &
    \begin{cases}
        (t_{\nu, 1 - \alpha}, \infty) & \mkern-36mu \textrm{under a \emph{greater/right-tailed} test } (H_0 \!: \mu_X \leq \theta_0, H_1 \!: \mu_X > \theta_0) \\
        (-\infty, t_{\nu, \alpha}) & \textrm{under a \emph{less/left-tailed} test } (H_0 \!: \mu_X \geq \theta_0, H_1 \!: \mu_X < \theta_0) \\
        (-\infty, t_{\nu, \alpha / 2}) \cup (t_{\nu, 1 - \alpha / 2}, \infty) & \mkern34mu \textrm{under a \emph{two-sided} test } (H_0 \!: \mu_X = \theta_0, H_1 \!: \mu_X \neq \theta_0) \,,
    \end{cases}
    \label{eq:statstest_ttest_critregion_def}
    \\
    p = &
    \begin{cases}
        1 - T_{\nu}(t_{\textrm{obs}}) & \textrm{under a \emph{greater/right-tailed} test \hspace*{120pt}} \\
        T_{\nu}(t_{\textrm{obs}}) & \textrm{under a \emph{less/left-tailed} test} \\
        2 \cdot (1 - T_{\nu}\left(|t_{\textrm{obs}}|\right)) & \textrm{under a \emph{two-sided} test} \,,
    \end{cases}
    \label{eq:statstest_ttest_pvalue_def}
\end{align}
where $t_{\textrm{obs}}$ is the observed test statistic, $t_{\nu, q}$ is the $q^{\textrm{th}}$ quantile of a Student's $t$-distribution with $\nu$ degrees of freedom, and~$T_{\nu}(\cdot)$ is the CDF of a Student's $t$-distribution with $\nu$ degrees of freedom.


\paragraph{Practical usage}
As mentioned at the beginning of the section, $t$-tests are popular in digital experiments. However, what digital experimenters call a $t$-test is usually not the pure Student's/Welch's $t$-test defined above. The Student's/Welch's $t$-test was initially developed for when the number of available samples is small. The test also impose normality assumptions on the data. Digital experimentation, on the other hand, often features a large number of responses that are unlikely to be well modelled by normal distributions.

Readers may recall that an approximate $z$-test, introduced in Section~\ref{sec:statstest_nhstexamples_ztest}, is suited to experiments with many arbitrarily distributed responses. However, it requires the knowledge of the population variances ($\sigma^2_X$, $\sigma^2_Y$), which we will never have in practice.\footnote{This is perhaps one of the reasons why the name ``$t$-test'' sticks in the digital experimentation community: It serves as a reminder that we are dealing with the sample variances instead of the population variances.}
To obtain the $t$-test digital experimenters use in practice, or the ``practical $t$-test'' in short, we need to extend the approximate $z$-test to take sample variances ($s^2_X$, $s^2_Y$) while keeping its other features. Fortunately, given enough samples, this is as simple as using the sample variance estimates as the population variances, or in other words, ``plugging in'' the sample variances.\footnote{This is known as the \emph{plug-in} principle. Why the plug-in principle works is a topic in its own right -- see, e.g., Chapters 4 \& 5 of~\cite{efron93introduction} for further discussions.} This results in
\begin{align}
    T \,\overset{\textrm{approx.}}{\sim}\, \mathcal{N}(0, 1) \,,
\end{align}
which enables experimenters to work with the normal distributions, instead of the more complicated Student's $t$-distributions, during experiment design and analysis (see Section~\ref{sec:statstest_nhstconcepts}). 

By delving into the underlying statistical theory, we can understand how the shift from Welch's $t$-test to the practical $t$-test (i.e., an approximate $z$-test with plug-in variance estimates) happens. Readers interested only in practical details can skip to the next titled paragraph. The shift can happen in two ways. The first maintains the normality assumptions imposed by Welch's t-test. Thus, the test statistic follows the Student's $t$-distribution with $\nu$ degrees of freedom. As the sample size increases, so would $\nu$, resulting in the test statistic distribution converging to a standard normal distribution:
\begin{align}
    T \,|_{X, Y \sim \mathcal{N}(\cdot, \cdot)} \sim t_{\nu}  \rightarrow \mathcal N(0, 1) \;\textrm{ as }\; \nu \rightarrow \infty \,.
    \label{eq:statstest_ttest_convergence_normal}
\end{align}

The second way removes the normality assumptions altogether. Using the same one-sample test statistic as Welch's $t$-test (see Equation~\eqref{eq:statstest_tstatistic_def_onesample}), we observe that the numerator, including~$\sqrt{n}$, converges in distribution to a normal distribution with a zero mean by the central limit theorem:
\begin{align}
    \sqrt{n}(\bar{X} - \mu_X) \,\overset{D}{\rightarrow}\, \mathcal{N}(0, \sigma^2_X) \,.
    \label{eq:statstest_ttest_convergence_nonnormal_CLT}
\end{align}
At the same time, the denominator converges in probability to the population standard deviation by the law of large numbers and the continuous mapping theorem:
\begin{align}
    s^2_X \,\overset{\mathbb{P}}{\rightarrow}\, \sigma^2_X  \,\implies\,  \sqrt{s^2_X} \,\overset{\mathbb{P}}{\rightarrow}\, \sigma_X \,.
    \label{eq:statstest_ttest_convergence_nonnormal_LLN}
\end{align}
Thus, by Slutsky's theorem, the test statistic converges in distribution to the quotient of the respective limits -- in somewhat sloppy mathematical notation:
\begin{align}
   T = \frac{\sqrt{n}(\bar{X} - \mu_X)}{\sqrt{s^2_X}}  
   \,\overset{D}{\rightarrow}\, \frac{\textrm{``}\mathcal{N}(0, \sigma^2_X)\textrm{''}}{\sigma_X} \equiv \mathcal{N}(0, 1) \,.
   \label{eq:statstest_ttest_convergence_nonnormal_final}
\end{align}
This formulation is a special case (square root version of the one-parameter test statistic) of the Wald test~\cite{wasserman2004all}. The exposition for the two-sample case follows in an identical fashion.

\paragraph{What constitutes a large enough sample size?}
Equations~\eqref{eq:statstest_ttest_convergence_normal} and~\eqref{eq:statstest_ttest_convergence_nonnormal_final} show that the test statistic distribution converges (in other words, eventually gets arbitrarily close) to the standard normal, provided the sample size is large. However, they make no claim on how quickly the convergence is happening or at what sample size the test statistic distribution will be practically indistinguishable from the standard normal.

Readers may have come across many resources suggesting $n=30$ as the ``magic number'' where beyond that, the test statistic is effectively normal~\cite{chang2006determination}. Some might go further and suggest one should switch from a $t$-test to a $z$-test beyond that point, regardless of the underlying distribution of the data~\cite{sullivan2017hypothesis}. These suggestions may be acceptable for pedagogical purposes but are misleading or even harmful in digital experimentation. Firstly, they indicate a clear boundary exists where things are normal on one side and not on the other, whereas in reality, no such boundary exists -- the $t_{29}$ (from $n=30$) and $t_{30}$ (from $n=31$) distributions are not much different from each other.\footnote{The issue of having arbitrary thresholds for decision-making is pervasive in statistical testing. An obvious example would be the convention for one to reject the null hypothesis when the $p$-value drops below a certain threshold (see Section~\ref{sec:statstest_nhstconcepts_pvalue}), which has led to years of bitter debates between statisticians.} 

Secondly, $n=30$ is usually insufficient for the test statistic distribution to practically resemble the standard normal. This is the case even when the normality assumption for the data holds. Consider the $t_{30}$ distribution, the distribution that the one-sample Welch's $t$-test statistic follows when $n=31$. While its PDF may look similar to the standard normal distribution on a plot, the tails of the two distributions are still substantially different. We can see this via the 95\% $z$- and $t$-quantiles:
\begin{align}
    z_{0.95} = t_{30,\, 0.94478...} \textrm{ and } z_{0.95517...} = t_{30,\, 0.95} \,.
\end{align}
The left equation above tells us that a $t_{30}$-distributed test statistic has a $5.52\%$ chance to exceed the 95\% $z$-quantile (the lower bound of the critical region for a greater/right-tailed $z$-test, see Equation~\eqref{eq:statstest_ztest_critregion_def}). Thus, if we assume the $t_{30}$-distributed test statistic is sufficiently normal and decide whether to reject based on the normal (instead of $t_{30}$) quantiles, we will inflate the Type I error from $5\%$ to $5.52\%$, or $10\%$ relatively. Such bias is further amplified at lower significance levels -- for the 99\% $z$- and $t$-quantiles,
\begin{align}
    z_{0.99} = t_{30,\, 0.98652...} \textrm{ and } z_{0.99299...} = t_{30,\, 0.99} \,.
\end{align}
In this case, rejecting at a $1\%$ significance level using the normal (instead of $t_{30}$) quantiles will inflate the Type I error from $1\%$ to $1.35\%$, or $35\%$ relatively. One will require hundreds (if not thousands) of samples to effectively limit such bias,\footnote{For example, to limit such bias to under $1\%$ relative to the significance level, we require at least 315 samples (per group) for a test at a $5\%$ significance level and 1,000 samples for a test at a $1\%$ significance level.} which, fortunately, is easily achievable in digital experiments.

It is worth reiterating that the bias above occurs even when the normality assumption in the data holds. A separate source of bias comes from when we deviate from the normality assumption. In this case, \cite{kohavi14sevenrules} provided a rule of thumb on the minimum sample size required per group for the test statistic to be sufficiently normal-like to run a practical $t$-test. It is based on the underlying distribution's moment coefficient of skewness:
\begin{align}
    355 \times s^2 , \textrm{ where }\, s = \frac{\mathbb{E}\left[(X - \mu_X)^3\right]}{\sigma^3_X} \,,
    \label{eq:statstest_ttest_RoTnonnormal}
\end{align}
In practice, one usually plugs the empirical moment coefficient of skewness into Equation~\eqref{eq:statstest_ttest_RoTnonnormal} to estimate the required sample size. The estimated sample size is most useful when $|s|>1$.
Needless to say, experimenters should ensure they have a sufficient sample size that can address both sources of bias. This is in addition to having enough samples to achieve sufficient test power (see Section~\ref{sec:statstest_nhstconcepts_samplesize}).

\subsection{The binomial test of proportions}
\label{sec:statstest_nhstexamples_binomial}

Another popular statistical test deals with binary responses. By statistical convention, the response of interest is referred to as success and the other as failure. The successes and failures are often modelled by a Bernoulli r.v. The test decides whether the proportion of successes from the observed responses equals a pre-specified probability.

\paragraph{Definitions} Let $X_1, \cdots, X_n$ be independent, identically, and Bernoulli distributed responses with parameter~$\pi_X$. We also let $B$ be the test statistic formed by summing the responses. We observe that the test statistic follows the binomial distribution:
\begin{align}
    B = \sum_{i=1}^{n} X_i \sim \textrm{Bin}(n, \pi_X) \,.
\end{align}
Unlike the $z$-test or $t$-test, the test statistic distribution depends on the parameter of interest~$\pi_X$. Thus, one should define their statistical hypotheses if they have not yet done so. The direction of the null hypothesis does not matter at this stage -- the test statistic follows the same distribution under the two-sided ($H_0: \pi_X = \theta_0$) and one-sided null hypotheses ($H_0: \pi_X \leq \theta_0$ or $H_0: \pi_X \geq \theta_0$):
\begin{align}
B \sim \textrm{Bin}(n, \theta_0) \,.
\end{align}

\paragraph{$p$-value \& critical region}
Suppose one then proceeds with the test statistic calculation and obtains $k$ (i.e., they have observed $k$ successes amongst $n$ responses). Its associated $p$-value, the probability that $B$ attains $k$ or more extreme values (see Section~\ref{sec:statstest_nhstconcepts_pvalue}), is
\begin{align}
  p = 
  \begin{cases}
    \sum_{i=k}^{n} \,\mathbb{P}(B = i) &  
    \textrm{under a \emph{greater/right-tailed} test } (H_0 \!: \mu_X \leq \theta_0, H_1 \!: \mu_X > \theta_0) \\
    \sum_{i=0}^{k} \,\mathbb{P}(B = i) & \textrm{under a \emph{less/left-tailed} test }(H_0 \!: \mu_X \geq \theta_0, H_1 \!: \mu_X < \theta_0) \\
    \sum_{i \textrm{ s.t. } \mathbb{P}(B = i) \leq \mathbb{P}(B = k)} \mathbb{P}(B = i) & \textrm{under a \emph{two-sided} test }(H_0 \!: \mu_X = \theta_0, H_1 \!: \mu_X \neq \theta_0)
    \,,
  \end{cases}
  \label{eq:statstest_binomtest_pvalue_def}
\end{align}
where $\mathbb{P}(B = i) = \binom{n}{i} (\pi_X)^i (1 - \pi_X)^{n-i}$.
Note how the summation indices change depending on the statistical hypothesis. For two-sided $p$-values, we obtain the summation indices by taking values that are equally or less likely to occur (as ``more extreme'') than $k$. Such an elaborate definition is required as most binomial distributions are asymmetric. This renders the approach used by $z$- and $t$-tests to obtain the $p$-value, i.e., taking the probability of the relevant tail and multiplying by two, infeasible.

We also note that a binomial r.v. is discrete. Thus, we define a critical \emph{set} instead:
\begin{align}
  \mathbb{T}_{\textrm{crit}} = 
  \begin{cases}
    \left\{\,i \:\Big|\: i \in \mathbb{N}^0,\, F^{-1}_{\textrm{Bin}(n, \theta_0)}(1 - \alpha) < i \leq n \right\} &
    \textrm{under a \emph{greater/right-tailed} test} \\
    \left\{\,i \:\Big|\: i \in \mathbb{N}^0,\, 0 \leq i < F^{-1}_{\textrm{Bin}(n, \theta_0)}(\alpha) \right\} & \textrm{under a \emph{less/left-tailed} test}
    \,,
  \end{cases}
  \label{eq:statstest_binomial_critset_def}
\end{align}
where $F^{-1}_{\textrm{Bin}(n, \theta_0)}(\cdot)$ is the quantile function of a binomial. Two observations arise: Firstly, we exclude the binomial quantiles from the critical sets as it ensures the probability of the test statistic attaining a value in the critical set would not exceed the significance level $\alpha$. Secondly, excluding the binomial quantiles means the critical region can be an empty set, usually when $n$ is small. In such a scenario, the test is considered poorly designed as it is impossible to reject the null hypothesis.

The critical set for a two-sided binomial test is more complicated than that for a one-sided test: one can use the $\alpha/2$ and $1 - \alpha/2$ binomial quantiles, but the resultant set rarely agrees with what one would obtain using the $p$-values, especially when~$n$ is small. This brings unnecessary confusion; thus, we omit it from Equation~\eqref{eq:statstest_binomial_critset_def}. In practice, when $n$ is small, it is easier for one to decide whether to reject using just the $p$-values. For large $n$, one generally uses the critical region provided by a practical $t$-test, as shown below.

\paragraph{Practical usage} The test described above is \emph{exact} -- it gives exact probabilities provided all the assumptions are met. It generally works well for small $n$, though it is inadequate for the sample size featured in digital experiments due to the need to calculate the probability mass for each possible value $B$ can take.

Instead, digital experimenters employ approximation methods. They appeal to the normal approximation to the binomial distribution (a special case of the CLT), which states that if the sample size $n$ is large and the success rate $\pi_X$ is not too close to zero or one,\footnote{The literature usually states the requirements as $n\pi_X > 5$ and $n(1-\pi_X) > 5$~\cite{schader1989tworules}. Some texts recommend that they should be greater than 10 to be safe.}
\begin{align}
    B \,\overset{\textrm{approx.}}{\sim}\, \mathcal{N}\left(n\pi_X,\, n\pi_X(1 - \pi_X)\right) \,.
\end{align}
Another way to represent the same approximation is to divide the test statistic by $n$. The resultant test statistic is the mean of $n$ Bernoulli r.v.'s, representing the proportion of successes:
\begin{align}
    B' \triangleq \frac{B}{n} \,\overset{\textrm{approx.}}{\sim}\, \mathcal{N}\left(\pi_X,\, \frac{\pi_X(1 - \pi_X)}{n}\right) \iff
    \frac{B' - \pi_X}{\sqrt{\frac{\pi_X(1 - \pi_X)}{n}}} \,\overset{\textrm{approx.}}{\sim}\, \mathcal{N}(0, 1) \,.
\end{align}
This enables one to use the practical $t$-test described in Section~\ref{sec:statstest_nhstexamples_ttest}.
Since there is no straightforward equivalent for the exact binomial test for two-sample comparison, digital experimenters almost always use the practical $t$-test with the test statistic defined in Equation~\eqref{eq:statstest_tstatistic_def_twosample}.


\section{Key Quantities in NHST}
\label{sec:statstest_nhstconcepts}

We introduce some quantities in NHST that are key to the design and analysis of digital experiments.

\subsection{\texorpdfstring{$p$}{p}-value}
\label{sec:statstest_nhstconcepts_pvalue}

We start with $p$-values, one of the most commonly reported measures\footref{footnote:measure} in NHSTs. Its construction is straightforward, with the resultant formulas/routines for different tests readily implemented in statistical software packages. However, perhaps due to the widespread availability and ease of access, it is also one of the most misconstrued quantities in NHSTs. The section aims to highlight the pitfalls surrounding its usage.

As alluded to in Section~\ref{sec:statstest_nhst}, the $p$-value of a test is often  descriptively defined as
\begin{align}
    p \triangleq \mathbb{P}(T \textrm{ attaining } t_{\textrm{obs}} \textrm{ or more extreme values} \,|\, H_0 \textrm{ is true}).
    \label{eq:statstest_pvalue_definition}
\end{align}
A more mathematically precise definition of $p$-value depends on the hypotheses and test, as those affect what constitutes extreme. See Equations~\eqref{eq:statstest_ztest_pvalue_def},~\eqref{eq:statstest_ttest_pvalue_def}, and~\eqref{eq:statstest_binomtest_pvalue_def} in the previous section for examples. One can also find a more abstract definition of $p$-value in statistical theory literature, e.g., Definition 8.3.26 in~\cite{casella2002statistical}.

The definition above applies to a simple null hypothesis, e.g., a nil hypothesis within a comparison test. Under a composite null hypothesis, e.g., a one-sided test,
\begin{align}
    p = \underset{h \in H_0}{\sup}\; \mathbb{P}(T \textrm{ attaining } t_{\textrm{obs}} \textrm{ or more extreme values} \,|\, h \textrm{ is true}).
    \label{eq:statstest_pvalue_definition_composite}
\end{align}
Each possible null hypothesis leads to a different test statistic distribution and, hence, a different $p$-value candidate for the supremum. At first glance, it seems like much work to derive all the test statistic distributions and calculate the $p$-value under every possible null hypothesis. In reality, virtually all named one-sided tests used in digital experimentation attain the supremum at the hypothesis boundary (e.g., parametric tests with $H_0: \Delta \leq \theta_0, H_1: \Delta > \theta_0$ attain the supremum at $\Delta = \theta_0$). Thus, computation of the test statistic and $p$-value always focuses on this single hypothesis as a simple hypothesis~\cite{huber2011justification}.

The widespread usage of $p$-values may give the impression that they give extensive information about the experiment. They do not. They have a fairly narrow scope: To the fullest extent of the interpretation, $p$-values can be characterised as a measure of the degree of surprise\footnote{Some prefer a degree of \emph{discrepancy}~\cite{bayarri1998quantifying} or \emph{incompatibility}~\cite{asa2016pvalue} instead.} for the data in hand under the assumptions imposed by the specific null hypothesis. Writing such characterisation as $\mathbb{P}(\textrm{Data} \,|\, H_0)$, we can see that changing the null hypothesis, which includes the underlying statistical model and experiment assumptions, will lead to a different, incomparable measure.

Crucially, $p$-values, by themselves and as characterised above, tell us nothing about the null hypothesis and the data-generating process. Ideally, one would decide whether to reject the null hypothesis based on $\mathbb{P}(H_0)$ or $\mathbb{P}(H_0 \,|\, \textrm{Data})$, i.e., the chance that the null hypothesis is actually true, with or without observing the data, respectively. In reality, both probabilities are generally unavailable under the NHST framework. Using $p$-values as a substitute for the same ideal is where one can get tripped up. Clearly, $\mathbb{P}(\textrm{Data} \,|\, H_0)$ and $\mathbb{P}(H_0 \,|\, \textrm{Data})$ are different. The former -- $p$-values -- do not tell us the probability that the null hypothesis is true, as the entire $p$-value calculation assumes that the null hypothesis is the preferred state of affairs, the ground truth. 

It is possible to circumvent the restriction above by making assumptions about $\mathbb{P}(H_0)$ or $\mathbb{P}(H_0 \,|\, \textrm{Data})$. Indeed,~\cite{shi2021reconnecting} showed that under some prior belief on how the parameters involved in $H_0$ may behave, $\mathbb{P}(\textrm{Data} \,|\, H_0)$ and $\mathbb{P}(H_0 \,|\, \textrm{Data})$ can be equivalent for certain statistical models and hypotheses.\footnote{More specifically,~\cite{shi2021reconnecting} showed that under non-informative priors, the $p$-value of a one-sided NHST and the posterior probability of a corresponding Bayesian test are (asymptotically) equivalent. Such equivalence applies for binary data with Jeffrey's prior and normal data with an improper flat prior. For two-sided tests, they demonstrated equivalence between the two concepts by reformulating a two-sided test as two one-sided tests and constructing a ``two-sided posterior probability'' that matches the $p$-value.} That said, such assumptions would bring us into Bayesian territory (see Section~\ref{sec:statstest_bayesian}), and we recommend readers who are new to $p$-values and their role in the frequentist vs Bayesian debate to revert to such equivalence only after familiarising themselves with the two perspectives.

In addition, $p$-values also tell us very little about the effect size of treatment and whether it is important in practice. Again, $p$-values merely represent how (in)compatible the data is with the specific null hypothesis assumptions. With carefully crafted hypotheses that seek to demonstrate specific effect sizes, e.g., those in non-inferiority or equivalence tests, one can indeed infer a non-negative or zero effect size solely from a small $p$-value. However, most digital experiments do not seek to demonstrate specific effect sizes (beyond, e.g., ``not zero''). Thus, their null hypothesis specification and $p$-value calculations are generally ignorant to the underlying effect size despite the latter usually being encoded in the data. To appreciate the extent of decoupling between $p$-values and effect sizes in these experiments, we note that for any given effect size, it is possible to construct statistical tests that yield an arbitrarily chosen $p$-value.\footnote{One can do so by changing, e.g., the sample size or the null hypothesis parameter(s).} One usually needs to report quantities discussed in Section~\ref{sec:statstest_nonexamples_effectsize} to communicate effect sizes. In all cases, they would also require domain expertise to appreciate the effect size in the relevant context and, more importantly, to decide the next steps based on the results.

Despite what $p$-values do not tell us, it remains useful in statistical testing to identify and potentially sift out specific statistical hypotheses and assumptions that the data is incompatible with. It is just not the one quantity to rule them all. In fact, no quantities in statistical testing assume such a role.
Experiment analysis is nuanced, and we require multiple sources of signal, including but not limited to the concepts introduced in Section~\ref{sec:statstest_nonexamples}, to understand the big picture and make well-informed decisions. This cannot happen if we over-emphasise reporting a single quantity that does not represent the experiment results. The quantity could also be prone to manipulation, both intentionally and unintentionally,\footnote{For $p$-value, it is known as data dredging~\cite{smith2002data} or ``p-hacking''~\cite{asa2016pvalue}.} making results less interpretable and thus less trustworthy.

To summarise the above, it is perhaps best to use the (slightly rearranged) words of the American Statistical Association~\cite{asa2016pvalue}:
\begin{quote}
    ``$p$-values can indicate how incompatible the data are with a specified statistical model [(one used by the null hypothesis)].''\\
    ``[They] do not measure the probability that the [scientific/statistical] hypothesis is true, or the probability that the data were produced by random chance alone.''\\
    ``[They also] do not measure the size of an effect or the importance of a result.''

    ``Proper inference requires full reporting and transparency.''\\
    ``By itself, a $p$-value does not provide a good measure of evidence regarding a model or hypothesis.''\\
    ``[One should not base] scientific conclusions and business or policy decisions […] only on whether a $p$-value passes a specific threshold.''\\
\end{quote}

\subsection{Type I and II errors}
\label{sec:statstest_nhstconcepts_errors}

The rest of the quantities stem from  decisions made in an NHST. Here, we consider what happens when we decide between two competing statistical hypotheses. 

Treating NHST as a binary decision problem enables us to analyse the different kinds of errors we commit. Of course, we do not know which of the null hypothesis ($H_0$) or the alternate hypothesis ($H_1$) is actually true -- otherwise, we do not need a statistical test at all. We can, however, consider what would have happened had either of the hypotheses been true. This gives us four possible scenarios, as shown below:
\vspace*{-0.8\baselineskip}
\begin{center}
\begin{tabular}{c||c|c}
    & $H_0$ is true & $H_1$ is true \\ \hline
  Test does not reject $H_0$ & 
  Correct decision / True negative &
  Type II error / False negative \\
  Test rejects $H_0$ & 
  Type I error / False positive & 
  Correct decision / True positive \\
\end{tabular}
\end{center}

Knowing how often each of the four scenarios happens enables us to quantify the performance of a decision rule. We are usually interested in the Type I and Type II error rates in statistical testing. The former is defined as the chance of committing a Type I error when $H_0$ is true (see Section~\ref{sec:statstest_nhstconcepts_significance}), and the latter is defined as the chance of committing a Type II error when $H_1$ is true (see Section~\ref{sec:statstest_nhstconcepts_power}). A good experiment design always controls for these two types of error, though we will show in Section~\ref{sec:statstest_nhstconcepts_power} that it is more of a trade-off.

It is worth noting that many other quantities exist to evaluate different aspects of a decision rule. To add to the complication, different communities may refer to the same quantity differently. For example, readers with a clinical background may recognise \emph{specificity} (one minus the Type~I error rate) and \emph{sensitivity} (one minus the Type~II error rate), while those with a computer science or machine learning background may be used to \emph{precision} (one minus the false positive risk, see~Section~\ref{sec:statstest_nhstconcepts_significance}) and \emph{recall} (another name for sensitivity). For consistency, we will stick to the terms presented in the table above for the remainder of the section.

\vspace*{-0.25\baselineskip}

\subsection{Significance level} 
\label{sec:statstest_nhstconcepts_significance}

The significance level, as commonly referred to in digital experimentation,\footnote{Strictly speaking, in statistical theory, Equation~\eqref {eq:statstest_significance_level_def} describes the \emph{size} of a statistical test, with the significance level being a property of a class of tests~\cite{marques2020size}. Having that said, in nearly all practical cases in digital experimentation, the size and level of a test are equal, and there is little practical impact in confusing the terms in our context.} of an NHST is defined as
\begin{align}
    \alpha \triangleq \mathbb{P}(\textrm{test rejects } H_0 \,|\, H_0 \textrm{ is true}).
    \label{eq:statstest_significance_level_def}
\end{align}
It is usually specified as a parameter (as opposed to a calculated property) of a test and determines the specific form of the rejection rules.

Similar to the $p$-value, the definition of the significance level is slightly different when a composite null hypothesis (e.g., in a one-sided test) is involved:
\begin{align}
   \alpha = \underset{h \in H_0}{\sup}\; \mathbb{P}(\textrm{test rejects } H_0 \,|\, h \textrm{ is true}) \,.
\end{align}
Under the same arguments for $p$-values, computation upon the specification of $\alpha$ always focuses on the boundary hypothesis as if it is a simple hypothesis. That said, given that a decision is now involved, experimenters should also take particular care when interpreting a test that failed to reject the null -- clearly, only a particular (but not the full range of the) null hypothesis can be true. 

The parameter signifies the \emph{false positive ratio/rate} experimenters are willing to take. While we are probably more interested in controlling the confusingly similarly named \emph{false positive risk} (i.e.,~${\mathbb{P}(H_0 \textrm{ is true} \,|\, \textrm{test rejects } H_0)}$, with the events conditioned the other way round), it is simply not a feature of NHSTs.

There are no rules on what $\alpha$ should be. Common sense dictates that it should be small -- a test is no good if it rejects the null hypothesis more often than not when the null hypothesis is true. Digital experimenters generally take $\alpha = 10\%$, $5\%$ or $1\%$, which stemmed from the values used in clinical trials decades ago, themselves chosen more as a matter of convenience (e.g., access to statistical tables presented with arbitrarily decided thresholds) rather than principle~\cite{lehmann1993fishernp}. 

It is also worth noting that some, including a large digital A/B testing platform provider~\cite{optimizelystatisticalsignificance}, refer to the complement of the $p$-value (i.e., $1-p$) as ``significance''. Fortunately, the two usages usually yield values on different extremes and thus create little confusion once the experiment context is clear.

\subsection{Power}
\label{sec:statstest_nhstconcepts_power}

The power of an NHST is defined as
\begin{align}
    1 - \beta \triangleq \mathbb{P}(\textrm{test rejects } H_0 \,|\, H_1 \textrm{ is true}),
\end{align}
where $\beta$ is the probability of committing a Type II error / false negative.

In the case where $H_1$ is a composite hypothesis (which is the case for all one-sided and two-sided tests), such probability is difficult to obtain. To illustrate, in parametric tests utilising a model with parameters $\theta \in \Theta$, we have to consider the test rejection probability under every parameter combination to obtain $1-\beta$, i.e., we have to calculate
\begin{align}
\int_{\theta \in \Theta} \underbrace{\mathbb{P}(\textrm{test rejects } H_0 \,|\, \theta)}_{\triangleq 1 - \beta_\theta}\, f(\theta \,|\, H_1 \textrm{ is true}) \,\textrm{d}\theta.
\end{align}
The second term of the integrand, which describes the distribution of the parameters under~$H_1$, is usually unknown in an NHST.\footnote{Imposing any assumptions on the parameter distribution will bring us into Bayesian territory (see Section~\ref{sec:statstest_bayesian}). However, most Bayesian testing approaches do not feature the concept of power as they do not involve a decision.} Thus, we generally refer to the first term of the integrand, i.e., the probability of the test rejecting $H_0$ \emph{given a specific alternate hypothesis}, when we talk about the power.

Unlike the significance level ($\alpha$), one can specify the power as a parameter \textit{a priori} or calculate the \textit{a posteriori}/\emph{post hoc} power.
The \textit{a priori} power,\footnote{Also known as \emph{prospective} power~\cite{ellis10essential}.} commonly denoted $\pi_{\textrm{min}}$, signifies the minimum power one desires for the upcoming test. Equivalently, via $\beta$, it represents the false negative ratio/rate they are willing to take. Clearly, we want the power to be close to one, and it is common to set $\pi_{\textrm{min}} = 80\%$ or $90\%$ in digital experiments. As shown below, specifying the power will affect what is possible with other test parameters, such as sample size (see Section~\ref{sec:statstest_nhstconcepts_samplesize}) and effect size (see Section~\ref{sec:statstest_nhstconcepts_mde}).


One can calculate the \textit{a posteriori}/\emph{post hoc} power before or after running the test. Both approaches answer, ``Given all the other test parameters, what is the chance that the test detects a certain effect size if it actually exists?'' The difference comes from the timing: in pre-test calculations, one often explores a range of potential test parameters based on their experience. In contrast, in post-test calculations, one often uses the observed effect size, response variance, and sample size as if they are the true values.

Calculating the power post-test using the observed effect size, known as a \emph{retrospective} or \emph{observed} power analysis, is controversial~\cite{faul2007gpower,hoenig2001abuse,kohavi2022abtestingintuition}.\footnote{The terminology surrounding power analysis is neither accurate nor precise. Some uses have semantics that deviate from the dictionary definition, and there is no consensus on the names of different approaches. Some consider an \textit{a posteriori}, \textit{post hoc}, and retrospective/observed analysis synonymous~\cite{althouse2021posthoc,schutz2022power}, others differentiate between an \textit{a posteriori}/\textit{post hoc} analysis and a retrospective/observed analysis~\cite{faul2007gpower,wuensch2020oversivew}.} This is because it relies on the unlikely assumption that the observed effect size, derived from a sample, is identical to the underlying population effect size. Its use is nonetheless common in other applications and, thus, not unheard of in digital experiments as experimenters move fields~\cite{kohavi2022abtestingintuition}.

\paragraph{Example: $z$-test}
Consider a two-sample, one-sided $z$-test with $H_0: \Delta = \theta_0$ and $H_1: \Delta > \theta_0$. The test statistic under $H_0$ is $z = \left((\bar{y} - \bar{x}) - \theta_0 \right) / \sqrt{\frac{\sigma^2_X}{n} + \frac{\sigma^2_Y}{m}}$, and we would reject the null hypothesis if $z > z_{1 - \alpha}$. The power for a specific $H_1: \Delta = \theta$ is thus
\begin{align}
    & \, 1 - \beta_\theta 
    = \mathbb{P}(Z > z_{1 - \alpha} \,|\, \Delta = \theta)
    = 1 - \mathbb{P}(Z < z_{1 - \alpha} \,|\, \Delta = \theta)
    \nonumber\\
    = &\,   1 - \mathbb{P}\left(\frac{\bar{Y}-\bar{X} - \theta_0}{\sqrt{\frac{\sigma^2_X}{n} + \frac{\sigma^2_Y}{m}}} < z_{1 - \alpha} \,\Bigg|\, \Delta = \theta \right) 
    & \textrm{(substituting def. of $Z$)} \nonumber\\
    = & \, 1 - \mathbb{P}\left(\frac{\bar{Y}-\bar{X} - \theta + \theta - \theta_0}{\sqrt{\frac{\sigma^2_X}{n} + \frac{\sigma^2_Y}{m}}} < z_{1 - \alpha} \,\Bigg|\, \Delta = \theta \right)
    & \textrm{(inserting self-cancelling $\theta$ pair)} \nonumber\\
    = & \, 
    1 - \mathbb{P}\left(\frac{\bar{Y}-\bar{X} - \theta}{\sqrt{\frac{\sigma^2_X}{n} + \frac{\sigma^2_Y}{m}}} < z_{1 - \alpha} - \frac{\theta - \theta_0}{\sqrt{\frac{\sigma^2_X}{n} + \frac{\sigma^2_Y}{m}}} \,\Bigg|\, \Delta = \theta \right) & \textrm{(moving const. terms to RHS)}. \label{eq:statstest_power_ztest_onesided_derivation}
\end{align}
We note that under this specific $H_1$, the r.v. on the RHS of the inequality follows the standard normal. Thus, the power evaluates to
\begin{align}
    1 - \beta_\theta = 1 - \Phi\left(z_{1 - \alpha} - \frac{\theta - \theta_0}{\sqrt{\frac{\sigma^2_X}{n} + \frac{\sigma^2_Y}{m}}}\right) \,.
    \label{eq:statstest_power_ztest_onesided_final}
\end{align}

In a two-sample, \emph{two}-sided $z$-test, we can obtain the power using the same procedure as Equations~\eqref{eq:statstest_power_ztest_onesided_derivation} and~\eqref{eq:statstest_power_ztest_onesided_final}, noting that we would reject the null hypothesis if $|z| > z_{1 - \alpha/2}$ instead and taking particular care when expanding the absolute function:
\begin{align}
    1 - \beta_\theta = \underbrace{\Phi\left(z_{\alpha/2} - \frac{\theta - \theta_0}{\sqrt{\frac{\sigma^2_X}{n} + \frac{\sigma^2_Y}{m}}}\right)}_{\textrm{Left-tail contribution}} + \underbrace{1 - \Phi\left(z_{1 - \alpha/2} - \frac{\theta - \theta_0}{\sqrt{\frac{\sigma^2_X}{n} + \frac{\sigma^2_Y}{m}}}\right)}_{\textrm{Right-tail contribution}} \,,
    \label{eq:statstest_power_ztest_twosided_final}
\end{align}
where the right-tail contribution is of the same form as that in Equation~\eqref{eq:statstest_power_ztest_onesided_final} (albeit with a different $z$-quantile).

There are also two alternate formulations for Equation~\eqref{eq:statstest_power_ztest_twosided_final}. The first, which we use in Section~\ref{sec:rade_abv}, assumes one uses a nil hypothesis as the null. In this case, $\theta_0 = 0$ and
\begin{align}
    1 - \beta_\theta  = \underbrace{\Phi\left(z_{\alpha/2} - \frac{\theta}{\sqrt{\frac{\sigma^2_X}{n} + \frac{\sigma^2_Y}{m}}}\right)}_{\textrm{Left-tail contribution}} + \underbrace{1 - \Phi\left(z_{1 - \alpha/2} - \frac{\theta}{\sqrt{\frac{\sigma^2_X}{n} + \frac{\sigma^2_Y}{m}}}\right)}_{\textrm{Right-tail contribution}} \,.
    \label{eq:statstest_power_ztest_twosided_nilhyp}
\end{align}
Another formulation exploits that in Equation~\eqref{eq:statstest_power_ztest_twosided_nilhyp}, the left-tail contribution is negligible when the ratio between $\theta$ and the standard error $\sqrt{\sigma^2_X/n + \sigma^2_Y/m}$ is large (i.e., $\theta/\sqrt{\cdots} \gg 0$), the right-tail contribution is negligible when the ratio is negatively large (i.e., $\theta/\sqrt{\cdots} \ll 0$), and the functions representing the two contributions are the horizontal reflection of each other along $\theta = 0$.\footnote{What matters is the magnitude of the entire ratio instead of its components ($\theta$ or the standard error). In digital experiments, one usually deals with small effect sizes and hence $\theta$ (see Section~\ref{sec:statstest_nhstconcepts_mde} and the "Improving sensitivity" paragraphs in Section~\ref{sec:rade_oce_statsmethods} for further discussions). However, the same experiments often involve a large sample size ($n$ and $m$), leading to a workable ratio.} This enables one to approximate the power using only the right-tail contribution term (similar to that in Equation~\eqref{eq:statstest_power_ztest_onesided_final}):
\begin{align}
    1 - \beta_\theta \approx 1 - \Phi\left(z_{1 - \alpha/2} - \frac{|\theta|}{\sqrt{\frac{\sigma^2_X}{n} + \frac{\sigma^2_Y}{m}}}\right) \,.
    \label{eq:statstest_power_ztest_twosided_absapprox}
\end{align}
This form is attractive as it eases downstream algebraic manipulation -- one only needs to deal with one normal CDF when making other parameters the subject of the equation (see Sections~\ref{sec:statstest_nhstconcepts_samplesize} and~\ref{sec:statstest_nhstconcepts_mde}). However, it does require the ratio between the absolute value of~$\theta$ and the standard error to be large (i.e., $|\theta|/\sqrt{\cdots} \gg 0$).\footnote{In practice, we can obtain good approximations when the ratio is greater than $z_{1 - \alpha / 2}$ + 1. At $\alpha = 5\%$, the quantity is approximately equal to three.}
This formulation is used in the following sections and Chapter~\ref{chap:pse}, where the ratio is assumed to be large enough to benefit from this approximation.

\paragraph{Determining factors}

In general, the power depends on many factors:
\begin{enumerate}
    \item The significance level ($\alpha$) - power increases when one increases $\alpha$ as $\beta_\theta$, the Type II error rate, decreases;
    \item The sampling uncertainty ($\sigma^2$) - power increases when $\sigma^2$ decreases as one is more likely to observe the same treatment effect under a lower noise level;
    \item The sample size ($n$/$m$) - power increases when $n$/$m$ increases due to the same reason above; and
    \item The effect size (e.g., $\theta$, assuming $\theta_0 = 0$ as in common practice) - power increases when $\theta$ increases as one is more likely to observe a larger effect size under the same noise level.
\end{enumerate}

Returning to our $z$-test examples, we observe from Equations~\eqref{eq:statstest_power_ztest_onesided_final} and~\eqref{eq:statstest_power_ztest_twosided_final} that power depends entirely on the above factors. In other words, assuming we can specify the value of $1 - \beta_\theta$ alongside the other parameters above, specifying any four out of the five (groups of) parameters will fix the final one. This gives experimenters multiple options during the power analysis stage that were alluded to in Section~\ref{sec:statstest_nhst}.

\subsection{Required sample size}
\label{sec:statstest_nhstconcepts_samplesize}

One common approach to conducting a power analysis seeks to obtain the sample size required for the statistical test and experiment.
This is done by either solving for the sample size ($n$ or $m$) in a known power formula given all other parameters. Alternatively, one finds the sample size that leads to a desired power value while holding other parameters constant in a power simulation. The approach is popular in medical experiments as experimenters have a higher influence on sample size (compared to sampling uncertainty/response variance or the effect size). It also appeals to digital experimenters as the number informs how long it takes to collect the samples, a sizeable portion of an experiment's overall duration.\footnote{Translating the required sample size to the number of days required to collect the samples can be either straightforward or tricky, depending on the experimental units. If the responses are based on website visits, the translation is as simple as dividing the required sample size by the average daily number of visits. On the other hand, if the responses are based on visitors who can return if and when they please, we require more advanced methods to estimate the time taken to observe a certain number of visitors~\cite{richardson22bayesian,zhou2023samplesize}.}

\paragraph{Example: $z$-test} 

We can obtain the required sample size for a $z$-test by specifying the desired power as $\pi_{\textrm{min}} < 1 - \beta_\theta$, substituting Equation~\eqref{eq:statstest_power_ztest_onesided_final} into the inequality, and rearranging the terms to make $n$ or $m$ the subject. Usually, one assumes $n = m$, i.e., the two groups have an equal sample size, as that generally yields the highest power and, thus, the lowest sample size requirement (assuming the two samples have similar variance).\footnote{Let $\textrm{se}(n) = \sqrt{\frac{\sigma^2_X}{n} + \frac{\sigma^2_Y}{n_{\textrm{tot}} - n}}$ be a function of the standard error to the sample size of the first group, where $\sigma^2_X$ and $\sigma^2_Y$ are the population variances, and $n_{\textrm{tot}}$ is the total number of experimental units available to be split into the two groups. Minimising $\textrm{se}(n)$ will maximise the power in Equations~\eqref{eq:statstest_power_ztest_onesided_final} and~\eqref{eq:statstest_power_ztest_twosided_absapprox}.
One can show (via, e.g., a derivative test) that $\textrm{se}(n)$ is minimised when $n = \frac{n_{\textrm{tot}}}{\sigma_Y / \sigma_X + 1}$. When $\sigma^2_X = \sigma^2_Y$, $\textrm{se}(n)$ is minimised when $n = n_{\textrm{tot}} / 2$, i.e., when the two groups have equal sample size.} This yields
\begin{align}
\pi_{\textrm{min}} < 1 - \Phi\left(z_{1 - \alpha} - \frac{\theta - \theta_0}{\sqrt{\frac{\sigma^2_X}{n} + \frac{\sigma^2_Y}{n}}}\right)
\;\iff\; n > \frac{(z_{1 - \alpha} - z_{1 - \pi_\textrm{min}})^2 (\sigma^2_X + \sigma^2_Y)}{(\theta - \theta_0)^2} \,.
\label{eq:statstest_samplesize_ztest}
\end{align}
We use the power definition in Equation~\eqref{eq:statstest_power_ztest_twosided_absapprox} instead for a two-sided test instead; otherwise, the procedure is identical.

A helpful rule of thumb~\cite{vanbelle2008statistical,kohavi20trustworthy,miller10hownot} is to express the minimum required sample size as
\begin{align}
    n = \frac{16 \sigma^2}{\theta^2} \,,
    \label{eq:statstest_samplesize_ruleofthumb}
\end{align}
where $\sigma^2$ is the population variance of \emph{one} group, and $\theta$ is the effect size as defined and used throughout this section (assuming $\theta_0 = 0$). The factor of 16 comes from rounding up the product of $(z_{1 - \alpha} - z_{1 - \pi_\textrm{min}})^2$ and the factor of two extracted from $\sigma^2_X + \sigma^2_Y$ in Equation~\eqref{eq:statstest_samplesize_ztest} (assuming $\alpha=5\%$, $\pi_\textrm{min}=80\%$, and $\sigma^2_X \approx \sigma^2_Y$ as per common practice).

One can also observe from the equations above that the required sample size $n$ is inversely proportional to the square of the effect size $\theta$. In other words, all other things being equal, one requires four times the number of samples to detect half the effect size. This highlights the importance of having a sensible effect size estimate. One can either draw on results from previous experiments or, without prior knowledge, use the minimum practically significant effect size -- one that would matter in practice.

\subsection{Minimum detectable effect}
\label{sec:statstest_nhstconcepts_mde}

Another useful approach to power analysis is to compute the required effect size $\theta$ based on other experiment parameters and see whether it is practically achievable. As power increases when the effect size increases, we are interested in the minimum effect size that yields the desired power, in other words, can be detected (via rejecting the null hypothesis) reliably (at the specified power level). Similar to the required sample size, this can be done by solving a known power formula (see below) or by simulation.

The approach appeals to those with little influence on other experiment parameters (e.g., sample size and response variance) due to operational constraints. Given the fact that digital experiments often feature small effect sizes~\cite{kohavi14sevenrules,xie16improving}, one may leverage a power analysis that requires an unrealistically large effect size to request better resources for their planned experiment, e.g., more participants or investment in methods that improve experiment sensitivity (see Section~\ref{sec:rade_oce_statsmethods}). 

We will also discuss how the minimum detectable effect (MDE) of a test comes into play when we evaluate competing digital experiment designs in Chapter~\ref{chap:pse}.

\paragraph{Example: $z$-test}

Similar to how we obtain the required sample size, we can obtain the MDE by specifying the desired power as $\pi_{\textrm{min}} < 1 - \beta_\theta$, substituting Equation~\eqref{eq:statstest_power_ztest_onesided_final} into the inequality, and rearranging the terms to make $\theta$ the subject. The only difference is that we do not need to assume anything about the sample size. This yields
\begin{align}
\pi_{\textrm{min}} < 1 - \Phi\left(z_{1 - \alpha} - \frac{\theta - \theta_0}{\sqrt{\frac{\sigma^2_X}{n} + \frac{\sigma^2_Y}{m}}}\right)
\;\iff\; 
\theta - \theta_0 > (z_{1 - \alpha} - z_{1 - \pi_\textrm{min}}) \sqrt{\frac{\sigma^2_X}{n} + \frac{\sigma^2_Y}{m}} \,.
\label{eq:statstest_mde_ztest}
\end{align}
The MDE, commonly denoted $\theta^*$, is then the minimum $\theta - \theta_0$ that satisfies Inequality~\eqref{eq:statstest_mde_ztest}, i.e., that specified by the RHS of the inequality. Some variants of Inequality~\eqref{eq:statstest_mde_ztest} may replace $\theta - \theta_0$ and $z_{1 - \alpha}$ with, respectively, $|\theta - \theta_0|$ and $z_{1 - \alpha/2}$ (by assuming a two-sided test and using Equation~\eqref{eq:statstest_power_ztest_twosided_absapprox} instead), or leave out $\theta_0$ (by assuming~$\theta_0 = 0$), or both.

\section{Non-parametric Tests}
\label{sec:statstest_nonparametric}

So far, we have discussed NHSTs comparing statistical hypotheses on the value of one or more statistical model parameters. In some cases, such hypotheses may not suit an experimenter's need -- they may not want to assume the data is generated by a specific distribution, or they may feel the test assumptions are too restrictive.

Instead of inferring specific parameter(s) in a statistical model, a non-parametric test focuses on the overall distributions. In a one-sample setting, the statistical hypotheses often have the following form:
\begin{align}
  H_0: F_X(\cdot) \equiv F(\cdot), \; H_1: F_X(\cdot) \not\equiv F(\cdot) \,,
  \label{eq:statstest_nonparametric_hypothesis_onesample}
\end{align}
i.e., whether a sample $X$ is generated from a specified distribution. Whereas in a two-sample setting, the hypotheses often have the following form:
\begin{align}
  H_0: F_X(\cdot) \equiv F_Y(\cdot), \; H_1: F_X(\cdot) \not\equiv F_Y(\cdot) \,,
  \label{eq:statstest_nonparametric_hypothesis_twosample}
\end{align}
i.e., whether two samples $X$ and $Y$ are generated by the same distribution. Similar to parametric tests, one-sided non-parametric tests exist and are often used, though different tests may have different definitions of what constitutes ``greater'' or ``less''.

This section will introduce two non-parametric tests commonly used in digital experiments: the Mann-Whitney $U$ test (Section~\ref{sec:statstest_mannwhitney}) and the $\chi^2$ goodness-of-fit test (Section~\ref{eq:statstest_chisq}).
Other examples, which we omit for brevity, include the Wilcoxon signed-rank~\cite{wilcoxon1945individual},\linebreak Kruskal-Wallis~\cite{kruskal1952useofranks}, and Kolmogorov–Smirnov tests~\cite{dodge2008kstest}.

\subsection{Mann-Whitney \texorpdfstring{$U$}{U} test}
\label{sec:statstest_mannwhitney}

A Mann-Whitney $U$ test (also known as a Wilcoxon rank-sum test~\cite{lehmann1975nonparametrics}) is a two-sample test determining whether the underlying distributions are equivalent~\cite{mann1947onatest}. Here, the null hypothesis is similar to that in Expression~\eqref{eq:statstest_nonparametric_hypothesis_twosample}.
In contrast, the greater (less than) alternative hypothesis is that one is more (less) likely to see a randomly chosen observation from a population $Y$ exceeding in value than a randomly chosen observation from the other population $X$ and, therefore, $Y$ is greater (less) than $X$ overall.\footnote{More formally, the greater (less than) alternative hypothesis says $Y$ is \emph{stochastically greater (less)} than $X$, defined as the probability that an observation from $Y$ exceeds that from population $X$ is greater (less) than the probability in the reverse case, i.e., observation from $X$ exceeds that from $Y$. This is defined mathematically as $\mathbb{P}(Y > X) > \mathbb{P}(X > Y)$ ($\mathbb{P}(Y > X) < \mathbb{P}(X > Y)$). The two-sided alternative hypothesis states that the two probabilities are unequal.}

Let $X_1, \cdots, X_n$ and $Y_1, \cdots, Y_n$ be our samples. We make no assumptions on the distribution of $X$ and $Y$ except that they are at least ordinal (i.e., there is some notion of a response being greater/less than another), and each copy is i.i.d. The test statistic is defined as
\begin{align}
    U \triangleq \sum_{i=1}^{n} \sum_{j=1}^{m} S(X_i, Y_j), \quad \textrm{where }S(X, Y)=
    \left\{\begin{array}{ll}
    1 & \textrm{if $Y < X$,}\\
    1/2 & \textrm{if $Y = X$,}\\
    0 & \textrm{if $Y > X$.}\\
    \end{array}\right.
    \label{eq:statstest_mannwhitney_teststatistic_def}
\end{align}
For large $n$ and $m$, a rank-based method that generates the same test statistic while avoiding the inefficient double sum is available. Statistical software packages should handle all this automatically upon presenting the responses.

When the sample size is large, $U$ approximately follows a normal distribution:
\begin{align}
    \frac{U - \mu_U}{\sigma_U} \,\overset{\textrm{approx.}}{\sim}\, \mathcal{N}(0, 1),
    \textrm{ where }
    \mu_U = \frac{n\cdot m}{2},\,
    \sigma_U = \sqrt{\frac{n\cdot m\cdot (n+m+1)}{12}} \,.
\end{align}
This enables us to use the $p$-values and critical regions of an approximate $z$-test (see Section~\ref{sec:statstest_nhstexamples_ztest}) to make a reject/not reject decision. In cases with many ties, one would adjust $\sigma_U$ to account for the $1/2$ contributions to the test statistic~\cite{lehmann1975nonparametrics}.

\paragraph{Practical usage}

The Mann-Whitney test is often employed as an alternative to the (practical) $t$-test. This may be due to a severe deviation from the normality assumptions in the data or an insufficient sample size for one to rely on the CLT or both. 

That said, one should not mindlessly use a Mann-Whitney test in lieu of a $t$-test. Strictly speaking, switching from a $t$-test to a Mann-Whitney test removes one's ability to produce statements on the effect size of a treatment described in Section~\ref{sec:statstest_nonexamples_effectsize}. This is due to the statistical hypotheses involved -- a $t$-test infers on the (difference in) population means while a Mann-Whitney test infers the overall distributions. At best, one can make claims like ``overall engagement has improved'' or ``users are spending more in general'' directly off the result of the test, but not ``average engagement has improved by 0.2 units'' or ``mean spend per user has increased by 83 pence''.\footnote{Many digital experimenters make the latter two claims off a Mann-Whitney test regardless.} Moreover, a Mann-Whitney test can be less efficient, or in other words, requires more samples to reject a nil-as-null hypothesis, than a $t$-test given the same effect size if the underlying distributions are sufficiently normal-like (see Section~\ref{sec:statstest_nhstexamples_ttest} on relevant heuristics).

Some literature refers to the Mann-Whitney test as a test for difference in medians. This is only true if one is willing to assume, in addition to that stated at the start of the section, that (1)~the underlying probability distributions $X$ and $Y$ are continuous and (2)~the distribution of $Y$ is a location-shifted version of $X$, i.e. $F_X(x) = F_Y(x + \delta)$ for some value of $\delta$. Failing these assumptions, it is possible to construct r.v.'s $X$ and $Y$ such that the median of $Y$ is equal to / less than $X$, yet a Mann-Whitney test will reject the null hypothesis in favour of a greater alternative hypothesis (i.e., $Y$ is stochastically greater than $X$). Generally, a Mann-Whitney test considers both the location and shape of the two distributions we are comparing~\cite{hart2001mannwhitney}.

\subsection{\texorpdfstring{$\chi^2$}{Chi-squared} goodness-of-fit test}
\label{eq:statstest_chisq}

We also discuss the $\chi^2$ goodness-of-fit test, which deals with categorical responses. In particular, it is a test to decide whether an observed frequency distribution over multiple categories is the same as the theoretical frequency.

Suppose there are $|\textrm{Cats.}|$ categories, with the \textbf{e}xpected/theoretical and \textbf{o}bserved frequency for each category~$i$ being~$E_i$ and~$O_i$, respectively. The test statistic is defined as 
\begin{align}
    C^2 = \sum_{i=1}^{|\textrm{Cats.}|} \frac{(O_i - E_i)^2}{E_i} \,.
\end{align}
Under the null hypothesis, it follows the $\chi^2$ distribution with $\nu$ degrees of freedom (d.f.):
\begin{align}
    C^2 \sim \chi^2_{\nu} \,.
\end{align}
For a goodness-of-fit test, $\nu$ equals the number of categories with counts minus the number of restrictions.\footnote{The same $\chi^2$ test statistic can also be used to test for homogeneity and independence. They feature a different approach to obtaining the degrees of freedom in the test statistic distribution.} The restrictions can be in the form of fitted parameters required to produce the expected count (taking one d.f. away per parameter), or rules that mandate how different sets of expected counts should behave, or both. Usually, there is only one rule for one-dimensional counts: the expected counts should sum up to the observed count sum. However, one may encounter many copies of such a rule along the marginals for multi-dimensional counts, substantially reducing the d.f.

We observe that the test statistic $C^2$ evaluates to zero when the observed frequency distribution is identical to the theoretical frequency distribution and grows large as the discrepancy between the two distributions increases. Thus, the test only makes sense as a greater/right-tailed test, and we reject the null hypothesis if $C^2$ exceeds the $1-\alpha$ quantile of the $\chi^2_\nu$ distribution.

\paragraph{Practical usage}
The $\chi^2$ test is often employed to check for sample ratio mismatch (SRM) in digital experiments. SRM occurs when the ratio between the sample size of two or more groups in an experiment deviates from the designed split. It often indicates one or more problems in software code that trigger a user's entry to an experiment, perform experimental unit randomisation, and track events. These problems often introduce biases that can invalidate experiment results~\cite{dmitriev2017adirtydozen,kohavi20trustworthy}. Of course, not all deviations are necessarily problematic. Small deviations can occur purely due to randomness, but the same deviation becomes unlikely as the sample size grows.\footnote{If we perform a random 50/50 split on our stream of incoming users, we often see~105 users in one group and~95 in the other from the first~200 users. However, one would be rightly suspicious if we see~\numprint{1050000} users in one group and~\numprint{950000} in the other, which indicates systemic bias.} The $\chi^2$ goodness-of-fit test formalises this process by quantifying how likely the observed deviation would happen, assuming the split is correct.

The SRM test operates like other $\chi^2$ goodness-of-fit tests, with the expected frequency of each group obtained by multiplying the total number of samples by the designed split proportion. The d.f. is the number of groups minus one, reflecting the only restriction that the expected group counts should sum up to the total number of samples. Digital experimenters also often set a lower significance level, e.g., at $\alpha = 1\%$ or even $\alpha = 0.1\%$. This is done on practical grounds: investigating why a test rejects the null hypothesis (i.e., flagging a potential SRM) is often costly~\cite{fabijan2019diagnosing}, incentivising experimenters to limit the number of false positives.

\section{Sequential Tests}
\label{sec:statstest_sequential}

As stipulated in Section~\ref{sec:statstest_nhst}, NHSTs require one to decide and commit to the number of samples required prior to the start of the experiment. This prevents experimenters from continuously monitoring and stopping an experiment early -- also known as \emph{peeking}~\cite{johari17peeking} and often used to achieve a shorter time-to-insight -- as it inflates the Type I / false positive error rate due to the look-elsewhere effect.\footnote{The look-elsewhere effect is best described as ``The harder you look for something, the more likely you will find it.'' Peeking is closely related to the multiple comparison problem (see Section~\ref{sec:rade_oce}). The key difference is that the former concerns observing and making decisions based on successive (dependent) test statistics within the same statistical test. In contrast, the latter concerns making decisions based on test statistics across different tests that may or may not be dependent.\label{footnote:peeking_vs_multiple_comparison}}

Some experimenters turn to sequential tests to circumvent the restriction on peeking~\cite{schultzberg2023choosing}.
Sequential tests (a.k.a. sequential analysis or \emph{interim analysis}) concern statistical tests that do not fix their sample size in advance. Instead, these tests evaluate the responses as they are observed. In addition to having a decision rule for the test outcome (e.g., reject/not reject~$H_0$), a sequential test also features a \emph{stopping rule}, which dictates whether an experimenter should continue or stop collecting new data. The decision rule in a sequential test may also depend on the stopping time or other elements of the stopping rule.

Many examples of sequential tests exist and are used in digital experimentation~\cite{miller2015simple}. Below, we introduce two sub-classes, including (mixed) sequential probability ratio tests and group sequential tests. Readers can refer to, e.g.,~\cite{schultzberg2023choosing}, for a critique on the pros and cons of different sequential tests.

\paragraph{(Mixed) sequential probability ratio test}

One of the earliest tests in the sequential\linebreak paradigm is the sequential probability ratio test (SPRT)~\cite{wald1945sequential}. It is a one-sample test that assumes the i.i.d. data $X_1, \cdots, X_n, \cdots, X_{n'}$ follows a parametric distribution and compares two simple hypotheses on a parameter of interest $\theta$, i.e., $H_0: \theta = \theta_0$ and $H_1: \theta = \theta_1$. The test statistic~$S_n$ is the cumulative sum of the log-likelihood ratio up until the latest observed data point~$X_n$. More specifically, $S_0 \triangleq 0$ and
\begin{align}
    S_n = S_{n-1} + \log \left(\frac{f(X_n \,|\, \theta_1)}{f(X_n \,|\, \theta_0)}\right) \,.
\end{align}
The stopping and decision rules are jointly specified as follows. Upon observing $X_n$ and calculating $S_n$:
\begin{itemize}
    \item If $S_n \geq \log \left(\frac{1-\beta}{\alpha}\right)$, accept $H_1$;
    \item Else if $S_n \leq \log \left(\frac{\beta}{1 - \alpha}\right)$, accept $H_0$;\footnote{Readers may notice the test refers to actions such as accepting $H_0$, something not supported in a classical NHST. This is not a typo. The SPRT stems from the (Neyman--Pearsonian) hypothesis test framework, an ``ancestor'' of the NHST framework. Instead of giving the null hypothesis a preferred status like that in an NHST, a hypothesis test decides between two competing hypotheses without any prior judgment and thus can accept either hypothesis. The test also does not have the notion of $p$-value, a feature of a (Fisherian) significance test~\cite{lew2020reckless}.}
    \item Else (i.e., when $\log \left(\frac{\beta}{1 - \alpha}\right) < S_n < \log \left(\frac{1-\beta}{\alpha}\right)$), continue to take observations;
\end{itemize}
where $\alpha$ and $1 - \beta$ correspond to the significance level and power (see Section~\ref{sec:statstest_nhstconcepts}) specified by the experimenter during the design stage, respectively.\footnote{We refer readers to the original work~\cite{wald1945sequential} for a full justification of \emph{why} these specific decision/stopping boundaries will control the Type I and II errors at the specified level.}

Many extensions of SPRT have been developed since. The mixture SPRT (or mSPRT~\cite{robbins1970statistical}) supports the now more commonly used hypothesis pair $H_0: \theta = \theta_0$ and $H_1: \theta \neq \theta_0$ by using a mixture distribution $H$ to specify likely values of $\theta$ among all possibilities. In other words, it specifies $H_1: \theta \sim H$. One then integrates the likelihood ratio across the PDF of~$H$, denoted~$f_H(\theta)$, when calculating the test statistic. This is defined, upon observing the first $n$ samples, as
\begin{align}
    \Lambda^{H, \theta_0}_{n} \triangleq \int \prod_{i=1}^{n} \frac{f(X_i \,|\, \theta)}{f(X_i \,|\, \theta_0)} \, f_H(\theta) \,\textrm{d}\theta .
\end{align}

The version of mSPRT commonly used in digital experimentation is specified in~\cite{johari17peeking}, which adapted the test into a two-sample setting. It assumes the samples $X_1, \cdots, X_n, \cdots, X_{n'}$ and $Y_1, \cdots, Y_n, \cdots, Y_{n'}$ are normally distributed with known variances $\sigma^2_X$ and $\sigma^2_Y$, respectively. The normality assumptions mean one can reduce the two samples into a single sample $(Y_1 - X_1), \cdots, (Y_n - X_n), \cdots, (Y_{n'} - X_{n'})$, observe the resultant sample follow a normal distribution with mean $\theta$ (to be inferred) and variance~$\sigma^2_X + \sigma^2_Y$,\footnote{In \cite{johari17peeking}, the text states that the resultant sample has a variance of $2\sigma^2$. This arises from assuming $\sigma^2_X = \sigma^2_Y = \sigma^2$.} and apply the one-sample mSPRT on the resultant sample. The authors of~\cite{johari17peeking} also noted that by taking $H$ as a normal distribution centred at the null hypothesis (i.e., $H = \mathcal{N}(\theta_0, \tau^2)$, where $\tau^2$ is a hyperparameter that we specify or learn from data), the test statistic would have the following closed-form upon observing the first $n$ samples:
\begin{align}
    \tilde{\Lambda}^{H, \theta_0}_{n}
    \triangleq \sqrt{\frac{\sigma^2_X + \sigma^2_Y}{\sigma^2_X + \sigma^2_Y + n\tau^2}}
    \exp{\left(\frac{n^2\tau^2(\bar{Y}_n - \bar{X}_n - \theta_0)^2}{2(\sigma^2_X + \sigma^2_Y)(\sigma^2_X + \sigma^2_Y + n\tau^2)}\right)} \,,
    \label{eq:statstest_sequential_msprt_teststatistic}
\end{align}
where $\bar{X}_n = n^{-1}\sum_{i=1}^{n}X_i$ and $\bar{Y}_n = n^{-1}\sum_{j=1}^{n}Y_j$ represent the sample mean of the two samples up to sample $n$. The test rejects the null hypothesis if $\tilde{\Lambda}^{H, \theta_0}_{n} \geq \frac{1}{\alpha}$. However, it can also seek new responses indefinitely if the inequality is not met.

\paragraph{Group sequential tests}
Another popular sub-class of sequential tests is group sequential tests. Instead of specifying a new test statistic, a group sequential test reuses the statistical test that one would have used if the experiment is fixed-horizon multiple times during the test duration. Instead, it adjusts each interim test's significance level (and thus critical value) to ensure the overall Type I error rate is equal to what has been specified.\footnote{The procedure resembles family-wise error rate control procedures such as Bonferroni correction, which we will cover in Section~\ref{sec:rade_oce}. That said, there are fundamental differences between the two sets of procedures -- see Footnote \ref{footnote:peeking_vs_multiple_comparison}.} 

The test procedure is then as follows. Many steps featured below are similar to that in an NHST, as described in Section~\ref{sec:statstest_nhst}:
\begin{enumerate}
    \item Determine the statistical hypotheses, the appropriate test \emph{for each interim analysis}, \emph{overall} significance level, power, \emph{maximum} sample size, and test statistic distribution under the null hypothesis $H_0$ (cf. Steps 1--4 in Section~\ref{sec:statstest_nhst});
    \item Upon collecting the required number of responses for the subsequent interim analysis:
    \begin{enumerate}
        \item Obtain the adjusted significance level (see below for examples),
        \item Calculate the test statistic based on responses observed so far (cf. Step 5),
        \item Obtain the adjusted critical region or calculate the $p$-value (cf. Steps 6/6A),
        \item Decide to reject/not reject $H_0$ \emph{within the context of the interim analysis}, using the adjusted critical region or significance level (cf. Steps 7/7A),
        \item If the interim analysis rejects $H_0$, decide the same for the overall test; \\ Otherwise, continue to take observations;
    \end{enumerate}
    \item Upon reaching the maximum sample size specified in Step 1, stop taking further observations and do not reject $H_0$.
\end{enumerate}

There are many ways to adjust the significance level in each interim test. Early approaches set a fixed discount on its value based on the number of \emph{groups} (i.e., the number of times the experimenter intends to peek) and apply the discounted value uniformly to every interim test~\cite{pocock1977group}. While experimenters can expect to use the same adjusted significance level for each interim test, such approaches require them to specify and commit to the number of times they intend to peek. 

More recent approaches feature an alpha spending function specified before starting an experiment. It describes how the overall significance level $\alpha$, which can be characterised as a budget of ``chance to commit Type I error'' in this context, is spent across multiple interim analyses~\cite{lan1983discrete}. Any monotonically increasing function that maps the proportion of responses observed so far to significance levels bounded above by $\alpha$, i.e., $[0, 1] \rightarrow [0, \alpha]$, can be used as an alpha spending function with various effectiveness. While these approaches require non-trivial calculations to obtain the values of adjusted significance level and critical region for each interim analysis,\footnote{That said, software packages that perform such calculations are readily available. See, e.g.,~\cite{casper2023ldbounds}.} they do not require experimenters to specify the number of times or when they intend to peek during the experiment~\cite{demets1994interim}.

\section{Bayesian Tests}
\label{sec:statstest_bayesian}

We also discuss statistical testing from a Bayesian perspective. Recall that we introduced $\mathbb{P}(H \,|\, \textrm{Data})$, the probability of a statistical hypothesis $H$ being true given the data, in Section~\ref{sec:statstest_nhstconcepts_pvalue}.\footnote{The actual hypothesis featured in Section~\ref{sec:statstest_nhstconcepts_pvalue} was, in fact, the null hypothesis $H_0$. The switch to a general statistical hypothesis $H$ is intentional, as all hypotheses are treated equally in the Bayesian regime.} We observed that this probability is often considered the more intuitively correct heuristic when making data-driven decisions than, e.g., $p$-values~\cite{cohen1994earth}. In 
Section~\ref{sec:statstest_nhstconcepts_pvalue}, ``Data'' was shorthand for the specific event ``test statistic attaining the observed or more extreme values''. However, the observation still holds if we take the term's general definition of ``the probability of the samples taking their observed value''.

The Bayesian perspective enables one to obtain $\mathbb{P}(H \,|\, \textrm{Data})$ by treating it as a posterior belief. Under such an approach, one first specifies their prior belief on the hypothesis as a probability, denoted as $\mathbb{P}(H)$. They then update the prior belief with the observations, summarised by the likelihood under the chosen statistical model (denoted as $f(\textrm{Data} \,|\, H)$), using Bayes' theorem to obtain the posterior belief.

We outline two approaches in the Bayesian setting. The first focuses on the posterior probability of a single hypothesis being true. By applying Bayes' theorem directly:
\begin{align}
    \mathbb{P}(H \,|\, \textrm{Data}) = 
    \frac{f(\textrm{Data} \,|\, H)\mathbb{P}(H)}{f(\textrm{Data})} \,.
    \label{eq:statstest_bayesian_bayestheorem}
\end{align}
In cases where the chosen statistical model is parametric, the hypothesis is usually a statement of plausible values of the parameter of interest $\theta$. The likelihood function has to account for all such parameter values:
\begin{align}
    f(\textrm{Data} \,|\, H) = \int f(\textrm{Data} \,|\, \theta, H) f(\theta \,|\,  H) \,\textrm{d}\theta \,.
    \label{eq:statstest_bayesian_likelihood_marginal}
\end{align}

The second approach arises from the need to compare and select statistical models. Of course, we already have those models in hand -- that specified by the null ($H_0$) and alternative hypotheses ($H_1$). Unlike the NHST regime, where there is a preferred/assumed state of matter and hence an unequal status balance between the two hypotheses, the models specified by the two hypotheses are treated equally in the Bayesian regime. Both are treated as candidate models, with updates done in the same way using the same observations.\footnote{One can still favour a particular hypothesis via the specification of the prior belief.}

The comparison and selection process utilises the Bayes factor~\cite{kass95bayes}. 
We first substitute $H_0$ and $H_1$ into Equation~\eqref{eq:statstest_bayesian_bayestheorem} separately to obtain two posterior probability formulas. Then, we observe that the quotient between the two posterior probabilities is
\begin{align}
    \underbrace{\frac{\mathbb{P}(H_1 \,|\, \textrm{Data})}{\mathbb{P}(H_0 \,|\, \textrm{Data})}}_{\textrm{Posterior odds}} = 
    \frac{f(\textrm{Data} \,|\, H_1) \mathbb{P}(H_1)}{f(\textrm{Data})} \cdot
    \frac{f(\textrm{Data})}{f(\textrm{Data} \,|\, H_0) \mathbb{P}(H_0)} = 
    \underbrace{\frac{f(\textrm{Data} \,|\, H_1)}{f(\textrm{Data} \,|\, H_0)}}_{\textrm{Bayes factor }(\triangleq \textrm{BF}_{10})}
    \cdot
    \underbrace{\frac{\mathbb{P}(H_1)}{\mathbb{P}(H_0)}}_{\textrm{Prior odds}} \,,
    \label{eq:statstest_bayesian_bf_definition}
\end{align}
with the expansion in Equation~\eqref{eq:statstest_bayesian_likelihood_marginal} applied where appropriate.
The Bayes factor $\textrm{BF}_{10}$ (the second-from-right fraction in Equation~\eqref{eq:statstest_bayesian_bf_definition}) represents the strength of evidence provided by the observations in favour of $H_1$. There is no universal scale with descriptors, though~\cite{kass95bayes} considers a Bayes factor of $\sqrt{10} \approx 3.162$, $10$, and $100$ to be the lower thresholds for ``substantial'', ``strong'', and ``decisive'' evidence.

An advantage of the Bayes factor over NHST is that the former can also represent the strength of evidence \emph{in favour} of $H_0$, the ``null'' hypothesis that one can only accumulate evidence \emph{against} in an NHST (Sections~\ref{sec:statstest_nhst}--\ref{sec:statstest_nhstconcepts}). We can obtain that by taking the reciprocal of the Bayes factor $\textrm{BF}_{10}$ (or the entire Equation~\eqref{eq:statstest_bayesian_bf_definition}) to obtain $\textrm{BF}_{01}$.
One can also recover the posterior probability of each of the two hypotheses from the posterior odds quickly if $H_0$ and $H_1$ cover the entire event space. This is usually the case if we use the standard one- and two-sided hypotheses featured in the previous sections.

\paragraph{Applications in digital experimentation} Bayesian testing has gained traction in digital experiments in recent years~\cite{deng15objective,deng16continuous}. It is often employed in lieu of sequential testing (see section above) to address the continuous monitoring/peeking problem -- see Section~\ref{sec:oced_introduction} for a more detailed account of the problem and~\cite{deng16continuous} for a discussion on how Bayesian and sequential testing differs in practice.

Many Bayesian tests in digital experimentation work off the statistical models used in a Wald/practical $t$-test. Consider a two-sample setting, with $X_1, \cdots, X_n, \cdots, X_{n'}$\linebreak and~$Y_1, \cdots, Y_m, \cdots, Y_{m'}$ being i.i.d. samples from the two groups.
We recall from Section~\ref{sec:statstest_nhstexamples_ttest} that given a large sample, the Wald/practical $t$-test statistic approximately follows the standard normal distribution. The Bayesian perspective allows one to update their belief with incoming observations. Thus, it makes sense to define a test statistic upon observing the first~$n$ samples from $X$ and $m$ samples from $Y$:
\begin{align}
    \sqrt{W_{n, m}} \triangleq \frac{\bar{Y}_m - \bar{X}_n - \Delta}{\sqrt{\left(\frac{s^2_X}{n} + \frac{s^2_Y}{m}\right)}} \,\overset{\textrm{approx.}}{\sim}\,\mathcal{N}(0, 1) \,,
\end{align}
where $\bar{X}_n = \frac{1}{n}\sum_{i=1}^{n} X_i$ and $\bar{Y}_m = \frac{1}{m}\sum_{j=1}^{m} Y_j$ are the running sample means. $s^2_X$ and $s^2_Y$ are the plug-in estimates for the population variances $\sigma^2_X$ and $\sigma^2_Y$, respectively.

Unlike the practical $t$-test, where one usually states hypotheses on the unstandardised effect size $\Delta$, we prefer inferring the effect size standardised by the pooled variance in a Bayesian test. This is because standardising the effect size will make it scale-independent, easing the specification or learning of the prior. To obtain the standardised effect size, we note $\sqrt{W_{n, m}}$ can be expanded as
\begin{align}
    \sqrt{W_{n, m}} = 
    \Bigg(
    \underbrace{\frac{\bar{Y}_m - \bar{X}_n}{\sqrt{\big(\frac{s^2_X}{n} + \frac{s^2_Y}{m}\big) / \left(\frac{1}{n}+ \frac{1}{m}\right)}}}_{\triangleq\, \delta_{n, m}} -
    \underbrace{\frac{\Delta}{\sqrt{\big(\frac{s^2_X}{n} + \frac{s^2_Y}{m}\big) / \left(\frac{1}{n}+ \frac{1}{m}\right)}}}_{\triangleq\, \Delta^{\!\circ}}
    \Bigg)
    \underbrace{\frac{1}{\sqrt{\frac{1}{n} + \frac{1}{m}\vphantom{\big(\frac{s^2_X}{n}\big)}}}}_{\triangleq\sqrt{E_{n, m}}} \;,
    \label{eq:statstest_bayesian_wald_decomposition}
\end{align}
where $\delta_{n, m}$, $\Delta^{\!\circ}$, and $E_{n,m}$ are the test statistic, standardised effect size and the effective sample size of the test, respectively. Given that $\sqrt{W_{n, m}}$ approximates the standard normal distribution, $\delta_{n, m}$, a linear transformation of $\sqrt{W_{n, m}}$, approximates a (non-standard) normal distribution:
\begin{align}
    \delta_{n, m} = 
    \frac{\bar{Y}_m - \bar{X}_n}{\sqrt{\big(\frac{s^2_X}{n} + \frac{s^2_Y}{m}\big) / \left(\frac{1}{n}+ \frac{1}{m}\right)}} = 
    \frac{\sqrt{W_{n, m}}}{\sqrt{E_{n, m}}} + \Delta^{\!\circ} 
    \, \overset{\textrm{approx.}}{\sim}\, \mathcal{N}\left(\Delta^{\!\circ}, \frac{1}{E_{n, m}}\right)\;.
\end{align}
This enables us to assume a normal likelihood for the standardised effect size.

Next, we specify the statistical hypothesis to calculate the Bayes factor. Similar to NHSTs, many hypothesis combinations exist (see Section~\ref{sec:statstest_considerations}).\footnote{That said, one should take extra care when specifying the statistical hypotheses in a Bayesian test, particularly the alternative hypothesis. In NHSTs, the alternative hypothesis is generally not involved in test calculations and can be lazily specified as the complement of the null hypothesis. This is different for Bayes factor calculations, which involve both hypotheses heavily.} A commonly used hypothesis pair is $H_0: \Delta^{\!\circ} = \theta_0$ and $H_1: \Delta^{\!\circ} \sim \mathcal{N}(\theta_0, V^2)$, where $V^2$ is a hyperparameter that we specify or learn from data. We observe that $H_0$ is identical to that used in NHSTs. Meanwhile, the prior specified in $H_1$ is a reasonable approximation to the distribution of (standardised) effect sizes (see Section~\ref{sec:vem_mathematical_formulation_distributions}) that also has nice mathematical properties -- the resultant model under~$H_1$ is a hierarchical normal model. Given the above, the Bayes factor is
\begin{align}
    (BF_{10})_{n, m} = \frac{f(\delta_{n,m} \,|\, H_1)}{f(\delta_{n,m} \,|\, H_0)} = \frac{\phi\left(\delta_{n,m};\, \theta_0,\, V^2 + \frac{1}{E_{n, m}}\right)}{\phi\left(\delta_{n,m};\, \theta_0,\, \frac{1}{E_{n, m}}\right)} \;,
    \label{eq:statstest_bayesian_bf_normal_example}
\end{align}
where $\phi(\,\cdot\,; \mu, \sigma^2)$ is the PDF of a normal distribution with mean $\mu$ and variance $\sigma^2$.
One can also obtain the posterior odds shown in Equation~\eqref{eq:statstest_bayesian_bf_definition} if they specify~$\mathbb{P}(H_0)$ and~$\mathbb{P}(H_1)$, the prior belief on the probability that each hypothesis is true.

Like every other Bayesian analysis, determining a good prior distribution for $\Delta^{\!\circ}$ is difficult. If one prioritises removing experimenters' biases from prior specification, they may consider leveraging past experiment data to learn the prior using an empirical Bayes approach~\cite{deng15objective}.\footnote{In~\cite{deng15objective,deng16continuous}, both prominent works on Bayesian testing in digital experimentation, the authors called the prior distribution obtained from the empirical Bayes ``objective''. We believe this is used in the general sense to contrast with a \emph{subjective prior} elicited from a domain expert. In Bayesian analysis, \emph{objective priors} usually mean \emph{non-informative priors}, default priors constructed by some formal or structural rules that are largely ignorant to the context surrounding the experiment~\cite{consonni2018prior,kass1996selection}, which is a different approach to empirical Bayes priors~\cite{upton2014dictionary}.} In our running example, this means using the expectation-maximisation algorithm to learn the parameter~$V^2$ and the prior probability~$\mathbb{P}(H_0)$. In the case where the speed in obtaining a large Bayes factor is essential, they may also consider using a \emph{non-local prior} for $H_1$, which is perhaps best described as drilling hole(s) on the probability density near the parameter(s) covered by $H_0$~\cite{johnson10ontheuse}. By unnesting $H_0$ from $H_1$, we avoid evidence in favour of $H_0$ to also contribute to $H_1$ as well and thus enable a quicker convergence had~$H_0$ been the ground truth.

\section{A Brief Recap}

We provided an introduction to statistical testing in digital experimentation in this chapter. Starting from the beginning of a statistical test -- the specification of one or more statistical hypotheses -- we introduced four (overlapping) classes of statistical tests. They are NHSTs, non-parametric tests, sequential tests, and Bayesian tests. For each class of tests, we described one or more examples that are popular among digital experimenters in detail. We also outlined three alternatives to formal statistical tests: the inter-ocular trauma test, effect size, and confidence intervals.

We have aimed to balance the theoretical foundations and the practical applications. For the former, we covered, among many topics, the distributional results that underpin the practical $t$-tests and the derivation of the required sample size formulas from first principles using the concept of power. For the latter, we have outlined the situations where a specific test is applied in digital experiments and the pitfalls in interpreting quantities (e.g., $p$-values) featured in a test.
In addition, we have carefully collated and linked the terminologies used in different communities. This is done with the intention for the guide to be accessible to a diverse yet mathematically-inclined audience. 

That said, the expansive nature of statistical testing means we have inevitably omitted the details of many examples and (primarily theoretical) concepts. We provided further details and pointers for interested readers in the footnotes and references throughout the chapter.

\cleardoublepage
\chapter{Datasets for Digital Experiments}
\label{chap:oced}

The chapter is adapted from the research paper ``\textit{Datasets for Online Controlled Experiments}'', presented at the \textit{35th Conference on Neural Information Processing Systems (NeurIPS '21)} \cite{liu2021datasets}.

\section{Motivation}
\label{sec:oced_introduction}

We move on and discuss data, another necessary ingredient for running digital experiments. Data plays a central role in developing the digital economy, especially in digital content, social networks, digital advertising and e-commerce. Its generation, processing, use, and storage have been the topic of interest for many software engineering, databases, and machine learning researchers and practitioners. That said, we feel data is less rigorously explored in the digital experimentation context than statistical tests (see Chapter~\ref{chap:statstest}). Its existence is often taken for granted by researchers or practitioners in digital experimentation or treated as merely one of the means to tackle the research/business problem (instead of the object of interest). 

To illustrate,
the ability to run experiments on the Web allows one to interact with many subjects within a short time frame and collect many responses. This, together with the scale of experimentation carried out by tech organisations, should lead to a wealth of datasets describing the result of an experiment. However, there are not many publicly available datasets describing digital experiments, and we believe they were never systematically reviewed nor categorised. This is in contrast to the machine learning field, which also enjoyed its application boom in the past decade yet already has established archives and detailed categorisations for its datasets~\cite{dua2019UCI,OpenML2013}.

We argue that the lack of relevant datasets arising from real experiments hinders the further development of digital experimentation and measurement methods (e.g., new statistical tests, bias correction, and variance reduction methods; see Chapter~\ref{chap:rade}). Many statistical tests proposed relied on simulated data that impose restrictive distributional assumptions and thus may not represent the real-world scenario. Moreover, it may be difficult to understand how methods differ and assess their relative strengths and weaknesses without a common dataset to compare them on.

To address this problem, we present the first ever survey for online controlled experiments (OCE) datasets, together with a taxonomy that can be generalised to all digital experiments.\footnote{Online controlled experiments, or online randomised controlled trials, are a subset of digital experiments. They are characterised by the experimenters' ability to perform random assignments and apply scientific control. See Section~\ref{sec:rade_classes} for further details.} Our survey identified 13 datasets, including standalone experiment archives, accompanying datasets from scholarly works, and demo datasets from online courses on the design and analysis of experiments. We also categorise these datasets based on dimensions such as the number of experiments each dataset contains, how granular each data point is time-wise and subject-wise, and whether it includes results from real experiment(s). 

The taxonomy enables us to discuss the data requirements for an experiment by systematically mapping out which data dimension is required for which statistical test and learning the hyperparameter(s) associated with the test. We also recognise that, in practice, data are often used for purposes beyond what it is originally collected for~\cite{kerr1998harking}. Hence, we posit the mapping is equally useful in allowing one to understand their options when choosing statistical tests given the format of data they possess. Together with the survey, the taxonomy helps us to identify what types of datasets are required for commonly used statistical tests yet are missing from the public domain. 

One of the gaps identified by the survey and taxonomy is datasets that can support the design and running of experiments with \emph{adaptive stopping} (a.k.a. continuous monitoring / optional stopping). We motivate their use below.
Traditionally, experimenters analyse experiments using null hypothesis significance tests (NHSTs), e.g., a Student's $t$-test. These tests require one to calculate and commit to a required sample size based on some expected treatment effect size, all prior to starting the experiment. Making extra decisions during the experiment, be it stopping the experiment early due to seeing favourable results or extending the experiment as it teeters ``on the edge of statistical significance''~\cite{munroe2015pvalues}, is discouraged as they risk one having more false discoveries than intended~\cite{greenland2016statistical,munroe2011significant}.

Clearly, the restrictions above are incompatible with modern decision-making processes.
Businesses operating online are incentivised to deploy beneficial features and roll back damaging changes as quickly as possible.  
In the ``free delivery'' banner example in Chapter~\ref{chap:introduction}, the business may have calculated that they require four weeks to observe enough users based on an expected 1\% change in the decision metric. If the experiment shows, two weeks in, that the banner is leading to a 2\% improvement,  it will be unwise not to deploy the banner to all users simply due to the need to run the experiment for another two weeks. Likewise, if the banner is shown leading to a 2\% loss, it makes every sense to immediately terminate the experiment and roll back the banner to stem further losses.

As a result, more experimenters are moving away from NHSTs and adopting adaptive stopping techniques. Experiments with adaptive stopping allow one to decide when to stop an experiment (i.e. stopping it earlier or prolonging it) based on the sample responses observed so far without compromising the statistical validity of false positive/discovery rate control. 
To encourage further development in this area, both in methods and data, we release the ASOS Digital Experiments Dataset, which contains daily checkpoints of decision metrics from multiple real OCEs run on the global online fashion retail platform. 

The dataset design is guided by the requirements identified by the mapping between the taxonomy and statistical tests, and to the best of our knowledge, is the first public dataset that can support the end-to-end design and running of digital experiments with adaptive stopping.
We demonstrate it can indeed do so by (1) running a sequential test and a Bayesian hypothesis test on all the experiments in the dataset and (2) estimating the value of hyperparameters associated with the tests. While the notion of ground truth does not exist in real digital experiments, we show that the dataset can also act as a quasi-benchmark for statistical tests by comparing results from the tests above with that of a $t$-test.

To summarise, our contributions are:
\begin{enumerate}
    \item (Sections~\ref{sec:oced_taxonomy} \&~\ref{sec:oced_survey}) We create, to the best of our knowledge, the first ever taxonomy on digital experiment datasets and apply it to publicly available online controlled experiment datasets;
    \item (Section~\ref{sec:oced_mapping}) We map the relationship between the taxonomy and statistical tests commonly used in experiments by identifying the minimally sufficient set of statistics and dimensions required in each test. The mapping, which also applies to offline experiments, enables experimenters to quickly identify the data collection requirements for their experiment design (and conversely, the test options available given the data availability); and
    \item (Section~\ref{sec:oced_dataset}) We make available, to the best of our knowledge, the first real, multi-experiment time series dataset, enabling the design and running of experimentation with adaptive stopping.\footnote{Link to the dataset and accompanying datasheet: \url{https://osf.io/64jsb/}. \label{footnote:oced_dataset_link}}
\end{enumerate}

\section{A Taxonomy for Digital Experiment Datasets}
\label{sec:oced_taxonomy}

We begin by presenting a taxonomy on digital experiment datasets, which is necessary to characterise and understand the results of a survey. To the best of our knowledge, there are no surveys nor taxonomies specifically on this topic prior to this work. While there is a large volume of work concerning the categorisation of datasets in machine learning~\cite{dua2019UCI,OpenML2013}, research work in the online randomised controlled experiment methods~\cite{auer18current,auer21controlled,fabijan2018online,ros18continuous}, and general experiment design \cite{hopewell2010quality,liu2020evaluation}, our search on Google Scholar and Semantic Scholar using combinations of the keywords ``online controlled experiment''/``A/B test'', ``dataset'', and ``taxonomy''/``categorization''/``categorisation'' yields no relevant results.

The taxonomy focuses on the following four main dimensions:

\paragraph{Experiment count} A dataset can contain the data collected from a \emph{single} or \emph{multiple} experiments. Results from a single experiment are useful for demonstrating how a test works, though any learning should ideally involve multiple experiments. Two closely related but relatively minor dimensions are the \emph{variant count} (number of control/treatment groups in the experiment) and the \emph{metric count} (number of performance metrics the experiment is tracking). Having an experiment with multiple variants and metrics enables the demonstration of methods such as false discovery rate control procedures (e.g.,~\cite{benjamini1995controlling}, see Section~\ref{sec:rade_oce_statsmethods} for more details) and learning the correlation structure within an experiment.

\paragraph{Response granularity} Depending on the experiment analysis requirements and constraints imposed by the digital experimentation platform, the dataset may contain data aggregated to various levels. Consider the ``free delivery'' banner example in Chapter~\ref{chap:introduction}, where the website users are randomly allocated to the treatment (showing the banner) and control (not showing the banner) groups to understand whether the banner changes the proportion of users who bought something. Each individual user is considered a randomisation unit~\cite{kohavi20trustworthy}. 

A dataset may contain, for each experiment, only summary statistics on the \emph{group} level, e.g., the proportion of users who have bought something in the control and treatment groups, respectively. It can also record one response per \emph{randomisation unit}, with each row containing the user ID and whether the user bought something. The more detailed activity logs at a \emph{sub-randomisation unit} level can also have each row containing information about a particular page view from a particular user.
    
\paragraph{Time granularity} An experiment can last between a week and many months~\cite{kohavi20trustworthy}, providing many possibilities for recording the result. A dataset can opt to record the \emph{overall result only}, showing the end state of an experiment. If the dataset contains multiple randomisation units or experiments, it may or may not have a \emph{timestamp} for each instance. It can also record \emph{intermediate checkpoints} for the decision metrics, ideally at regular intervals such as \emph{daily} or \emph{hourly}. These checkpoints can either be a \emph{snapshot} of the interval (recording activities between time~$t$ and~$t+1$, time~$t+1$ and~$t+2$, etc.) or \emph{cumulative} from the start of the experiment (recording activities between time~$0$ and~$1$, time~$0$ and~$2$, etc.).

\paragraph{Syntheticity} A dataset can record data generated from a \emph{real} process. It can also be \emph{synthetic} -- generated via simulations with distributional assumptions applied. A dataset can also be \emph{semi-synthetic} if generated from a real-life process and subsequently augmented with synthetic data.

Note that we can also describe datasets arising from any experiments (including offline and non-randomised controlled experiments) using these four dimensions. We will discuss in Section~\ref{sec:oced_mapping} how these dimensions map to common statistical tests used in digital experimentation.

In addition, we also record the \emph{application domain}, \emph{target demographics}, and the \emph{temporal coverage} of the experiment(s) featured in a dataset. In an age when data are often reused, one must understand the underlying context and that learnings from a dataset created under a specific context may not translate to another context. 
We also see the surfacing of such context as a way to promote considerations in fairness and transparency for experimenters as more experiment datasets become available~\cite{jiang2019guineapig}. For example, having target demographics information on a meta-level helps experimenters identify who were involved, or perhaps more importantly, who were \emph{not} involved in experiments and could be adversely impacted by an effectively untested treatment. 

Finally, the datasets can also differ in their medium of \emph{documentation} and the presence/absence of a \emph{data management / long-term preservation plan}. The latter includes the hosting location, the presence/absence of a DOI, and the type of license. We record these attributes for the datasets surveyed below for completeness.

\section{Public Online Controlled Experiment Datasets}
\label{sec:oced_survey}

Here, we discuss our approach to produce the first ever survey on OCE datasets and present its results.
The survey is compiled via two search directions, which we describe below. For both directions, we conduct a first round search in May 2021, with follow up rounds in August and October 2021 to ensure we have the most updated results.

We first searched on the vanilla Google search engine using the keywords ``Online controlled experiment ``dataset'''', ``A/B test ``dataset'''', and ``Multivariate test ``dataset''''. For each keyword, we inspect the first 10 pages of the search result (top 100 results) for scholarly articles, web pages, blog posts, and documents that may host or describe a publicly available OCE dataset (or both).
The search term ``dataset'' is in double quotes to limit the search results to those explicitly mentioning dataset(s).
We also searched on specialist data search engines/hosts, namely Google Dataset Search (GDS) and Kaggle, using the keywords ``Online controlled experiment(s)'' and ``A/B test(s)''. We inspect the metadata and description for all the results returned (except for GDS, where we inspect the first 100 results for ``A/B test(s)'') for relevant datasets as defined below.\footnote{Searching for the keyword ``Online controlled experiment'' on GDS and Kaggle returned 42 and 7 results, respectively, and that for ``A/B test'' returned ``100+'' and 286 results, respectively. Curiously, replacing ``experiment'' and ``test'' in the keywords with their plural form changes the number of results, with the former returning 6 and 10 results on GDS and Kaggle, respectively, and the latter returning ``100+'' and 303 results, respectively.}

A dataset must record the result arising from a randomised controlled experiment run online to be included in the survey. 
The criterion excludes experimental data collected from offline experiments, e.g. those in agriculture~\cite{wright20agridatdataset}, medicine~\cite{doyle2019clinical}, and economics~\cite{dupas2016targeting}.
It also excludes datasets used to perform quasi-experiments and observational studies, e.g. the LaLonde dataset used in econometrics~\cite{dehejia99causal} and datasets constructed for uplift modelling tasks~\cite{diemert2018largescale,hillstrom2008minethatdata}.\footnote{Uplift modelling (UM) tasks for online applications often start with an OCE~\cite{radcliffe2007using}, and thus we can consider UM datasets as OCE datasets with extra randomisation unit level features. The nature of the tasks is different, though: OCEs are concerned with validating the average treatment effect across the population using a statistical test, whereas UM is concerned with modelling the conditional average treatment effect for each individual user, making it more a general causal inference task that is outside the scope of this survey.}

The result is presented in Table~\ref{tab:oced_survey}. We place the 13 OCE datasets identified in this exercise along the four taxonomy dimensions defined in Section~\ref{sec:oced_taxonomy} and record the additional features.
These datasets include two standalone archives, one for online media experiments~\cite{matias2021upworthydataset} and the other for experiments on an online education platform~\cite{selent2016assistmentsdataset}, plus two accompanying datasets for peer-reviewed research articles~\cite{tenorio2017meututordataset,young2014librarydataset}. 
There are also tens of Kaggle datasets, blog posts, and code repositories that describe or duplicate one of the six example datasets used in five massive open online courses on online controlled experiment design and analysis~\cite{baathmobileproject,camposabtestingcourse,grimesabtestingcourse,grossmanabtestingcourse,udacityanalystcourse}. Finally, we identify three standalone datasets hosted on Kaggle with relatively light documentation~\cite{ay2020syntheticalabdataset,emmanuel2020addataset,klimonova2020grocerydataset}.

The table shows several gaps in OCE dataset availability, the most obvious being the lack of datasets that record responses at a sub-randomisation unit level.
In the sections below, we will identify more of these gaps and discuss their implications for digital experiment analysis.

\pagebreak

\begin{landscape}
\begin{table}
\vspace*{-5pt}
\hspace*{-6pt}
\renewcommand{\arraystretch}{1.05}
\begin{adjustbox}{width=24.8cm}
\begin{tabular}{l|l||l|l|l|l||l|l|l||l|l}
Dataset Name                                             & 
Ref.                                                     & \begin{tabular}[t]{@{}l@{}}
Experiment Count\\- Variant / Metric\\$\;\;\,$Count
\end{tabular}                                            &
\begin{tabular}[t]{@{}l@{}}
Response\\Granularity
\end{tabular}                                            &
Time Granularity                                         &
Syntheticity                                             &
\begin{tabular}[t]{@{}l@{}}
Application\\Domain
\end{tabular}                                            &
\begin{tabular}[t]{@{}l@{}}
Target\\Demographic
\end{tabular}                                            &
\begin{tabular}[t]{@{}l@{}}
Temporal\\Coverage
\end{tabular}                                            &
Documentation                                            &
\begin{tabular}[t]{@{}l@{}}
Data management / long- \\ term preservation plan
\end{tabular}                              \\[1.5em]\hline
Upworthy Research Archive                                & \cite{matias2021upworthydataset}                         & \begin{tabular}[c]{@{}l@{}}
Multiple (\numprint{32487})\\ - 2--14 / 2 (C)
\end{tabular}                                            &
Group                                                    &
\begin{tabular}[c]{@{}l@{}}
Overall result only\\ Timestamp per expt.
\end{tabular}                                            &
Real                                                     &
\begin{tabular}[c]{@{}l@{}}
Media \& Ads\\ Copy/Creative
\end{tabular}                                            &
\begin{tabular}[c]{@{}l@{}}
Mostly English-\\speaking users \\ in USA
\end{tabular}                                            &
\begin{tabular}[c]{@{}l@{}}
Jan 2013 -- \\ Apr 2015
\end{tabular}                                            &
\begin{tabular}[c]{@{}l@{}}
Peer reviewed\\ data article
\end{tabular}                                            & \begin{tabular}[c]{@{}l@{}}
Host: Open Science Framework\\ DOI: $\cmark$ (see \cite{matias2021upworthydataset})\\ Licence: CC BY 4.0
\end{tabular}                              \\[1.5em]\hline
\begin{tabular}[c]{@{}l@{}}
ASSISTments Dataset from \\ Multiple Randomized \\ Controlled Experiments\end{tabular}                                 &
\cite{selent2016assistmentsdataset}                      & \begin{tabular}[c]{@{}l@{}}
Multiple (22)\\- 2 / 2 (BC)
\end{tabular}                                            & 
Rand. Unit                                               &
\begin{tabular}[c]{@{}l@{}}
Overall result only\\ $\xmark$ timestamp
\end{tabular}                                            &
Real                                                     &
\begin{tabular}[c]{@{}l@{}}
Education\\ Teaching model
\end{tabular}                                            &
\begin{tabular}[c]{@{}l@{}}
Mostly middle school \\
students (age 11-14) \\ 
in/near MA, USA
\end{tabular}                                            &
\begin{tabular}[c]{@{}l@{}}
2013 --\\ 2015
\end{tabular}                                            &
\begin{tabular}[c]{@{}l@{}}
Peer reviewed\\ data article 
\end{tabular}                                            &
\begin{tabular}[c]{@{}l@{}}
Host: Author website \\ DOI: Unknown\\ Licence: Unknown
\end{tabular}                              \\[1.5em]\hline
\begin{tabular}[c]{@{}l@{}}
A/B Testing Web Analytics Data \\ (From \cite{young2014librarypaper}) 
\end{tabular}                                            &
\cite{young2014librarydataset}                           &
\begin{tabular}[c]{@{}l@{}}
Single\\- 5 / 2 (C)
\end{tabular}                                            &
Group                                                    &
\begin{tabular}[c]{@{}l@{}}
Overall result only \\ $\checkmark$ timestamp
\end{tabular}                                            &
Real                                                     &
\begin{tabular}[c]{@{}l@{}}
Education\\ UX/UI change
\end{tabular}                                            &
\begin{tabular}[c]{@{}l@{}}
Mostly English-\\speaking university \\ library users
\end{tabular}                                            &
\begin{tabular}[c]{@{}l@{}}
May 2013 --\\ Jun 2013
\end{tabular}                                            &
\begin{tabular}[c]{@{}l@{}}
Accompanying\\ dataset to\\ peer-reviewed\\ research article
\end{tabular}                                            &
\begin{tabular}[c]{@{}l@{}}
Host: University library\\ DOI: $\cmark$ (see \cite{young2014librarydataset})\\ Licence: CC BY-SA 4.0
\end{tabular}                              \\[1.5em]\hline
\begin{tabular}[c]{@{}l@{}}
Dataset of two experiments of the\\ application of gamified peer assessment\\ model into online learning environment\\  MeuTutor (From \cite{tenorio2016meututorpaper})
\end{tabular}                                            &
\cite{tenorio2017meututordataset}                        &
\begin{tabular}[c]{@{}l@{}}
Multiple (2)\\- 3+2 / 3+3 (R+C)
\end{tabular}                                            &
Rand. Unit                                               &
\begin{tabular}[c]{@{}l@{}}
Overall result only\\ $\xmark$ timestamp
\end{tabular}                                            &
Real                                                     &
\begin{tabular}[c]{@{}l@{}}
Education\\ Teaching model
\end{tabular}                                            &
\begin{tabular}[c]{@{}l@{}}
High school students \\ in Brazil taking \\ ENEM (age 17)
\end{tabular}                                            &
\begin{tabular}[c]{@{}l@{}}
Jul 2015 --\\ Aug 2015
\end{tabular}                                            &
\begin{tabular}[c]{@{}l@{}}
Peer reviewed\\ data article 
\end{tabular}                                            &
\begin{tabular}[c]{@{}l@{}}
Host: Journal website\\ DOI: $\cmark$ (see \cite{tenorio2017meututordataset})\\ Licence: CC BY 4.0
\end{tabular}                              \\[1.5em]\hline
\begin{tabular}[c]{@{}l@{}}
Udacity Free Trial Screener Experiment\\(From Udacity A/B Testing \\ Course - Final Project \cite{grimesabtestingcourse})
\end{tabular}                                            &
\begin{tabular}[c]{@{}l@{}}
See e.g.\\ \cite{nahar2021udacityabtestdataset,shen2021udacityabtestdataset,wan2019abtestingudacity}
\end{tabular}                                            &
\begin{tabular}[c]{@{}l@{}}
Single\\ - 2 / 4 (C)
\end{tabular}                                            &
Group                                                    &
\begin{tabular}[c]{@{}l@{}}
Daily checkpoint\\ Snapshot
\end{tabular}                                            &
Real                                                     &
\begin{tabular}[c]{@{}l@{}}
Education\\ UX/UI change
\end{tabular}                                            &
\begin{tabular}[c]{@{}l@{}}
Mostly English-\\speaking users
\end{tabular}                                            &
\begin{tabular}[c]{@{}l@{}}
Unknown
\end{tabular}                                            &
\begin{tabular}[c]{@{}l@{}}
Blog posts \&\\
Kaggle notebooks\\
e.g. \cite{messerschmied2019udacityabtesting,rotem2018udacityabtesting,yu2017abtestingforudacity}
\end{tabular}                                            &
\begin{tabular}[c]{@{}l@{}}
Host: Kaggle / GitHub (multiple) \\ DOI: Unknown\\ Licence: Unknown
\end{tabular}                              \\[1.5em]\hline
\begin{tabular}[c]{@{}l@{}}
``Analyse A/B Test Results'' Dataset \\ (From Udacity Online Data \\ Analyst Course - Project 3 \cite{udacityanalystcourse})
\end{tabular}                                            & \begin{tabular}[c]{@{}l@{}}
See e.g.\\ \cite{alreemi2019udacityDANDabtestdataset,dawoud2021udacityDANDabtestdataset,patience2019udacityDANDabtestdataset}
\end{tabular}                                            &
\begin{tabular}[c]{@{}l@{}}
Single\\ - 2 / 1 (B)
\end{tabular}                                            & 
Rand. Unit                                               &
\begin{tabular}[c]{@{}l@{}}
Overall result only \\ Timestamp per RU
\end{tabular}                                            &
Unknown                                                  &
\begin{tabular}[c]{@{}l@{}}
E-commerce\\ UX/UI change
\end{tabular}                                            &
\begin{tabular}[c]{@{}l@{}}
Unknown
\end{tabular}                                            &
\begin{tabular}[c]{@{}l@{}}
Unknown
\end{tabular}                                            &
\begin{tabular}[c]{@{}l@{}}
Blog posts \&\\Kaggle notebooks\\
e.g.~\cite{chen2020udacityDANDabtest,malesova2020udacityDANDabtest,raza2020udacityDANDabtest}
\end{tabular}                                            &
\begin{tabular}[c]{@{}l@{}}
Host: Kaggle / GitHub (multiple) \\ DOI: Unknown\\ Licence: Unknown
\end{tabular}                              \\[1.5em]\hline
\begin{tabular}[c]{@{}l@{}}
Mobile Games A/B Testing \\ with Cookie Cats \\ (DataCamp project \cite{baathmobileproject})
\end{tabular}                                            & 
\begin{tabular}[c]{@{}l@{}}
See e.g.\\ \cite{arpit20cookiecatsdataset,sui19cookiecatdataset,yarkin21cookiecatdataset}
\end{tabular}                                            & 
\begin{tabular}[c]{@{}l@{}}
Single\\- 2 / 3 (BC)
\end{tabular}                                            &
Rand. Unit                                               &
\begin{tabular}[c]{@{}l@{}}
Overall result only\\ $\xmark$ timestamp
\end{tabular}                                            &
Real                                                     &
\begin{tabular}[c]{@{}l@{}}
Gaming\\ Design change
\end{tabular}                                            &
\begin{tabular}[c]{@{}l@{}}
Unknown (likely \\ Facebook users)
\end{tabular}                                            &
\begin{tabular}[c]{@{}l@{}}
Unknown
\end{tabular}                                            &
Kaggle notebook                                          &
\begin{tabular}[c]{@{}l@{}}
Host: Kaggle / Course website\\  DOI: Unknown\\ Licence: Unknown
\end{tabular}                              \\[1.5em]\hline
\begin{tabular}[c]{@{}l@{}}
Experiment Dataset (From \\ DataCamp A/B Testing \\ in R Course \cite{camposabtestingcourse})
\end{tabular}                                            & \cite{camposabtestingexptdataset}                        &
\begin{tabular}[c]{@{}l@{}}
Single\\ - 2 / 1 (B)
\end{tabular}                                            &
Rand. Unit                                               &
\begin{tabular}[c]{@{}l@{}}
Overall result only\\ Timestamp per RU
\end{tabular}                                            &
Synthetic                                                &
\begin{tabular}[c]{@{}l@{}}
Tech\\ UX/UI Change
\end{tabular}                                            &
\begin{tabular}[c]{@{}l@{}}
N/A
\end{tabular}                                            &
\begin{tabular}[c]{@{}l@{}}
N/A
\end{tabular}                                            &
\begin{tabular}[c]{@{}l@{}}
Course notes\\ Blog posts
\end{tabular}                                            &
\begin{tabular}[c]{@{}l@{}}
Host: Course website\\ DOI: Unknown\\ Licence: Unknown
\end{tabular}                              \\[1.5em]\hline
\begin{tabular}[c]{@{}l@{}}
Data Visualization Website - \\ April 2018 (From DataCamp \\ A/B Testing in R Course \cite{camposabtestingcourse})
\end{tabular}                                            &
\cite{camposabtestingwebsitedataset}                     &
\begin{tabular}[c]{@{}l@{}}
Single\\ - 2 / 4 (BR)
\end{tabular}                                            &
Rand. Unit                                               &
\begin{tabular}[c]{@{}l@{}}
Overall result only\\ Timestamp per RU
\end{tabular}                                            &
Synthetic                                                &
\begin{tabular}[c]{@{}l@{}}
Tech\\ UX/UI Change
\end{tabular}                                            &
\begin{tabular}[c]{@{}l@{}}
N/A
\end{tabular}                                            &
\begin{tabular}[c]{@{}l@{}}
N/A
\end{tabular}                                            &
Course notes                                             &
\begin{tabular}[c]{@{}l@{}}
Host: Course website\\ DOI: Unknown\\ Licence: Unknown\\
\end{tabular}                              \\[1.5em]\hline
\begin{tabular}[c]{@{}l@{}}
AB Testing Result (From \\ Customer Analytics and \\ A/B Testing in Python~\cite{grossmanabtestingcourse})
\end{tabular}                                            &
\cite{grossmanabtestingdataset}                        &
\begin{tabular}[c]{@{}l@{}}
Single\\- 2 / 2 (CR)
\end{tabular}                                            &
Rand. Unit                                               &
\begin{tabular}[c]{@{}l@{}}
Overall result only\\ Timestamp per RU
\end{tabular}                                            &
Unknown                                                  &
\begin{tabular}[c]{@{}l@{}}
Tech\\UX/UI Change
\end{tabular}                                            &
\begin{tabular}[c]{@{}l@{}}
Unknown
\end{tabular}                                            &
\begin{tabular}[c]{@{}l@{}}
Jan 2014 --\\ Jan 2018
\end{tabular}                                            &
\begin{tabular}[c]{@{}l@{}}
Course notes 
\end{tabular}                                            &
\begin{tabular}[c]{@{}l@{}}
Host: Course website\\ DOI: Unknown\\ Licence: Unknown\\
\end{tabular}                              \\[1.5em]\hline
Grocery Website Data for AB Test                         & \cite{klimonova2020grocerydataset}                       &
\begin{tabular}[c]{@{}l@{}}
Single\\ - 2 / 1 (B)
\end{tabular}                                            &
Rand. Unit                                               &
\begin{tabular}[c]{@{}l@{}}
Overall result only\\ $\xmark$ timestamp
\end{tabular}                                            &
Unknown                                                  &
\begin{tabular}[c]{@{}l@{}}
E-commerce\\ UX/UI change
\end{tabular}                                            &
\begin{tabular}[c]{@{}l@{}}
Unknown
\end{tabular}                                            &
\begin{tabular}[c]{@{}l@{}}
Unknown
\end{tabular}                                            &
Kaggle notebook                                          &
\begin{tabular}[c]{@{}l@{}}
Host: Kaggle\\ DOI: Unknown\\ Licence: Unknown\\
\end{tabular}                              \\[1.5em]\hline
\begin{tabular}[c]{@{}l@{}}
Ad A/B Testing\\ (aka SmartAd AB Data)
\end{tabular}                                            &
\cite{emmanuel2020addataset}                             &
\begin{tabular}[c]{@{}l@{}}
Single\\ - 2 / 2 (B)
\end{tabular}                                            &
Rand. Unit                                               &
\begin{tabular}[c]{@{}l@{}}
Overall result only\\ Timestamp per RU
\end{tabular}                                            & 
Unknown                                                  &
\begin{tabular}[c]{@{}l@{}}
Media \& Ads\\ Display ads
\end{tabular}                                            &
\begin{tabular}[c]{@{}l@{}}
Unknown
\end{tabular}                                            &
\begin{tabular}[c]{@{}l@{}}
Jul 2020
\end{tabular}                                            &
\begin{tabular}[c]{@{}l@{}}
Kaggle notebook
\end{tabular}                                            &
\begin{tabular}[c]{@{}l@{}}
Host: Kaggle\\ DOI: N/A\\ Licence: CC BY-SA 3.0\\
\end{tabular}                              \\[1.5em]\hline
\begin{tabular}[c]{@{}l@{}}
Synthetical A/B-Tests
\end{tabular}                                            &
\cite{ay2020syntheticalabdataset}                        &
\begin{tabular}[c]{@{}l@{}}
Multiple (\numprint{25856})\\- 2 / 1 (R)
\end{tabular}                                            &
Group.                                                   &
\begin{tabular}[c]{@{}l@{}}
Overall result only\\ $\xmark$ timestamp
\end{tabular}                                            &
Synthetic                                                &
\begin{tabular}[c]{@{}l@{}}
N/A
\end{tabular}                                            &
\begin{tabular}[c]{@{}l@{}}
N/A
\end{tabular}                                            &
\begin{tabular}[c]{@{}l@{}}
N/A
\end{tabular}                                            &
\begin{tabular}[c]{@{}l@{}}
Kaggle notebook 
\end{tabular}                                            &
\begin{tabular}[c]{@{}l@{}}
Host: Kaggle\\ DOI: Unknown \\ Licence: CDLA-Sharing-1.0
\end{tabular}                              \\[1.5em]
\end{tabular}
\end{adjustbox}

\caption[Results from the first ever survey of OCE datasets.]{Results from the first ever survey of OCE datasets. The 13 datasets identified are placed on the four taxonomy dimensions defined in Section~\ref{sec:oced_taxonomy}, together with the additional attributes recorded. In the Experiment Count column, a second-line value (corresponding to Variant/Metric Count) of ``$x$ / $y$  (BCR)'' means the dataset features $x$ variants and $y$ metrics in each experiment, with the metrics based on \textbf{B}inary, \textbf{C}ount, and \textbf{R}eal-valued responses. In the Response Granularity and Time Granularity columns, Randomisation Unit is abbreviated as RU or Rand. Unit.
Many resources are accessed via links to non-scholarly articles -- blog plots, Kaggle dataset pages, and GitHub repositories. These resources may not persist over time.}
\label{tab:oced_survey}
\end{table}
\end{landscape}

\section{Matching Dataset Taxonomy with Statistical Tests}
\label{sec:oced_mapping}

Specifying the data requirements (or structure) and performing statistical tests are perhaps two of the most common tasks carried out by data scientists. However, the link between the two processes is seldom mapped out explicitly. It is all too common to consider from scratch the question, ``I need to run this statistical test, how should I format my dataset?'' (or more controversially, ``I have this dataset, what statistical tests can I run?''~\cite{kerr1998harking}) for every new project/application, despite the list of possible dataset dimensions and statistical tests remaining essentially the same.

We aim to speed up the process above by describing
what summary statistics are required to perform common statistical tests in digital experiments and link the statistics back to the taxonomy dimensions defined in Section~\ref{sec:oced_taxonomy}. The exercise is similar to identifying the sufficient statistic(s) for a statistical model~\cite{fisher1922mathematicalfoundations}. However, the identification is done for the encapsulating statistical inference procedure, with a practical focus on data dimension requirements. We do so by stating the formula used to calculate the corresponding effect sizes and test statistics and observing the summary statistics required in common. The general approach also enables one to apply the resultant mapping to any experiments that involve a two-sample statistical test, including offline experiments and experiments without a randomised
control. We will refrain from discussing the full model assumptions and their applicability for brevity. Instead, we point readers to the relevant work in the literature.

\subsection{Effect size and Welch's \texorpdfstring{$t$}{t}-test}
\label{sec:oced_mapping_ttest}

We first look at the requirements to calculate effect sizes (using Cohen's $d$) and perform the most basic statistical test (using Welch's $t$-statistic). Recall from Sections~\ref{sec:statstest_nonexamples_effectsize} and~\ref{sec:statstest_nhstexamples_ttest} that the formulas to calculate these two quantities are
\begin{align}
    d &\, = \frac{\bar{Y} - \bar{X}}{\sqrt{\frac{(n - 1) s^2_X + (m - 1) s^2_Y}{n + m - 2}}} \,,
    \tag{*\ref{eq:statstest_cohens_d_def}} \\[0.5em]
    T &\, = \frac{\bar{Y} - \bar{X} - \Delta}{ \sqrt{\frac{s^2_X}{n} + \frac{s^2_Y}{m}}} \,,
    \tag{*\ref{eq:statstest_tstatistic_def_twosample}}
\end{align}
where $(\bar{X}, \bar{Y})$, $(s^2_X, s^2_Y)$, and $(n, m)$ are the means, variances, and counts of the samples $X$ and~$Y$, respectively.

We observe that we require only six \emph{group-level} summary statistics to calculate the two quantities above: two means, two variances, and two counts. 
Besides the counts, the exact list of quantities required depends on the distributional assumptions used in a test, which dictates what the sufficient statistics are.\footnote{Welch's $t$-test assumes normally distributed data and hence requires the sample mean and variance for both samples as they are a joint sufficient statistic for normal distributions with unknown variance. For decision metrics derived from binary responses, we only require the sample mean as it is a sufficient statistic for a Bernoulli distribution.}
We call these quantities \emph{Dimension Zero (D0)} quantities, as they are the bare minimum required to run a statistical test. These quantities will be expanded along the taxonomy dimensions defined in Section~\ref{sec:oced_taxonomy}.

\paragraph{Cluster randomisation / dependent data}
The sample variance estimates ($s^2_X$, $s^2_Y$) may be biased when cluster randomisation is involved. Using the ``free delivery'' banner example again, instead of randomly assigning each individual user to the control and treatment groups, the business may randomly assign postcodes to the two groups, with all users from the same postcode getting the same version of the website. In this case, user responses may become correlated, violating the independence assumptions in statistical tests. Common workarounds, including the use of bootstrap~\cite{bakshy13uncertainty} and the Delta method~\cite{deng18applying} (see Sections~\ref{sec:rade_oce_statsmethods} and~\ref{sec:rade_abv}), generally require access to \emph{sub-randomisation unit} responses.

\subsection{Experiments with adaptive stopping}
\label{sec:oced_mapping_adaptive_stopping}

As discussed in Section~\ref{sec:oced_introduction}, experiments with adaptive stopping are getting increasingly popular among the digital experimentation community. Here, we motivate the data requirement for statistical tests in this domain by looking at the quantities required to calculate the test statistics for a mixture sequential probability ratio test (mSPRT, see Section~\ref{sec:statstest_sequential} or~\cite{johari17peeking}) and a Bayesian hypothesis test using Bayes factor (see Section~\ref{sec:statstest_bayesian} or~\cite{deng16continuous}), two popular approaches in digital experimentation. 
Many other tests support adaptive stopping~\cite{miller10hownot,wald1945sequential}, though the data requirements should be largely identical in terms of the dimensions defined in Section~\ref{sec:oced_taxonomy}.

We first recall from Section~\ref{sec:statstest_sequential} that running a mSPRT with a normal mixing distribution ${H=\mathcal{N}(\theta_0, \tau^2)}$ involves calculating the following test statistic upon observing the first~$n$ samples from $X$ and $Y$:
\begin{align}
    \tilde{\Lambda}^{H, \theta_0}_{n}
    = \sqrt{\frac{\sigma^2_X + \sigma^2_Y}{\sigma^2_X + \sigma^2_Y + n\tau^2}}
    \exp{\left(\frac{n^2\tau^2(\bar{Y}_n - \bar{X}_n - \theta_0)^2}{2(\sigma^2_X + \sigma^2_Y)(\sigma^2_X + \sigma^2_Y + n\tau^2)}\right)} \,,
    \tag{*\ref{eq:statstest_sequential_msprt_teststatistic}}
\end{align}
where $\bar{X}_n = n^{-1}\sum_{i=1}^{n}X_i$ and $\bar{Y}_n = n^{-1}\sum_{j=1}^{n}Y_j$ represent the sample mean of $X$ and $Y$ up to sample $n$, respectively.

For Bayesian hypothesis tests, we recall from Section~\ref{sec:statstest_bayesian} that the test statistic~$\delta_{n, m}$ and effective sample size $E_{n,m}$ can be obtained by decomposing the Wald test statistic upon observing the first~$n$ samples from~$X$ and $m$ samples from $Y$:
\begin{align}
    \sqrt{W_{n, m}} = 
    \Bigg(
    \underbrace{\frac{\bar{Y}_m - \bar{X}_n}{\sqrt{\big(\frac{s^2_X}{n} + \frac{s^2_Y}{m}\big) / \left(\frac{1}{n}+ \frac{1}{m}\right)}}}_{\triangleq\, \delta_{n, m}} -
    \underbrace{\frac{\Delta}{\sqrt{\big(\frac{s^2_X}{n} + \frac{s^2_Y}{m}\big) / \left(\frac{1}{n}+ \frac{1}{m}\right)}}}_{\triangleq\, \Delta^{\!\circ}}
    \Bigg)
    \underbrace{\frac{1}{\sqrt{\frac{1}{n} + \frac{1}{m}\vphantom{\big(\frac{s^2_X}{n}\big)}}}}_{\triangleq\sqrt{E_{n, m}}} \;,
    \tag{*\ref{eq:statstest_bayesian_wald_decomposition}}
\end{align}
where $\Delta^{\!\circ}$ is the standardised effect size. $\delta_{n, m}$ and $E_{n,m}$ are then used to calculate the Bayes factor under the hypothesis pair $H_0: \Delta^{\!\circ} = \theta_0$ and ${H_1: \Delta^{\!\circ} \sim \mathcal{N}(\theta_0, V^2)}$:
\begin{align}
    (BF_{10})_{n, m} = \frac{f(\delta_{n,m} \,|\, H_1)}{f(\delta_{n,m} \,|\, H_0)} = \frac{\phi\left(\delta_{n,m};\, \theta_0,\, V^2 + \frac{1}{E_{n, m}}\right)}{\phi\left(\delta_{n,m};\, \theta_0,\, \frac{1}{E_{n, m}}\right)} \;,
    \tag{*\ref{eq:statstest_bayesian_bf_normal_example}}
\end{align}
where $\phi(\,\cdot\,; \mu, \sigma^2)$ is the PDF of a normal distribution with mean $\mu$ and variance $\sigma^2$.

During an experiment with adaptive stopping, we calculate the test statistics stated above many times for different $n$ and $m$. This means a dataset can only support the running of such experiments if it contains \emph{intermediate checkpoints} for the counts ($n$, $m$) and the means ${(\bar{X}_n, \bar{Y}_m)}$, ideally \emph{cumulative} from the start of the experiment. One often also requires the variances at the same time points (see below). The only exception to the dimensional requirement above is where the dataset contains responses at a \emph{randomisation unit} or finer level of granularity and, despite recording the \emph{overall results only}, has a \emph{timestamp per randomisation unit}. Under this special case, we can still construct the cumulative means ($\bar{X}_n$, $\bar{Y}_m$) for all relevant values of $n$ and $m$ by ordering the randomisation units by their associated timestamps.

\paragraph{Learning the effect size distribution (hyper)parameters}
The two tests introduced above feature some hyperparameters ($\tau^2$ and $V^2$) that must be specified or learned from data. These parameters characterise the prior belief of the effect size distribution, which will be the most effective if it ``matches
the distribution of true effects across the experiments a user runs''~\cite{johari17peeking}. Common parameter estimation procedures~\cite{adams18empirical,azevedo19empirical,guo20empirical} require results from \emph{multiple related experiments}.

\paragraph{Estimating the response variance}
In the equations above, the response variance of the two samples, $\sigma^2_X$ and $\sigma^2_Y$ (or $s^2_X$ and $s^2_Y$), are assumed to be known before we observe all the samples. In practice, we often use the plug-in empirical estimates $(s_X^2)_n$ and $(s_Y^2)_m$ -- the sample variances for the first~$n$ samples from $X$ and $m$ samples from $Y$, respectively. Thus, the data dimensional requirement is identical to that of the counts and means as discussed above. We will also require a \emph{sub-randomisation unit} response granularity when the plug-in estimates are biased due to dependent data (see  Section~\ref{sec:oced_mapping_ttest}).

\subsection{Non-parametric tests}

We also briefly discuss the data requirements for non-parametric tests.
Recall from Section~\ref{sec:statstest_nonparametric} that one of the most commonly used non-parametric tests in digital experimentation, the Mann-Whitney $U$-test, calculates the following test statistic:
\begin{align}
    U = \sum_{i=1}^{n} \sum_{j=1}^{m} S(X_i, Y_j), \quad \textrm{where }S(X, Y)=
    \left\{\begin{array}{ll}
    1 & \textrm{if $Y < X$,}\\
    1/2 & \textrm{if $Y = X$,}\\
    0 & \textrm{if $Y > X$.}\\
    \end{array}\right.
    \tag{*\ref{eq:statstest_mannwhitney_teststatistic_def}}
\end{align}
While a rank-based method is available for large $n$ and $m$, both methods require knowledge of all the $X_i$ and $Y_j$. This requirement applies to many other non-parametric tests (see Section~\ref{sec:statstest_nonparametric}). The observation suggests that a dataset can only support a non-parametric test if it at least provides responses at a \emph{randomisation unit} level.

We conclude by showing how we can combine the individual data requirements above to obtain the requirement to design or run experiments for more complicated statistical tests. This is possible due to the orthogonal design of the taxonomy dimensions. Consider an experiment with adaptive stopping using Bayesian non-parametric tests (e.g., with a P\'{o}lya Tree prior~\cite{chen2014bayesian,holmes2015twosample}). It involves a non-parametric test, requiring responses at a \emph{randomisation unit} level. It computes multiple Bayes factors for adaptive stopping and hence requires \emph{intermediate checkpoints} for the responses (or a timestamp for each randomisation unit). Finally, it needs to learn the hyperparameters of the P\'{o}lya Tree prior, which requires \emph{multiple related experiments}. The substantial data requirement along 3+ dimensions perhaps explains the lack of relevant digital experiment datasets and the tendency for experimenters to use simpler statistical tests for the day-to-day design or running of digital experiments (or both).

\section{A Novel Dataset for Experiments with Adaptive Stopping}
\label{sec:oced_dataset}

We finally introduce the ASOS Digital Experiments Dataset, which we believe is the first public dataset that supports the end-to-end design and running of digital experiments with adaptive stopping. We motivate why this is the case, provide a light description of the dataset (and a link to the more detailed accompanying datasheet),\footref{footnote:oced_dataset_link} and showcase the capabilities of the dataset via a series of experiments. We also discuss the ethical implications of releasing this dataset.

Recall from Section~\ref{sec:oced_mapping_adaptive_stopping} that in order to support the end-to-end design and running of experiments with adaptive stopping, we require a dataset that 
\begin{enumerate}
    \item Includes \emph{multiple} related experiments;
    \item Is \emph{real}, so that any parameters learned are reflective of the real-world scenario;
\end{enumerate}
and either
\begin{enumerate}[leftmargin=22pt]
    \item[3a.] Contains \emph{intermediate checkpoints} for the summary statistics during each experiment (i.e., time-granular), or
    \item[3b.] Contains responses at a \emph{randomisation unit} granularity with a timestamp for each randomisation unit (i.e. response-granular with timestamps).
\end{enumerate}

None of the datasets surveyed in Section~\ref{sec:oced_survey} meet all three criteria. While the Upworthy~\cite{matias2021upworthydataset}, ASSISTments~\cite{selent2016assistmentsdataset}, and MeuTutor~\cite{tenorio2017meututordataset} datasets meet the first two criteria, they all fail to meet the third.\footnote{All three report the overall results only and hence are not time-granular. The Upworthy dataset reports group-level statistics and hence is not response-granular. The ASSISTments and MeuTutor datasets are response-granular but lack the timestamp to order the samples.} The Udacity Free Trial Screener Experiment dataset meets the last two criteria by having results from a real experiment with daily snapshots of the decision metrics (and hence time-granular), which supports the running of an experiment with adaptive stopping. However, the dataset only contains a single experiment, which does not help in learning the effect size distribution (the design).

The ASOS Digital Experiments Dataset contains results from OCEs run by a business unit within ASOS.com, a global online fashion retail company. In terms of the taxonomy defined in Section~\ref{sec:oced_taxonomy}, 
the dataset contains \emph{multiple} (78) \emph{real} experiments, with two to five variants in each experiment and four decision metrics based on binary, count, and real-valued responses. The results are aggregated on a \emph{group} level, with \emph{daily or 12-hourly checkpoints} of the metric values \emph{cumulative} from the start of the experiment. The dataset design meets all the three criteria stated above and thus differentiates itself from other public datasets.

We provide readers with an accompanying datasheet (based on~\cite{gebru2020datasheets}) that provides further information about the dataset.  We also host the dataset on Open Science Framework to ensure it is easily discoverable and can be preserved long-term.\footref{footnote:oced_dataset_link} It is worth noting that the dataset is released with the intent to support development in the statistical methods required to run digital experiments. The experiment results shown in the dataset are not representative of ASOS.com's overall business operations, product development, or experimentation program operations, and no conclusion of such should be drawn from this dataset.

\subsection{Potential use cases}

\begin{figure}
  \includegraphics[width=0.243\textwidth, trim = 2.5mm 2.5mm 2.5mm 2.5mm, clip]{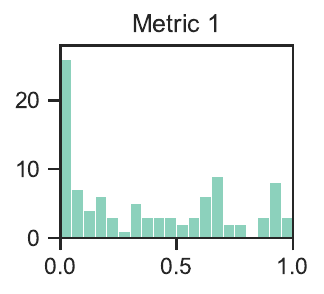}
  \includegraphics[width=0.243\textwidth, trim = 2.5mm 2.5mm 2.5mm 2.5mm, clip]{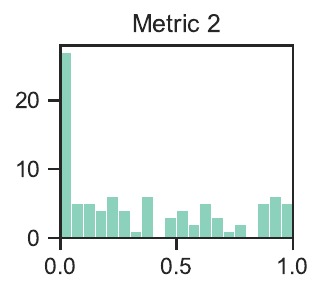}
  \includegraphics[width=0.243\textwidth, trim = 2.5mm 2.5mm 2.5mm 2.5mm, clip]{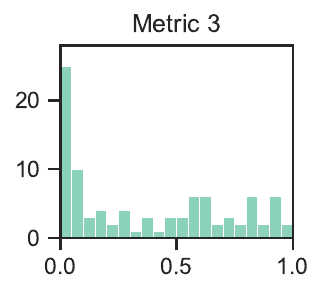}
  \includegraphics[width=0.243\textwidth, trim = 2.5mm 2.5mm 2.5mm 2.5mm, clip]{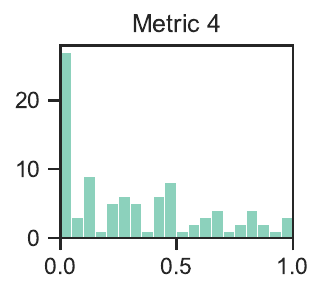}
  \caption[Distribution of $p$-values attained by the 99 OCEs in the ASOS Digital Experiment Dataset using Welch's $t$-tests, split by decision metrics.]{Distribution of $p$-values attained by the 99 OCEs in the ASOS Digital Experiment Dataset using Welch's $t$-tests, split by decision metrics. The leftmost bar in each histogram represents experiments with $p < 0.05$. Here, we treat OCEs with multiple variants as multiple independent OCEs.}
  \label{fig:oced_experiment_ttest_pvalue}
  \vspace*{\baselineskip}
\end{figure}

\paragraph{Meta-analyses}
The multi-experiment nature of the dataset enables one to perform meta-analyses. A simple example is characterising the distribution of $p$-values (under Welch's $t$-test) across all experiments (see Figure~\ref{fig:oced_experiment_ttest_pvalue}). We observe that roughly a quarter of experiments in this dataset attain~$p < 0.05$ and attribute this to the fact that what we experiment in digital experiments is often guided by what domain experts think may have an impact. That said, we invite external validation on whether there is evidence for data dredging using, e.g.,~\cite{miller2020metaanalytic}.

\paragraph{Design and running of experiments with adaptive stopping} 
We also demonstrate that the dataset can indeed support digital experiments with adaptive stopping by performing a mixed sequential probability ratio test (mSPRT) and a Bayesian hypothesis test via the Bayes factor for each experiment and metric. This requires learning the hyperparameters $\tau^2$ and $V^2$. We learn, for each metric, a na\"{i}ve estimate for~$V^2$ by collating the~$\delta_{n, m}$ (see Equation~\eqref{eq:statstest_bayesian_wald_decomposition}) at the end of each experiment and taking their sample variance. This yields the estimates \texttt{1.30e-05}, \texttt{1.07e-05}, \texttt{6.49e-06}, and \texttt{5.93e-06} for the four metrics respectively. 

For $\tau^2$, we learn near-identical na\"{i}ve estimates by collating the value of Cohen's~$d$ (see Equation~\eqref{eq:statstest_cohens_d_def}) instead. However, as $\tau^2$ captures the spread of unstandardised effect sizes, we specify in each test $\tau^2 = d \cdot (s^2_X)_n$, where~$(s^2_X)_n$ is the sample variance of all responses up to the $n^{\textrm{th}}$ observation in that particular experiment. 
The Bayesian tests also require a prior belief in the null hypothesis being true ($\mathbb{P}(H_0)$) -- we set it to $0.75$ based on what we observed in the $t$-tests above.

\begin{figure}
  \begin{center}
  \includegraphics[height=0.22\textwidth, trim =2.5mm 2.5mm 2.5mm 2.5mm, clip]{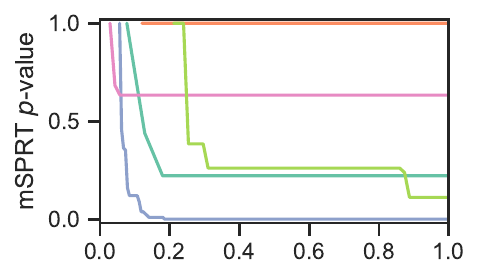}
  \includegraphics[height=0.221\textwidth, trim =2.5mm 2.5mm 2.5mm 2.5mm, clip]{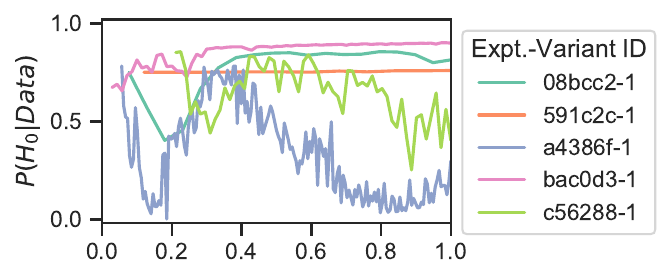}
  \end{center}
  \vspace*{-0.5em}
  \caption[Change in $p$-values in a mixed sequential probability ratio test and posterior belief in the null hypothesis in a Bayesian hypothesis test during the experiment for five experiments selected at random.]{Change in (left) $p$-values in a mixed sequential probability ratio test ($\tau^2 = \texttt{5.92e-06}$) and (right) posterior belief in the null hypothesis ($\mathbb{P}(H_0 | \textrm{data})$) in a Bayesian hypothesis test ($V^2 = \texttt{5.93e-06}$, $\mathbb{P}(H_0)=0.75$) during the experiment for five experiments selected at random. The experiment duration (x-axis) is normalised by its overall runtime. Only results for Metric~4 are shown.}
  \label{fig:oced_experiment_msprt_pvalue_bf}
  \vspace*{\baselineskip}
\end{figure}

We then calculate the $p$-value in mSPRT and the posterior belief in the null\linebreak hypothesis~($\mathbb{P}(H_0|\textrm{data})$) in the Bayesian test for each experiment and metric at each daily/12-hourly checkpoint, following the procedures stated in~\cite{johari17peeking} 
and~\cite{deng16continuous,kass95bayes}, respectively. We plot the results for five experiments selected at random in Figure~\ref{fig:oced_experiment_msprt_pvalue_bf}. It shows that the $p$-value for an mSPRT is monotonically non-increasing. At the same time, the posterior belief for a Bayesian test can fluctuate depending on the effect size observed so far.

\paragraph{A quasi-benchmark for adaptive stopping methods}
Real digital experiments, unlike machine learning tasks, generally do not have a notion of ground truth. Nonetheless, the dataset can still be used as a quasi-ground truth to compare two hyperparameter settings of the same adaptive stopping method or between two adaptive stopping methods. 
For example, we can treat the reject / not reject verdict from a Welch’s $t$-test at the end of an experiment as the “ground truth” and compare it to the reject / not reject verdict of an mSPRT at different experiment stages.
This yields many ``confusion matrices'' over different stages of individual experiments, where a “Type I error” corresponds to cases in which a Welch’s $t$-test fails to reject the null hypothesis and an mSRPT rejects the null hypothesis. A confusion matrix for the end of each experiment can be seen in  Table~\ref{fig:oced_experiment_comparison}. As the dataset was collected without early stopping, it allows us to perform sensitivity analysis and optimisation on the hyperparameters of mSPRT under what can be construed as a “precision-recall” trade-off of statistically significant treatments.

\paragraph{Other use cases} The time series nature of this dataset enables one to detect bias (of the estimator) across time, e.g., those caused by concept drift or feedback loops. In the context of digital experiments,~\cite{chen2019automatic} described several methods to detect invalid experiments over time that may be run on this dataset. 
Moreover, with the time series spanning multiple experiments and decision metrics, one can also learn the correlation structure across experiments, decision metrics, and time~\cite{pougetabadie2019variance,wang2020causal}.

\begin{table}
   \onehalfspacing
   \caption{Comparing the number of reject / not reject verdicts given by Welch's $t$-test and an mSPRT at the end of an experiment for all four metrics.}
   \label{fig:oced_experiment_comparison}
   \begin{center}
   \begin{tabular}{c|cccc}
    $t$-test & 
      \multicolumn{2}{c}{Rejects $H_0$} & 
      \multicolumn{2}{c}{Does not reject $H_0$}  \\\hline
    mSPRT & 
      Rejects $H_0$ & 
      Does not reject $H_0$ & 
      Rejects $H_0$ & 
      Does not reject $H_0$ \\\hline
    Metric 1 & 19 & 7  & 2  & 71 \\
    Metric 2 & 20 & 7  & 18 & 49 \\
    Metric 3 & 16 & 9  & 6  & 63 \\
    Metric 4 & 16 & 11 & 4  & 63 
  \end{tabular}
  \end{center}
  \vspace*{\baselineskip}
\end{table}

\subsection{Ethical considerations}

We finally discuss the ethical implications of releasing the dataset, touching on data protection and anonymisation, potential misuse, and the ethical considerations for running digital experiments in general.

\paragraph{Data protection and anonymisation}
The dataset records aggregated activities of hundreds of thousands or millions of website users for business measurement purposes. Hence, it is impossible to identify a particular user. Moreover, to minimise the risk of disclosing business sensitive information, all experiment context is either removed or anonymised. One should not be able to tell who is in an experiment, when the experiment is run, what treatment the experiment involves, and what decision metrics are used. We refer readers to the accompanying datasheet\footref{footnote:oced_dataset_link} for further details in this area.

\paragraph{Potential misuses}
A digital experiment dataset, no matter how anonymised it is, reflects the behaviour of its participants under a specific application domain and time. We urge potential users of this dataset to exercise caution when attempting to generalise the learnings. 
It is important to emphasise that the learnings differ from the statistical methods and processes demonstrated on this dataset. We believe the latter are generalisable, i.e., they can be applied to other datasets with similar data dimensions regardless of the datasets' application domain, target demographics, and temporal coverage, and appeal for potential dataset users to focus on such.

One example of generalising the learnings is using this dataset as a full performance benchmark. As discussed above, this dataset does not have a notion of ground truth, and any quasi-ground truths constructed are themselves a source of bias to estimators. Thus, experiment design comparisons need to be considered at a theoretical level~\cite{liu2020evaluation}. 
Another example will be directly applying the value of hyperparameter(s) obtained while training a model on this dataset to another dataset. While this may work for similar application domains, the less similar they are, the less likely the hyperparameters learned will transfer. This may risk introducing bias, both in the estimator and in fairness.

\paragraph{Running digital experiments in general}
The dataset is released to support experiments with adaptive stopping, which will enable a faster experimentation cycle. The ethical concerns will naturally mount as we run more digital experiments, which are ultimately human subjects research. We reiterate the importance of the following three principles when we design and run experiments~\cite{united1978belmont,us2018office}: respect for persons, beneficence (properly assess and balance the risks and benefits), and justice (ensure participants are not exploited), and refer readers to Chapter 9 of~\cite{kohavi20trustworthy} and its references for further discussions in this area.

\section{A Brief Recap}
\label{sec:oced_conclusion}

Digital experiments are a powerful tool for online organisations to assess the impact of their digital products and services. To safeguard future methodological development in the area, it is vital that we have access to and a systematic understanding of relevant datasets arising from real experiments.
We described the result of the first ever survey on publicly available OCE datasets. We also provided a dimensional taxonomy that links the data collection and statistical test requirements for digital experiments in general. In addition, we released the first ever dataset that can support digital experiments with adaptive stopping, which design is grounded on a theoretical discussion between the taxonomy and statistical tests. Via extensive experiments, we also showed that the dataset is capable of addressing the identified gap in the literature. 

Our work on surveying, categorising, and enriching the publicly available digital experiment datasets is just the beginning, and we invite the community to join in the effort. As discussed above, we have yet to see a dataset that can support methods dealing with correlated data due to cluster randomisation, or the end-to-end design and running of experiments with adaptive stopping using Bayesian non-parametric tests. We also see ample opportunity to generalise the survey to cover datasets arising from uplift modelling tasks, quasi-experiments, and observational studies. Finally, we can further expand the taxonomy, which already supports datasets from all experiments, with extra dimensions (e.g., number of features to support stratification, control variate, and uplift modelling methods) as the area matures.
\cleardoublepage
\chapter{Recent Advances in Digital Experimentation Methods}
\label{chap:rade}

Part of the chapter (Section~\ref{sec:rade_abv}) is adapted from the research paper ``\textit{Measuring e-Commerce Metric Changes in Online Experiments}'', presented at \textit{ACM Web Conference 2023 (WWW '23)}~\cite{liu2023measuring}.

\section{Motivation}

After establishing the statistical tests and datasets required to run a digital experiment in Chapters~\ref{chap:statstest} and~\ref{chap:oced}, we can now describe what a digital experiment entails. To begin, we outline the causal reasoning behind experiments. Establishing causation (i.e., making conclusions such as ``this treatment causes that business metric to change'') is a step further than establishing correlation\footnote{Also known as \emph{association} within the causal inference community~\cite{hernan2020causal}.} that a statistical test may do. We thus require additional theoretical grounding to represent the concept of causation. This chapter provides an initial view of the potential outcomes framework / Rubin causal model and its associated language.\footnote{Other tools and frameworks exist, e.g., directed acyclic graphs and do-calculus, which are arguably more generalisable. Here, we prioritise the ease of understanding.}

Using the potential outcomes framework, we can describe different classes of experiments. The classes, primarily characterised by the level of control an experimenter has, are randomised controlled trials (RCTs), quasi-experiments, and natural experiments. Having a classification of digital experiments enables us to highlight the different methodological and implementation challenges faced by digital experimenters on each front. 

Given the availability of excellent and up-to-date review articles that deal with established and emerging approaches in each area, it is perhaps not the most effective for us to cover yet another systemic review of digital experimentation methods. Thus, we provide a more lightweight, narrative overview beyond introducing the review articles. We will describe the major challenges and highlight several prominent works for each challenge as entry points for readers interested in further exploration.

As we continue to place equal emphasis on the theory and practice of digital experimentation methods, we also enrich the overview with case studies. These case studies provide spotlights on digital randomised controlled trials with dependent responses and quasi-experiments that assign treatments based on geographical regions. In addition to providing evidence on the extent of these challenges, the case studies also detail the practical considerations when implementing the methods to address them.

The rest of the chapter is organised as follows. We introduce the basics of the potential outcomes framework in Section~\ref{sec:rade_pof}. Using its notation and rules, we outline the three classes of experiments in Section~\ref{sec:rade_classes}. We then interleave the reviews on advances in each class of experiments (RCTs in Section~\ref{sec:rade_oce} and quasi-experiments/natural experiments in Section~\ref{sec:rade_causalinference}) with spotlights on methods that deal with dependent responses (Section~\ref{sec:rade_abv}) and geo-experiments (Section~\ref{sec:rade_causalinference_geoexperiments}).

\section{From Correlation to Causation -- The Potential Outcomes Framework}
\label{sec:rade_pof}

Readers may recall that in Chapter~\ref{chap:statstest}, we noted that statistical tests alone are insufficient to establish causal relationships between treatments and changes in responses. In this section, we bridge the gap between correlation/association thinking (as featured in statistical tests) and causal thinking (as featured in experiments and causal inference) by introducing the most basic concepts of the potential outcomes framework. In subsequent sections, we will refer to and extend these concepts.

This introduction extends the notations in Chapter~\ref{chap:statstest} and takes a more leisurely pace than other introductions to background materials in the thesis. This is deliberate -- in the author's experience, it is a big jump to fully appreciate the framework's premises, rules, and notations, especially for those who are used to dealing only with randomised control trials.
Readers familiar with the potential outcomes framework and the $Y_i(0)$-style notation in causal inference may consider skimming through the section. They may also find other expositions on the framework~\cite{deng2021causal,hernan2020causal,imbens2015causal} useful.

\paragraph{Observed responses} 
In Chapter~\ref{chap:statstest}, we often deal with two sets of responses: $X_1, \cdots, X_n$ and $Y_1, \cdots, Y_m$. In the general context of statistical testing, the responses can arise from two arbitrary sources, e.g., the height of males vs females in a survey or ice cream sales vs the number of drowning incidents~\cite{mumford2013correlation}. 
In the context of an experiment, the two sets of responses would have come from experimental units that are assigned and subjected to two different treatments.\footnote{We focus on the two-treatment case for the rest of the chapter. The numerical designation of the treatment assignments, i.e., 0, 1, $\cdots$, eases the extension of the framework to cover multiple treatments.} Formally, by modelling the treatment \textbf{a}ssignment using a binary r.v. $\mathcal{A}$ and denoting the two possible assignments $\mathcal{A} = 0$ and $\mathcal{A} = 1$, we can represent the responses under the additional context as
\begin{align}
X_i \,|\, \mathcal{A} = 0  \quad\textrm{ and }\quad Y_j \,|\, \mathcal{A} = 1 \,.
\label{eq:rade_pof_response_given_treatment}
\end{align}
Assuming all $X_i$ and $Y_j$ are identically distributed copies of their respective populations, by taking the sample mean of the two sets of responses, we obtain an unbiased estimator of the expected response of the overall populations $X$ and $Y$ given the corresponding treatments:
\begin{align}
\mathbb{E}(X \,|\, \mathcal{A} = 0) \quad\textrm{ and }\quad \mathbb{E}(Y \,|\, \mathcal{A} = 1) \,.
\label{eq:rade_pof_expected_response_given_treatment}
\end{align}

\paragraph{Counterfactual responses} 
Note $X_i$ and $Y_j$ in Expression~\eqref{eq:rade_pof_response_given_treatment} are merely the \emph{observable} responses from their associated experimental units. We do not observe how the experimental unit responding $X_i$ under~${\mathcal{A} = 0}$ would have responded had ${\mathcal{A} = 1}$, nor do we observe how the unit responding $Y_j$ under $\mathcal{A} = 1$ would have responded had $\mathcal{A} = 0$. These ``would haves'' are known as the \emph{counterfactual} responses and are essential to causal reasoning -- understanding the impact of a treatment requires understanding both the outcomes when the treatment is and is not prescribed.

The potential outcome framework unifies the observed and counterfactual responses from the same experimental unit by treating all equally as potential responses (or \emph{outcomes}) and giving each potential response a separate designation. In the two-treatment case, they are denoted as
\begin{align}
  X_i \triangleq
  \begin{cases}
    X_i(0) & \textrm{ if } \mathcal{A} = 0 \\
    X_i(1) & \textrm{ if } \mathcal{A} = 1 \,,
  \end{cases}
  \qquad
  Y_j \triangleq
  \begin{cases}
    Y_j(0) & \textrm{ if } \mathcal{A} = 0 \\
    Y_j(1) & \textrm{ if } \mathcal{A} = 1 \,.
  \end{cases}
  \label{eq:rade_pof_counterfactual_response}
\end{align}
More generally, we can write $X_i = X_i(a)$ if~$\mathcal{A} = a$ and $Y_j = Y_j(a)$ if $\mathcal{A} = a$, where~${a \in \{0, 1\}}$.\footnote{In~\cite{hernan2020causal}, the same quantities are denoted $X_i^{a}$ and $Y_j^{a}$, sometimes with the numeric value of $a$ explicitly specified, e.g., $Y_j^{a=0}$ for disambiguation.} The framework features a missing data problem -- some potential responses (generally one max per experimental unit) are observable, and others must be estimated. 

Using the framework, we can define the causal impact of a treatment over another more formally. In a two-treatment setting, the treatment effect (of treatment 1 over 0) on a single experimental unit that responds $X_i$ or $Y_j$ is
\begin{align}
    \Delta_{X_i} = X_i(1) - X_i(0), \quad
    \Delta_{Y_j} = Y_j(1) - Y_j(0) \,.
\end{align}
Assuming again that all $X_i$ and $Y_j$ are identically distributed copies of their respective populations,
we can also define the average treatment effect (ATE) across the overall populations $X$ and $Y$ as
\begin{align}
    \Delta_{X} = \mathbb{E}(X(1) - X(0)), \quad
    \Delta_{Y} = \mathbb{E}(Y(1) - Y(0)) \,.
    \label{eq:rade_pof_ate}
\end{align}

The framework also enables us to formalise the adage ``correlation does not imply causation''. So far, we have encountered the expected observed response given a treatment $\mathbb{E}(X \,|\, \mathcal{A} = a)$ (see Expression~\eqref{eq:rade_pof_expected_response_given_treatment}) and the expected potential response under the said treatment $\mathbb{E}(X(a))$. They may appear to be descriptions of the same quantity but are not the same. This is because a third variable (e.g., a confounder) may affect both~$X$ and $\mathcal{A}$. 

To illustrate, we use the ``ice cream consumption causes drowning'' fallacy~\cite{mumford2013correlation} -- let $X$ be the number of (observed) drowning incidents and $\mathcal{A}$ be a binary r.v., representing high ($\mathcal{A} = 1$) and low ($\mathcal{A} = 0$) ice cream sales. It is established that the number of drowning incidents increases along with ice cream sales. In other words, $\mathbb{E}(X \,|\, \mathcal{A} = 1)$ is high, $\mathbb{E}(X \,|\, \mathcal{A} = 0)$ is low, and the difference $\mathbb{E}(X \,|\, \mathcal{A} = 1) - \mathbb{E}(X \,|\, \mathcal{A} = 0)$ is positive. However, it is also established that ice cream consumption does not cause drowning, nor does drowning incidents cause increased ice cream sales.\footnote{It is conceivable for drowning incidents to cause \emph{decreased} ice cream sales due to emergency responses.} Instead, the correlation/association in the expected observed responses arises due to a third factor: weather. Hot and sunny spells generally raise the demand for ice cream and, unfortunately, the number of drowning incidents as more people visit various water bodies and thus are exposed to the risk of drowning.

The expected potential responses better reflect the lack of a causal relationship between ice cream sales and drowning incidents. Given the potential number of drowning incidents under high ice cream sales $\mathbb{E}(X(1))$, one would expect the counterfactual response $\mathbb{E}(X(0))$, i.e., the number of drowning incidents had the ice cream sales been low \emph{and all else equal}, will be similar if not the same. This leads to a negligible causal impact $\mathbb{E}(X(1)) - \mathbb{E}(X(0))$.

\section{Classes and Examples of Digital Experiments}
\label{sec:rade_classes}

\begin{figure}
  \begin{center}
  \includegraphics[width=0.9\textwidth, trim = 17mm 45mm 20mm 45mm, clip]{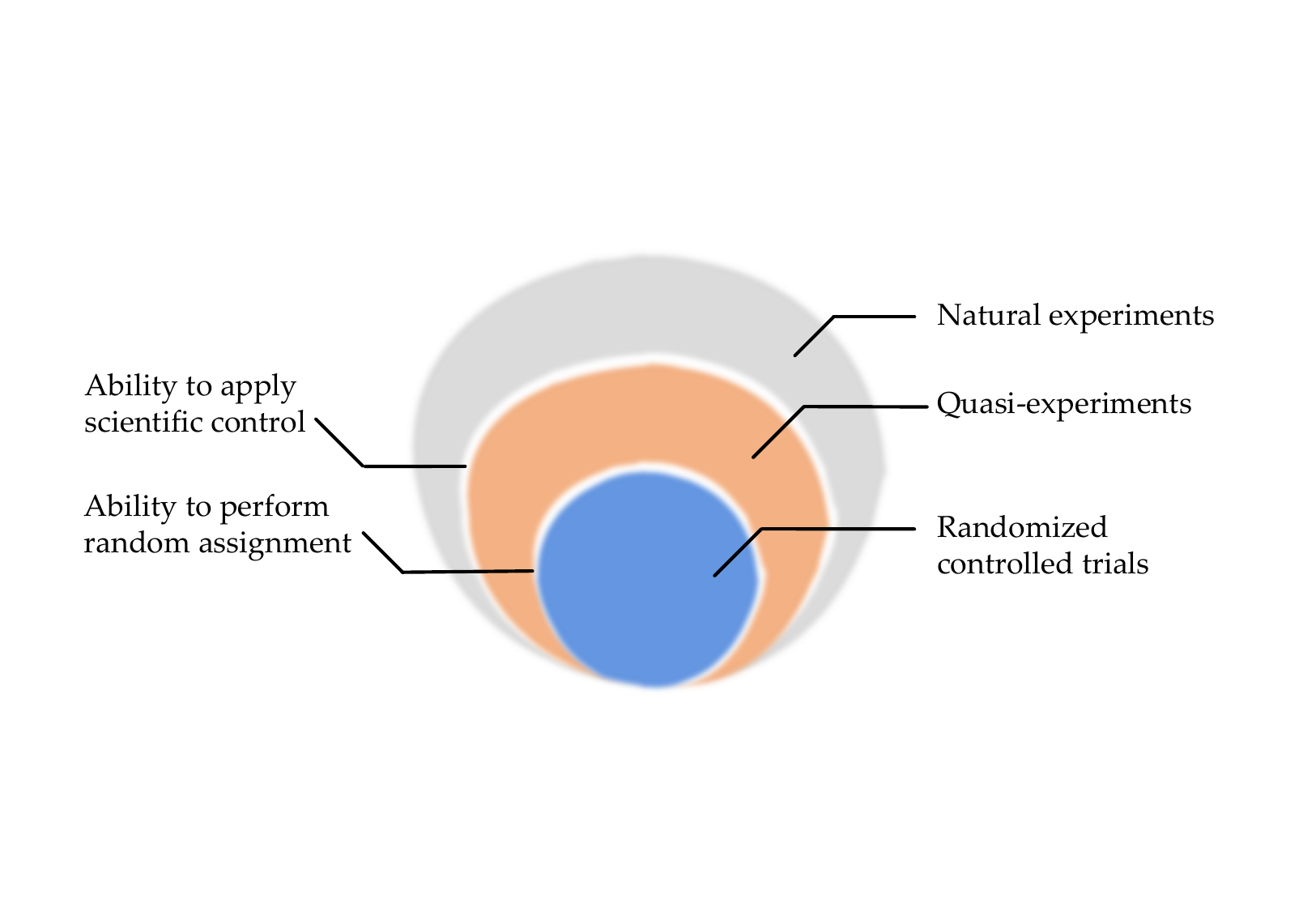}
  \end{center}
  \caption[The onion model for different types of experiments.]{The onion model for different types of experiments, in decreasing degree of control from the experimenter (inner to outer): Randomised controlled trials, quasi-experiments, and natural experiments. Quasi-experiments and randomised controlled trials are differentiated by the experimenter's ability to assign experiment participants randomly into different groups for different treatments. Natural experiments and quasi-experiments are differentiated by the experimenter's ability to vary the treatment (or lack thereof) prescribed to an experiment participant.}
  \label{fig:intro_exptypes_onionmodel_basic}
  \vspace*{\baselineskip}
\end{figure}

To aid the discussion in the rest of the chapter, we provide an overview of different classes of experiments defined in the traditional statistics literature and further examples within digital experimentation. Figure~\ref{fig:intro_exptypes_onionmodel_basic} shows the classes of experiments and what differentiates a class from another. 

\subsection{Randomised controlled trials}
\label{sec:rade_classes_rct}

The class of experiments where experimenters have the most control is randomised controlled trials (RCTs). They are also known as online controlled experiments (OCEs) in the digital experimentation community~\cite{kohavi20trustworthy}. The free delivery banner experiment described at the beginning of the thesis (see Figure~\ref{fig:intro_exp_abtestillustration_freedelivery}) belongs to this class. These experiments are characterised by the experimenter's ability to apply a scientific control and randomly assign experimental units to different groups.\footnote{Unlike in medical trials, the issue of \emph{blinding}, where information that may influence the result is withheld from the experimenter or experiment participants (or both) to minimise biases, is rarely discussed in digital experimentation. It is often addressed by having the Internet between the experimenter and participants: the overwhelming majority of participants are unaware of what treatment(s) are in place when they visit a website. Moreover, experimenters would only know who the participants are during post-experiment analysis as they delegate the task of assigning and tracking participants to automated computer systems.} For the free delivery banner example, this corresponds to the ability to show a version of the website without the ``free delivery'' banner and randomly assign users to see different website versions, respectively.

Most of the value of an RCT comes from the ability to perform random assignments, which has two major implications. Here, we describe the implications using the potential outcomes framework. Firstly, randomisation generally implies exchangeability.\footnote{Also known as \emph{unconfoundedness} or \emph{ignorability}. While there are randomisation schemes that lead to non-exchangeable potential responses, in practice, almost all online controlled experiments rely on randomisation schemes (e.g., simple randomisation) that yield some form of exchangeability~\cite{savje2021randomization}.} This means the potential responses and the actual treatment are independent. We denote such exchangeability as $X(0), X(1) \indep \mathcal{A}$ and $Y(0), Y(1) \indep \mathcal{A}$ and observe that if the potential responses are exchangeable,
\begin{align}
    \mathbb{E}(X(a))  & = \mathbb{E}(X(a) \,|\, \mathcal{A} = 0) = \mathbb{E}(X(a) \,|\, \mathcal{A} = 1) \, \textrm{ and } \nonumber\\
    \mathbb{E}(Y(a))  & = \mathbb{E}(Y(a) \,|\, \mathcal{A} = 0) = \mathbb{E}(Y(a) \,|\, \mathcal{A} = 1) \,, \; a\in \{0, 1\} \,.
    \label{eq:rade_pof_mean_exchangability}
\end{align}
Secondly, as experimental units in both groups can be viewed as a simple sample of the same underlying population under randomisation, the r.v.'s representing the potential responses of \emph{both} groups under the same treatment, i.e.,
\begin{align}
    X_1(a), \cdots, X_n(a), Y_1(a), \cdots, Y_m(a) \textrm{ for } a\in \{0, 1\} \,,
    \label{eq:rade_pof_interchangeability}
\end{align}
are interchangeable.\footnote{Also known as an \emph{exchangeable sequence of r.v.'s}~\cite{chow1997probability}. There are a few subtle differences between the two exchangeability definitions. Thus, we use a different name to disambiguate these two concepts.} This means they can be swapped within and across the groups without consequences.

The two implications mean we can obtain an unbiased estimate/estimator on the average treatment effect using only the observed responses under randomisation. To illustrate, we can express the ATE defined in Equation~\eqref{eq:rade_pof_ate} as
\begin{align}
  \Delta_X \,& = \mathbb{E}(X(1) - X(0))  & \textrm{(By definition -- Eq.~\eqref{eq:rade_pof_ate})} \nonumber\\
  & = \mathbb{E}(X(1)) - \mathbb{E}(X(0)) & \textrm{(Linearity of expectation)}  \nonumber\\
  & = \mathbb{E}(Y(1)) - \mathbb{E}(X(0)) & \textrm{(Interchangeability between $X(1)$ and $Y(1)$)} \nonumber\\
  & = \mathbb{E}(Y(1) \,|\, \mathcal{A} = 1) - \mathbb{E}(X(0) \,|\, \mathcal{A} = 0) & \textrm{\hspace*{2em}(Exchangeability of $Y(1)$ and $X(0)$ -- Eq.~\eqref{eq:rade_pof_mean_exchangability})} \nonumber\\
  & = \mathbb{E}(Y \,|\, \mathcal{A} = 1) - \mathbb{E}(X \,|\, \mathcal{A} = 0)  &  \textrm{(By definition -- Eq.~\eqref{eq:rade_pof_counterfactual_response})} \,.
  \label{eq:rade_pof_randomisation}
\end{align}
We can also show that $\Delta_Y$ is equal to the last line of Equation~\eqref{eq:rade_pof_randomisation} using the same arguments (plus the interchangeability between $X(0)$ and $Y(0)$). This should not come as a surprise as $X$ and $Y$ are based on experimental units from the same population and thus should share the same ATE. As stated in Section~\ref{sec:rade_pof}, estimating the last line of Equation~\eqref{eq:rade_pof_randomisation} is simply a matter of taking two sample means over the two groups of observed responses. This has the effect of reducing the ATE estimation problem into a much more straightforward two-sample testing problem. Experimenters can use the many readily available methods in statistical testing (see Chapter~\ref{chap:statstest}) and avoid estimating expected counterfactual responses, which may attract considerable bias.

To sum up, the results above show that randomisation generally implies that any difference in the responses can be attributed to the causal effect of the treatment and not to other measured or unmeasured confounding. Such a desirable property perhaps explains why RCTs are considered the ``gold standard'' in medical trials~\cite{jones2015history} and are the most common type of digital experiment. We will outline further statistical challenges to running RCTs online in Section~\ref{sec:rade_oce}.

\subsection{Quasi-experiments}
\label{sec:rade_classes_quasi}
Taking away the ability to randomly assign experiment units in randomised controlled trials leaves us with a quasi-experiment.
Consider the example in~\cite{xu2016evaluating}, where a digital organisation developed a mobile application downloadable from app stores such as Apple App Store and Google Play. The organisation is interested in whether a new app feature shipped within a new app version will improve a business metric relative to the existing app version. For the new feature to reach app users, developers usually upload the new version onto the app stores. The app stores then have to review the new app version, and once approved, app users have to update their app.

In this scenario, the ability to assign who sees the new or existing version of the app (let alone randomly) is taken away from the experimenters and put in the hands of the users, who decide when to update the app, and the app stores, who decide when to allow the new version of the app to be downloaded.

The inability to randomise treatment assignments generally marks the beginning of causal inference's remit. Experimenters can no longer expect the exchangeability mentioned in the previous subsection to automatically hold and have to employ other methods to estimate the counterfactual responses from one or more observed responses. In the language of the potential outcomes framework, we want to estimate $X_i(1)$ and $Y_j(0)$ using $X_i \,|\, \mathcal{A} = 0$ and $Y_j \,|\, \mathcal{A} = 1$, where $X_i$ represents the responses from users who have yet to download the new app version, $Y_j$ represents the responses from users who have seen the new version (and presumably interacted with the new feature), and $\mathcal{A}$ be the binary treatment assignment indicator. We will introduce several causal inference approaches and how they apply to quasi-experiments in Section~\ref{sec:rade_causalinference}.

\subsection{Natural experiments}
\label{sec:rade_classes_natural}
If we further remove the ability to apply scientific control,\footnote{Also known as \emph{manipulating} the treatment or independent variables~\cite{rogers2019experimental}.} we end up with natural experiments. While experimenters cannot apply different treatments (in this case, different versions of a digital product or experience), participants may still encounter scenarios that include one or more of the said treatments by the act of nature. This means it remains possible to draw inferences on the causal effect of the treatments from observational data. 

An experimenter may opt not to apply a scientific control and run a natural experiment instead due to various operational, reputational, legal and ethical considerations. For example, experiments aiming to measure the value added by existing call centre support may be prevented from actively blocking some users’ access to the call centre as it negatively impacts user experience. Other experiments seeking to understand the price elasticity of demand\footnote{The relationship between the price of a product/service sold by an organisation and its demand.} may be keen to avoid engaging in the controversial practice of price discrimination\footnote{The act of offering different users a different price for the same digital product/service.} to avoid negative press and risk of legal actions~\cite{oecd2018personalised,wild2018aprice}. Lastly, testing different news headlines simultaneously in front of readers may also damage editorial credibility or even ``fan the flames'' of partisan political beliefs in the long run~\cite{jiang2019guineapig}. Experimenters may instead leverage responses from users who have never contacted the call centre, sales on similar digital products/services priced differently, and reactions on past news headlines on similar topics to understand the respective causal impact.

Most natural experiments share the same causal inference approaches as quasi-experiments. We will cover a few applications in Section~\ref{sec:rade_causalinference}.

\vspace*{1.1\baselineskip}

\subsection{Special case -- Experiments for personalisation strategies}
\label{sec:rade_classes_pse}

It is worth noting that the boundaries between each experiment class introduced above are not clear-cut. Depending on the situation, experimenters may have various degrees of control over random assignments.

Consider a follow-up experiment to our free delivery banner experiment introduced above.
Suppose the organisation is now interested in whether varying the \emph{timing} of showing the ``free delivery'' banner will lead to further improvement. One can run an experiment as simple as comparing ``showing the banner after a user has viewed \emph{three} shop products'' vs ``showing the banner after a user has viewed \emph{four} shop products'' and see which timing rule leads to a better outcome.

In the experiment, the treatment (timing rule) a user is subjected to is \emph{jointly} determined by the assignment from experimenters and the user's actions. While the experimenters can randomly assign which timing rule to put the user through, the user must also qualify under the rules by seeing three or four shop products. Only by doing so, the user would see the ``free delivery'' banner, without which the notion of treatment effect would not make sense.
This means experimenters have some, but not complete, control in random assignments -- more than that in the new app feature experiment described in Section~\ref{sec:rade_classes_quasi}, but certainly less than the vanilla free delivery banner experiment discussed in Section~\ref{sec:rade_classes_rct}.\footnote{Some literature describes this scenario as \emph{one-sided non-compliance}~\cite{waisman2023multicell}.}

The banner display timing experiment is one of many experiments for \emph{personalisation strategies} -- complex sets of targeted user interactions common in e-commerce and digital marketing.\footnote{The example provided is grossly simplified to facilitate discussion. In practice, display timing strategies, in true personalisation spirit, usually specify how different users are shown their ``free delivery'' banner at a different stage of their website visit, depending on their browsing behaviour. Thus, an experiment that compares two display timing strategies is essentially comparing the execution of two \emph{sets} of display timing rules generated from two meta-principles.} We will focus on the design of these experiments and further elaborate on their unique challenges in Chapter~\ref{chap:pse}.

\vspace*{1.1\baselineskip}

\section{Advances in Online Controlled Experiments}
\label{sec:rade_oce}

Here, we provide a brief review of the advances in OCEs. Since its inception in the late 1990s~\cite{kohavi17encyclopedia}, the field has seen tremendous growth in research output and its economic and societal impact~\cite{thomke20experimentation}. More recently, we have also seen some comprehensive and up-to-date reviews in the area:
\begin{itemize}
  \item \cite{kohavi20trustworthy} provided a narrative, general-purpose guide to the field while collating many relevant works.\footnote{The author reviewed part of the manuscript before the book's publication in 2020.}
  \item \cite{auer21controlled} and \cite{quin2023ab} provided systematic literature reviews. The former built a taxonomy based on the process-infrastructure distinction, type of methodological approach, and the challenges/benefits involved. The latter categorised the works by the subject of an experiment,\footnote{Defined in~\cite{quin2023ab} as the application domain and the element subjected to experimental manipulation.} choices made in the experiment design/execution processes, and the role of stakeholders.
  \item \cite{gupta19topchallenges} and \cite{larsen2023statistical} described the challenges. The former provided a general view by including data, engineering, and organisational culture topics. The latter focused on the statistical aspect and provided a more technical treatment to each topic presented.
\end{itemize}

This review aims to be a gateway to the multitude of research works in the area by unifying the many nomenclatures, providing a concise problem statement for each challenge, and suggesting works accessible to those new to a particular line of work.
For a fuller exposition of the challenges and research in this area, readers should refer to the five high-quality reviews listed above.

Most of the section is dedicated to advanced statistical methods, building on concepts and notations covered in Chapter~\ref{chap:statstest} (on statistical testing) and Sections~\ref{sec:rade_pof} and~\ref{sec:rade_classes_rct} (on the potential outcomes framework). We will also briefly outline works addressing other topics in running successful OCEs, as mentioned in the second research question in Section~\ref{sec:introduction_research_q}. These include choosing decision metrics / evaluation criteria, scalable computer systems, and fostering a growth mindset organisational culture.


\subsection{Advanced statistical methods}
\label{sec:rade_oce_statsmethods}

We begin with the statistical challenges in OCEs that necessitate the development of advanced methods. These challenges included in the section are the multiple testing/comparison problem, post-selection inference, heterogeneous treatment effects, dependent responses, stable unit treatment value assumption (SUTVA) violations, and improving experiment sensitivity.

\paragraph{Multiple testing/comparisons}
Recall from Section~\ref{sec:statstest_nhstconcepts_significance} that the probability of rejecting the null $H_0$ in an NHST when $H_0$ is true (i.e. the false positive rate) is $\alpha$. The false positive rate is usually small so that when a test rejects $H_0$, we have reasonable conviction to hail the experiment result as a discovery. That said, the probability of committing a false positive accumulates as we perform additional comparisons. 
For $N_{\textrm{test}}$ independent NHSTs, we expect $(\alpha \cdot N_{\textrm{test}})$ tests to reject $H_0$ when $H_0$ is true. This means that for every 20 NHSTs run under $\alpha = 0.05$, we expect the tests to reject $H_0$ once on average, even when all the $H_0$ are true~\cite{munroe2011significant}. Hailing all test rejections as discoveries without accounting for the inflated false positive rate will lead to biased decisions.\footnote{This is closely related to the continuous monitoring/peeking problem discussed in Section~\ref{sec:statstest_sequential} and Section~\ref{sec:oced_introduction}. That said, comparisons in the continuous monitoring/peeking problem are often correlated, leading to the risk of committing a false positive accumulating differently.}

The problem is increasingly prevalent due to the increasing number of experiments, treatment pairs, and user segments to compare (see paragraph on heterogeneous treatment effect below). 
Usually, experimenters seek to control the family-wise error rate (FWER) or the false discovery rate (FDR), defined over many experiments as
\begin{align}
    \textrm{FWER} & \triangleq  \mathbb{P}(\textrm{\# Type I error} \geq 1) = 1 - \mathbb{P}(\textrm{\# Type I error} = 0), \\[0.5em]
    \textrm{FDR} & \triangleq \mathbb{E}\left(\frac{\textrm{\# Type I error}}{\textrm{\# Test rejecting } H_0}\right) \,.
\end{align}
The most common techniques that control for FWER include Bonferonni correction~\cite{dunn1961multiple} and \v{S}id\'{a}k correction~\cite{sidak1967rectangular}, whereas those for FDR include the Benjamini-Hochberg~\cite{benjamini1995controlling} and knockoff procedures~\cite{foygelbarber2015controlling}.

\paragraph{Post-selection inference}

Having multiple comparisons over experiments, treatment pairs and sub-populations will remain a potential source of bias even when we move away from binary decisions and on to effect size estimation. Consider a two-sample, right-tailed $t$-test (and all its associated notation, see Section~\ref{sec:statstest_nhstexamples_ttest}). In addition to deciding to reject/not reject $H_0: \Delta = \theta_0$, we also estimate the treatment effect~$\Delta$. We know $\bar{Y} - \bar{X}$ is an unbiased estimator for $\Delta$. However, with many statistical tests, we generally ignore all tests that fail to reject~$H_0$ -- assuming $\Delta = \theta_0$ as per the null hypothesis -- and focus on tests that reject~$H_0$. The treatment effect estimate after such test selection now overestimates $\Delta$:
\begin{align}
    \mathbb{E}\left(\bar{Y} - \bar{X} \,|\, T > t_{\nu, 1-\alpha}\right) 
    = \mathbb{E}\left(\bar{Y} - \bar{X} \,\bigg|\, \bar{Y} - \bar{X} > \theta_0 + t_{\nu, 1-\alpha}\sqrt{\frac{s^2_X}{n} + \frac{s^2_Y}{m}}\right)
    > \mathbb{E}\left(\bar{Y} - \bar{X}\right)
    = \Delta \,.
\end{align}
Another way to put this in the context of multiple comparisons is that the test selection process above favours ``lucky'' OCEs, some of which are merely false discoveries. Inferring quantities with the test selection process is an example of \emph{post-selection inference}~\cite{deng2021postselection}, and the estimation bias it introduces is also known as the \emph{winner's curse}~\cite{lee2018winner}.

The best way to detect the existence of the winner's curse and correct the bias is via replication~\cite{deng2021postselection}. In practice, it is often operationally challenging. It also seems unnecessary if one only seeks the aggregated treatment effect from multiple successful OCEs; in this case one can employ a maximum likelihood approach to estimate the expected bias from the test selection process~\cite{lee2018winner}. To estimate the treatment effects of individual OCEs, one can perform pseudo-replication by splitting an experiment into two sub-experiments with the same treatments and build a regression model on the expected treatment effect of the second sub-experiment (without any selection) using that of the first sub-experiment (which goes through test selection)~\cite{coey2019improving}. They can also use empirical Bayes methods~\cite{azevedo19empirical} such as Tweedie's formula and the James–Stein estimator~\cite{efron2011tweedie}. Finally, \cite{deng2021postselection} combined elements of the approaches above to improve the precision of individual treatment effect estimates.

\paragraph{Heterogeneous treatment effects}

Many experiments seek to measure the average treatment effect across the population. That said, it is valuable for experimenters to understand whether there are subgroups whose treatment effect may deviate from that of the entire population~\cite{duivesteijn17haveitbothways}.
To detect such heterogeneous treatment effects, experimenters often estimate the conditional average treatment effect (CATE) instead, i.e.
\begin{align}
    \mathbb{E}(Y_j(1) - Y_j(0) \,|\, \mathcal{C}_j),
\end{align}
where $\mathcal{C}_j$ represents the covariates for experimental unit $j$. Many works assume conditional interchangeability and exchangeability, such that they can estimate CATE using the conditional observed responses $\mathbb{E}(X_i \,|\, \mathcal{C}_i = c)$ and $\mathbb{E}(Y_j \,|\, \mathcal{C}_j = c)$ from an OCE. See Section~\ref{sec:rade_causalinference_cate} for further discussions on CATE estimation and 
Section 3 of~\cite{larsen2023statistical} for a summary of works in the area.


\paragraph{Dependent responses}

Many OCEs feature statistical tests (see Section~\ref{sec:statstest_nhstexamples}) that assume i.i.d. responses. While many OCEs do have observations that are approximately i.i.d.~\cite{deng17trustworthy}, such assumptions are not always justified. We have (implicitly) covered deviations from the identically distributed assumptions when we discussed heterogeneous treatment effects above. Below, we look at deviations from the independence assumptions. 

One can find dependent responses in many experiment designs. A significant portion features dependent treatment assignments. This could be explicitly imposed by randomly assigning pre-existing groups of experimental units (instead of individuals) to treatment groups, also known as \emph{cluster randomisation}, as briefly discussed in Section~\ref{sec:oced_mapping_ttest}. It could also implicitly arise as one randomises individual experimental units independently yet observes and infers based on more granular responses, effectively forming a cluster per experimental unit.\footnote{Digital experimenters often differentiate an experimental unit, a \emph{randomisation unit} (one subjected to independent randomisation), and an analysis unit (one that yields a response). These units do not necessarily coincide~\cite{deng2011choice,deng17trustworthy}.} In Section~\ref{sec:rade_abv}, we will present an extensive case study on dependent responses in e-commerce OCEs, which arises from experimenters randomising by users and measuring changes to business metrics based on more granular transaction- or item-based responses.

Having dependent responses can affect treatment effect estimates and decisions from statistical tests.
Consider a one-sample $t$-test (see Section~\ref{sec:statstest_nhstexamples_ttest} for all its associated notation). We recall from Equation~\eqref{eq:statstest_tstatistic_def_onesample} that its test statistic is $(\bar{X} - \mu_X)/\sqrt{s_X^2/n}$. Moreover, the test statistic's denominator is an estimator of the standard error of the mean (SE):
\begin{align}
  \sqrt{\textrm{Var}(\bar{X})}
  & = \sqrt{\frac{1}{n^2} \big(\textstyle\sum_{i=1}^{n} \textrm{Var}(X_i) + 2 \textstyle\sum_{i<j} \textrm{Cov}(X_i, X_j) \big)} \;.
  \label{eq:rade_se_def}
\end{align}
Under i.i.d. assumptions, $\textrm{Cov}(X_i, X_j) = 0 \;\;\forall i \neq j$ and thus ${\textrm{Var}(\bar{X}) = \sum_i \textrm{Var}(X_i)/n^2 = \sigma^2_X/n}$. This implies $\sqrt{s_X^2/n}$ is an unbiased estimator of the SE. However, if $X_i$ are dependent, then $\textrm{Cov}(X_i, X_j) \neq 0$, which makes $\sqrt{s_X^2/n}$ a biased estimator. In the case study below, we will demonstrate how the bias affects the power and confidence interval calculations and, thus, the decisions made in statistical tests and experiments.

Digital experimenters use the delta method~\cite{deng18applying} or one-way/block bootstrap~\cite{bakshy13uncertainty} to mitigate the bias. We will provide a critique of these methods in the case study below. Another interesting work in the area is \cite{deng17trustworthy}, which explores scenarios where an experimenter can treat responses as i.i.d. before providing a variance formula that exploits the relationship between the experimental/randomisation/analysis units and accommodates a wide range of randomisation mechanisms in practical settings.

\paragraph{Stable unit treatment value assumption (SUTVA) violations}

Responses can also become dependent by violating the \emph{stable unit treatment value assumption} (SUTVA), an assumption we made implicitly when we defined the potential outcomes in Section~\ref{sec:rade_pof} / Equation~\eqref{eq:rade_pof_counterfactual_response}.\footnote{Also known as having \emph{spillover effects}~\cite{sinclair2012detecting,sobel2006whatdo}, in the sense that a treatment spills over to experimental units in another treatment group (due to, e.g., interactions between experimental units in different groups) and induces unwanted treatment effects.} More precisely, SUTVA is an assumption in two parts -- in the words of~\cite{imbens2015causal}:
\begin{enumerate}
    \item (No interference~\cite{cox1958planning,rosenbaum2007interference}) -- ``The potential outcomes for any [experimental] unit do not vary with the treatments assigned to other units, and,'' 
    \item (No hidden variations of treatments) -- ``[F]or each [experimental] unit, there are no different forms or versions of each treatment level, which lead to different potential outcomes.''
\end{enumerate}

In OCEs, interference mainly arises due to the network effects. As experimental units interact with each other in the network, their potential response no longer depends on what treatment they have received but also on what treatments their connections have received. In other words, a treatment may spill over to an experimental unit's connections. The effect can be found directly in social networks (where a user interacts with their connections)~\cite{karrer2021network} or indirectly via marketplaces (where suppliers/buyers compete for shared resources)~\cite{li2022interference} and recommender systems (where the system recommend multiple items/users to the same user/item)~\cite{pougetabadie2019variance}. 

While dependent treatment assignment, which breaks the independence assumption but maintains SUTVA, generally affects the variance estimate, a SUTVA violation can bias the treatment effect estimates. \cite{bakshy13uncertainty} have shown that interaction amongst users and, separately, the items recommended can result in an increased false positive rate in experiments when applied to Facebook data, likely due to bias in both the treatment effect and variance estimates.

A common way to mitigate the spillover effect in social networks is to randomise treatment group assignments over clusters\footnote{In network speak, sub-graphs that are highly connected within.} of experimental units (e.g. communities of users) instead of the units themselves~\cite{backstrom11network}. Clustering by non-overlapping communities means one can hold SUTVA at a cluster level, as the units generally interact heavily within the same cluster but sparsely between clusters~\cite{liu2016overlapping}.
However, the choice of clusters is non-trivial; this means a cluster-based treatment assignment may only be justified when the network effect is apparent. \cite{saveski17detecting}~developed a framework that detects whether network effects exist for a particular OCE by simultaneously running the same experiment under individual-based and cluster-based assignment and testing if the treatment effect differs (and hence if SUTVA holds up reasonably to justify individual-based assignment). Section 6 of~\cite{larsen2023statistical} describes many more relevant works, including those focusing on indirect network effects.

The second part of SUTVA (no hidden variations of treatments) is seldom explicitly discussed in OCEs. That said, it is conceivable that many ``imperfections [...] hidden  throughout the
engineering stack or in the design process''~\cite{wong2022addressing} are capable of breaking this part of SUTVA and thus bias the (potential) responses. Many examples are presented as pitfalls in OCE operations, which we will cover in Section~\ref{sec:rade_oce_others} / Table~\ref{tab:rade_experimentation_pitfall_meta}.

\paragraph{Improving sensitivity}

We look beyond challenges that lead to biases in our treatment effect estimator.
Recall from Section~\ref{sec:statstest_nhstconcepts_power} that the power of a test depends on many factors, including the significance level, sampling uncertainty, sample size and effect size. While we have discussed the role of sample and effect sizes in Sections~\ref{sec:statstest_nhstconcepts_samplesize} and~\ref{sec:statstest_nhstconcepts_mde}, respectively, little regarding the sampling uncertainty has been said so far in the thesis. Some works look into increasing the sensitivity of OCEs by reducing sampling uncertainty, which enables an experimenter to reduce the sample size required for an experiment or the minimum detectable effect (or both). The works are motivated by observations where \cite{kohavi14sevenrules} noted that ordinarily, successful experiments in technology companies only improve metrics by a fraction of a per cent; \cite{xie16improving} also described the need to detect small effects as huge customer bases often translate them into substantial gains in revenue and profit. 

The methods described below rely on decomposing the variance in a metric and attempting to eliminate unnecessary components. Controlled-experiment Using Pre-Existing Data (CUPED) is a control variates method that measures a metric modified by a linear function of the covariates~\cite{deng13improving}.\footnote{Also known as \emph{regression adjustment} in econometrics~\cite{liou2020varianceweighted}.} The idea is to eliminate the effect of correlated covariates on the metric under measurement. \cite{deng15diluted} later apply the idea to reformulate dilution\footnote{The act of translating the treatment effect from users who have been exposed to a treatment (``triggered'' users) to the general user base. The metric is diluted as we incorporate untriggered users (who contribute zero).} as a variance reduction problem, resulting in a more easy to implement and less error-prone formula. A similar idea is used by~\cite{poyarkov16boosted}, which subtracted the predicted value of a metric using boosted decision trees as predictors. In the same spirit, stratified sampling is used to eliminate variance between strata. Stratified sampling can be applied pre- or post-experiment, and while the theoretical bounds are better for pre-experiment stratification, post-stratification is often preferable for large-scale OCEs~\cite{xie16improving}.

\subsection{Other topics}
\label{sec:rade_oce_others}

To complete our review of OCEs, we also review topics that we did not cover in the statistical tests and data chapters, as well as the advanced methods section above.

\paragraph{Anatomy of an OCE} 
A mature OCE system/platform often features complex processes and infrastructure, and many seek to provide an overview or describe specific parts of the ecosystem. \cite{kohavi09controlled} provided a definitive guide on setting up, running, reporting, and scaling OCEs. Several papers described the scalable, concurrent testing infrastructure and its associated challenges at major technology companies~\cite{kohavi13online,tang10overlapping,xu15frominfrastructure}. Others described the OCE lifecycle~\cite{fabijan20online} and characterised the elements defining an experiment~\cite{auer19characteristics}.

\paragraph{State and progress of OCEs' usage}
A class of works aim to capture the state and track the progress of the use of OCEs (or DEM capabilities in general) across many organisations. Some propose models that represent the maturity/growth of DEM capabilities~\cite{fabijan17evolution,widerfunnel18thestate}. Others focus on experiment outcomes and perform meta-analyses on results from multiple OCEs -- see \cite{brooksbell20thestate,browne17whatworks,johnson2017online} for results across industries, in e-commerce, and in display advertising, respectively. \cite{ghosh20effects} bridged the two topics above by linking experiment performance to organisational structure.

\paragraph{Pitfalls in OCE operations}
Despite the increasing adoption and maturity of the use of OCE, running a successful OCE after another is, by the admission of the biggest players in the industry, difficult~\cite{kohavi20trustworthy}. This is due to many non-obvious challenges and pitfalls experimenters face while running an OCE end-to-end. In addition to that related to statistical testing (see Chapter~\ref{chap:statstest}) and requiring advanced statistical methods (see Section~\ref{sec:rade_oce_statsmethods}), many accounts~\cite{crook09sevenpitfalls,dmitriev2017adirtydozen,estellercucala19experimentation,kohavi11unexpected,kohavi12trustworthy} describe pitfalls related to experimentation infrastructure, decision metric selection and interpretation~\cite{deng16datadriven}, and external effects~\cite{hohnhold15focusing}. We summarise these challenges and pitfalls in Table~\ref{tab:rade_experimentation_pitfall_meta}.
To mitigate the potential problems, \cite{lu14separation} proposed techniques that separate problematic user responses affected by some of the external effects (namely dilution, carry-over effect, and novelty effect) from valid ones. \cite{chen2019automatic,fabijan2019diagnosing} also developed a series of diagnostic tests to identify the root cause(s) of a sample ratio mismatch, a common symptom for tests that are potentially invalidated due to encountering one or more of the pitfalls  (see ``practical usage'' paragraph of Section~\ref{eq:statstest_chisq}).

\begin{landscape}
  \begin{table}
    \centering
    \caption[Pitfalls in running an OCE end-to-end as described in nine selected works.]{Pitfalls in running an OCE end-to-end as described in Crook et al.~\cite{crook09sevenpitfalls}, Kohavi and Longbotham~\cite{kohavi11unexpected}, Kohavi et al.~\cite{kohavi12trustworthy}, Lu and Liu~\cite{lu14separation}, Hohnhold et al.~\cite{hohnhold15focusing}, Dmitriev et al.~\cite{dmitriev2017adirtydozen}, Fabijan et al.~\cite{fabijan2019diagnosing}, Chen et al.~\cite{chen2019automatic}, and Esteller-Cucala et al.~\cite{estellercucala19experimentation}. The value in each cell represents the section number where the pitfall is described in each work.}
    \label{tab:rade_experimentation_pitfall_meta}
    \includegraphics[width=1.45\textwidth, trim=18mm 30mm 19mm 10mm, clip]{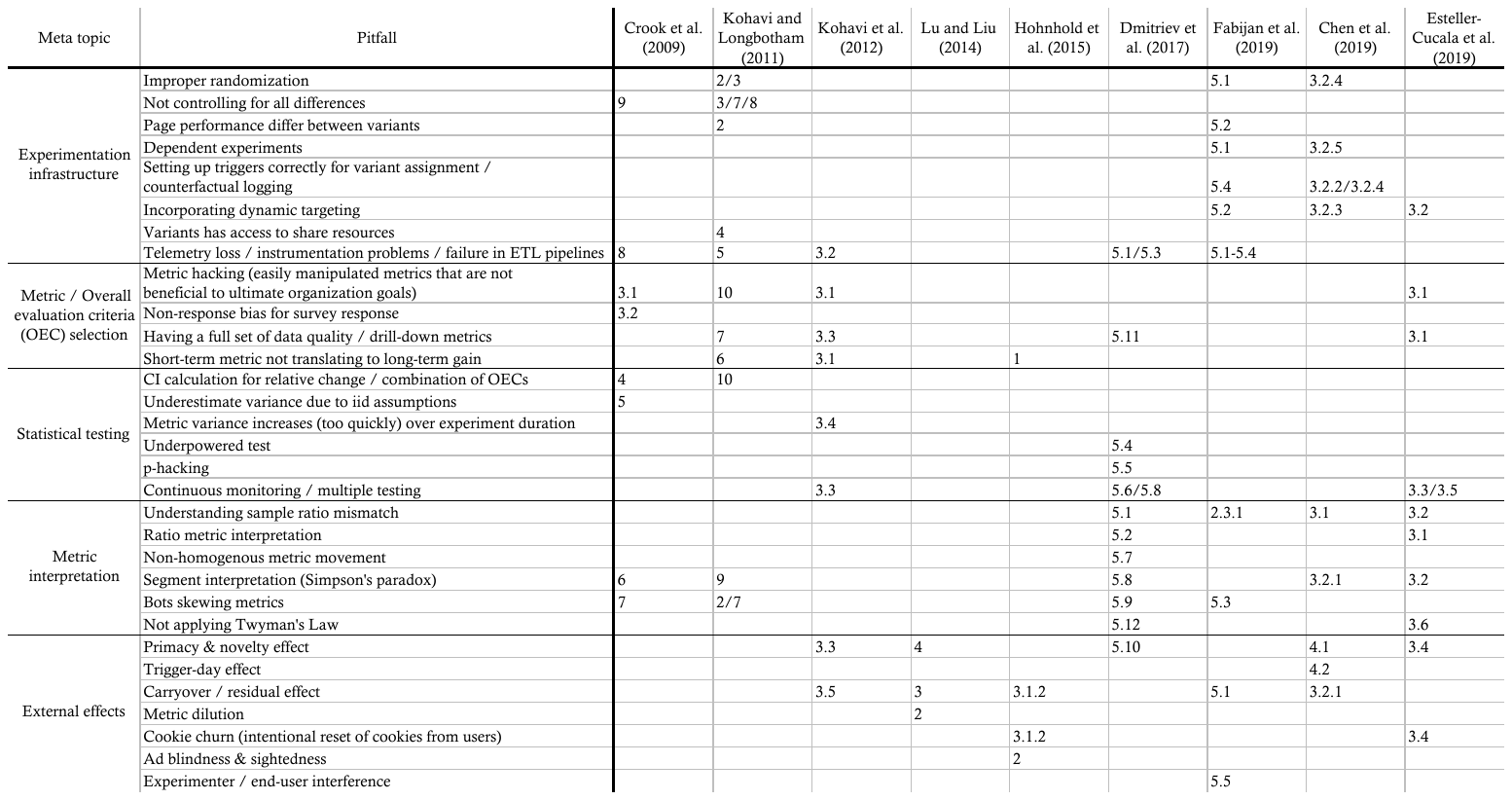}
  \end{table}
\end{landscape}

\section{A Spotlight on Dependent Responses in e-Commerce Experiments}
\label{sec:rade_abv}

To complement our review of advances in OCE, we present a case study on dependent responses in e-commerce OCEs. We intend for the case study to be a short but crucial piece of evidence that drives awareness of the pitfall among experimenters in e-commerce and hence encourages the adoption of established mitigation approaches.

Like many technology organisations, e-commerce organisations aim to provide a consistent user experience; hence, randomisation in OCEs is generally done on a per-user basis. On the other hand, e-commerce organisations are unique as they feature physical inventories and thus track a set of business metrics that are transaction- or item-based. These metrics include \emph{average basket value} (ABV),\footnote{Average (mean) amount spent in each transaction/basket by a user.} \emph{average basket size} (ABS),\footnote{Average (mean) number of items purchased in each transaction/basket by a user.} and \emph{average selling price} (ASP).\footnote{Average (mean) price per item sold.}

Experiments that measure changes to ABV, ABS, and ASP often feature dependent responses.
In these experiments, the responses (or \emph{analysis units}) are at a more granular transaction or item level than the \emph{randomisation units}, which are at a user level. Given that a user can make many transactions and purchase many items during an experiment, the value of these transactions and items will likely correlate based on the user's preference. This creates a local dependence structure between users and transactions/items, which violates the usual i.i.d. assumptions in common statistical tests that OCEs employ (e.g., a Student's $t$-test). If left unmitigated, it risks experimenters having wrong estimates of sampling uncertainty and making wrong conclusions from a statistical test.

Having dependent responses in experiments due to differences in granularity between randomisation and analysis units is not a recent revelation~\cite{kohavi20trustworthy}. In some sense, one may regard such a setup as a cluster-randomised controlled experiment, where an experimenter randomly assigns clusters of transactions or items belonging to the same user to the same treatment groups.\footnote{In such a setup, the size of a cluster can be zero and would only be determined after the experiment ends. This is because experimenters cannot control how much a user buys on the website.} 
In terms of obtaining an unbiased estimate for the sampling uncertainty, many viable approaches already exist: experimenters in medicine, economics, and social sciences often employ a crossed random effects model~\cite{baayen08mixedeffects} or cluster robust standard errors~\cite{cameron11robust}. Meanwhile, those in digital technology prefer using bootstrap~\cite{bakshy13uncertainty} or the delta method~\cite{deng18applying} to estimate the sample variance and the standard error due to their ability to scale to large datasets and relative lack of model assumptions.

Despite the existing literature dedicated to dependent responses in digital experiments, there is little published evidence specifically in the context of OCEs in e-commerce. Most published research is based on the context of OCEs in digital advertising and content, where experimenters randomise by users and analyse by sessions or page views~\cite{bakshy13uncertainty,deng18applying}; or social networks, where experimenters both randomise and analyse by users whose responses may correlate via their social connections~\cite{eckles17design,gui15network}. We believe the lack of evidence contributes to insufficient awareness of the issue from experimenters in this area.

This case study's contribution is evidence from real-life e-commerce experimentation operations. In particular:
\begin{enumerate}[topsep=0em]
    \item We show, using three e-commerce transaction datasets (with two publicly available), a positive correlation between the value/size of transactions and items from the same user. This leads to inflation in measurement uncertainty (Section~\ref{sec:rade_abv_correlation});
    \item We quantify the extent of the said inflation, which can be anywhere between 1x to $>$100x the usual estimate and is dependent on the business metric, experiment duration, and organisation (Section~\ref{sec:rade_abv_restimation});
    \item We highlight the impact of the said inflation on test power and confidence interval coverage in null hypothesis significance tests (Section~\ref{sec:rade_abv_impact}); and
    \item We share lessons learned while incorporating some of the established mitigation approaches into an e-commerce company's internal experiment analysis platform, including a critique of some popular approaches and practical tradeoffs during implementation (Section~\ref{sec:rade_abv_deployment}).
\end{enumerate}

\subsection{Beware of Hidden Inflation in Measurement Uncertainty}
\label{sec:rade_abv_correlation}

We first motivate why we may obtain inaccurate estimates of the sampling uncertainty when measuring changes to e-commerce metrics based on transactions or items in OCEs. Recall from the previous section (see Equation~\eqref{eq:rade_se_def}) that $\sqrt{s_X^2/n}$, the vanilla sample standard error (SE), is an unbiased estimator for the SE under i.i.d. assumptions yet becomes biased when the responses are dependent.

The i.i.d. assumption is the prevailing assumption when experimenters randomise by users and analyse by user~\cite{deng17trustworthy}. However, it is unlikely to hold when experimenters randomise by users and analyse by transactions or items. Many users tend to make a transaction that is similarly sized and valued to their previous transaction(s). They also tend to purchase items that are at a similar price point.
To demonstrate the claims above, we utilise three online retail/e-commerce transaction datasets. The UCI Online Retail II~\cite{chen12datamining,dua2019UCI} and Olist Brazilian e-Commerce~\cite{olist18braziliandataset} datasets are publicly available.\footnote{The experiment code and results on the two publicly available datasets are available on GitHub: \url{https://github.com/liuchbryan/oce-ecomm-abv-calculation}.} We also use a proprietary dataset from ASOS.com, a global online fashion retail company, that records transactions within a specific two-month period in 2022 on a particular mobile platform.\footnote{The data described are not representative of ASOS.com's overall business operations, and one should not draw any such conclusion from the dataset.} We summarise the three datasets in Table~\ref{tab:ecomm_dataset_summary}.

\begin{table}
  \onehalfspacing
  \caption{Summary of the three online retail/e-commerce transaction datasets featured in this paper (post data cleaning).}
  \label{tab:ecomm_dataset_summary}
  \begin{adjustbox}{width=17cm}
  \begin{tabular}{c|ccccc}
    Dataset & 
    \begin{tabular}[t]{@{}c@{}}
      \# Users/\\Customers
    \end{tabular} & 
    \begin{tabular}[t]{@{}c@{}}
      \# Transactions/\\Orders
    \end{tabular}& 
    \begin{tabular}[t]{@{}c@{}}
      \# Items/\\Units
    \end{tabular} & 
    \begin{tabular}[t]{@{}c@{}}
      \# Products/\\SKUs
    \end{tabular}& 
    \begin{tabular}[t]{@{}c@{}}
      Time span
    \end{tabular}
    \\[2em]
    \hline
    ASOS (Proprietary) &
    $\sim$\numprint{4180000} &
    $\sim$\numprint{9680000} &
    $\sim$\numprint{32900000} &
    $\sim$\numprint{330000} &
    \begin{tabular}[t]{@{}c@{}}
      62 days\\(2 months)
    \end{tabular}
    \\[2em]
    \begin{tabular}[t]{@{}c@{}}
      UCI Online Retail II\\\cite{dua2019UCI,chen12datamining}
    \end{tabular} &
    \numprint{5852} &
    \numprint{36594} &
    \numprint{10690447} &
    \numprint{4621} & 
    \begin{tabular}[t]{@{}c@{}}
      739 days\\(2 years)
    \end{tabular}
    \\[2em]
    \begin{tabular}[t]{@{}c@{}}
      Olist Brazilian\\e-Commerce~\cite{olist18braziliandataset}
    \end{tabular} &
    \numprint{94983} &
    \numprint{98199} &
    \numprint{112101} &
    \numprint{32729} & 
    \begin{tabular}[t]{@{}c@{}}
      729 days\\(2 years)
    \end{tabular}\\[2em]
  \end{tabular}
  \end{adjustbox}
  \vspace*{\baselineskip}
\end{table}

\begin{figure}
  \begin{center}
    \includegraphics[height=0.295\textwidth, trim = 0mm 2mm 2mm 2mm, clip]{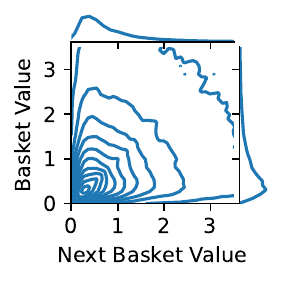}
    \includegraphics[height=0.295\textwidth, trim = 10mm 2mm 2mm 2mm, clip]{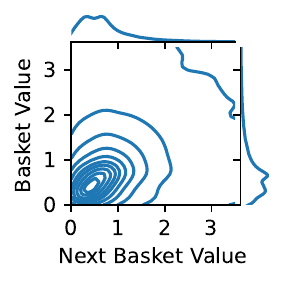}
    \includegraphics[height=0.295\textwidth, trim = 10mm 2mm 2mm 2mm, clip]{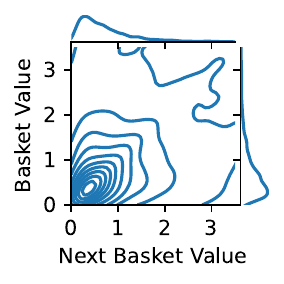}\\
    \includegraphics[height=0.295\textwidth, trim = 0mm 2mm 2mm 2mm, clip]{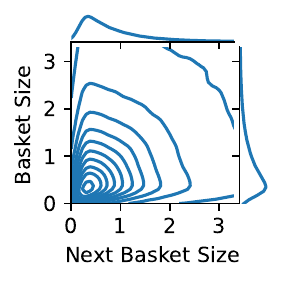}
    \includegraphics[height=0.295\textwidth, trim = 10mm 2mm 2mm 2mm, clip]{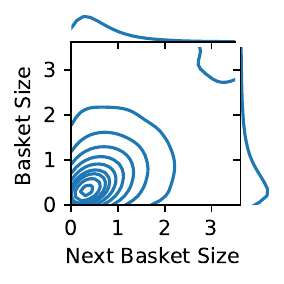}
    \includegraphics[height=0.295\textwidth, trim = 10mm 2mm 2mm 2mm, clip]{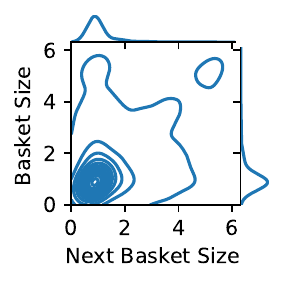}\\
  \end{center}
  \vspace*{-0.5em}
  \caption[Kernel density estimation plots showing a user's transaction value/size against that of their next transaction in three online retail transaction datasets.]{Kernel density estimation plots showing a user's transaction (Top row) value / (Bottom row) size against that of their next transaction -- if it exists -- in the (Left) ASOS, (Middle) UCI Online Retail II, and (Right) Olist Brazilian e-Commerce datasets. We normalise the transaction values/sizes by the ABV/ABS of respective datasets.}
  \label{fig:abv_EDA_BasketValue_True}
  \vspace*{\baselineskip}
\end{figure}

For each dataset, we plot the value and size of a user's transaction against that of their next transaction (if it exists) in Figure~\ref{fig:abv_EDA_BasketValue_True}. We observe from the figures a high empirical density along the $x=y$ diagonal and strong directionality in the kernel density estimates. This suggests a positive correlation between transaction values/sizes from the same user.
For brevity, we do not show the plots for item values, yet contend that a perfect correlation arises when a customer purchases multiple units of the same product.

As responses from the same user become positively correlated, we have $\textrm{Cov}(X_i, X_j) > 0$ for some~$i$ and~$j$ in Equation~\eqref{eq:rade_se_def} and thus a higher SE than the vanilla sample SE.

\subsection{Re-estimating the SE using bootstrap}
\label{sec:rade_abv_restimation}

\begin{figure}
  \begin{center}
    \includegraphics[width=0.8\textwidth, trim = 2mm 2mm 2mm 2mm, clip]{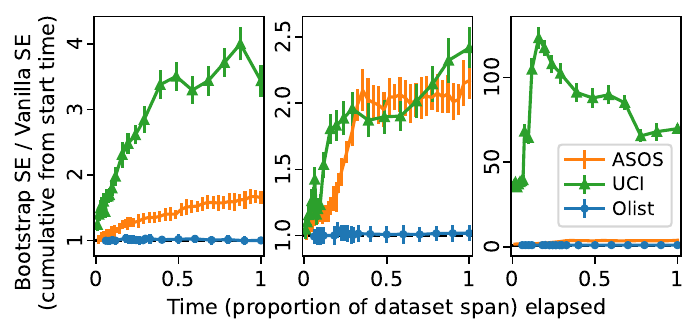}
  \end{center}
  \vspace*{-0.5em}
  \caption[The ratio between the one-way bootstrap standard error (SE) estimates and the vanilla sample SE estimates for different business metrics.]{The ratio between the one-way bootstrap standard error (SE) estimates~\cite{bakshy13uncertainty} and the vanilla sample SE estimates for (Left) ABV, (Middle) ABS, and (Right) ASP. The error bars represent the 95\% confidence interval for the one-way bootstrap SE estimates.}
  \label{fig:abv_SE_ABV}
  \vspace*{\baselineskip}
\end{figure}

We now explore how much the SE can inflate. We apply the bootstrap procedure described in Section~2.2 of~\cite{bakshy13uncertainty} to the datasets described above to re-estimate the SE. 
Bootstrapping generates a resample by sampling the original set of responses with replacement (or applying a random weight in this case). This yields a different sample mean. By repeating the process many times and taking many bootstrap means, we can estimate the SE by calculating the standard deviation of the bootstrap.

We use a one-way bootstrap (a.k.a. block bootstrap)
to account for the dependency between transactions/items and users. Instead of generating the resample by sampling each transaction/item individually, we sample clusters of transactions/items belonging to the same user, mirroring our randomisation process.\footnote{For ASP, in addition to the dependency between items and users, there may be additional dependence between items and products/stock-keeping units (SKUs). One may account for both using a two-way bootstrap described in Section~\ref{sec:rade_abv_deployment} or Section 2.3 of~\cite{bakshy13uncertainty}. \label{footnote:rade_asp}} Here, we use random weights generated from a Poisson(1) distribution to reweight our samples before calculating a bootstrap mean. The random weights emulate the number of times a transaction/item was sampled with replacement (along with other transactions/items associated with the same user). Finally, we estimate the SE from 500 bootstrap means.


Figure~\ref{fig:abv_SE_ABV} shows how the one-way bootstrap SE differs from the vanilla sample SE in successive expanding windows, corresponding to different potential OCE durations. We observe that the bootstrap SE is significantly higher than the vanilla sample SE, with the exact ratio between the one-way bootstrap SE and vanilla sample SE heavily dependent on the dataset/audience, business metric, and duration. For example, the bootstrap SE for ASP in the UCI Online Retail~II dataset is $>$100x the vanilla SE as many of their users buy tens or hundreds of the same product. Meanwhile, the ratios for ABV and ABS in the Olist Brazilian e-Commerce dataset are practically one. This is due to only 3.2\% of transactions coming from returning users. Given such differences, experimenters should strive to obtain a reasonably accurate estimate.

\begin{figure}
    \begin{center}
    \includegraphics[width=0.8\textwidth, trim = 2mm 2mm 2mm 2mm, clip]{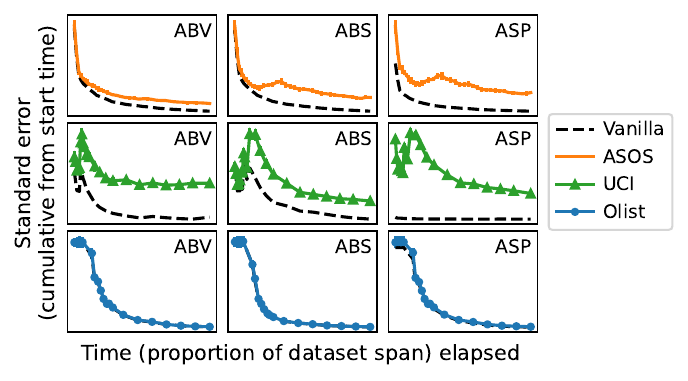}
    \end{center}
    \vspace*{-0.5em}
    \caption[Trajectories of the one-way bootstrap standard error (SE) estimates and the vanilla SE estimates under different dataset-metric combinations.]{Trajectories of the one-way bootstrap standard error (SE) estimates  (solid colored lines) and the vanilla SE estimates (dashed black lines) under different dataset-metric combinations. Estimates are cumulative from day one of the datasets. All x- and y-axes are on different scales.}
    \label{fig:abv_svbse_trajectory}
    \vspace*{\baselineskip}
\end{figure}

We also observe from Figure~\ref{fig:abv_svbse_trajectory} that the bootstrap SE may no longer drop as more transactions are involved over time and may go up again in some cases. This is likely due to returning users making further transactions, thus creating more and larger clusters of transaction/item values and sizes. It leads to the response variance increasing, sometimes more quickly than the increase in sample size. The observation suggests that lengthening an experiment solely to collect more responses (hence lowering the~SE) may backfire. In addition to established practices on sample size estimation~\cite{richardson22bayesian}, experimenters should also consider how the variance of their metrics evolves when designing OCEs with dependent data.

\subsection{Impact on OCE decisions}
\label{sec:rade_abv_impact}

We finally discuss how an inflated SE due to dependent data can affect decisions made from a NHST. Firstly, it reduces the power of the test, making any potential treatment effect harder to detect. Consider a two-tailed Student's $t$-test with significance level~$\alpha$ and $\nu$ d.f. Its power (adapted from Equation~\eqref{eq:statstest_power_ztest_twosided_nilhyp}) is
\begin{align}
     1 - T_{\nu}\big(t_{\nu, 1 - \alpha/2} - \theta/ \textrm{SE}\big) + T_{\nu}\big(-t_{\nu, 1 - \alpha/2} - \theta / \textrm{SE}\big),
     \label{eq:abv_test_power}
\end{align}
where $\theta$ is the effect size, $T_{\nu}(\cdot)$ is the CDF, and $t_{\nu, q}$ is the $q^{\textrm{th}}$ quantile of a $t$-distributed r.v. If the SE increases, then both the standardised effect size $\theta / \textrm{SE}$ and the test power in Expression~\eqref{eq:abv_test_power} decrease.

Secondly, having an inflated SE without knowing so will lead to tests producing confidence intervals that are too narrow, risking more false positives. In the test above, the $(1-\alpha)$ confidence interval~(CI) is 
$\big(\bar{x} \,\pm\, t_{\nu, 1 - \alpha/2} \cdot \textrm{SE}\big)$ -- see Equation~\eqref{eq:statstest_confidence_interval_normal}.
Fixing the CI bounds, we observe that $t_{\nu, 1 - \alpha/2}$ must decrease (and cease to be a $(1 - \alpha/2)$~quantile) when the SE increases. This means that the said CI will no longer have a $(1-\alpha)$ but a lower coverage, i.e., there is now a lower chance the interval will contain the actual effect size.

\begin{figure}
  \begin{center}
    \includegraphics[width=0.4\textwidth, trim = 0mm 2mm 2mm 2mm, clip]{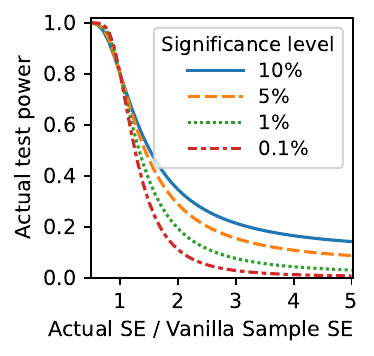}
    \includegraphics[width=0.4\textwidth, trim = 0mm 2mm 2mm 2mm, clip]{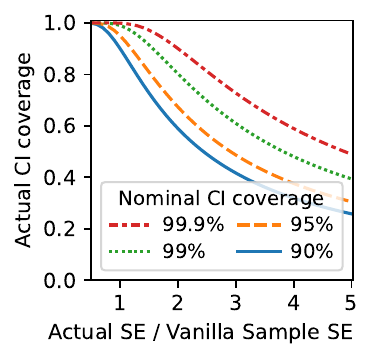} 
  \end{center}
  \vspace*{-0.5em}
  \caption[The actual test power of and the coverage of the centred confidence interval from a two-tailed \texorpdfstring{$z$}{z}-test against different standard error ratios.]{(Left) The actual test power of and (Right) the coverage of the centred confidence interval (CI) from a two-tailed $z$-test against different standard errors (SE), expressed as a multiple of the vanilla sample SE $\sqrt{s^2/n}$. We choose the test parameters such that the test power $=80\%$ and the CI coverage $=$ the nominal values when the SE multiples are one.}
  \label{fig:abv_actual_test_power_CI_coverage}
  \vspace*{\baselineskip}
\end{figure}

To show the full extent of the issues above, we plot the test power and CI coverage of a two-tailed $z$-test, essentially a Student's $t$-test with large degrees of freedom, against different SEs in Figure~\ref{fig:abv_actual_test_power_CI_coverage}. We observe that the power of a $z$-test with a 5\% significance level (dashed line) tumbles from 80\% to around~29\% when we merely double the vanilla sample SE. Moreover, the centred 95\% CI calculated using the vanilla sample SE would only have roughly 67\% coverage. These results reinforce the importance of having an accurate estimate of the SE.

\subsection{Lessons Learned During Deployment}
\label{sec:rade_abv_deployment}

We finally share some lessons learned when incorporating the calculations of e-commerce metrics into ASOS.com's experiment analysis platform, which has been in production since Sep 2019. The analysis platform extends the functionality of a third-party experimentation platform, namely splitting, logging, and alerting. It adds value to the business by computing and performing statistical tests on bespoke business metrics that use internal datasets.

As mentioned at the beginning of the section, experimenters in digital technology prefer the use of bootstrap (see Section~2.2 of~\cite{bakshy13uncertainty}) and the delta method (see Section~3 of~\cite{deng18applying}) to re-estimate the SE, which acts as a plug-in estimate for subsequent statistical tests. We described and implemented the bootstrap approach in Section~\ref{sec:rade_abv_restimation} when we quantified the extent of the SE inflation.
The delta method approach expresses a transaction-/item-based metric as the quotient of two user-based metrics. For example, we can express ABV as the quotient of the business metrics ``mean spend per user (across all baskets)'' and ``mean number of baskets per user.'' Given that both metrics in the quotient are asymptotically normal and based on randomisation unit-based responses, which are commonly assumed to be i.i.d.~\cite{deng17trustworthy}, we can use the delta method to estimate the variance of the quotient. The resultant formula has an explicit term for the dependency between users and transactions/items.

\paragraph{Delta method - a straightforward formula with hidden complexity and inflexibility}
The delta method approach is more compute efficient than the bootstrap approach. It requires passing only group-level statistics around instead of analysis/sub-randomisation unit-level responses required by bootstrap methods.\footnote{That said, to calculate the group-level statistics, which include the covariance between users and transactions/items, one still requires \emph{access} to sub-randomisation unit responses initially (see Section~\ref{sec:oced_mapping_ttest}).} However, it is unclear how the results in~\cite{deng18applying}, developed with a one-way dependency between randomisation and analysis units in mind, would apply to business metrics that feature responses dependent on other types of units.\footref{footnote:rade_asp} The approach also requires considerable statistical mastery to understand and implement, risking decreased engagement from engineering teams and non-technical stakeholders due to general apprehensiveness towards ``black box'' algorithms.

\paragraph{Multi-way bootstrap may be overkill by being overly conservative}

The multi-way bootstrap described in Section 2.3 of~\cite{bakshy13uncertainty} aims to address dependencies between different types of units. One may apply such an approach to item-based metrics (e.g., ASP) calculations, where dependence between items and products exists alongside dependence between items and users.

Recall from Section~\ref{sec:rade_abv_restimation} that in a one-way reweighting bootstrap, we apply a random weight drawn from Poisson(1) to each user and all their associated items before calculating a single bootstrap mean. In a two-way bootstrap, we apply a random weight to each user and a separate random weight to each product. The final weight for each item is the product of the user weight and the product weight.\footnote{In the general case where there are dependencies between multiple types of units, we apply a random weight to each unit within a single unit type, repeat for other unit types with a separate set of random weights, and take the final weight of each link/interaction/dependency between the unit types as the product of the weight of their associated units.} Similar to the one-way bootstrap, the final weight emulates the number of times the item will be selected if we separately sample the users and products with replacement. The calculations for both the bootstrap means and SE remain identical post-reweighting.

As a two-way bootstrap preserves the correlation structure between items, users, and products during sampling, one would expect the method to generate less biased (if not unbiased) SE estimates. However, the SE estimates from a two-way bootstrap are impractically large (see Table~\ref{tab:ecomm_asp_oneway_twoway_comparison}). This appears to match the results reported in~\cite{bakshy13uncertainty}, where two-way bootstrap SE estimates generally yield CIs that are overly conservative in A/A tests.\footnote{A/A tests are tests featuring two supposedly identical treatments and, thus, produce results that are consistent with a nil null hypothesis. See Section~\ref{sec:statstest_considerations} for further discussions. \label{footnote:rade_aa_test}} While the true SE will always be unknown, we believe overestimating it is just as bad as underestimating it, as experimenters will then struggle to design any experiments with sufficient power.

The analysis platform currently uses a one-way, user-based bootstrap to estimate the SE of e-commerce metrics. The approach mimics the randomisation process, and there is extensive evidence (both in~\cite{bakshy13uncertainty} and within the organisation) that it produces CIs with the right coverage in A/A tests.
The team responsible constantly evaluates whether a multi-way bootstrap will become necessary, especially for experiments with product-treatment interactions.\footnote{Section 4 of~\cite{bakshy13uncertainty} shows that a one-way, user-based bootstrap may underestimate the SE of item-based metrics when there are product-treatment interactions. Note that ``products'' in this case study are referred to as ``items'' in~\cite{bakshy13uncertainty} due to different business contexts.}

\begin{table}
  \onehalfspacing
  \caption[The ratio between the bootstrap standard error (SE) estimates and the vanilla sample SE estimates for ASP.]{The ratio between the bootstrap standard error (SE) estimates and the vanilla sample SE estimates for ASP. Each entry represents the mean and standard deviation of the ratios across multiple experiment runs.}
  \label{tab:ecomm_asp_oneway_twoway_comparison}
  \begin{center}
  \begin{tabular}{cccc}
    \hline
    Dataset & Duration & One-way bootstrap & Two-way bootstrap\\
    \hline
    ASOS &
    First 30d &
    3.51 $\pm$ 0.13 &
    17.77 $\pm$ 0.56 \\
    \hline
    \multirow{3}{*}{UCI} &
    First 10\%&
    64.45 $\pm$ 1.94 &
    72.83 $\pm$ 2.20 \\
    &
    First 50\% &
    87.95 $\pm$ 2.82 &
    105.15 $\pm$ 3.33 \\
    &
    All &
    69.82 $\pm$ 1.37 &
    94.33 $\pm$ 2.28 \\
    \hline
    \multirow{3}{*}{Olist} &
    First 20\% &
    1.104 $\pm$ 0.035 &
    2.083 $\pm$ 0.075 \\
    &
    First 50\% &
    1.058 $\pm$ 0.036 &
    2.531 $\pm$ 0.066 \\
    &
    All &
    1.076 $\pm$ 0.037 &
    3.364 $\pm$ 0.109 \\
    \hline
  \end{tabular}
  \end{center}
  \vspace*{\baselineskip}
\end{table}

\paragraph{Further consideration - error vs runtime/cost tradeoff}
Bootstrap methods introduce their error (more precisely, \emph{variance component}) to SE estimates during the resampling process. This is in addition to the (inflated) sampling uncertainty and generally scales along $O\big(1/\sqrt{B}\big)$, where $B$ is the number of bootstrap resamples~\cite{efron93introduction}. Given the runtime of bootstrap methods scales along~$O(B)$, at some point, it is no longer economically practical for a production system to draw further resamples to reduce the error to SE estimates. 

Deciding where the tipping point is more an art than science and depends more on individual needs. Unlike many technology giants which perform analysis in a streaming fashion, our analysis platform features batch-processing pipelines due to temporal constraints from upstream data sources. As a result, our priority is to keep the processing jobs' runtime (and hence compute cost) low. The team finds $B \in [500, 1000]$ works well as it leads to a coefficient of variation (error of SE estimate divided by mean SE estimate) of~$<5\%$. That said, we believe the tradeoff also applies to streaming systems as the compute latency increases with a larger~$B$, which impacts system availability.





\section{Advances in Quasi-experiments and Natural Experiments}
\label{sec:rade_causalinference}

We move on to works related to quasi-experiments and natural experiments. While they differ regarding whether an experimenter can impose scientific control, they involve the same set of causal inference techniques to estimate the treatment effect. We thus combine the discussion on related methods and examples for the two experiment classes.

To date, few works are dedicated to reviewing the use of causal inference techniques in digital experimentation. Chapter 11 of~\cite{kohavi20trustworthy} briefly outlined commonly used causal inference methods with examples, and~\cite{deng2021causal} is a work-in-progress discussing its application in the online industry. That said, causal inference is an established discipline, and many excellent introductions and reviews exist:
\begin{itemize}
  \item \cite{hernan2020causal} and~\cite{imbens2015causal} provide an accessible introduction to the area and review some models and techniques. Both books stem from courses the respective authors taught in the past decades and are geared towards researchers and practitioners in the medical and social sciences, the latter sharing some scientific goals with digital experimentation.
  \item \cite{pearl09causal}~focuses on the theoretical constructs (or representations) in causal inference -- structural equation models, the potential outcomes framework (see Section~\ref{sec:rade_pof} for the basics), and graphical models. The paper also proposed the structural causal model, which is seen as either a unifying or an alternative representation to those above.
  \item \cite{yao2021survey} extensively surveys the methods, datasets, software packages, and applications in (general) causal inference. Sections of note include  the applications in advertising and recommendations (Sections 6.1--6.2). The paper is tailored to the computer science and data mining communities, which substantially overlap with the digital experimentation community.
  \item \cite{kaddour2022causal} explores works in the intersection of causal inference and machine learning \linebreak (CausalML), which has flourished in the past decade. At the same time,~\cite{varian2016causal} outlines many of its applications in economics and marketing. Many experiments described in the latter are run offline, though most techniques can be directly translated to an online setting.
\end{itemize}

We keep our review close to digital experimentation and cover the following in the remainder of the section. We first revisit the conditional average treatment effect (CATE) and conditional exchangeability concepts in Section~\ref{sec:rade_causalinference_cate}. Many causal inference methods seek to estimate the former and assume the latter. We then list several classes of popular methods in causal inference, followed by examples applying the said methods in digital experiments in Section~\ref{sec:rade_causalinference_methods}. We finally provide a mini-spotlight in geo-experiments, a form of quasi-experiments that assign treatments based on the geographical region a user falls within, in Section~\ref{sec:rade_causalinference_geoexperiments}.

\subsection{CATE and Conditional Exchangeability Revisited}
\label{sec:rade_causalinference_cate}

Many causal inference techniques seek to estimate the conditional average treatment effect (CATE), the average treatment effect among those sharing certain covariates $\mathcal{C}$ (those on which the treatment assignment $\mathcal{A}$ does not affect):\footnote{Here, we separate the covariates~$\mathcal{C}$ and the treatment assignment indicator~$\mathcal{A}$. Some in the causal inference community lump these two objects together when discussing CATE. Others may have a strict separation between these objects, with additional terminologies such as average treatment effect on the treated (ATT, i.e., CATE on ${\mathcal{A} = 1}$) and average treatment effect on the control/untreated (ATC/ATU, i.e. CATE on~$\mathcal{A} = 0$)~\cite{greifer2023choosing}.}
\begin{align}
  \mathbb{E}(Y(1) - Y(0) \,|\, \mathcal{C}) = \mathbb{E}(Y(1) \,|\, \mathcal{C}) - \mathbb{E}(Y(0) \,|\, \mathcal{C}) \,.
\end{align}
The availability of covariates means one can produce a more expressive model $Y = g(\mathcal{A},\, \mathcal{C})$ to capture the complex dynamics between the treatments and observed responses, achieved by drawing on the wealth of potentially complex statistical and machine learning models~\cite{johnson2007applied}. Of course, such a model may still only capture correlation/association instead of causation, as its dependent/target variables are not the potential responses.

If, in addition, the covariates identify and account for all confounders, we noted that the potential responses are conditionally exchangeable, i.e., $Y(0), Y(1) \indep \mathcal{A} \,|\, \mathcal{C}$.\footnote{Also known as \emph{conditional unconfoundedness} or \emph{strong ignorability}~\cite{deng2021causal,rosenbaum1983central,savje2021randomization}.} Similar to its unconditional counterpart in Equation~\eqref{eq:rade_pof_mean_exchangability}, if the potential responses are conditionally exchangeable, we have
\begin{align}
  \mathbb{E}(Y(a) \,|\, \mathcal{C}) = \mathbb{E}(Y(a) \,|\, \mathcal{A}=1,\, \mathcal{C}) = \mathbb{E}(Y(a) \,|\, \mathcal{A}=0,\, \mathcal{C}),\; a\in \{0, 1\} \,.
  \label{eq:rade_cate_cond_exchangeability}
\end{align}
This, combined with the definition of potential outcomes in Equation~\eqref{eq:rade_pof_counterfactual_response}, enables us to estimate the potential responses using the observed responses~$Y$ conditioned on the covariates~$\mathcal{C}$ and, thus, develop a model that captures the complex dynamics between the treatments and potential responses, i.e., causation.

At first glance, conditional exchangeability appears to be a huge assumption. This is mostly true but not insurmountable in practice. 
We live in a complex and dynamic environment; thus, it is difficult to identify and account for all confounders, which biases the treatment effect estimate as discussed in Section~\ref{sec:rade_pof}.
On the other hand, many environmental variables have a tenuous link to our treatments and responses, and the bias they introduce may be practically insignificant. Identifying confounders is one of the main tasks in causal discovery and remains an active research topic, together with the study of the impact of hidden confounders~\cite{greenland1999confounding,pearl2009simpsons}.

It is also worth emphasising that the decision to estimate CATE instead of ATE and the ability to perform random assignments are two orthogonal concepts. One can estimate CATE in randomised controlled trials (as in the case of heterogeneous treatment effect detection and uplift modelling discussed in Sections~\ref{sec:rade_oce_statsmethods} and~\ref{sec:oced_survey}). They can also recover the population-level ATE from various CATE estimates by applying the law of total expectation:
\begin{align}
  \mathbb{E}(Y(1) - Y(0)) = \mathbb{E}_{\mathcal{C}}(\mathbb{E}(Y(1) - Y(0) \,|\, \mathcal{C})).
\end{align}
That said, it might be difficult in practice as the distribution of $\mathcal{C}$ is often unknown. While one can use the empirical distribution of $\mathcal{C}$, it may incur bias.

\subsection{Methods and Examples}
\label{sec:rade_causalinference_methods}

Given the underlying goal, we are in a position to list and briefly describe some popular causal inference techniques. \cite{angrist2014mastering} called the following approaches the ``furious five'' in econometrics, most of which are translatable to digital experiments:
\begin{itemize}[itemsep=0.25em]
    \item Randomisation: This has been extensively discussed in the previous sections.
    \item (Outcome) regression: A model that maps the relationship between the treatment assignment indicator, other covariates, and the responses. By manipulating the treatment assignment indicator within the fitted model, one can obtain the CATE for a particular covariate combination, provided the covariates present are sufficient for conditional exchangeability. See Chapter 9 of~\cite{gelman2006data} for further discussions.
    \item Instrumental variables: Variables that only affect the responses via their effect on the treatment assignment indicator or other covariates (but not directly). They must also be independent of the error term. These variables are often employed when the treatment assignment indicator or other covariates correlate with the unobserved error terms in a regression model.
    \item Regression discontinuity design: Specifies a cut-off to a continuous eligibility index and assigns treatment based on which side of the cut-off an experimental unit falls on. By exploiting the assumption that there is no meaningful difference in experimental units' characteristics or activity around the cut-off point before the treatment, one can estimate the local treatment effect as all others are roughly equal.
    \item Difference-in-differences: Calculates, per treatment, the difference in average response pre- and post-treatment and, then, another difference in the differences arising from different treatments. Under the parallel trend assumptions, i.e., the non-treatment variables affect the average response in both groups similarly across time, one is constructing a counterfactual when taking the per-treatment difference and estimating the treatment effect by comparing the counterfactual to the observed when taking the final difference.
\end{itemize}

\paragraph{Propensity scores} To address hidden confounders between treatment and outcomes, some experimenters match the experimental units or reweigh the responses. This is usually done via the \emph{propensity score}, the probability that an experimental unit is assigned a certain treatment given the covariates associated with the unit~\cite{rosenbaum1983central}. The score is most straightforwardly estimated via a logistic regression model. Once calculated, it is used to match a unit receiving a treatment with other units receiving another treatment (known as \emph{propensity score matching})~\cite{rosenbaum1983central}. The estimated treatment effect is the difference in responses between the unit and its matched counterparts, averaged over all matches. Alternatively, one can reweigh the responses using the score and take the difference between the reweighted average response to obtain the treatment effect (known as \emph{inverse probability/propensity weighting})~\cite{hernan2006estimating}.

Another increasingly popular technique related to propensity scores is \emph{doubly robust estimation}~\cite{funk2011doubly}. It combines outcome regression and propensity score models and yields a valuable property: the treatment effect estimates are consistent when at least one of the two models is correctly specified.
This approach is taken by~\cite{xu2016evaluating} to estimate the effect of changes to LinkedIn's mobile app, where the ability to perform random assignments is surrendered to the app store and users (as discussed in Section~\ref{sec:rade_classes_quasi}).

\paragraph{Mediation analyses} Digital experimenters also turn to mediation analyses. They feature the use of mediators, variables directly influenced by the treatment assignment, and, in turn, directly influence the responses.\footnote{Not to be confused with instruments. Explaining \emph{why} and \emph{how} a mediator and an instrument are different involves an introduction to graphical models in causal inference, which we have omitted for brevity. See, e.g., Chapters 2 \& 3 of~\cite{pearl2016causal} for more details.} Using such intermediary variables enables one to attribute it to different causal pathways between the treatment assignment and change in responses. It also enables one to produce more accurate treatment effect estimates if the variables are unaffected by the same confounders that affect the treatment assignment indicator and responses. This is the case in~\cite{hill15measuring}, who, seeking to measure the effect of a display advert targeting algorithm on certain user behaviour, uses ad viewability\footnote{Defined in~\cite{hill15measuring} as whether at least half of the advert has come to a user's view for longer than a second. Treatment in display advertising is usually defined as serving an advert, i.e., the advert is planned/scheduled to appear on a user's screen as they scroll through the page. While such a definition leads to much easier tracking from an experimenter's perspective, the ad may not come into users' view for many reasons  (e.g., the user stops scrolling or has the wrong browser dimensions).~\cite{hill15measuring} observed that 45\% of adverts have not come to view and hence cannot influence any change in user behaviour.} as a mediating event. Moreover,~\cite{wang2020causal}, in the quest to measure the impact of algorithmic changes to, e.g., predictive models or recommender systems, uses evaluation metrics from online A/B tests as mediators. They bridge the algorithmic changes’ effect on evaluation metrics in offline experiments to that on business KPIs.



\subsection{A mini spotlight on geo-experiments}
\label{sec:rade_causalinference_geoexperiments}

A class of quasi-experiments gaining traction in the past decade is geo-experiments. They are heavily used in display advertising~\cite{barajas20advertising,vaver11measuring} -- the author also led a team at ASOS.com to build and maintain such a geo-experimentation platform to help its marketing function manage tens of millions of spend~\cite{benbaccar2023geo,mccabe2021accelerating}.

A geo-experiment assigns treatments at a geographical region level. For example, it may assign all users based in Oxfordshire to control and all users based in Cambridgeshire to treatment. Such treatment assignment is necessitated by the inability to target and track individual users (due to privacy and data protection concerns or third-party platform restrictions) and, sometimes, by interference between individual users (see Section~\ref{sec:rade_oce_statsmethods}). The treatment assignment may or may not be random -- while a randomised assignment will simplify the experiment to a straightforward cluster randomised OCE (see Section~\ref{sec:rade_oce_statsmethods}), experimenters may decline to do so due to various operational constraints (e.g., cross-region user activities, minimum advertising budget/coverage requirements).

We estimate the treatment effect by comparing the responses from regions receiving the treatment to a counterfactual, constructed using the responses from control regions, on how the same regions would have performed had there been no treatment. More precisely and in the language of potential outcomes, consider a simplified case where we have two regions, one with response $X$ and the other with response $Y$. We call the former the control region and the latter the treatment region. We also assume we can access some covariates~$\mathcal{C}$ shared between the two regions that address all confounding, i.e., conditional exchangeability $Y(0), Y(1) \indep \mathcal{A} \,|\, \mathcal{C}$ holds. 

Before the experiment-proper, we assign both regions to $\mathcal{A} = 0$ (i.e., control) and observe
\begin{align}
    X \,|\, \mathcal{A}=0,\, \mathcal{C}  \quad\textrm{ and }\quad Y \,|\, \mathcal{A}=0,\, \mathcal{C} \,.
    \label{eq:rade_geo_training_observed}
\end{align}
We know from the definition of potential outcomes (see Section~\ref{sec:rade_pof} / Equation~\eqref{eq:rade_pof_counterfactual_response}) that Expression~\eqref{eq:rade_geo_training_observed} is equivalent to
\begin{align}
   X(0) \,|\, \mathcal{A}=0,\, \mathcal{C}  \quad\textrm{ and }\quad Y(0) \,|\, \mathcal{A}=0,\, \mathcal{C} \,.
   \label{eq:rade_geo_training_potential}
\end{align}
Suppose we have a mapping $g(\cdot)$ between the two observations under the same treatment assignment using the covariates, i.e., $Y = g(X, \mathcal{A}=0,\, \mathcal{C}) $. Given that the observations in Expressions~\eqref{eq:rade_geo_training_observed} and~\eqref{eq:rade_geo_training_potential} are equivalent, the following must also be true:
\begin{align}
  Y(0) = g(X, \mathcal{A}=0,\, \mathcal{C}) \,.
  \label{eq:rade_geo_mapping}
\end{align}

During the experiment, we assign the control region to $\mathcal{A} = 0$ and the treatment region to $\mathcal{A} = 1$ and observe\footnote{Readers may notice that it seems possible to observe responses under \emph{both} treatments ($Y \,|\, \mathcal{A}=0,\, \mathcal{C}$ in Expression~\eqref{eq:rade_geo_training_observed} and $Y \,|\, \mathcal{A}=1,\, \mathcal{C}$ in Expression~\eqref{eq:rade_geo_testing_observed}), which contradicts the premise of counterfactual reasoning. There is merely an artefact of oversimplifying the notations for illustration purposes -- the two observed responses happened at different points in time and, thus, ought to be represented by different r.v.'s. Here, we pretend $Y \,|\, \mathcal{A}=0,\, \mathcal{C}$ was never observed when we constructed the counterfactual.}
\begin{align}
    X \,|\, \mathcal{A}=0,\, \mathcal{C}  \quad\textrm{ and }\quad Y \,|\, \mathcal{A}=1,\, \mathcal{C} \,.
    \label{eq:rade_geo_testing_observed}
\end{align}
To estimate $Y(0)$, we require constructing a counterfactual using the observation $X$ and the mapping $g(\cdot)$, i.e. $g(X, \mathcal{A}=0,\, \mathcal{C}) \,|\, \mathcal{A}=0,\, \mathcal{C}$. This enables us to take the expected difference between the observed $Y$ and the counterfactual to obtain the CATE:
\begin{align}
    &  \mathbb{E}(Y \,|\, \mathcal{A}=1,\, \mathcal{C}) -
      \mathbb{E}(g(X, \mathcal{A}=0,\, \mathcal{C}) \,|\, \mathcal{A}=0,\, \mathcal{C}) &  \nonumber\\
    = \, & \mathbb{E}(Y \,|\, \mathcal{A}=1,\, \mathcal{C}) -
      \mathbb{E}(Y(0) \,|\, \mathcal{A}=0,\, \mathcal{C}) & \textrm{(RHS: By Eq.~\eqref{eq:rade_geo_mapping})}  \nonumber\\
    = \, & \mathbb{E}(Y(1) \,|\, \mathcal{A}=1,\, \mathcal{C}) -
      \mathbb{E}(Y(0) \,|\, \mathcal{A}=0,\, \mathcal{C}) & \textrm{(LHS: By definition -- Eq.~\eqref{eq:rade_pof_counterfactual_response})} \nonumber\\
    = \, & \mathbb{E}(Y(1) \,|\, \mathcal{C}) -
      \mathbb{E}(Y(0) \,|\, \mathcal{C}) & \textrm{(Conditional exchangeability -- Eq.~\eqref{eq:rade_cate_cond_exchangeability})} \nonumber\\
    = \, & \mathbb{E}(Y(1) - Y(0) \,|\, \mathcal{C}) & \textrm{(Linerarity of conditional expectation)}\,.
\end{align}

The example above is grossly simplified. In reality, there are many copies of $X$ and $Y$, both along the geographical (responses from multiple regions) and time dimensions (responses at multiple time points). Given the many variables involved, experimenters often prefer using the more expressive structural equation model (instead of the potential outcomes framework) to express causal assumptions~\cite{barajas20advertising,vaver11measuring}.

Many models have been proposed for the mapping $g(\cdot)$, with the two most common being geo-based regression (GBR) and time-based regression (TBR) models. GBR uses the aggregated metric value of the regions before and during the experiment, plus the advertising spend as covariates~\cite{vaver11measuring,vaver12periodic}. Thus, it allows experimenters to easily extract the return on ad spend (ROAS) metric from the fitted model. However, the statistical power of GBR is derived from the number of regions involved, which is problematic when the number of regions available for an experiment is small. The challenge is mitigated by TBR, which uses the value of the covariates at each time interval. TBR also employs Bayesian regression techniques, more precisely, Bayesian structural time series. This allows access to the full posterior distribution of the counterfactual and, thus, the calculation of a credible interval~\cite{chen19robust,kerman17estimating}. 

That said, similar to many other Bayesian models, one needs to take care when specifying the prior distributions. The team at ASOS.com found during development that non-informative priors generate credible intervals (associated with the posterior distribution of the counterfactual response or, more generally, the posterior predictive) that are too narrow based on A/A tests.\footref{footnote:rade_aa_test} The team thus switched to informative priors that enabled them to calibrate the variance of the posterior predictive based on simulated experiments.

There are many other considerations in designing a geo-experiment. They include (1) the creation of regions and (2) the selection of regions into control and treatment groups.
Often, regions are handcrafted based on expert input, though~\cite{rolnick2019randomized} showed it can be learnt from clustering past user queries using clickstream data. The optimal allocation of the regions into control and treatment groups is studied in \cite{barajas20advertising} and~\cite{au18atimebased}, with the former focusing on finding the best region pair (i.e. one region each for control and treatment) and the latter finding the best disjoint subsets of regions available\footnote{Consider the case where the experimenter can assign treatments over regions $\{1, 2, 3, \cdots, 7\}$. The optimal allocation may assign $\{2, 4, 7\}$ to control and $\{3, 5\}$ to treatment, leaving $\{1, 6\}$ unused.} using a hill-climbing algorithm. 

Finally, an experimenter also needs to consider the duration for different stages of a geo-experiment~\cite{benbaccar2023geo,mccabe2021accelerating}. Unlike most digital OCEs, geo-experiments feature multiple stages: two -- the period before the experiment where the data arising from two identical treatments is used to train the model and the period where the treatments differ -- or more if an experimenter imposes one or more ``cooldown'' periods to avoid conflating effect from multiple treatments across time. The test power of geo-experiments is also less dependent on the sample size but more on the model fit, which reflects the variance of the counterfactual responses. This complicates power, sample size (or duration), and minimum detectable effect calculations. Often, one performs a grid search across all possible combinations of stage durations to determine the optimal experiment runtime.

\section{A Brief Recap}
\label{sec:rade_recap}

We presented a brief review of recent advances in digital experimentation. These include advanced statistical methods that address the many sources of bias (e.g., post-selection inference, heterogeneous treatment effects, SUTVA violations) to experiments, improve experiment sensitivity, and produce treatment effect estimates using general causal inference techniques when one cannot perform a randomised controlled trial. To organise the research reported in the chapter, we have also introduced the potential outcomes framework and classified experiments into online randomised controlled trials, quasi-experiments, and natural experiments depending on the level of control an experimenter has on performing random assignments and imposing scientific control.

Interleaved between the reviews is an original case study on dependent responses in\linebreak e-commerce OCEs, which arises as experimenters randomise by users to ensure a consistent user experience yet analyse by a more granular transaction- or item-based decision metric as is common in the industry. Using three real online retail/e-commerce transaction datasets, we provided evidence of the extent of the dependent responses in practice. We then highlighted its impact on test power and confidence interval coverage. Finally, we shared the lessons learned when incorporating relevant methods in the review above into the internal experiment analysis platform of ASOS.com, a global online fashion retail company. We also provided a spotlight on geo-experiments, an increasingly popular experimental design to measure the effect of display advertising.
\cleardoublepage
\chapter{An Evaluation Framework for Personalisation Strategy Experiment Designs}
\label{chap:pse}

This chapter is adapted from the research paper ``\textit{An Evaluation Framework for Personalization Strategy Experiment Designs}'', presented at and awarded Best Student Paper of \textit{AdKDD 2020 Workshop (in conjunction with SIGKDD '20)} \cite{liu2020evaluation}.

\section{Motivation}
\label{sec:pse_introduction}

Without mentioning experimental design, no discussion on statistical challenges in digital experimentation and measurement capabilities is complete. As alluded to in Section~\ref{sec:rade_classes_pse}, in this chapter, we focus on the design of experiments that compare \emph{personalisation strategies} -- complex sets of targeted user interactions executed by e-commerce and digital marketing organisations that aim to create an individualised experience for every user to their websites. In addition to that mentioned in Section~\ref{sec:rade_classes_pse}, examples of personalisation strategies include the scheduling, budgeting, and ordering of marketing activities directed at a user based on their browsing and purchase history.

Compared to one's usual online controlled experiment (see Section~\ref{sec:rade_classes_rct} for examples), 
experiments for personalisation strategies face two unique challenges. Firstly, strategies are often only applicable to a small fraction of the user base, leading to many simple experiment designs suffering from either a lack of test power due to low sample size or a diluted treatment effect due to the inclusion of irrelevant samples~\cite{deng15diluted}.

Secondly, users are not randomly assigned \emph{a priori} but must qualify to be treated with a strategy via their actions or attributes. This leads to a one-sided non-compliance problem where experimenters do not have complete control over applying scientific control~\cite{johnson2023inferno}. At best, they can randomise a user's eligibility to be treated and treat those who are both eligible and qualified~\cite{johnson2023inferno,waisman2023multicell}. Under such a situation, groups of users subjected to different strategies cannot be assumed statistically equivalent and, hence, are not directly comparable.

While several variance reduction techniques, including stratification and control variates~\cite{deng13improving,poyarkov16boosted} (see Section~\ref{sec:rade_oce}), partially address the challenges, the strata and control variates can vary dramatically from one personalisation strategy experiment to another, requiring many \emph{ad hoc} adjustments. As a result, such techniques may not scale well when organisations design and run hundreds or thousands of experiments at any given time.

We argue that personalisation strategy experiments should focus on the assignment of users from the strategies they qualified for to the treatment/analysis groups. We call this mapping process an \emph{experiment setup}. Identifying the best experiment setup increases the chance of detecting any treatment effect. An experimentation framework can also quickly reuse and switch between different setups with little custom input, ensuring the operation can scale. More importantly, the process does not hinder the subsequent application of variance reduction techniques, meaning we can still apply the techniques \textit{post hoc} if required.

To date, many experiment setups exist to compare personalisation strategies. An increasingly popular approach is to compare the strategies using multiple control groups -- Quantcast calls it a dual control~\cite{quantcast}, and Facebook calls it a multi-cell lift study~\cite{liu2018designing,waisman2023multicell}. In the two-strategy case, this involves running two experiments on two random partitions of the user base in parallel, with each experiment further splitting the respective partition into treatment/control and measuring the \emph{incrementality} (the change in a business metric as compared to the case where we do nothing) of each strategy. We then compare the incrementalities of the two strategies against each other.

Despite the setup above gaining traction in display advertising, there is a lack of literature on whether it is a better setup -- one with a higher sensitivity, presents a higher effect size, or both. While~\cite{liu2018designing} noted that multi-cell lift studies require many users, they did not discuss how the number compares to other setups.\footnote{A single-cell lift study is often used to measure the incrementality of a single personalisation strategy and, hence, is not a representative comparison.} Identifying and adopting a better experiment setup can reduce the required sample size and enable more cost-effective experimentation.

We address the gap in the literature by introducing an evaluation framework that compares experiment setups given two personalisation strategies. The framework is designed to be flexible in dealing with a wide range of baselines and changes in user responses presented by any pair of strategies (\emph{situations} hereafter). However, we also recognise the need to quickly compare typical setups and provide some simple rules of thumb for situations where one setup will be better. In particular, we outline the desirable conditions for employing treatment effect dilution and a multi-cell setup.

To summarise, our contributions are: 
\begin{enumerate}
    \item (Section~\ref{sec:pse_eval_framework}) We develop a flexible evaluation framework for personalisation strategy experiments, where one can compare two experiment setups given the situation presented by two competing strategies;
    \item (Section~\ref{sec:pse_comparison}) We provide simple rules of thumb to enable experimenters who do not require the full flexibility of the framework to compare typical setups quickly; and 
    \item (Section~\ref{sec:pse_experiments}) We make our results useful to practitioners by making the code that performs empirical verification publicly available.\footnote{The code is available on GitHub: \url{https://github.com/liuchbryan/experiment\_design\_evaluation}. \label{footnote:pse_repo_link}}
\end{enumerate}

\section{Evaluation Framework}
\label{sec:pse_eval_framework}

We first present our evaluation framework for personalisation strategy experiments. The experiments compare two personalisation strategies, which we refer to as Strategy~1 and Strategy~2. 
Often, one is the existing strategy, and the other is a new strategy we intend to test and learn from. 
In this section, we introduce
\begin{enumerate}
    \item How users qualifying themselves into strategies creates non-statistically equivalent groups,
    \item How experimenters usually assign the users, and
    \item When we consider an assignment to be better.
\end{enumerate}
    
\subsection{User grouping}
\label{sec:pse_user_grouping}

\begin{figure}
\begin{center}
    \includegraphics[width=0.61\textwidth, trim = 0 0 0 0]{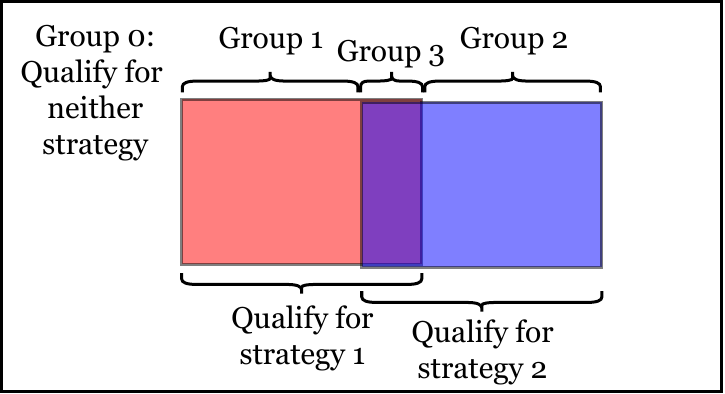}
\end{center}
\vspace*{-0.5\baselineskip}
\caption[Venn diagram of the user groups in our evaluation framework for personalisation strategy experiment designs.]{Venn diagram of the user groups in our evaluation framework. The outer, left inner (red), and right inner (blue) boxes represent the entire user base, those who qualify for Strategy 1, and those who qualify for Strategy 2, respectively.}
\label{fig:pse_ME_groups}
\vspace*{\baselineskip}
\end{figure}

As users qualify themselves into the two strategies, four disjoint groups emerge: those who qualify for neither strategy, those who qualify only for Strategy 1, those who qualify only for Strategy 2, and those who qualify for both strategies. We denote these groups (user) Groups~0, 1, 2, and 3, respectively (see Figure~\ref{fig:pse_ME_groups}). It is perhaps obvious that we cannot assume those in different user groups are statistically equivalent and compare them directly.

We assume Groups 0, 1, 2, and 3 have $n_0$, $n_1$, $n_2$, and $n_3$ users, respectively. We also assume responses from users (which we aggregate, often by taking the mean, to obtain our business metric) are distributed differently between groups, and within the same group, between the scenario where the group is subjected to the treatment associated with the corresponding strategy and where nothing is done (baseline). We list all group-scenario combinations in Table~\ref{tab:pse_group_id} and denote the mean and variance of the responses $(\mu_G, \sigma^2_G)$ for a combination~$G$.\footnote{For example, responses for Group~1 without any interventions have mean and variance ($\mu_{C1}$, $\sigma^2_{C1}$), and that for Group~2 with the treatment prescribed under Strategy~2 have mean and variance ($\mu_{I2}$, $\sigma^2_{I2}$).}

\begin{table}
\caption[All group-scenario combinations in our evaluation framework for personalisation strategy experiments.]{All group-scenario combinations in our evaluation framework for personalisation strategy experiments. The columns represent the groups described in Figure~\ref{fig:pse_ME_groups}. The baseline represents the scenario where we do nothing. We assume those who qualify for both strategies (Group 3) can only receive treatment(s) associated with either strategy.}
\label{tab:pse_group_id}
\begin{center}
\onehalfspacing
\begin{tabular}{c|c|c|c|c}
    &  Group 0 & Group 1  & Group 2  & Group 3 \\\hline
 Baseline (\textbf{C}ontrol) & $C0$ & $C1$ & $C2$ & $C3$ \\\hline
 Under treatment (\textbf{I}ntervention) & $/$ & $I1$ & $I2$ & $\begin{array}{ll} \textrm{Under Strategy 1:} \; I\phi \\ \textrm{Under Strategy 2:}  \; I\psi \end{array} $\\
\end{tabular}
\end{center}
\vspace*{\baselineskip}
\end{table}

\begin{figure}
\begin{center}
    \includegraphics[width=0.85\textwidth, trim = 0 0 0 0, clip]{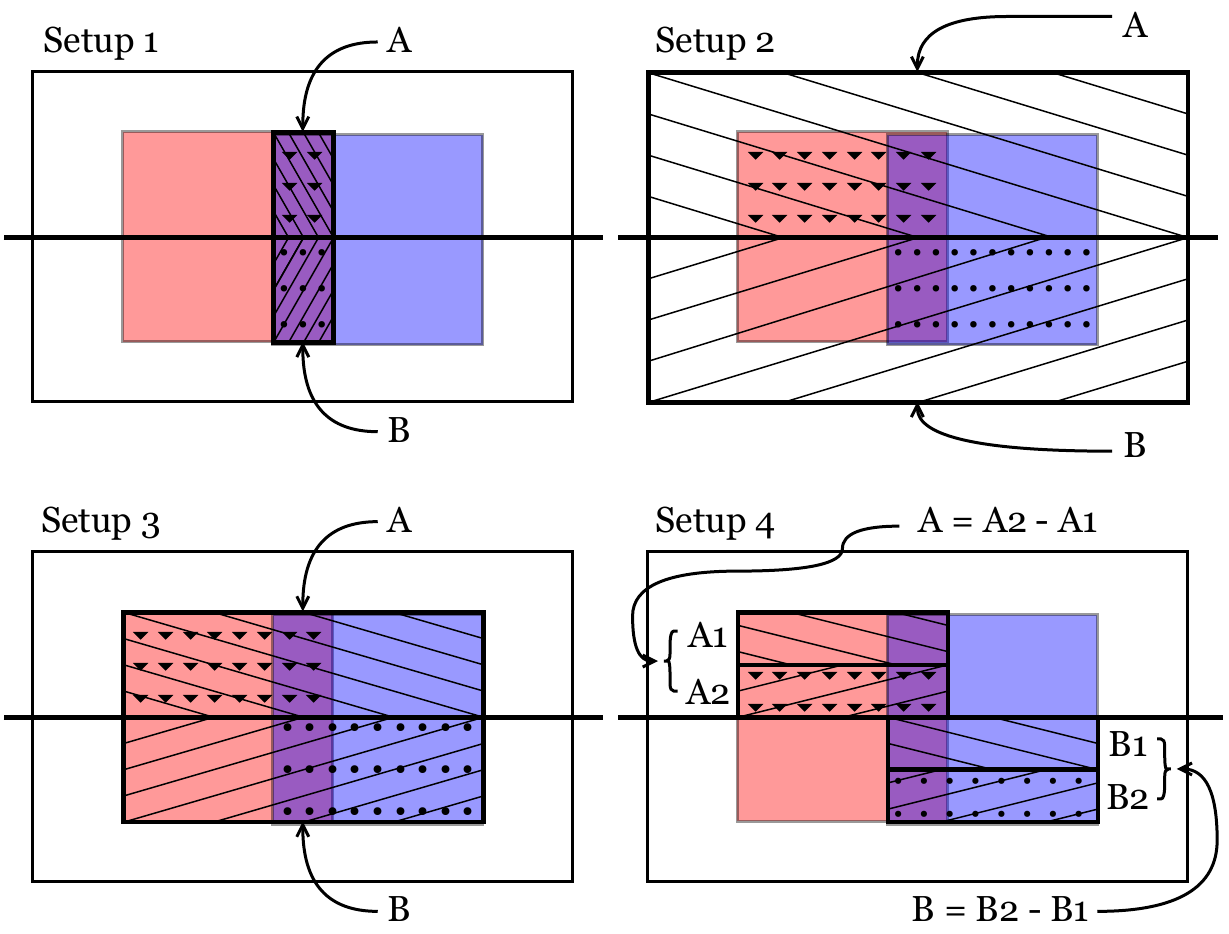}
\end{center}
\vspace*{-0.5em}
\caption[Experiment setups overlaid on the user grouping Venn diagram in Figure~\ref{fig:pse_ME_groups}.]{Experiment setups overlaid on the user grouping Venn diagram in Figure~\ref{fig:pse_ME_groups}. The hatched boxes cover users included in the analysis, and the downward triangles and dots cover users subjected to the treatment(s) prescribed under Strategies 1 and 2, respectively. See Section~\ref{sec:pse_experiment_setup} for a detailed description.}
\label{fig:pse_ME_designs}
\end{figure}

\subsection{Experiment setups}
\label{sec:pse_experiment_setup}
Many experiment setups exist and are in use in different organisations. Here, we introduce four typical setups of various sophistication, which we also illustrate in Figure~\ref{fig:pse_ME_designs}.

\paragraph{Setup 1 (Users in the intersection only)} The setup considers only users who qualify for both strategies. The said users are randomly split (usually 50/50) into two (analysis) groups, $A$ and $B$, and are prescribed the treatment specified by Strategies 1 and 2, respectively. The setup is easy to implement, though it is difficult to translate any learnings from the experiment to other user groups (e.g., those who qualify for one strategy only)~\cite{dmitriev2017adirtydozen}.

\paragraph{Setup 2 (All samples)} The setup is a simple A/B test that considers all users, regardless of whether they qualify for any strategy. The users are randomly split into two analysis groups, $A$ and $B$, and are prescribed the treatment specified by Strategy 1(2) if (i) they qualify under the strategy and (ii) they are in Group $A$($B$). This setup is easiest to implement but usually suffers severely from dilution in treatment effect~\cite{deng15diluted}.

\paragraph{Setup 3 (Qualified users only)} The setup is similar to Setup 2, except only those who qualified for at least one strategy (``triggered'' users in some literature~\cite{deng15diluted}) are included in the analysis groups. The setup sits between Setup 1 and Setup 2 in terms of user coverage. It has the advantage of capturing the greatest  number of relevant samples yet having the least treatment effect dilution. However, the setup also prevents one from telling the incrementality of a strategy itself, but only the difference in incrementalities between the two strategies.

\paragraph{Setup 4 (Dual control / multi-cell lift test)} As described in Section~\ref{sec:pse_introduction}, the setup first split the users randomly into two randomisation groups. For the first randomisation group, we consider those who qualify for Strategy 1 and split them into Analysis Groups $A1$ and $A2$. Group $A2$ receives the treatment prescribed under Strategy~1, and Group $A1$ acts as a control. The incrementality for Strategy~1 is then the difference in the business metric between Groups $A2$ and $A1$. Next, We apply the same process to the second randomisation group, with Strategy~2 and Analysis Groups $B1$ and $B2$ in place. Finally, we compare the incrementality for Strategies 1 and 2. The setup allows one to obtain the incrementality of each strategy and minimises treatment effect dilution. However, it also leaves many samples unused and creates additional analysis groups, and hence generally suffers from a low test power~\cite{liu2018designing}.

\subsection{Evaluation criteria}
\label{sec:pse_evaluation_criteria}
There are several considerations when one evaluates competing experiment setups. These include technical considerations, such as the complexity of setting up the setups on an OCE framework, and business considerations, such as whether the incrementality of individual strategies is required.

Here, we focus on the statistical aspect and propose two evaluation metrics: 
\begin{enumerate}[topsep=0pt]
    \item The actual average treatment effect size as presented by the two analysis groups in an experiment setup, and
    \item The sensitivity of the experiment setup, represented by the minimum detectable effect (MDE) under a pre-specified test power.
\end{enumerate}
Both evaluation metrics are necessary: the former indicates whether a setup suffers from treatment effect dilution, whereas the latter indicates whether the setup suffers from a lack of power or sample size.\footnote{The evaluation metric on experiment sensitivity pertains to the \emph{efficiency} of a statistical test~\cite{ems2011efficiency}. However, existing literature on this topic often compares the minimum sample size required by competing statistical tests instead. As shown earlier in Section~\ref{sec:statstest_nhstconcepts}, under a given significance level $\alpha$ and power threshold $\pi_{\textrm{min}}$, fixing the effect size will determine the sample size required and vice versa. Therefore, we opt to compare the MDE of competing personalisation strategy experiment designs because (1) it is generally easier to estimate the sample size and (2) it is on the same scale as the other evaluation metric.} An ideal setup should yield a high actual effect size and a high sensitivity (i.e., a low MDE),\footnote{We will use the terms ``high(er) sensitivity'' and ``low(er) MDE'' interchangeably.} though as we observe in the next section, it is usually a trade-off.

We formally define the two evaluation metrics. Let $A$ and $B$ be the two analysis groups in an experiment setup, with user responses randomly distributed with independent mean and variance $(\mu_A, \sigma^2_A)$ and $(\mu_B, \sigma^2_B)$, respectively. We also assume the sample size, namely~$n_A$ and $n_B$ for Groups $A$ and $B$, respectively, is sufficient such that the mean response of the two groups approximately follows the normal distribution by the central limit theorem. This enables us to run a two-sample approximate $z$-test.

The actual effect size is simply the difference in the mean of the two analysis groups:
\begin{align}
    \Delta \triangleq \mu_B - \mu_A \,.
    \label{eq:pse_ae_def}
\end{align}
It is worth noting that, unlike the means and variances of the user groups, the response means of the analysis groups, thus $\Delta$, vary depending on the experiment setup. We will demonstrate this property in Section~\ref{sec:pse_actual_mde_size}.

As covered in Section~\ref{sec:statstest_nhstconcepts_mde}, the MDE of a two-sample, two-sided $z$-test is
\begin{align}
    \theta^* \triangleq \left(z_{1 - \frac{\alpha}{2}} - z_{1 - \pi_{\textrm{min}}}\right)\,\sqrt{\frac{\sigma^2_A}{n_A} + \frac{\sigma^2_B}{n_B}} \,.
    \label{eq:pse_mde_def}
\end{align}

We finally define what it means to be better under these evaluation metrics. WLOG, we assume the actual effect sizes of the two competing experiment setups are positive\footnote{We swap the analysis groups if both actual effect sizes are negative. It is likely an error if the actual effect sizes are of opposite signs.} and say
a setup $S$ is superior to another setup $R$ if, all else being equal,
\begin{enumerate}
    \item $S$ produces a higher actual effect size ($\Delta_S > \Delta_R$) \emph{and} a lower minimum detectable effect size ($\theta^*_S < \theta^*_R$), or
    \item The gain in actual effect is greater than the loss in sensitivity:
    \begin{align}
        \Delta_S - \Delta_R > \theta^*_S - \theta^*_R \,,
    \end{align}
    which means an actual effect still stands a higher chance to be observed under $S$.
\end{enumerate}

\section{Comparing Experiment Setups}
\label{sec:pse_comparison}

Having described the evaluation framework above, in this section, we use the framework to compare the typical experiment setups described in Section~\ref{sec:pse_experiment_setup}. We will first derive the actual effect size and MDE for each setup in Section~\ref{sec:pse_actual_mde_size} and use the result to create rules of thumb on (1) whether diluting the treatment effect by including users who qualify for neither strategy is beneficial (Section~\ref{sec:pse_expt_comparison_dilution}) and (2) if dual control is a better setup for personalisation strategy experiments (Section~\ref{sec:pse_expt_comparison_dual_control}), two questions that are often discussed among e-commerce and marketing-focused experimenters.

\subsection{Actual and minimum detectable effect sizes}
\label{sec:pse_actual_mde_size}

We first present the four experiment setups' actual effect size and MDE. For each setup, we first compute each analysis group's sample size, mean response, and response variance, which arises as a mixture of user groups described in Section~\ref{sec:pse_user_grouping}. We then substitute the quantities computed into the definitions of~$\Delta$ (see Equation~\eqref{eq:pse_ae_def}) and~$\theta^*$ (see Equation~\eqref{eq:pse_mde_def}) to obtain the setup-specific actual effect size and MDE. We assume all random splits are done 50/50 in these setups to maximise the test power.

\paragraph{Setup 1 (Users in the intersection only)}
We recall the setup, which considers only users who qualify for both personalisation strategies (i.e. the intersection), randomly splits User Group~3 into two analysis groups, $A$ and $B$, each with the following number of samples:
\begin{align}
    n_A = n_B = \frac{n_3}{2} \,. \nonumber
\end{align}
Users in Analysis Group $A$ are provided treatment under Strategy~1, and users in Analysis Group $B$ are provided treatment under Strategy~2. This leads to the groups' responses having the following mean and variance: 
\begin{align}
    \mu_A = \mu_{I\phi} \,,\;\; 
    \mu_B = \mu_{I\psi} \,,\;\;
    \sigma^2_A = \sigma^2_{I\phi} \,,\;\;
    \sigma^2_B = \sigma^2_{I\psi} \,. \nonumber
\end{align}

The actual effect size and MDE for Setup 1 are, hence,
\begin{align}
    \Delta_{S1} &= \mu_{I\psi} - \mu_{I\phi} \,, \label{eq:pse_Delta_S1} \\
    \theta^*_{S1} &= (z_{1-\frac{\alpha}{2}} - z_{1-\pi_{\textrm{min}}})
    \sqrt{\frac{\sigma^2_{I\phi}}{\frac{n_3}{2}} + \frac{\sigma^2_{I\psi}}{\frac{n_3}{2}}} \,. \label{eq:pse_theta_star_S1}
\end{align}

\paragraph{Setup 2 (All samples)}
This setup also contains two analysis groups, $A$ and $B$, each taking half of all available users (regardless of whether they qualify for any strategy):
\begin{align}
    n_A = n_B = \frac{n_0 + n_1 + n_2 + n_3}{2} \,. \nonumber
\end{align}
The mean response and response variance for Groups $A$ and $B$ are the weighted mean response and response variance of the constituent user groups, respectively, weighted by the constituent groups' size. As we only provide treatment to those who qualify for Strategy~1 in Group~$A$ and those who qualify for Strategy~2 in Group~$B$, this leads to different responses in different constituent user groups:
\begin{align}
    \mu_A = \frac{n_0\mu_{C0} + n_1\mu_{I1} + n_2\mu_{C2} + n_3\mu_{I\phi}}{n_0 + n_1 + n_2 + n_3} \,, \;\;
    \mu_B = \frac{n_0\mu_{C0} + n_1\mu_{C1} + n_2\mu_{I2} + n_3\mu_{I\psi}}{n_0 + n_1 + n_2 + n_3} \,; \nonumber\\[0.5em]
    \sigma^2_A = \frac{n_0\sigma^2_{C0} + n_1\sigma^2_{I1} + n_2\sigma^2_{C2} + n_3\sigma^2_{I\phi}}{n_0 + n_1 + n_2 + n_3} \,, \;\;
    \sigma^2_B = \frac{n_0\sigma^2_{C0} + n_1\sigma^2_{C1} + n_2\sigma^2_{I2} + n_3\sigma^2_{I\psi}}{n_0 + n_1 + n_2 + n_3} \,. \nonumber
\end{align}
We then substitute the above into the actual effect size and MDE definitions and simplify the resultant expressions to obtain
\begin{align}
    \Delta_{S2} = \, &   \frac{n_1(\mu_{C1} - \mu_{I1}) + n_2(\mu_{I2} - \mu_{C2}) + n_3(\mu_{I\psi} - \mu_{I_\phi})}{n_0 + n_1 + n_2 + n_3} \,,
    \label{eq:pse_Delta_S2}\\[0.5em]
    \theta^*_{S2} = \, &  (z_{1-\frac{\alpha}{2}} - z_{1-\pi_{\textrm{min}}})
    \sqrt{\frac{
    2\left(n_0(2\sigma^2_{C0}) + n_1(\sigma^2_{I1} \!+\! \sigma^2_{C1}) +
    n_2(\sigma^2_{C2} \!+\! \sigma^2_{I2}) + n_3(\sigma^2_{I\phi} \!+\! \sigma^2_{I\psi})\right)
    }{(n_0 + n_1 + n_2 + n_3)^2}} \,.
    \label{eq:pse_theta_star_S2}
\end{align}

\paragraph{Setup 3 (Qualified users only)}
The setup is very similar to Setup~2, with members from User Group~0 excluded:
\begin{align}
    n_A = n_B = \frac{n_1 + n_2 + n_3}{2} \,. \nonumber
\end{align}
The absence of members from User Group 0 means they are not featured in the weighted mean response and response variance of the two analysis groups:
\begin{align}
    \mu_A = \frac{n_1\mu_{I1} + n_2\mu_{C2} + n_3\mu_{I\phi}}{n_1 + n_2 + n_3} \,, \;\;
    \mu_B = \frac{n_1\mu_{C1} + n_2\mu_{I2} + n_3\mu_{I\psi}}{n_1 + n_2 + n_3} \,; \nonumber\\[0.5em]
    \sigma^2_A = \frac{n_1\sigma^2_{I1} + n_2\sigma^2_{C2} + n_3\sigma^2_{I\phi}}{n_1 + n_2 + n_3} \,, \;\;
    \sigma^2_B = \frac{n_1\sigma^2_{C1} + n_2\sigma^2_{I2} + n_3\sigma^2_{I\psi}}{n_1 + n_2 + n_3} \,. \nonumber
\end{align}

This leads to the following actual effect size and MDE for Setup 3:
\begin{align}
    \Delta_{S3} = \, &   \frac{n_1(\mu_{C1} - \mu_{I1}) + n_2(\mu_{I2} - \mu_{C2}) + n_3(\mu_{I\psi} - \mu_{I_\phi})}{n_1 + n_2 + n_3} \,, 
    \label{eq:pse_Delta_S3}\\[0.5em]
    \theta^*_{S3} = \, &  (z_{1-\frac{\alpha}{2}} - z_{1-\pi_{\textrm{min}}}) 
    \sqrt{\frac{2\left(n_1(\sigma^2_{I1} + \sigma^2_{C1}) + n_2(\sigma^2_{C2} + \sigma^2_{I2}) + n_3(\sigma^2_{I\phi} + \sigma^2_{I\psi})\right)}{(n_1 + n_2 + n_3)^2}} \,.
    \label{eq:pse_theta_star_S3}
\end{align}

\paragraph{Setup 4 (Dual control)}
Setup 4 is unique amongst the experiment setups introduced as it has four analysis groups. Two analysis groups ($A1$ and $A2$) are drawn from those who qualified for Strategy 1 and are allocated into the first randomisation group. The other two ($B1$ and $B2$) are drawn from those who qualified for Strategy 2 and are allocated into the second randomisation group:
\begin{align}
    n_{A1} = n_{A2} = \frac{n_1 + n_3}{4}\,,\;\; n_{B1} = n_{B2} = \frac{n_2 + n_3}{4} \,.
    \nonumber
\end{align}
Each analysis group's mean response and response variance are the weighted mean response and response variance of the user groups involved, respectively:
\begin{align}
    \mu_{A1} = \frac{n_1\mu_{C1} + n_3\mu_{C3}}{n_1 + n_3} \,, \;\,
    \mu_{A2} = \frac{n_1\mu_{I1} + n_3\mu_{I\phi}}{n_1 + n_3} \,, \;\,
    \mu_{B1} = \frac{n_2\mu_{C2} + n_3\mu_{C3}}{n_2 + n_3} \,, \;\,
    \mu_{B2} = \frac{n_2\mu_{I2} + n_3\mu_{I\psi}}{n_2 + n_3} \,; \nonumber\\[0.5em]
    \sigma^2_{A1} = \frac{n_1\sigma^2_{C1} + n_3\sigma^2_{C3}}{n_1 + n_3} \,, \;\;
    \sigma^2_{A2} = \frac{n_1\sigma^2_{I1} + n_3\sigma^2_{I\phi}}{n_1 + n_3} \,, \;\;
    \sigma^2_{B1} = \frac{n_2\sigma^2_{C2} + n_3\sigma^2_{C3}}{n_2 + n_3} \,, \;\;
    \sigma^2_{B2} = \frac{n_2\sigma^2_{I2} + n_3\sigma^2_{I\psi}}{n_2 + n_3} \,. \nonumber
\end{align}

As the setup takes the difference of differences in the mean responses (i.e., the difference between the mean response for Groups~$B2$ and~$B1$ and the difference between the mean response for Groups~$A2$ and~$A1$),\footnote{Not to be confused with the difference-in-differences method, which captures the mean responses for the control and treatment groups at both the experiment's beginning (pre-intervention) and end (post-intervention). Here, we only capture the mean responses once at the end of the experiment.} the actual effect size is specified, post-simplification, as follows:
\begin{align}
    \Delta_{S4} & = (\mu_{B2} - \mu_{B1}) - (\mu_{A2} - \mu_{A1}) \nonumber\\
    & = \frac{n_2(\mu_{I2} - \mu_{C2}) + n_3(\mu_{I\psi} - \mu_{C3})}{n_2 + n_3} - \frac{n_1(\mu_{I1} - \mu_{C1}) + n_3(\mu_{I\phi} - \mu_{C3})}{n_1 + n_3} \,.
    \label{eq:pse_Delta_S4}
\end{align}
The MDE for Setup 4 is similar to that specified in the RHS of Equation~\eqref{eq:pse_mde_def}, albeit with more analysis groups featured in the square root term:
\begin{align}
    \theta^*_{S4} = \  & (z_{1-\frac{\alpha}{2}} - z_{1-\pi_{\textrm{min}}})
    \sqrt{\frac{\sigma^2_{A1}}{n_{A1}} + \frac{\sigma^2_{A2}}{n_{A2}} + \frac{\sigma^2_{B1}}{n_{B1}} + \frac{\sigma^2_{B2}}{n_{B2}}} \nonumber\\
    = \,&  2 \cdot (z_{1-\frac{\alpha}{2}} - z_{1-\pi_{\textrm{min}}}) \sqrt{\frac{n_1\left(\sigma^2_{C1} + \sigma^2_{I1}\right) + n_3\left(\sigma^2_{C3} + \sigma^2_{I\phi}\right)}{(n_1 + n_3)^2} +
          \frac{n_2\left(\sigma^2_{C2} + \sigma^2_{I2}\right) + n_3\left(\sigma^2_{C3} + \sigma^2_{I\psi}\right)}{(n_2 + n_3)^2}} .
    \label{eq:pse_theta_star_S4}
\end{align}

\subsection{Is dilution always bad?}
\label{sec:pse_expt_comparison_dilution}

The use of responses from users who do not qualify for any of the strategies we are comparing, an act known as treatment effect dilution, has stirred countless debates in experimentation teams. On the one hand, responses from these users make any treatment effect less pronounced by contributing exactly zero; on the other hand, it might be necessary as one does not know who actually qualifies for a strategy~\cite{liu2018designing}, or it might be desirable as they can be leveraged to reduce the variance of the treatment effect estimator~\cite{deng15diluted}.
Here, we are interested in whether we should engage in the act of dilution given the assumed user responses prior to an experiment. This can be clarified by understanding the conditions where Setup~3 would emerge superior (as defined in Section~\ref{sec:pse_evaluation_criteria}) to Setup~2. We relegate most of the intermediate algebraic work to Appendix~\ref{sec:miscmath_metric_dilution} for brevity.

By inspecting Equations~\eqref{eq:pse_Delta_S2} and~\eqref{eq:pse_Delta_S3}, it is clear that $\Delta_{S3} > \Delta_{S2}$ if~${n_0 > 0}$. Thus, Setup~3 is superior to Setup~2 under the first criterion if $\theta^*_{S3} < \theta^*_{S2}$, which is the case if $\sigma^2_{C0}$, the response variance of users who qualify for neither strategy, is large. This can be shown by substituting Equations~\eqref{eq:pse_theta_star_S2} and~\eqref{eq:pse_theta_star_S3} into $\theta^*_{S3} < \theta^*_{S2}$ and rearranging the terms to obtain
\begin{align}
    \frac{\left(n_1(\sigma^2_{I1} + \sigma^2_{C1}) + n_2(\sigma^2_{C2} + \sigma^2_{I2}) + n_3(\sigma^2_{I\phi} + \sigma^2_{I\psi})\right) \cdot (n_0 + 2n_1 + 2n_2 + 2n_3)}{2 (n_1 + n_2 + n_3)^2} < \sigma^2_{C0} \,.
    \label{eq:pse_setup23_criterion1}
\end{align}
If we assume the response variances are similar across groups with users who qualified for at least one strategy, i.e. $\sigma^2_{I1} \approx \sigma^2_{C1} \approx \cdots \approx \sigma^2_{I\psi} \approx \sigma^2_G$, Inequality~\eqref{eq:pse_setup23_criterion1} can then be simplified as
\begin{align}
    \sigma^2_G \left(\frac{n_0}{n_1 + n_2 + n_3} + 2\right) < \sigma^2_{C0} \,,
\end{align}
which can be used quickly to determine if one should consider dilution.

If Inequality~\eqref{eq:pse_setup23_criterion1} does not hold (i.e., $\theta^*_{S3} \geq \theta^*_{S2}$), we should then consider when the second criterion (i.e., $\Delta_{S3} - \Delta_{S2} > \theta^*_{S3} - \theta^*_{S2}$) holds. Writing
\begin{align}
    \eta & = n_1(\mu_{C1} - \mu_{I1}) + n_2(\mu_{I2} - \mu_{C2}) + n_3(\mu_{I\psi} - \mu_{I\phi}) \,, \nonumber\\
    \xi & = n_1(\sigma^2_{C1} + \sigma^2_{I1}) + n_2(\sigma^2_{I2} + \sigma^2_{C2}) + n_3(\sigma^2_{I\psi} + \sigma^2_{I\phi}) \,, \textrm{ and} \nonumber\\
    z & = z_{1-\frac{\alpha}{2}} - z_{1 - \pi_{\textrm{min}}} \,, 
    \label{eq:pse_setup23_maths_shorthand}
\end{align}
we can substitute Equations~\eqref{eq:pse_Delta_S2}, \eqref{eq:pse_theta_star_S2}, \eqref{eq:pse_Delta_S3}, and \eqref{eq:pse_theta_star_S3} into $\Delta_{S3} - \Delta_{S2} > \theta^*_{S3} - \theta^*_{S2}$ and rearrange the terms to obtain
\begin{align}
    \frac{n_1 + n_2 + n_3}{n_0}\sqrt{2n_0\sigma^2_{C0} + \xi} \,>\, 
    \frac{n_0 + n_1 + n_2 + n_3}{n_0}\sqrt{\xi} - \frac{\eta}{\sqrt{2}z} \,.
    \label{eq:pse_setup23_criterion2_initial}
\end{align}
As the LHS of Inequality~\eqref{eq:pse_setup23_criterion2_initial} is always positive, Setup 3 is superior if the RHS $\leq 0$. Noting 
\begin{align}
    \Delta_{S3} = \frac{\eta}{n_1 + n_2 + n_3}\,\textrm{ and }\,
    \theta^*_{S3} = \frac{\sqrt{2}\cdot z\cdot\sqrt{\xi}}{n_1 + n_2 + n_3} \,, \nonumber
\end{align}
the trivial case is satisfied if
\begin{align}
\frac{n_0 + n_1 + n_2 + n_3}{n_0} \, \theta^*_{S3} \leq \Delta_{S3} \,.
    \label{eq:pse_setup23_criterion2_strong}
\end{align}

If the RHS of Inequality~\eqref{eq:pse_setup23_criterion2_initial} is positive, we can safely square both sides and use the identities for $\Delta_{S3}$ and $\theta^*_{S3}$ to get
\begin{align}
    \frac{2\sigma^2_{C0}}{n_0} >
    \frac{\left(\theta^*_{S3} - \Delta_{S3} + \frac{n_1 + n_2 + n_3}{n_0} \theta^*_{S3}\right)^2 - \left(\frac{n_1 + n_2 + n_3}{n_0} \theta^*_{S3}\right)^2 }{2z^2} \,.
    \label{eq:pse_setup23_criterion2_final}
\end{align}
As the LHS is always positive, the second criterion is met if
\begin{align}
    \theta^*_{S3} \leq \Delta_{S3} \,.
    \label{eq:pse_setup23_criterion2_weak}
\end{align}
Note that this is a weaker, and thus more easily satisfiable condition than that introduced in Inequality~\eqref{eq:pse_setup23_criterion2_strong}.
This suggests that an experiment setup is always superior to a diluted alternative if the experiment is already adequately powered -- introducing any dilution will make things worse.

Failing the condition in Inequality~\eqref{eq:pse_setup23_criterion2_weak}, we can always fall back to Inequality~\eqref{eq:pse_setup23_criterion2_final}. While the inequality operates in squared space, it is essentially comparing the standard error of User Group~0 (LHS) -- those who qualify for neither strategy -- to the gap between the minimum detectable and actual effects ($\theta^*_{S3} - \Delta_{S3}$). The gap can be interpreted as the current noise level. Thus, a higher standard error means mixing in Group~0 users will introduce extra noise, and one is better off without them. Conversely, a smaller standard error means Group~0 users can lower the noise level, i.e. stabilise the treatment effect fluctuation, and one should take advantage of them.

To summarise, treatment effect dilution in a personalisation strategy experiment setup is \emph{not} helpful if 
\begin{enumerate}
    \item Users who do not qualify for any strategies have a large response variance (see Inequality~\eqref{eq:pse_setup23_criterion1}), or
    \item The experiment is already adequately powered (see Inequality~\eqref{eq:pse_setup23_criterion2_weak}).
\end{enumerate}
It could help if the experiment has yet to gain sufficient power and users who do not qualify for any strategy provide low-variance responses (i.e., when the complement of Inequality~\eqref{eq:pse_setup23_criterion2_final} holds), such that they exhibit stabilising effects when included in the analysis.

\subsection{When is dual control more effective?}
\label{sec:pse_expt_comparison_dual_control}
Often, when advertisers compare two personalisation strategies, the question of whether to use a dual control/multi-cell design comes up. Proponents of such an approach celebrate its ability to tell a story by making the incrementality of an individual strategy available, while opponents voice concerns about the complexity of setting up the design. Here, we are interested in whether Setup 4 (dual control) is superior to Setup 3 (a simple A/B test) under the prescribed evaluation criteria, and if so, under what circumstances.

We first observe $\theta^*_{S4} > \theta^*_{S3}$ always holds; hence, a dual control setup will never be superior to the more straightforward under the first criterion. This can be verified by substituting in Equations~\eqref{eq:pse_theta_star_S4} and~\eqref{eq:pse_theta_star_S3} and rearranging the terms to show $\theta^*_{S4} > \theta^*_{S3}$ is equivalent to
\begin{align}
    & 2\Bigg(
      \frac{n_1}{(n_1 + n_3)^2} (\sigma^2_{C1} + \sigma^2_{I1}) +
      \frac{n_2}{(n_2 + n_3)^2} (\sigma^2_{C2} + \sigma^2_{I2}) +
      \nonumber\\
    & \quad \frac{n_3}{(n_1 + n_3)^2} \sigma^2_{I\phi} +
      \frac{n_3}{(n_2 + n_3)^2} \sigma^2_{I\psi} +
      \left(\frac{n_3}{(n_1 + n_3)^2} + \frac{n_3}{(n_2 + n_3)^2}\right) \sigma^2_{C3} \Bigg) \nonumber\\
    > \; &
      \frac{n_1}{(n_1 + n_2 + n_3)^2} (\sigma^2_{C1} + \sigma^2_{I1}) +
      \frac{n_2}{(n_1 + n_2 + n_3)^2} (\sigma^2_{C2} + \sigma^2_{I2}) + 
    \nonumber \\
    & \frac{n_3}{(n_1 + n_2 + n_3)^2} \sigma^2_{I\phi} + 
      \frac{n_3}{(n_1 + n_2 + n_3)^2} \sigma^2_{I\psi} \;,
    \label{eq:pse_setup34_criterion1}
\end{align}
which, given all $n$-terms are non-negative and $\sigma^2$-terms are positive, always holds : not only the coefficients of the $\sigma^2$-terms are larger on the LHS than their RHS counterparts, the LHS also carries an additional~$\sigma^2_{C3}$ term with non-negative coefficient and a factor of two.

Moving on to the second evaluation criterion, we recall that Setup 4 is superior if $\Delta_{S4} - \Delta_{S3} > \theta^*_{S4} - \theta^*_{S3}$. Otherwise, Setup 3 is superior under the same criterion. We can see the full flexibility of the model by substituting Equations~\eqref{eq:pse_Delta_S3}, \eqref{eq:pse_theta_star_S3}, \eqref{eq:pse_Delta_S4}, and~\eqref{eq:pse_theta_star_S4} into $\Delta_{S4} - \Delta_{S3} > \theta^*_{S4} - \theta^*_{S3}$ and rearranging the terms to obtain
\begin{align}
    &\frac{n_1 \frac{n_2(\mu_{I2}-\mu_{C2}) + n_3(\mu_{I\psi}-\mu_{C3})}{n_2 + n_3} -
           n_2 \frac{n_1(\mu_{I1}-\mu_{C1}) + n_3(\mu_{I\phi}-\mu_{C3})}{n_1 + n_3}}
           {\sqrt{n_1(\sigma^2_{C1} + \sigma^2_{I1}) +
                  n_2(\sigma^2_{C2} + \sigma^2_{I2}) +
                  n_3(\sigma^2_{I\phi} + \sigma^2_{I\psi})}}
    \nonumber\\
    > & \sqrt{2}z 
    \vast[
      \sqrt{
        2 \cdot 
        \frac{
          \begin{array}{l}
            \big(1 + \frac{n_2}{n_1 + n_3}\big)^2 
              \big[n_1(\sigma^2_{C1} + \sigma^2_{I1}) +
              n_3(\sigma^2_{C3} + \sigma^2_{I\phi})\big] + \\[-0.1em]
            \quad \big(1 + \frac{n_1}{n_2 + n_3}\big)^2 
              \big[n_2(\sigma^2_{C2} + \sigma^2_{I2}) +
              n_3(\sigma^2_{C3} + \sigma^2_{I\psi})\big]
            \end{array}
        }
        {n_1(\sigma^2_{C1} + \sigma^2_{I1}) +
         n_2(\sigma^2_{C2} + \sigma^2_{I2}) +
         n_3(\sigma^2_{I\phi} + \sigma^2_{I\psi})}
        }
      - 1
    \vast] \,,
    \label{eq:pse_setup34_criterion2_full}
\end{align}
where $z = z_{1-\frac{\alpha}{2}} - z_{1 - \pi_{\textrm{min}}}$.

A key observation from inspecting Inequality~\eqref{eq:pse_setup34_criterion2_full} is that the LHS of the inequality scales along $O(\sqrt{n})$, where $n$ is the number of users, while the RHS remains constant. This leads to the insight that Setup 4 is more likely to be superior if the $n$-terms are large. Here, we assume the ratio $n_1:n_2:n_3$ remains unchanged when we scale the number of samples, an assumption that generally holds when an organisation increases their reach while maintaining its user mix. It is worth pointing out that our claim is stronger than that in previous work -- we have shown that having a large user base not only fulfils the requirement of running a dual control experiment as described in~\cite{liu2018designing}, but it also makes a dual control experiment a better setup than its simpler counterparts in terms of actual and minimum detectable effect sizes. 

We can see the scaling relationship more clearly by simplifying the $\sigma^2$- and $n$-terms. If we assume the response variances are similar across user groups (i.e., ${\sigma^2_{C1} \approx \sigma^2_{I1} \approx \cdots \approx \sigma^2_{I\psi} \approx \sigma^2_G}$), the RHS of Inequality~\eqref{eq:pse_setup34_criterion2_full} becomes
\begin{align}
    \sqrt{2} z 
    \left[ 
        \sqrt{\frac{n_1 + n_2 + n_3}{n_1 + n_3} + 
        \frac{n_1 + n_2 + n_3}{n_2 + n_3}} - 1 
    \right] \,,
    \label{eq:pse_setup34_criterion2_sigmaSimplified}
\end{align}
which remains a constant if the ratio $n_1:n_2:n_3$ remains unchanged. Separately, if we assume the number of users in Groups 1, 2, and 3 is similar (i.e., $n_1 \approx n_2 \approx n_3 \approx n$), the LHS of Inequality~\eqref{eq:pse_setup34_criterion2_full} becomes
\begin{align}
    \frac{\sqrt{n} \big((\mu_{I2} - \mu_{C2}) - (\mu_{I1} - \mu_{C1}) + \mu_{I\psi} - \mu_{I\phi}\big)}{2\sqrt{\sigma^2_{C1} + \sigma^2_{I1} + \sigma^2_{C2} + \sigma^2_{I2} + \sigma^2_{I\phi} + \sigma^2_{I\psi}}},
    \label{eq:pse_setup34_criterion2_nSimplified}
\end{align}
which clearly scales along $O(\sqrt{n})$.

We conclude the section by indicating what a large $n$ may look like. Suppose the response variances and the number of users are similar across user groups. In that case, we can rearrange Inequality~\eqref{eq:pse_setup34_criterion2_full} to make $n$ the subject:
\begin{align}
    n > \left(2\sqrt{12}\left(\sqrt{6} - 1\right)z\right)^2 \frac{\sigma^2_G}{\Delta^2},
    \label{eq:pse_setup34_criterion2_bothSimplified}
\end{align}
where $\Delta = (\mu_{I2} - \mu_{C2}) - (\mu_{I1} - \mu_{C1}) + \mu_{I\psi} - \mu_{I\phi}$ is the difference in actual effect sizes between Setups 4 and 3. Under a 5\% significance level and 80\% power, the first coefficient amounts to around 791, roughly 50 times the coefficient one would use to determine the sample size of a simple A/B test~\cite{miller10hownot}. This suggests that a dual control setup is perhaps a luxury accessible only to the largest advertising platforms and their top advertisers. For example, consider an experiment to optimise the conversion rate where the baselines attain 20\% (hence having a variance of $0.2(1 - 0.2) = 0.16$). If there is a 2.5\% relative (i.e. 0.5\% absolute) effect between the competing strategies, the dual control setup will only be superior if there are $>5$M users in each user group. 

\section{Empirical Verification}
\label{sec:pse_experiments}
Having performed theoretical calculations for the actual and detectable effects and conditions where an experiment setup is superior to another, we verify those calculations using simulation results.
We focus on the results presented in Section~\ref{sec:pse_actual_mde_size}, as the rest of the results presented followed those calculations.

In each experiment setup evaluation, we randomly select the value of the $\mu$- and $\sigma$-parameters from a uniform distribution,\footnote{Denoting $U(a,b)$ as a uniform distribution with bounds $a$ and $b$, we draw $\mu_{G} \sim U(-10, 10)$ and ${\sigma^2_G \sim U(1, 10)}$ for $G \in \{C0, C1, C2, C3, I1, I2, I\phi, I\psi\}$ (see Table~\ref{tab:pse_group_id} for how the groups are defined).} with the bounds chosen that are reflective of what advertisers usually see with decision metrics based on binary and count responses.\footnote{This is done to rule out any CLT convergence issues due to heavy-tailed distributed responses or low sample size. We acknowledge the limitations of such an approach in Section~\ref{sec:conclusion_future_work}. \label{footnote:pse_avoid_clt_convergence_issue}} We also select the $n$-parameters uniformly on a log scale, with the bounds chosen to represent usual user numbers in personalisation strategy experiments and to ensure the empirical verification runs within a practical timeframe.\footnote{We draw $n_0, n_1, n_2, n_3 \sim 5 \times 10^{U(1, 3.5)}$.} We favour sampling user numbers uniformly on a log scale (instead of a linear scale) due to the diminishing real-life experiment count and diminishing impact of having a single additional user as we increase the magnitude of user numbers. Each parameter is drawn independently from other parameters within the same evaluation and across different evaluations. We fix $\alpha = 5\%$ and $\pi_{\textrm{min}} = 80\%$ across all evaluations as per the norm in digital experiments.

We then take 1,000 actual effect samples, each by 
\begin{enumerate}
    \item Sampling normally distributed responses from the user groups under the specified parameters,\footref{footnote:pse_avoid_clt_convergence_issue}
    \item Computing the mean for the analysis groups, and
    \item Taking the difference of the means.
\end{enumerate}

We also take 100 MDE samples in separate evaluations, each by
\begin{enumerate}
    \item Sampling a critical value under the null hypothesis,
    \item Computing the test power under a large number of possible effect sizes, each using the critical value and sampled response means under the alternate hypothesis, and
    \item Searching the effect size space for the value that gives the predefined power.
\end{enumerate} 
As the power vs effect size curve is noisy, given the use of simulated power samples, we use the bisection algorithm provided by the \texttt{noisyopt} package to perform the search. The algorithm dynamically adjusts the number of samples taken from the same point on the curve to ensure the noise does not send us down the wrong search space. 

We expect the mean of the sampled actual effect and MDE to match the theoretical value.
To verify this, we perform~1,000 bootstrap resamplings on the above samples to obtain an empirical bootstrap distribution of the sample mean in each evaluation. The $95\%$ bootstrap resampling confidence interval (BRCI) should then contain the theoretical mean $95\%$ of the time. 
Furthermore, the histogram of the theoretical quantity's percentile rank, in relation to the bootstrap samples, across multiple evaluations should also follow a uniform distribution~\cite{talts2018validating}.

\begin{table}
\onehalfspacing
\caption[Number of evaluations where the theoretical value of the quantities falls between the 95\% bootstrap confidence interval for each experiment setup.]{Number of evaluations where the theoretical value of the quantities (columns) falls between the 95\% bootstrap confidence interval for each experiment setup (rows). See Section~\ref{sec:pse_experiments} for a detailed description of the empirical verification.}
\label{tab:pse_experiment_in_BRCI_result}
\begin{center}
\begin{tabular}{c|c|c}
 & Actual effect size & Minimum detectable effect \\\hline
 Setup 1 & 1049/1099 (95.45\%) & 66/81 (81.48\%) \\
 Setup 2 & 853/999 (85.38\%) & 87/106 (82.08\%) \\
 Setup 3 & 922/1099 (83.89\%) & 93/116 (80.18\%) \\
 Setup 4 & 240/333 (72.07\%) & 149/185 (80.54\%)\\
\end{tabular}
\end{center}
\vspace*{\baselineskip}
\end{table}

The result is shown in Table~\ref{tab:pse_experiment_in_BRCI_result}. One can observe that more evaluations have their theoretical quantity lying outside the BRCI than expected. 
Upon further investigation, we observed a characteristic $\cup$-shape from the histograms of the percentile ranks for the actual effects. 
This suggests that the bootstrap samples may be under-dispersed but otherwise centred on the theoretical quantities. 

We also observed the histograms for MDEs curving upward to the right. This suggests that the theoretical value is a slight overestimate (of $<1\%$ to the bootstrap mean in all cases). We believe this is likely due to a small bias in the bisection algorithm. The algorithm tests if the mean of the power samples is less than the target power to decide which half of the search space to continue along. Given that we can bisect up to 10 times in each evaluation, we will likely see a false positive even when we set the significance level for individual comparisons to 1\%. This leads to the algorithm favouring a smaller MDE sample. Nevertheless, we are satisfied with the theoretical quantities for experiment design purposes since we have tested for a wide range of parameters and the overall bias is small.

\section{A Brief Recap}

We have addressed the problem of comparing experiment designs for personalisation strategies by presenting an evaluation framework that allows experimenters to evaluate which experiment setup they should adopt. The flexible framework can be easily extended to compare setups that compare more than two strategies by adding more user groups (i.e. new sets to the Venn diagram in Figure~\ref{fig:pse_ME_groups}). A new setup can also be quickly incorporated as it is essentially a different weighting of user group-scenario combinations shown in Table~\ref{tab:pse_group_id}.
The framework also allows the development of simple rules of thumb, such as:
\begin{enumerate}
    \item Treatment effect dilution should never be employed if the experiment already has sufficient power, though it can be helpful if the experiment is under-powered and the non-qualifying users provide a ``stabilising effect''; and
    \item A dual control setup is superior to simpler setups only if one has access to the user base of the largest organisations.
\end{enumerate}
We have validated the theoretical results via simulations and made the code available\footref{footnote:pse_repo_link} so that practitioners can benefit from the results immediately when designing their upcoming experiments.

\cleardoublepage

\chapter{Conclusion}
\label{chap:conclusion}

\section{Summary of Thesis Achievements}

We presented and addressed several statistical and data challenges experimenters face when building digital experimentation and measurement (DEM) capabilities from the ground up. They include
\begin{enumerate}
    \item (Chapter~\ref{chap:vem}) Valuing DEM capabilities to justify related investments via a novel ranking under lower uncertainty model that quantifies the value gained and risk when one lowers the measurement uncertainty of the items they are prioritising;
    \item (Chapter~\ref{chap:statstest}) Understanding statistical tests and their alternatives to estimate the potential treatment impact via an introduction to statistical testing that aims to balance the theoretical foundations and the practical applications;
    \item (Chapter~\ref{chap:oced}) Recognising the availability of datasets to validate existing and develop new methods via the creation of the first ever taxonomy for digital experiment datasets, survey for publicly available online controlled experiment (OCE) datasets, and OCE dataset that can support experiments with adaptive stopping;
    \item (Chapter~\ref{chap:oced}) Identifying data collection requirements based on the statistical tests chosen (and vice versa) via a mapping between statistical tests and datasets, two concepts that are often studied in isolation;
    \item (Chapter~\ref{chap:rade}) Acquiring knowledge of advanced methods related to the design, running, analysis, and interpretation of digital experiments (while avoiding pitfalls) via a review of recent advances in methods and examples of OCEs, quasi-experiments, and natural experiments;
    \item (Chapter~\ref{chap:rade}) Implementing selected methods in the literature via a case study that critiques the effectiveness and practicalness of popular approaches that address dependent responses in OCEs (and geo-experiments to some extent); and
    \item (Chapter~\ref{chap:pse}) Designing experiments to compare competing personalisation strategies, known to suffer from low test power and lack of complete control in random assignments, via developing an evaluation framework that is both flexible and capable of generating simple rules of thumb.
\end{enumerate}


Digital experimentation is a multi-disciplinary field with researchers and practitioners from many backgrounds. This is reflected in the diversity of statistical disciplines involved in this thesis.
Addressing the challenges above requires the study of order statistics, statistical testing, data collection and processing, experimental design, and causal inference. In addition, the thesis also draws upon techniques in computational statistics and distributed computing to perform the large-scale simulations and empirical verification featured in the thesis.
The breadth of topics makes it likely that a researcher or practitioner in digital experimentation can take away something from the text.

The thesis also reached considerable depth within each research topic.
Most of the novel research contributions in the thesis have already been presented in and included in proceedings of highly-regarded conferences in applied statistics, machine learning, and data mining. These include
the IEEE International Conference on Data Mining (ICDM), AdKDD Workshop (in conjunction with ACM
SIGKDD Conference on Knowledge Discovery and Data Mining, KDD), Conference on Neural Information Processing Systems (NeurIPS), and ACM Web Conference (TheWebConf, formerly WWW). The research has been further recognised via the inclusion of
the extended ICDM paper into a journal special issue for ``Highly-rated Short Papers for ICDM 2019''
and the Best Student Paper award for AdKDD 2020.

\paragraph{Impact beyond academic research} The research in this thesis is motivated by the need for and has directly contributed to the development of DEM capabilities at ASOS.com (a global online fashion retail company) from its infancy:
\begin{itemize}
    \item Results from the ranking under lower uncertainty model presented in Chapter~\ref{chap:vem}, using parameters reflecting the company's measurement capability at the time, have formed one of the main arguments for the company to invest in initial DEM capabilities.
    \item Since then, the author has spent years engaging with and educating company colleagues from both technical and non-technical backgrounds on statistical testing, datasets, and experimental design in an online setting. Chapters~\ref{chap:statstest},~\ref{chap:oced}, and~\ref{chap:pse} reflect the distilled knowledge from that process. Striking a balance between theoretical rigour and applicability in practice also makes the text suitable for other pedagogical purposes. 
    \item The author led a team that incorporates advanced statistical techniques in ASOS.com's internal experiment analysis platform. These include the mSPRT, a sequential test, to address peeking (see Section~\ref{sec:statstest_sequential}) and one-way/block bootstrap to address dependent responses as described in the case study in Section~\ref{sec:rade_abv_deployment}. The author also led another team that has built and is maintaining a geo-experimentation framework in-house that utilises techniques mentioned in Section~\ref{sec:rade_causalinference_geoexperiments}.
\end{itemize}

\section{Future Work}
\label{sec:conclusion_future_work}

The research has also opened up many opportunities for further work in different statistical and data topics. Below, we outline them in the order they emerge in the thesis.

\paragraph{Ranking under lower uncertainty} We believe the general ranking under lower uncertainty problem, introduced in Chapter~\ref{chap:vem}, will interest many in the statistics and operations research community. Many exciting questions remain, not only for valuing DEM capabilities (e.g., understanding how the expected gain and risk change when we assume different item value distributions, see Section~\ref{sec:vem_extension_t_distributed}), but also for general prioritisation processes (e.g., the value gained when not all items in the prioritisation process can have their estimation noise reduced, see Section~\ref{sec:vem_extension_partial_noise_reduction}). It also yields an opportunity for us to gain a further understanding of order statistics via questions like the probability that two order statistics in the ranking under lower uncertainty model are generated from the same r.v. (see Appendix~\ref{sec:miscmath_prob_Ir_Js}).

\paragraph{Datasets for digital experiments} The work in Chapter~\ref{chap:oced} has highlighted the need for more publicly available digital experiment datasets, both in quantity and variety, a call echoed by~\cite{larsen2023statistical}. This is necessary to foster academic-industry collaboration and safeguard future methodological development. As more datasets become available, the survey, taxonomy, and mapping introduced in the chapter will require updating and extending.

\paragraph{Dependent responses in e-commerce experiments} In Section~\ref{sec:rade_abv_deployment}, we observed that the two-way bootstrap, which addresses responses dependent on multiple types of units, produces impractically large standard error estimates and unnecessarily conservative confidence intervals under online retail/e-commerce transactions datasets. Further work is required to understand how and why this is the case. This enables experimenters to adjust the models and methods to estimate the degree of dependence between responses accurately.

\paragraph{Evaluation framework for personalisation strategy experiment designs} The evaluation framework introduced in Chapter~\ref{chap:pse} assumes the responses from each user group-scenario combination are randomly distributed with the mean independent of the variance, with the evaluation criteria calculated assuming the decision metric is approximately normally distributed under the central limit theorem. While they are reasonable assumptions in practice, it begs the question of whether the evaluation framework and its results are robust to deviations from these assumptions, e.g., with binary responses (where the mean and variance correlate) and heavy-tailed distributed responses (where the sample mean converges to a normal distribution slowly).

\hspace*{0pt}

In addition to the extensions outlined above, we propose the following related, high-potential research direction involving sub-disciplines not covered in the thesis. These sub-disciplines include (1) automatic experimentation and optimisation capabilities via bandit and Bayesian optimisation algorithms~\cite{han2020contextual,optimizely2021multiarmed} and (2) combining experimental and observational data to provide more accurate treatment effect estimates~\cite{kallus2018removing}.

\paragraph{Automatic experiment prioritisation with decision makers in-the-loop}
It is increasingly common to deploy automatic experimentation and optimisation capabilities within individual business functions or digital products. However, the problem of finding the optimal sequencing of experimental interventions at a cross-functional level or across multiple digital products is often dealt with manually. We argue that a Bayesian optimisation framework, which balances exploitation (obtaining the highest valued intervention) and exploration (eliminating uncertainty in value estimates), is well suited to recommending the next interventions to be evaluated. The framework should be central to an iterative decision framework where we
\begin{enumerate}
  \item Measure interventions’ value via experiments,
  \item Estimate other candidate interventions’ value using causal inference techniques, and 
  \item Identify the next best interventions to be evaluated.
\end{enumerate}
This can be roughly linked to the points on the objective function, the surrogate model, and the acquisition function. 
With the objective function being largely unknown, we propose researching into constructing novel surrogate models and acquisition functions that can address other challenges presented by prioritising experiments in real life. These include quantifying the uncertainty of interventions in a mixed experimental-observational setting with potential unmeasured confounders, having concept drift erode the confidence in past effect estimates over time, and incorporating preference from decision makers in addition to modelled effect estimates.

\hspace*{0pt}

To sum up, the need to understand the cause and effect of decisions and test out different alternatives in the current digital era has led to a rich body of work in digital experimentation and measurement. The field has attracted substantial interdisciplinary interest in the past two decades and will remain fast-growing, as demonstrated by the open challenges above. This thesis has contributed to, and seeks to continue inspiring such growth.
\cleardoublepage


\cleardoublepage

\appendix

\counterwithout{equation}{chapter}
\counterwithin{equation}{section}

\chapter{Miscellaneous Mathematical Results}
\label{chap:miscmath}

\section[Probability in Generating Specific Order Statistic Pairs]{Ranking Under Lower Uncertainty: Probability in Generating Specific Order Statistic Pairs}
\label{sec:miscmath_prob_Ir_Js}

We consider the problem posed in Section~\ref{sec:vem_var_D}, namely the probability that two order statistics in the ranking under lower uncertainty problem, one from the high-noise set and the other from the low-noise set, are generated from the same random variable. 

This appendix is adapted from the appendix of the research paper “What is the Value of Experimentation and Measurement?” published in \textit{Data Science and Engineering}~\cite{liu20whatisthevalue}. It also incorporates relevant discussions on Cross Validated Stack Exchange~\cite{liu2020probability}.

\begin{figure}
    \begin{center}
        \includegraphics[width=0.95\textwidth, trim = 0 0 0 0, clip]{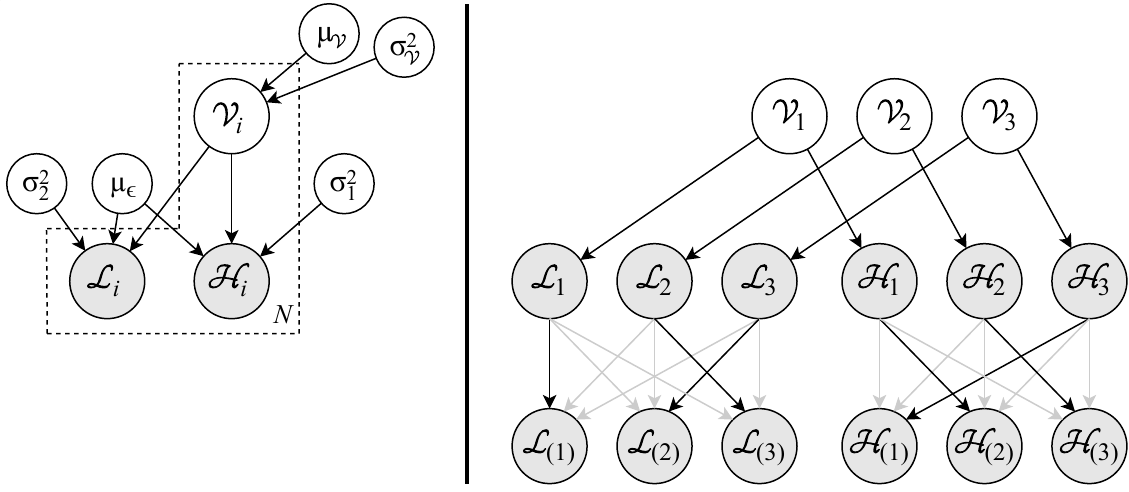}
    \end{center}
    \caption[Figure~\ref{fig:vem_rulu_generative_model_two_sample} and RHS of Figure~\ref{fig:vem_rulu_ranking_three_items} restated.]{(L) Figure~\ref{fig:vem_rulu_generative_model_two_sample} -- The generative model for the ranking under lower uncertainty problem in plate notation. (R) RHS of Figure~\ref{fig:vem_rulu_ranking_three_items} -- Relationship between different quantities in a three-item generative model. $\mathcal{V}_i$, $\mathcal{H}_i$/$\mathcal{L}_i$, and $\mathcal{H}_{(r)}$/$\mathcal{L}_{(s)}$, ${i, r, s \in \{1, 2, 3\}}$, represent the true value, the unranked noisy estimates, and the ranked noisy estimates of the items, respectively. Here $\mathcal{H}_{(r)}$ and~$\mathcal{L}_{(s)}$ may be generated by the same~$\mathcal{V}_i$ for some~$r$ and~$s$ combinations.}
    \label{fig:miscmath_rulu_gen_model_three_items}
\end{figure}

We restate the setup and problem formally, which is also re-illustrated in Figure~\ref{fig:miscmath_rulu_gen_model_three_items}. Let
\begin{align}
    \mathcal{V}_i \overset{\textrm{i.i.d.}}{\sim} F_{\mathcal{V}}(\cdot)\,, 
    & \textrm{ where } \mathbb{E}(\mathcal{V}_i) = \mu_{\mathcal{V}} \,, \textrm{Var}(\mathcal{V}_i) = \sigma^2_{\mathcal{V}} \,, \tag{*\ref{eq:vem_X_n_eps_n_modelling_general}}\\
    \epsilon_{1i} \overset{\textrm{i.i.d.}}{\sim} F_{\epsilon_1}(\cdot)\,,
    & \textrm{ where } \mathbb{E}(\epsilon_{1i}) = \mu_\epsilon \,, \textrm{Var}(\epsilon_{1i}) = \sigma^2_1 \,, \nonumber\\
    \epsilon_{2i} \overset{\textrm{i.i.d.}}{\sim} F_{\epsilon_2}(\cdot)\,,
    & \textrm{ where } \mathbb{E}(\epsilon_{2i}) = \mu_\epsilon \,, \textrm{Var}(\epsilon_{2i}) = \sigma^2_2 \,, \tag{*\ref{eq:vem_eps_1_eps_2_def}}
\end{align}
where $\mathcal{V}_i \perp \epsilon_{1j}$, $\mathcal{V}_i \perp \epsilon_{2j}$, and $\epsilon_{1i} \perp \epsilon_{2j}$ $\forall\, i, j$. $\mathcal{V}_i$ represents the true value of the $i^\textrm{th}$ item under consideration in the ranking under lower uncertainty problem, while $\epsilon_{1i}$ and $\epsilon_{2i}$ represents noise under different noise levels. We further let
\begin{align}
    \mathcal{H}_i = \mathcal{V}_i + \epsilon_{1i} \,, \quad
    \mathcal{L}_i = \mathcal{V}_i + \epsilon_{2i} \quad \forall i \,. \tag{*\ref{eq:vem_Yn_Zn_def}}
\end{align}
Here, $\mathcal{H}_i$ and $\mathcal{L}_i$ are the estimated values for $\mathcal{V}_i$ under high and low estimation noise, respectively. We rank $\mathcal{H}$ and $\mathcal{L}$ separately to obtain two sets of order statistics: $\mathcal{H}_{(r)}$ and $\mathcal{L}_{(s)}$. 
We then introduce two functions, $\mathcal{I}: \{1, ..., N\} \rightarrow \{1, ..., N\}$ and $\mathcal{J}: \{1, ..., N\} \rightarrow \{1, ..., N\}$, which we use to map the ranks of $\mathcal{H}$ and $\mathcal{L}$ back to their original indices (and thus the indices of $\mathcal{V}$), respectively.
We are interested in $\mathbb{P}(\mathcal{I}(r) = \mathcal{J}(s))$, the probability that the $r^\textrm{th}$ ranked~$\mathcal{H}$ and the $s^\textrm{th}$ ranked $\mathcal{L}$ are generated by the same~$\mathcal{V}_i$.

Before deriving the probability, we obtain some intuition on what we are dealing with via some simulations. In each run within a simulation, we:
\begin{enumerate}
    \item Given $N$, $\mu_{\mathcal{V}}$, $\mu_{\epsilon}$, $\sigma^2_{\mathcal{V}}$, $\sigma^2_{1}$, and $\sigma^2_{2}$, generate $\mathcal{V}_i$, $\mathcal{H}_i$, and $\mathcal{L}_i$ $\forall i \in \{1, ..., N\}$ as specified above.
    \item Obtain the rank for each realised $\mathcal{H}$ while preserving the order of the data array (i.e., the indices). Do the same for $\mathcal{L}$.
    \item For each index $i$, obtain the rank of $\mathcal{H}_i$ as $r$ and the rank of $\mathcal{L}_i$ as $s$. The pair $(r, s)$ is said to have gained a ``hit''.
\end{enumerate}
After many runs, we obtain an empirical distribution of $\mathbb{P}(\mathcal{I}(r) = \mathcal{J}(s))$ across all possible $(r, s)$ pairs by dividing the number of ``hits'' for each rank pair by the total number of simulation runs.\footnote{The statement ``the $r^\textrm{th}$ ranked $\mathcal{H}$ and the $s^\textrm{th}$ ranked $\mathcal{L}$ are generated by the same $\mathcal{V}_i$'' is equivalent to ``the same $\mathcal{V}_i$ generates the $r^\textrm{th}$ ranked $\mathcal{H}$ and the $s^\textrm{th}$ ranked $\mathcal{L}$'', and thus Step 3 above provides a more computationally efficient way to obtain the empirical distribution  --  we do not need to determine the index mapping functions $\mathcal{I}(\cdot)$ and $\mathcal{J}(\cdot)$).}
We observe that each $(r, s)$ pair gets $N$ chances to gain a ``hit'' within each simulation run, with the restriction that each $r$ (and each $s$) would ultimately gain one ``hit''. The latter owes to the fact that each rank within $\mathcal{H}$ and $\mathcal{L}$ should appear once and only once.

\begin{figure}
    \onehalfspacing
    \begin{center}
        \includegraphics[width=0.85\textwidth, trim = 0 0 0 0, clip]{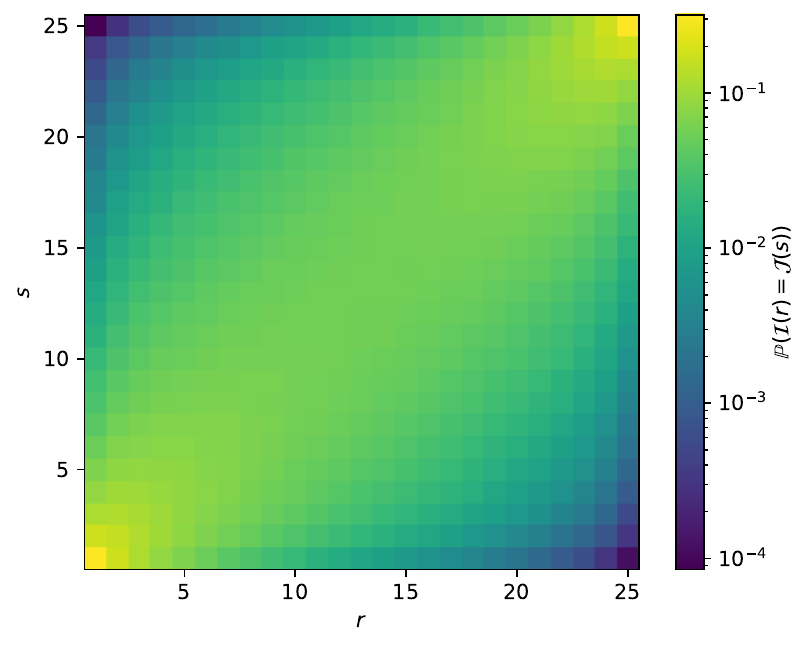}
    \end{center}
    \vspace*{-18pt}
    \caption[Heatmap showing the empirical distribution of $\mathbb{P}(\mathcal{I}(r) = \mathcal{J}(s))$ across different $r$ and $s$.]{Heatmap showing the empirical distribution of $\mathbb{P}(\mathcal{I}(r) = \mathcal{J}(s))$ across different $r$ and $s$, generated from \numprint{200000} simulation runs with $N = 25$, $\mu_{\mathcal{V}} = \mu_{\epsilon} = 0$, $\sigma^2_{\mathcal{V}} = 1$, $\sigma^2_1 = 0.5$, and $\sigma^2_{2} = 0.4$. Here, all $\mathcal{V}_i$, $\mathcal{H}_i$, and $\mathcal{L}_i$ follow the normal distribution. Every row and column in the heatmap sums to one.}
    \label{fig:miscmath_prob_Ir_eq_Js_2d}
\end{figure}

\begin{figure}
    \begin{center}
        \includegraphics[width=0.925\textwidth, trim = 0 0 0 0, clip]{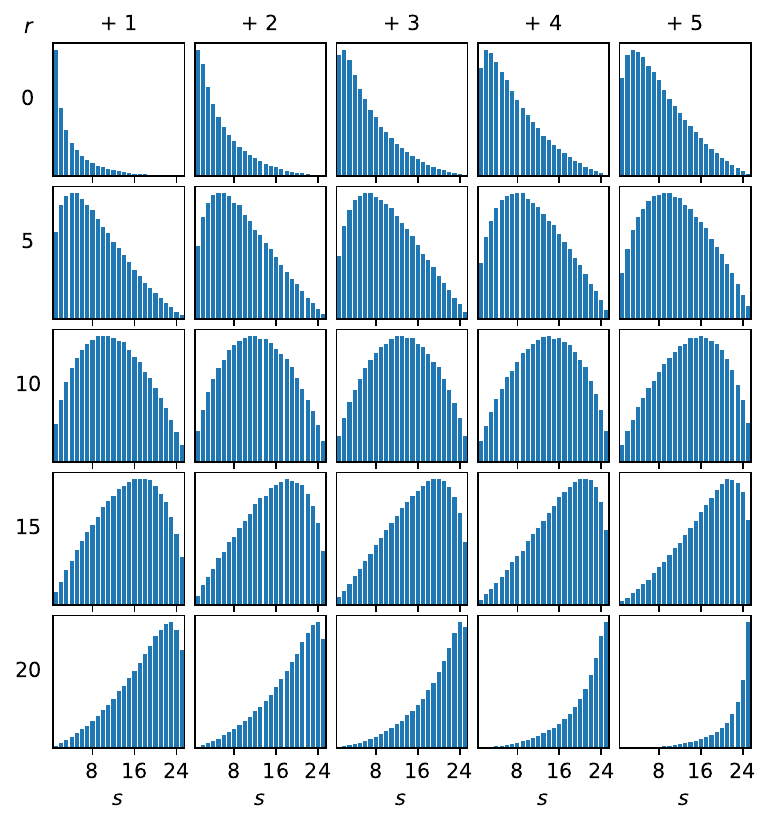}
    \end{center}
    \caption[The marginal distribution of $\mathbb{P}(\mathcal{I}(r) = \mathcal{J}(s))$ across different $s$ for a given $r$.]{
    The marginal distribution of $\mathbb{P}(\mathcal{I}(r) = \mathcal{J}(s))$ across different $s$ for a given $r$, generated from 200 000 simulation runs with $N = 25$, $\mu_{\mathcal{V}} = \mu_{\epsilon} = 0$, $\sigma^2_{\mathcal{V}} = 1$, $\sigma^2_1 = 0.5$, and $\sigma^2_{2} = 0.4$. Here, all $\mathcal{V}_i$, $\mathcal{H}_i$, and $\mathcal{L}_i$ follow the normal distribution.
    Each plot represents a column in the heatmap in Figure~\ref{fig:miscmath_prob_Ir_eq_Js_2d} and has a different scale on the y-axis to emphasise the relative difference in the probability across different $s$ for a given $r$.}
    \label{fig:miscmath_prob_Ir_eq_Js_r_marginals}
\end{figure}

Figures~\ref{fig:miscmath_prob_Ir_eq_Js_2d} and~\ref{fig:miscmath_prob_Ir_eq_Js_r_marginals} show the empirical $\mathbb{P}(\mathcal{I}(r) = \mathcal{J}(s))$ for a particular ranking under lower uncertainty model. We observe from Figure~\ref{fig:miscmath_prob_Ir_eq_Js_2d} that the probability is clearly not uniform and higher when $r$ and~$s$ are close to each other (i.e., along the $r=s$ diagonal), reaching the maximum when $r=s=1$ and $r=s=N$. This agrees with the intuition that if $\mathcal{H}_i$ is the largest (smallest) amongst $\mathcal{H}$, chances are that the associated $\mathcal{V}_i$ is quite large (small) as well, and thus the associated $\mathcal{L}_i$ stands a high chance of becoming the largest (smallest) amongst $\mathcal{L}$. On the other hand, the probability is lower when $r$ and $s$ are far apart, reaching the minimum when $(r, s) = (1, N)$ or $(N, 1)$. This also makes sense  --  for $\mathcal{V}_i$ to generate both $\mathcal{H}_{(N)}$ and $\mathcal{L}_{(1)}$, one requires a large realisation for~$\epsilon_{1i}$ and a large but negative realisation for $\epsilon_{2i}$. The two realisations rarely happen together.

We also note the similarity between the marginal distributions in Figure~\ref{fig:miscmath_prob_Ir_eq_Js_r_marginals} (i.e., the probability across $s$ for a given $r$) and the family of beta-binomial distributions. This is not surprising: beta distributions are closely connected to order statistics.
As shown below, the observation will shape our approach to obtaining the probability.

\subsection{Representing the probability as a certain number of successes in multiple Bernoulli trials}

To obtain $\mathbb{P}(\mathcal{I}(r) = \mathcal{J}(s))$, one can express it as follow:
\begin{align}
    \mathbb{P}\left(\mathcal{I}(r) = \mathcal{J}(s)\right)
    = &\, \sum_{k=1}^{N} \mathbb{P}\left(\mathcal{I}(r)=k \,\cap\, \mathcal{J}(s)=k\right) & \textrm{(law of total prob.)} \nonumber\\
    = &\, \sum_{k=1}^{N} \mathbb{P}\left(\mathcal{J}(s)=k \,|\, \mathcal{I}(r)=k \right) \mathbb{P}\left(\mathcal{I}(r)=k \right)  &  \textrm{(conditional prob.)}
    \label{eq:miscmath_PIr_Js_conditional_prob}
\end{align}

We first focus on the event
$\mathcal{J}(s) = k$. Given any $k$, the event
is equivalent to precisely $(s-1)$ $\{\mathcal{L}_j\}_{j\neq k}$ being less than~$\mathcal{L}_k$, which makes $\mathcal{L}_k$ the $s^{\textrm{th}}$ ranked $\mathcal{L}$. Thus, we can express Equation~\eqref{eq:miscmath_PIr_Js_conditional_prob} as
\begin{align}
    \mathbb{P}\left(\mathcal{I}(r) = \mathcal{J}(s)\right) 
    = \sum_{k=1}^{N} \mathbb{P}\left(\sum_{j=1,\, j\neq k}^{N} \mathbb{I}_{\left\{\mathcal{L}_j < \mathcal{L}_k\right\}} = s-1 \,\Bigg|\, \mathcal{I}(r) = k\right)
    \mathbb{P}\left(\mathcal{I}(r)=k \right) \,.
    \label{eq:miscmath_PIr_Js_sum_of_ind_var}
\end{align}
We then recognise:
\begin{enumerate}[itemsep=0pt]
    \item $\mathbb{P}(\mathcal{I}(r)=k) = \frac{1}{N} \,\forall k$. This is because all $\mathcal{V}_i$ and $\mathcal{H}_i$ are i.i.d. As a result, each $\mathcal{H}_i$ is equally likely to have generated the order statistic $\mathcal{H}_{(r)}$;
    \item The conditional probability within the outer summation does not depend on the index~$k$ but the rank $r$; and
    \item Given $k = \mathcal{I}(r)$, this particular $\mathcal{L}_k$ is now different from all the other $\{\mathcal{L}_j\}_{j\neq k}$, as the former now carries information on its possible values from $\mathcal{H}_{(r)}$ via $\mathcal{V}_{\mathcal{I}(r)}$. To stress the difference, we perform the index substitution here and refer to the random variable as $\mathcal{L}_{\mathcal{I}(r)}$ from now on.
\end{enumerate}
This enables us to further simplify Equation~\eqref{eq:miscmath_PIr_Js_sum_of_ind_var} as:
\begin{align}
    \mathbb{P}\left(\mathcal{I}(r) = \mathcal{J}(s)\right) 
    = & \, \sum_{k=1}^{N} \left(\mathbb{P}\left(\sum_{j=1,\, j\neq \mathcal{I}(r)}^{N} \mathbb{I}_{\left\{\mathcal{L}_j < \mathcal{L}_{\mathcal{I}(r)}\right\}} = s-1 \right)
    \cdot \frac{1}{N} \right)\nonumber\\
    = & \, \mathbb{P}\Bigg(\underbrace{\sum_{j=1,\, j\neq \mathcal{I}(r)}^{N} \mathbb{I}_{\left\{\mathcal{L}_j < \mathcal{L}_{\mathcal{I}(r)}\right\}}}_{\triangleq \, C} = s-1 \Bigg) \,,
    \label{eq:miscmath_PIr_Js_C_eq_sm1}
\end{align}
where we also define $C$ (the \textbf{c}ount) as the summation of the indicator variables.

From \eqref{eq:miscmath_PIr_Js_C_eq_sm1}, we observe that $C$ represents the number of successes in $(N-1)$ independent Bernoulli trials (i.e., $\mathcal{L}_j < \mathcal{L}_{\mathcal{I}(r)}$). Moreover, the success probability of each Bernoulli trial is not fixed but depends on the value of $\mathcal{L}_{\mathcal{I}(r)}$, itself a random variable. We know that for any given $l \in \mathbb{R}$, the probability of $\mathcal{L}_j$ being less than $l$ is, by definition, $F_{\mathcal{L}_j}(l)$. Thus, we can define another random variable $\mathcal{P} = F_{\mathcal{L}_j}(\mathcal{L}_{\mathcal{I}(r)})$ to show how the success \textbf{p}robability in the Bernoulli trials is distributed, with the probability density function (PDF) being
\begin{align}
    f_{\mathcal{P}}\left(p \,|\, r, ...\right)
    & \, = \frac{f_{\mathcal{L}_{\mathcal{I}(r)}}\left(F_{\mathcal{L}_j}^{-1}(p) \,|\, r, ...\right)}{f_{\mathcal{L}_j}\left(F_{\mathcal{L}_j}^{-1}(p) \,|\, ...\right)} \,,
\end{align}
where we omit other parameters in the ranking under lower uncertainty model such as $\mu_{\mathcal{V}}$, $\sigma^2_{\mathcal{V}}$, ..., for clarity.

This PDF arises from two standard results:
\begin{enumerate}
    \item (Transformation of random variables) If $Y = g(X)$ and $g^{-1}(\cdot)$ exists, then \\ $f_Y(y) = f_X\left(g^{-1}(x)\right) \left|\frac{\textrm{d}g^{-1}(x)}{\textrm{d}x}\right|$; and
    \item (Derivative of inverse functions) $\frac{\textrm{d}g^{-1}(x)}{\textrm{d}x} = \frac{1}{g'\left(g^{-1}(x)\right)}$ for all functions $g(\cdot)$ if both the inverse $g^{-1}(\cdot)$ and the derivative $g'(\cdot)$ exist (which they do for $F_{\mathcal{L}_j}(\cdot)$).
\end{enumerate}
$C$ then follows the following compound probability distribution:
\begin{align}
    f_{C}(c \,|\, N-1, r, ...) = \int_0^1 \textrm{Bin}(c \,|\, N-1, p) \,f_{\mathcal{P}}(p \,|\, r, ...) \,\textrm{d}p \,,
\end{align}
where $\textrm{Bin}(c \,|\, N-1, p)$ represents the PDF of the binomial distribution with $N-1$ trials and success probability $p$.

\subsection{Modelling the success count distribution using beta-binomials}
\label{sec:miscmath_prob_Ir_Js_modelling}

Deriving the exact distribution for $\mathcal{P}$ and $C$ is beyond the scope of this work, and we believe the distributions are analytically intractable in many cases. To estimate the covariance in~\eqref{eq:vem_cov_XIr_XJs}, we will model $\mathcal{P}$ as beta distributions, with the parameters~$\alpha_{\mathcal{P}}$ and~$\beta_{\mathcal{P}}$ obtained via the method of moments. We believe beta distributions are a natural choice for $\mathcal{P}$ as they are closely related to order statistics. Moreover, beta distributions are also conjugate priors to binomial distributions, which eases the computation of the probability masses for $C$.

To obtain the beta distribution parameters, we first require the mean and variance for $\mathcal{L}_{\mathcal{I}(r)}$ and $F_{\mathcal{L}_j}(\mathcal{L}_{\mathcal{I}(r)})$. We know that $\mathcal{L}_{\mathcal{I}(r)} = \mathcal{V}_{\mathcal{I}(r)} + \epsilon_{2(\mathcal{I}(r))}$, $\mathcal{V}_{\mathcal{I}(r)} \perp \epsilon_{2(\mathcal{I}(r))}$ from~\eqref{eq:vem_Yn_Zn_def}, and hence we have
\begin{align}
    \mathbb{E}(\mathcal{L}_{\mathcal{I}(r)})
    & = \mathbb{E}(\mathcal{V}_{\mathcal{I}(r)}) + \mathbb{E}(\epsilon_{2(\mathcal{I}(r))})
    \approx \, \mu_\mathcal{V} + \frac{\sigma^2_\mathcal{V}}{\sigma^2_\mathcal{V} + \sigma^2_1}\left(F_\mathcal{H}^{-1}\Big(\frac{r - \alpha}{N - 2\alpha + 1}\Big) - (\mu_\mathcal{V} + \mu_\epsilon)\right) + \mu_\epsilon \;, \label{eq:vem_E_ZIr_general} \\
    \textrm{Var}(\mathcal{L}_{\mathcal{I}(r)}) & =  \textrm{Var}(\mathcal{V}_{\mathcal{I}(r)}) + \textrm{Var}(\epsilon_{2(\mathcal{I}(r))})
    \nonumber\\ 
    & \approx \, \frac{\sigma^2_1 \sigma^2_\mathcal{V}}{\sigma^2_\mathcal{V} + \sigma^2_1} \, + \frac{\sigma^4_\mathcal{V}}{\left(\sigma^2_\mathcal{V} + \sigma^2_1\right)^2}  \frac{r(N-r+1)}{(N+1)^2 (N+2)}
    \frac{1}{\big(f_\mathcal{H}\big(F_\mathcal{H}^{-1}\big(\frac{r}{N+1}\big)\big)\big)^2} + \sigma^2_2 \;, \label{eq:vem_var_ZIr_general}
\end{align}
where $\mathbb{E}(\mathcal{V}_{\mathcal{I}(r)})$ and $\textrm{Var}(\mathcal{V}_{\mathcal{I}(r)})$ are obtained from~\eqref{eq:vem_E_XIr_general} and~\eqref{eq:vem_var_xir}, respectively.

We can then approximate the expected value and variance of $F_{\mathcal{L}_j}(\mathcal{L}_{\mathcal{I}(r)})$ using Taylor series expansion:
\begin{align}
    \mathbb{E}\left(F_{\mathcal{L}_j}(\mathcal{L}_{\mathcal{I}(r)})\right)
    \approx & \, F_{\mathcal{L}_j}\left(\mathbb{E}(\mathcal{L}_{\mathcal{I}(r)})\right) + 
    \frac{1}{2} f'_{\mathcal{L}_j}\left(\mathbb{E}(\mathcal{L}_{\mathcal{I}(r)})\right) \cdot \textrm{Var}(\mathcal{L}_{\mathcal{I}(r)}) + ... \;, \\
    \textrm{Var}\left(F_{\mathcal{L}_j}(\mathcal{L}_{\mathcal{I}(r)})\right) 
    \approx & \, \left(f_{\mathcal{L}_j}(\mathbb{E}(\mathcal{L}_{\mathcal{I}(r)}))\right)^2 \cdot \textrm{Var}(\mathcal{L}_{\mathcal{I}(r)}) \,+ \nonumber\\
    & \; \frac{1}{4}\left(f'_{\mathcal{L}_j}(\mathbb{E}(\mathcal{L}_{\mathcal{I}(r)}))\right)^2 \cdot \textrm{Var}\left(\left(\mathcal{L}_{\mathcal{I}(r)} - \mathbb{E}(\mathcal{L}_{\mathcal{I}(r)})\right)^2\right) + ... \;, 
\end{align}
where $f_{\mathcal{L}_j}(\cdot)$ and $f'_{\mathcal{L}_j}(\cdot)$ are the probability density function and its derivative for $\mathcal{L}_j$, respectively.
We observe that the first-order approximation (a special case of the delta method) is insufficiently accurate when compared against simulation results (see Section~\ref{sec:miscmath_prob_Ir_Js_validation}). This is likely due to $F_{\mathcal{L}_j}$ being non-linear. We thus recommend using a second or higher-order approximation.

Finally, we denote $\mu_{\mathcal{P}} \triangleq \mathbb{E}\left(F_{\mathcal{L}_j}(\mathcal{L}_{\mathcal{I}(r)})\right)$,
$\sigma^2_{\mathcal{P}} \triangleq  \textrm{Var}\left(F_{\mathcal{L}_j}(\mathcal{L}_{\mathcal{I}(r)})\right)$ and obtain the beta distribution parameters~$\alpha_{\mathcal{P}}$ and~$\beta_{\mathcal{P}}$ via the method of moments:
\begin{align}
    \alpha_{\mathcal{P}} = \left(\frac{1 - \mu_{\mathcal{P}}}{\sigma^2_{\mathcal{P}}} - \frac{1}{\mu_{\mathcal{P}}} \right) \mu_{\mathcal{P}}^2 \,, 
    \quad
    \beta_{\mathcal{P}} = \alpha_{\mathcal{P}}\left(\frac{1}{\mu_{\mathcal{P}}} - 1\right) \,.
    \label{eq:vem_beta_param_mom}
\end{align}

\subsection{Estimation under normal assumptions}
To complement the main text, we also discuss how the quantities derived above behave under normal assumptions. Firstly,~\eqref{eq:vem_E_ZIr_general} and~\eqref{eq:vem_var_ZIr_general} are now
\begin{align}
    \mathbb{E}(\mathcal{L}_{\mathcal{I}(r)})
    & \approx \mu_\mathcal{V} + \mu_\epsilon + \frac{\sigma^2_\mathcal{V}}{\sqrt{\sigma^2_\mathcal{V} + \sigma^2_1}} \Phi^{-1}\left(\frac{r - \alpha}{N - 2\alpha + 1}\right) \,, \label{eq:vem_E_ZIr_normal} \\[-0.3em]
    \textrm{Var}(\mathcal{L}_{\mathcal{I}(r)}) & \approx \frac{\sigma^2_1 \sigma^2_\mathcal{V}}{\sigma^2_\mathcal{V} + \sigma^2_1} \, +
    \frac{\sigma^4_\mathcal{V}}{\sigma^2_\mathcal{V} + \sigma^2_1}  \frac{r(N-r+1)}{(N+1)^2 (N+2)}
    \frac{1}{\big(\phi\big(\Phi^{-1}\big(\frac{r}{N+1}\big)\big)\big)^2} + \sigma^2_2 \,. \label{eq:vem_var_ZIr_normal}
\end{align}

We then recall from~\eqref{eq:vem_Yn_Zn_def} that $\mathcal{L}_j \overset{\textrm{i.i.d.}}{\sim} \mathcal{N}\left(\mu_\mathcal{V} + \mu_\epsilon, \sigma^2_\mathcal{V} + \sigma^2_2\right)$, and hence
\begin{align}
    F_{\mathcal{L}_j}(l) = \Phi\left(\frac{l - (\mu_\mathcal{V} + \mu_\epsilon)}{\sqrt{\sigma^2_\mathcal{V} + \sigma^2_2}}\right).
    \label{eq:vem_f_Zn_def_normal}
\end{align}
This opens up multiple pathways to estimate $\mu_{\mathcal{P}}$ and $\sigma^2_{\mathcal{P}}$.
Firstly, we can proceed with the Taylor series expansion approach, using the quantities derived above and noting that the first and second derivatives of~\eqref{eq:vem_f_Zn_def_normal} are
\begin{align}
    f_{\mathcal{L}_j}(l) & = \frac{1}{\sqrt{\sigma^2_\mathcal{L} + \sigma^2_2}} \,\phi\left(\frac{l - (\mu_\mathcal{V} + \mu_\epsilon)}{\sqrt{\sigma^2_\mathcal{V} + \sigma^2_2}}\right) \,, \textrm{ and} \\
    f'_{\mathcal{L}_j}(l) & = -\frac{l - (\mu_\mathcal{V} + \mu_\epsilon)}{(\sigma^2_\mathcal{V} + \sigma^2_2)^{\frac{3}{2}}}\, \phi\left(\frac{l - (\mu_\mathcal{V} + \mu_\epsilon)}{\sqrt{\sigma^2_\mathcal{V} + \sigma^2_2}}\right) \,.
\end{align}
However, while the estimate for $\mathbb{E}(F_{\mathcal{L}_j}(\mathcal{L}_{\mathcal{I}(r)}))$ under this approach is reasonably accurate, we find the estimates for $\textrm{Var}(F_{\mathcal{L}_j}(\mathcal{L}_{\mathcal{I}(r)}))$ unsatisfactory in some cases, even when we involve the \emph{sixth}-order Taylor polynomial (see Section~\ref{sec:miscmath_prob_Ir_Js_validation}). 

Instead, we estimate the quantities using Owen's work on integrals of Gaussian functions~\cite{owen80table}. We first define $\mathcal{L}^*$, being $\mathcal{L}_{\mathcal{I}(r)}$ normalised by the parameters of~$\mathcal{L}_j$:
\begin{align}
    \mathcal{L}^* \triangleq \frac{\mathcal{L}_{\mathcal{I}(r)} - (\mu_\mathcal{V} + \mu_\epsilon)}{\sqrt{\sigma^2_\mathcal{V} + \sigma^2_2}} \,.
    \label{eq:vem_Z*_def}
\end{align}
Since $\mathcal{L}_{\mathcal{I}(r)}$ is approximately normally distributed, $\mathcal{L}^*$ is also approximately normally distributed with mean and variance
\begin{align}
    \mu_{\mathcal{L}^*} = \frac{\mathbb{E}(\mathcal{L}_{\mathcal{I}(r)}) - (\mu_\mathcal{V} + \mu_\epsilon)}{\sqrt{\sigma^2_\mathcal{V} + \sigma^2_2}}, \;\;
    \sigma^2_{\mathcal{L}^*} = \frac{1}{\sigma^2_\mathcal{V} + \sigma^2_2} \textrm{Var}(\mathcal{L}_{\mathcal{I}(r)}) \,,
    \label{eq:vem_e_var_Z*}
\end{align}
where $\mathbb{E}(\mathcal{L}_{\mathcal{I}(r)})$ and $\textrm{Var}(\mathcal{L}_{\mathcal{I}(r)})$ are approximated in~\eqref{eq:vem_E_ZIr_normal} and~\eqref{eq:vem_var_ZIr_normal}, respectively. This allows us to represent $\mathcal{L}^*$ by scaling a standard normal r.v. $Z$:
\begin{align}
    \mathcal{L}^* \approx \mu_{\mathcal{L}^*} + \sigma_{\mathcal{L}^*}Z \,.
    \label{eq:vem_Z*_scaled_S}
\end{align}
We can then write $F_{\mathcal{L}_j}(\mathcal{L}_{\mathcal{I}(r)})$ as
\begin{align}
    F_{\mathcal{L}_j}\left(\mathcal{L}_{\mathcal{I}(r)}\right) = \Phi\left(\mathcal{L}^*\right) \approx \Phi\left(\mu_{\mathcal{L}^*} + \sigma_{\mathcal{L}^*}Z\right)
\end{align}
by substituting, in turn,~\eqref{eq:vem_f_Zn_def_normal}, \eqref{eq:vem_Z*_def} and~\eqref{eq:vem_Z*_scaled_S} into the LHS of the equation.

We then make use of the following identities provided by Owen~\cite{owen80table} (Equations 10010.8 and~20010.4):
\begin{align}
    \mathbb{E}\left(\Phi(\mu_{\mathcal{L}^*} + \sigma_{\mathcal{L}^*}Z)\right) & = \Phi\left(\frac{\mu_{\mathcal{L}^*}}{\sqrt{1 + \sigma^2_{\mathcal{L}^*}}}\right) \,, \label{eq:vem_E_Z*}\\
    \mathbb{E}\left((\Phi(\mu_{\mathcal{L}^*} + \sigma_{\mathcal{L}^*}Z))^2\right)
    & = \, \Phi\left(\frac{\mu_{\mathcal{L}^*}}{\sqrt{1 + \sigma^2_{\mathcal{L}^*}}}\right) - 2 \cdot T\left(\frac{\mu_{\mathcal{L}^*}}{\sqrt{1 + \sigma^2_{\mathcal{L}^*}}}, \frac{1}{\sqrt{1 + 2\sigma^2_{\mathcal{L}^*}}}\right) \,,
    \label{eq:vem_E_Z*_sq}
\end{align}
where $T(\cdot, \cdot)$ represents Owen's $T$ function~\cite{owen1956tables}, of which a fast numerical algorithm is available from~\cite{patefield2000fast}. While the original work does not provide an identity for the variance of~$\Phi({\mathcal{L}^*})$, we can obtain so from~\eqref{eq:vem_E_Z*} and~\eqref{eq:vem_E_Z*_sq} easily:
\begin{align}
    & \textrm{Var}\left(\Phi(\mu_{\mathcal{L}^*} + \sigma_{\mathcal{L}^*}Z)\right) \nonumber\\
    =\, & \mathbb{E}\left((\Phi(\mu_{\mathcal{L}^*} + \sigma_{\mathcal{L}^*}Z))^2\right) - \left(\mathbb{E}\left(\Phi(\mu_{\mathcal{L}^*} + \sigma_{\mathcal{L}^*}Z)\right)\right)^2 \nonumber\\
    =\, & \Phi\left(\frac{\mu_{\mathcal{L}^*}}{\sqrt{1 + \sigma^2_{\mathcal{L}^*}}}\right) \left(1 - \Phi\left(\frac{\mu_{\mathcal{L}^*}}{\sqrt{1 + \sigma^2_{\mathcal{L}^*}}}\right) \right) - 2\cdot T\left(\frac{\mu_{\mathcal{L}^*}}{\sqrt{1 + \sigma^2_{\mathcal{L}^*}}},\, \frac{1}{\sqrt{1 + 2\sigma^2_{\mathcal{L}^*}}}\right) \,.
    \label{eq:vem_var_Z*}
\end{align}

We finally substitute~\eqref{eq:vem_E_Z*} and~\eqref{eq:vem_var_Z*} into~\eqref{eq:vem_beta_param_mom} to obtain the beta distribution parameters under normal assumptions.

\subsection{Some basic validations}
\label{sec:miscmath_prob_Ir_Js_validation}

We close the investigation on the supposedly minor quantity in the ranking under lower uncertainty problem by validating the theoretical results derived above. We have accepted in Section~\ref{sec:miscmath_prob_Ir_Js_modelling} that the probability $\mathbb{P}(\mathcal{I}(r) = \mathcal{J}(s))$ and its associated distributions are analytically intractable in many cases. As a result, a full validation across many parameter combinations (similar to what we have done for quantities such as $\mathbb{E}(\mathcal{V}_{\mathcal{I}(r)})$ and $\textrm{Var}(\mathcal{W}_\mathcal{L})$ in the main text) for what are best estimates is overkill. Instead, we will perform visual checks on whether the fitted distributions are reasonable approximation to the empirical distribution for a handful of parameter combinations under normal assumptions. This is not to say we have fully addressed the original question posed at the top of this chapter  --  there is still plenty to uncover for the said probability in terms of estimation method and interpretation  --  it is just not the primary quantity of interest for our ranking under lower uncertainty problem.

We also calculate the mean Kullback-Leibler (KL) divergence between the empirical distribution and distributions produced from different modelling approaches to compare them quantitatively. Here, we define the mean KL divergence as
\begin{align}
    & \, \frac{1}{N} \sum_{r=1}^{N} D_{KL}\left(\hat{f}_{C \,|\, N-1, r, ...} \,||\, f_{\hat{C} \,|\, N-1, r, ...}\right) \nonumber\\
    = &\, \frac{1}{N} \sum_{r=1}^{N} \sum_{s=1}^{N} \hat{f}_{C}(s-1 \,|\, N-1, r, ...) \log\left(\frac{\hat{f}_{C}(s-1 \,|\, N-1, r, ...)}{f_{\hat{C}}(s-1 \,|\, N-1, r, ...)}\right)
    \,,
\end{align}
where $\hat{f}_{C}(\cdot)$ denotes the probability mass function (PMF) of the empirical marginal distribution and $f_{\hat{C}}(\cdot)$ denotes the PMF of the fitted distribution. In other words, we calculate the KL divergence of each marginal distribution shown in Figure~\ref{fig:miscmath_prob_Ir_eq_Js_r_marginals} and average it over the~$N$ marginal distributions.\footnote{We use the empirical marginal distribution as the target distribution for KL divergence as we believe the true distribution is analytically intractable in many cases. This means the average KL divergence may vary slightly from one simulation to another due to using fresh samples. We ensure such variance is negligible by performing hundreds of thousands of simulation runs when generating each empirical marginal distributions.}

\begin{figure}
    \begin{center}
        \includegraphics[width=0.925\textwidth, trim = 0 0 0 0, clip]{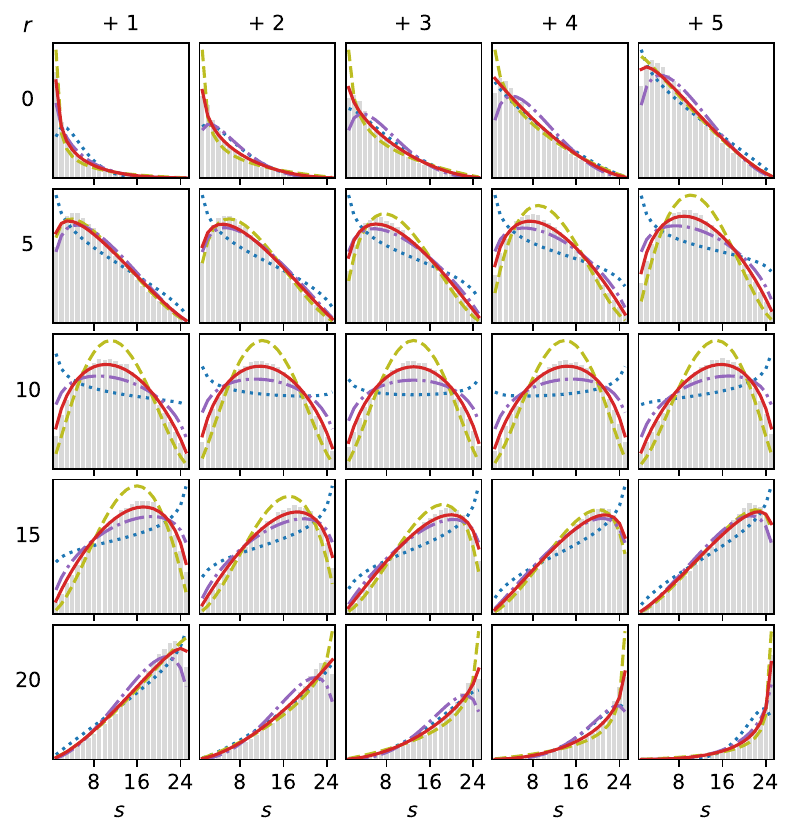}
    \end{center}
    \vspace*{-15pt}
    \caption[The fitted marginal distributions overlaid on Figure~\ref{fig:miscmath_prob_Ir_eq_Js_r_marginals}.]{The fitted marginal distributions  --  that using the Taylor series expansion up to the second-order (blue dotted line), fourth-order (olive dashed line), and sixth-order (purple dot-dashed line), as well as that using Owen's integrals of Gaussian functions (red solid line) -- overlaid on Figure~\ref{fig:miscmath_prob_Ir_eq_Js_r_marginals}, which shows the empirical marginal distributions of $\mathbb{P}(\mathcal{I}(r) = \mathcal{J}(s))$ (grey bars).}
    \label{fig:miscmath_prob_Ir_eq_Js_r_marginals_w_fits_N25}
\end{figure}

We observe that the approach using Owen's integrals on Gaussian functions produces marginal distributions produces better fit on the empirical distribution than the approaches using Taylor series expansion up to the sixth-order Taylor polynomial, both visually (see Figure~\ref{fig:miscmath_prob_Ir_eq_Js_r_marginals_w_fits_N25}) and in terms of the mean KL divergence (see Table~\ref{tab:miscmath_prob_Ir_Js_KL_div_N1025}). We also observe the fit using Owen's integrals is less ideal (yet still reasonable) in cases where:
\begin{enumerate}[itemsep=0pt]
    \item $r$ or $s$ is close to $1$ or $N$,
    \item $N$ is small, and
    \item $\sigma^2_X$ and $\sigma^2_1$ are similar in magnitude.
\end{enumerate}
These cases lead to a higher variance in order statistic and rank estimators and greater deviation from the normal assumptions, which make accurate estimation more difficult.

\begin{figure}
    \begin{center}
        \includegraphics[width=0.925\textwidth, trim = 0 0 0 0, clip]{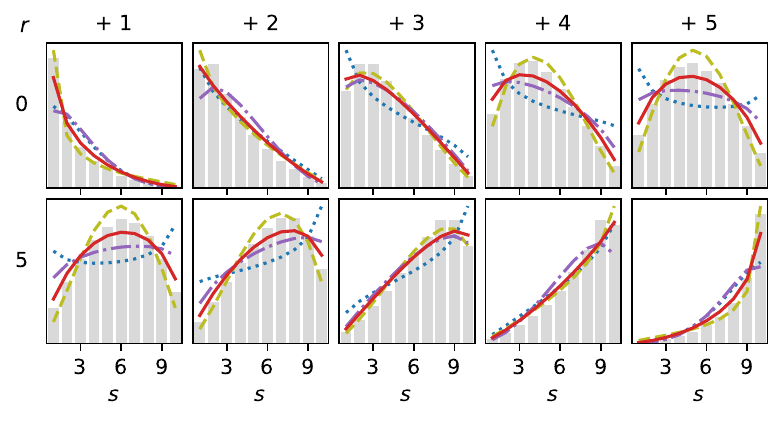}
    \end{center}
    \vspace*{-15pt}
    \caption[The fitted marginal distributions overlaid on the empirical marginal distributions of $\mathbb{P}(\mathcal{I}(r) = \mathcal{J}(s))$ ($N = 10$).]{The fitted marginal distributions -- that using the Taylor series expansion up to the second-order (blue dotted line), fourth-order (olive dashed line), and sixth-order (purple dot-dashed line), as well as that using Owen's integrals of Gaussian functions (red solid line) -- overlaid on the empirical marginal distributions of $\mathbb{P}(\mathcal{I}(r) = \mathcal{J}(s))$ (grey bars), with $N=10$, $\mu_\mathcal{V} = \mu_\epsilon = 0$, $\sigma^2_\mathcal{V} = 1$, $\sigma^2_1 = 0.5$, and $\sigma^2_2 = 0.4$.}
    \label{fig:miscmath_prob_Ir_eq_Js_r_marginals_w_fits_N10}
\end{figure}

\begin{table}
    \onehalfspacing
    \centering
    \begin{tabular}{c|c|c}
         Modelled distribution & \multicolumn{2}{c}{Mean KL divergence} \\
         & $N = 10$ & $N = 25$\\\hline
         Uniform & 0.24582 $\pm$ 0.00025 & 0.28358 $\pm$ 0.00018 \\
         Beta-binomial (2nd order Taylor SE) & 0.05924 $\pm$ 0.00017 & 0.06190 $\pm$ 0.00007 \\
         Beta-binomial (4th order Taylor SE) & 0.01070 $\pm$ 0.00004 & 0.01930 $\pm$ 0.00004 \\
         Beta-binomial (6th order Taylor SE) & 0.03414 $\pm$ 0.00013 & 0.02029 $\pm$ 0.00004 \\
         Beta-binomial (Owen's integrals) & \textbf{0.00784} $\pm$ 0.00006 & \textbf{0.00317} $\pm$ 0.00002 \\
    \end{tabular}
    \caption[Mean Kullback-Leibler (KL) divergence between the empirical marginal distribution of \texorpdfstring{$\mathbb{P}(\mathcal{I}(r) = \mathcal{J}(s))$}{P(I(r) = J(s))} and distributions produced from different modelling approaches.]{Mean Kullback-Leibler (KL) divergence between the empirical marginal distribution of $\mathbb{P}(\mathcal{I}(r) = \mathcal{J}(s))$ and distributions produced from different modelling approaches, with $\mu_\mathcal{V} = \mu_\epsilon = 0$, $\sigma^2_\mathcal{V} = 1$, $\sigma^2_1 = 0.5$, and $\sigma^2_2 = 0.4$. Each entry records the mean and standard deviation of the mean KL divergence across multiple simulations.}
    \label{tab:miscmath_prob_Ir_Js_KL_div_N1025}
\end{table}

\newpage

\section{Calculations Related to Dilution in Personalisation Strategy Experiment Designs}
\label{sec:miscmath_metric_dilution}

This appendix expands the exposition in Section~\ref{sec:pse_expt_comparison_dilution} by providing the detailed algebraic manipulations of the inequalities involved. In the section, we discussed, under the context of designing experiments for personalisation strategies, the conditions where an experiment setup with treatment effect dilution (aka Setup~2) will emerge superior to one without treatment effect dilution (aka Setup~3) and vice versa.

We recap the mathematical arguments presented in the section. Under the evaluation framework introduced in Chapter~\ref{chap:pse}, we can establish superiority of an experiment setup via two criteria. Setup~3 is superior to Setup~2 if at least one of the following inequalities hold: 
\begin{align}
    \textrm{First criterion: } &  \theta^*_{S3} < \theta^*_{S2} \textrm{ (as $\Delta_{S3} > \Delta_{S2}$ trivially holds)} , \nonumber \\
    \textrm{Second criterion: } & \Delta_{S3} - \Delta_{S2} > \theta^*_{S3} - \theta^*_{S2} \,, \nonumber
\end{align}
where $\Delta_{S2}$, $\theta^*_{S2}$, $\Delta_{S3}$, and $\theta^*_{S3}$ are that stated in Equations~\eqref{eq:pse_Delta_S2}, \eqref{eq:pse_theta_star_S2}, \eqref{eq:pse_Delta_S3}, and \eqref{eq:pse_theta_star_S3} respectively -- we will restate the equations below where appropriate.
If neither inequality holds (and both sides are not equal), we consider Setup~2 as superior to Setup~3 under the second criterion as the following holds:
\begin{align}
    \Delta_{S3} - \Delta_{S2} < \theta^*_{S3} - \theta^*_{S2} 
    \;\;\iff\;\;  \Delta_{S2} - \Delta_{S3} > \theta^*_{S2} - \theta^*_{S3}.
    \nonumber
\end{align}

This appendix only deals with the detailed algebraic manipulation. We refer readers back to Section~\ref{sec:pse_expt_comparison_dilution} in the main text for the interpretations of and the insights from the resultant inequalities.

\subsection{The first criterion}

We first show $\theta^*_{S3} < \theta^*_{S2}$, where
\begin{align}
    \theta^*_{S2} = \, &  (z_{1-\frac{\alpha}{2}} - z_{1-\pi_{\textrm{min}}})
    \sqrt{\frac{
    2\left(n_0(2\sigma^2_{C0}) + n_1(\sigma^2_{I1} + \sigma^2_{C1}) +
    n_2(\sigma^2_{C2} + \sigma^2_{I2}) + n_3(\sigma^2_{I\phi} + \sigma^2_{I\psi})\right)
    }{(n_0 + n_1 + n_2 + n_3)^2}} \,,
    \tag{*\ref{eq:pse_theta_star_S2}} \\[0.5em]
    \theta^*_{S3} = \,  &  (z_{1-\frac{\alpha}{2}} - z_{1-\pi_{\textrm{min}}}) 
    \sqrt{\frac{2\left(n_1(\sigma^2_{I1} + \sigma^2_{C1}) + n_2(\sigma^2_{C2} + \sigma^2_{I2}) + n_3(\sigma^2_{I\phi} + \sigma^2_{I\psi})\right)}{(n_1 + n_2 + n_3)^2}} \,,
    \tag{*\ref{eq:pse_theta_star_S3}}
\end{align}
is equivalent to 
\begin{align}
    \frac{\left(n_1(\sigma^2_{I1} + \sigma^2_{C1}) + n_2(\sigma^2_{C2} + \sigma^2_{I2}) + n_3(\sigma^2_{I\phi} + \sigma^2_{I\psi})\right) \cdot (n_0 + 2n_1 + 2n_2 + 2n_3)}{2 (n_1 + n_2 + n_3)^2} < \sigma^2_{C0} \,.
    \tag{*\ref{eq:pse_setup23_criterion1}}
\end{align}
This is the condition that will lead to Setup~3 being superior to Setup~2 under the first criterion in our evaluation framework. 

We start by writing the inequality $\theta^*_{S3} < \theta^*_{S2}$ in full:
\begin{align}
    & (z_{1-\frac{\alpha}{2}} - z_{1-\pi_{\textrm{min}}}) 
    \sqrt{\frac{2\left(n_1(\sigma^2_{I1} + \sigma^2_{C1}) + n_2(\sigma^2_{C2} + \sigma^2_{I2}) + n_3(\sigma^2_{I\phi} + \sigma^2_{I\psi})\right)}{(n_1 + n_2 + n_3)^2}} \nonumber\\
    <\, & (z_{1-\frac{\alpha}{2}} - z_{1-\pi_{\textrm{min}}})
    \sqrt{\frac{
    2\left(n_0(2\sigma^2_{C0}) + n_1(\sigma^2_{I1} + \sigma^2_{C1}) + n_2(\sigma^2_{C2} + \sigma^2_{I2}) + n_3(\sigma^2_{I\phi} + \sigma^2_{I\psi})\right)}{(n_0 + n_1 + n_2 + n_3)^2}} \,.
    \label{eq:pse_setup23_criterion1_full}
\end{align}
Canceling the $z_{1-\frac{\alpha}{2}} - z_{1 - \pi_{\min}}$ and $\sqrt{2}$ terms on both sides, then squaring both sides\footnote{The operation maintains the \emph{iff} relation as the radicand is positive.} yields
\begin{align}
    & \frac{n_1(\sigma^2_{I1} + \sigma^2_{C1}) + n_2(\sigma^2_{C2} + \sigma^2_{I2}) + n_3(\sigma^2_{I\phi} + \sigma^2_{I\psi})}{(n_1 + n_2 + n_3)^2} \nonumber\\
    <\, & \frac{
    n_0(2\sigma^2_{C0}) + n_1(\sigma^2_{I1} + \sigma^2_{C1}) +
    n_2(\sigma^2_{C2} + \sigma^2_{I2}) + n_3(\sigma^2_{I\phi} + \sigma^2_{I\psi})
    }{(n_0 + n_1 + n_2 + n_3)^2} \,.
    \tag{\ref{eq:pse_setup23_criterion1_full}a}
\end{align}

We then write $\xi = n_1(\sigma^2_{I1} + \sigma^2_{C1}) + n_2(\sigma^2_{C2} + \sigma^2_{I2}) + n_3(\sigma^2_{I\phi} + \sigma^2_{I\psi})$ and move the $\xi$ term on the RHS to the LHS:
\begin{align}
    \xi\left(\frac{1}{(n_1 + n_2 + n_3)^2} - \frac{1}{(n_0 + n_1 + n_2 + n_3)^2}\right) < \frac{n_0(2\sigma^2_{C0})}{(n_0 + n_1 + n_2 + n_3)^2} \,.
    \tag{\ref{eq:pse_setup23_criterion1_full}b}
\end{align}
As the partial fractions on the LHS can be consolidated as
\begin{align}
    & \frac{1}{(n_1 + n_2 + n_3)^2} - \frac{1}{(n_0 + n_1 + n_2 + n_3)^2} 
    =\, \frac{(n_0 + n_1 + n_2 + n_3)^2 - (n_1 + n_2 + n_3)^2}{(n_1 + n_2 + n_3)^2(n_0 + n_1 + n_2 + n_3)^2} \nonumber\\
    =\, & \frac{(n_0 + 2n_1 + 2n_2 + 2n_3)n_0}{(n_1 + n_2 + n_3)^2(n_0 + n_1 + n_2 + n_3)^2}\,,
    \nonumber
\end{align}
where the second step utilizes the identity $a^2 - b^2 = (a+b)(a-b)$, we can write Inequality~(\ref{eq:pse_setup23_criterion1_full}b) as
\begin{align}
    \xi\left(\frac{(n_0 + 2n_1 + 2n_2 + 2n_3)n_0}{(n_1 + n_2 + n_3)^2(n_0 + n_1 + n_2 + n_3)^2}\right) < \frac{n_0(2\sigma^2_{C0})}{(n_0 + n_1 + n_2 + n_3)^2} \,.
    \tag{\ref{eq:pse_setup23_criterion1_full}c}
\end{align}
We finally cancel the $n_0$ and $(n_0+n_1+n_2+n_3)^2$ terms on both sides, move the factor of two to the LHS, and write $\xi$ in its full form to arrive at
\begin{align}
    \frac{\left(n_1(\sigma^2_{I1} + \sigma^2_{C1}) + n_2(\sigma^2_{C2} + \sigma^2_{I2}) + n_3(\sigma^2_{I\phi} + \sigma^2_{I\psi})\right) \cdot  (n_0 + 2n_1 + 2n_2 + 2n_3)}{2 (n_1 + n_2 + n_3)^2} < \sigma^2_{C0} \,,
    \nonumber
\end{align}
which is identical to Inequality~\eqref{eq:pse_setup23_criterion1}.

\subsection{The second criterion}

\paragraph{The master inequality}
We also show that the inequality $\Delta_{S3} - \Delta_{S2} > \theta^*_{S3} - \theta^*_{S2}$,
where
\begin{align}
    \Delta_{S2} = \, &   \frac{n_1(\mu_{C1} - \mu_{I1}) + n_2(\mu_{I2} - \mu_{C2}) + n_3(\mu_{I\psi} - \mu_{I_\phi})}{n_0 + n_1 + n_2 + n_3} \,,
    \tag{*\ref{eq:pse_Delta_S2}}\\[0.5em]
    \Delta_{S3} = \, &   \frac{n_1(\mu_{C1} - \mu_{I1}) + n_2(\mu_{I2} - \mu_{C2}) + n_3(\mu_{I\psi} - \mu_{I_\phi})}{n_1 + n_2 + n_3} \,, 
    \tag{*\ref{eq:pse_Delta_S3}}\\[0.5em]
    \theta^*_{S2} = \, &  (z_{1-\frac{\alpha}{2}} - z_{1-\pi_{\textrm{min}}})
    \sqrt{\frac{
    2\left(n_0(2\sigma^2_{C0}) + n_1(\sigma^2_{I1} + \sigma^2_{C1}) +
    n_2(\sigma^2_{C2} + \sigma^2_{I2}) + n_3(\sigma^2_{I\phi} + \sigma^2_{I\psi})\right)
    }{(n_0 + n_1 + n_2 + n_3)^2}} \,,
    \tag{*\ref{eq:pse_theta_star_S2}} \\
    \theta^*_{S3} = \, &  (z_{1-\frac{\alpha}{2}} - z_{1-\pi_{\textrm{min}}}) 
    \sqrt{\frac{2\left(n_1(\sigma^2_{I1} + \sigma^2_{C1}) + n_2(\sigma^2_{C2} + \sigma^2_{I2}) + n_3(\sigma^2_{I\phi} + \sigma^2_{I\psi})\right)}{(n_1 + n_2 + n_3)^2}} \,,
    \tag{*\ref{eq:pse_theta_star_S3}}
\end{align}
is equivalent to
\begin{align}
    \frac{n_1 + n_2 + n_3}{n_0}\sqrt{2n_0\sigma^2_{C0} + \xi} > 
    \frac{n_0 + n_1 + n_2 + n_3}{n_0}\sqrt{\xi} - \frac{\eta}{\sqrt{2}z}
    \tag{*\ref{eq:pse_setup23_criterion2_initial}} \,,
\end{align}
where
\begin{align}
    \eta & = n_1(\mu_{C1} - \mu_{I1}) + n_2(\mu_{I2} - \mu_{C2}) + n_3(\mu_{I\psi} - \mu_{I\phi}) \,, \nonumber\\
    \xi & = n_1(\sigma^2_{C1} + \sigma^2_{I1}) + n_2(\sigma^2_{I2} + \sigma^2_{C2}) + n_3(\sigma^2_{I\psi} + \sigma^2_{I\phi}) \,, \textrm{ and} \nonumber\\
    z & = z_{1-\frac{\alpha}{2}} - z_{1 - \pi_{\textrm{min}}}\,. \tag{*\ref{eq:pse_setup23_maths_shorthand}}
\end{align}
We start by writing $\Delta_{S3} - \Delta_{S2} > \theta^*_{S3} - \theta^*_{S2}$ in terms of $\eta$, $\xi$, and $z$ as shown above: 
\begin{align}
    \frac{\eta}{n_1 + n_2 + n_3} - 
    \frac{\eta}{n_0 + n_1 + n_2 + n_3}
    > z\sqrt{\frac{2\xi}{(n_1 + n_2 + n_3)^2}} - 
    z\sqrt{\frac{2\big(n_0(2\sigma^2_{C0})+ \xi\big)}{(n_0 + n_1 + n_2 + n_3)^2}} \,.
    \label{eq:pse_setup23_criterion2_initial_full}
\end{align}
Pulling out the common factors on each side we have
\begin{align}
    & \eta\left(\frac{1}{n_1 + n_2 + n_3} - 
    \frac{1}{n_0 + n_1 + n_2 + n_3}\right)
    > \sqrt{2} z \Bigg(\frac{\sqrt{\xi}}{n_1 + n_2 + n_3} - \frac{\sqrt{2n_0\sigma^2_{C0}+ \xi}}{n_0 + n_1 + n_2 + n_3}\Bigg) \,.
    \tag{\ref{eq:pse_setup23_criterion2_initial_full}a}
\end{align}
Writing the partial fraction on the LHS of
Inequality~(\ref{eq:pse_setup23_criterion2_initial_full}a) as a composite fraction we have
\begin{align}
    & \eta\left(\frac{n_0}{(n_1+n_2+n_3)(n_0+n_1+n_2+n_3)}\right)
    >  \sqrt{2} z \Bigg(\frac{\sqrt{\xi}}{n_1 + n_2 + n_3} - \frac{\sqrt{2n_0\sigma^2_{C0}+ \xi}}{n_0 + n_1 + n_2 + n_3}\Bigg) \,.
    \tag{\ref{eq:pse_setup23_criterion2_initial_full}b}
\end{align}
We then move the composite fraction to the RHS and the $\sqrt{2}z$ term to the LHS:
\begin{align}
    \frac{\eta}{\sqrt{2}z} >
    \frac{(n_1+n_2+n_3)\cdot (n_0+n_1+n_2+n_3)}{n_0}
    \Bigg(\frac{\sqrt{\xi}}{n_1 + n_2 + n_3} - \frac{\sqrt{2n_0\sigma^2_{C0}+ \xi}}{n_0 + n_1 + n_2 + n_3}\Bigg) \,,
    \tag{\ref{eq:pse_setup23_criterion2_initial_full}c}
\end{align}
and expand the brackets, canceling terms that appear on both the denominator and the numerator of the resultant fractions in the RHS:
\begin{align}
    \frac{\eta}{\sqrt{2}z} >
    \frac{n_0 + n_1 + n_2 + n_3}{n_0} \sqrt{\xi} - \frac{n_1 + n_2 + n_3}{n_0} \sqrt{2n_0\sigma^2_{C0}+ \xi} \,.
    \tag{\ref{eq:pse_setup23_criterion2_initial_full}d}
\end{align}
Finally, we swap the position of the leftmost term with that of the rightmost term to arrive at
\begin{align}
    \frac{n_1 + n_2 + n_3}{n_0}\sqrt{2n_0\sigma^2_{C0} + \xi} > 
    \frac{n_0 + n_1 + n_2 + n_3}{n_0}\sqrt{\xi} - \frac{\eta}{\sqrt{2}z} \,,
    \nonumber
\end{align}
which is identical to Inequality~\eqref{eq:pse_setup23_criterion2_initial}.

\vspace*{-0.5em}
\paragraph{The trivial case: RHS $\leq 0$}
We observe that the LHS of Inequality~\eqref{eq:pse_setup23_criterion2_initial} is always positive, and hence
the inequality trivially holds if the RHS is non-positive.
Here we show RHS $\leq 0$ is equivalent to
\begin{align}
    \frac{n_0 + n_1 + n_2 + n_3}{n_0} \, \theta^*_{S3} \leq \Delta_{S3} \,,
    \tag{*\ref{eq:pse_setup23_criterion2_strong}}
\end{align}
where 
\begin{align}
    \Delta_{S3} = \, &   \frac{n_1(\mu_{C1} - \mu_{I1}) + n_2(\mu_{I2} - \mu_{C2}) + n_3(\mu_{I\psi} - \mu_{I_\phi})}{n_1 + n_2 + n_3} \,, 
    \tag{*\ref{eq:pse_Delta_S3}}\\[0.5em]
    \theta^*_{S3} = \, &  (z_{1-\frac{\alpha}{2}} - z_{1-\pi_{\textrm{min}}}) 
    \sqrt{\frac{2\left(n_1(\sigma^2_{I1} + \sigma^2_{C1}) + n_2(\sigma^2_{C2} + \sigma^2_{I2}) + n_3(\sigma^2_{I\phi} + \sigma^2_{I\psi})\right)}{(n_1 + n_2 + n_3)^2}} \,.
    \tag{*\ref{eq:pse_theta_star_S3}}
\end{align}
The can be done by writing RHS $\leq 0$ in full:
\begin{align}
    \frac{n_0 + n_1 + n_2 + n_3}{n_0}\sqrt{\xi} - \frac{\eta}{\sqrt{2}z} \leq 0 \,, 
    \label{eq:pse_setup23_criterion2_strong_full}
\end{align}
and moving the second term on the LHS to the RHS:
\begin{align}
    \frac{n_0 + n_1 + n_2 + n_3}{n_0}\sqrt{\xi} \leq \frac{\eta}{\sqrt{2}z} \,. \tag{\ref{eq:pse_setup23_criterion2_strong_full}a}
\end{align}
We then add a factor of $\sqrt{2}z / (n_1 + n_2 + n_3)$ on both sides to get
\begin{align}
    \frac{n_0 + n_1 + n_2 + n_3}{n_0}\frac{\sqrt{\xi}\sqrt{2}z}{n_1+n_2+n_3} \leq \frac{\eta}{n_1+n_2+n_3} \,. \tag{\ref{eq:pse_setup23_criterion2_strong_full}b}
\end{align}
Noting from Equations \eqref{eq:pse_Delta_S3} and \eqref{eq:pse_theta_star_S3} that
\begin{align}
    \Delta_{S3} = \frac{\eta}{n_1 + n_2 + n_3} \,\textrm{ and }\,
    \theta^*_{S3} = \frac{\sqrt{2}\cdot z\cdot\sqrt{\xi}}{n_1 + n_2 + n_3} \,, \nonumber
\end{align}
we finally replace the terms in Inequality~(\ref{eq:pse_setup23_criterion2_strong_full}b) with $\Delta_{S3}$ and $\theta^*_{S3}$ to arrive at
\begin{align}
    \frac{n_0 + n_1 + n_2 + n_3}{n_0}\theta^*_{S3} \, \leq \Delta_{S3} \,,
    \nonumber
\end{align}
which is identical to Inequality~\eqref{eq:pse_setup23_criterion2_strong}.

\paragraph{The non-trivial case: RHS $>0$}
We finally tackle the case where the RHS of the master inequality (Inequality~\eqref{eq:pse_setup23_criterion2_initial}) is greater than zero. We show in this non-trivial case, the master inequality
\begin{align}
    \frac{n_1 + n_2 + n_3}{n_0}\sqrt{2n_0\sigma^2_{C0} + \xi} > 
    \frac{n_0 + n_1 + n_2 + n_3}{n_0}\sqrt{\xi} - \frac{\eta}{\sqrt{2}z}
    \tag{*\ref{eq:pse_setup23_criterion2_initial}}
\end{align}
is equivalent to
\begin{align}
    \frac{2\sigma^2_{C0}}{n_0} >
    \frac{\left(\theta^*_{S3} - \Delta_{S3} + \frac{n_1 + n_2 + n_3}{n_0} \theta^*_{S3}\right)^2 - \left(\frac{n_1 + n_2 + n_3}{n_0} \theta^*_{S3}\right)^2}{2z^2} \; ,
    \tag{*\ref{eq:pse_setup23_criterion2_final}}
\end{align}
where
\begin{align}
    \Delta_{S3} = \, &   \frac{n_1(\mu_{C1} - \mu_{I1}) + n_2(\mu_{I2} - \mu_{C2}) + n_3(\mu_{I\psi} - \mu_{I_\phi})}{n_1 + n_2 + n_3} \,, 
    \tag{*\ref{eq:pse_Delta_S3}}\\[0.5em]
    \theta^*_{S3} = \, &  (z_{1-\frac{\alpha}{2}} - z_{1-\pi_{\textrm{min}}}) 
    \sqrt{\frac{2\left(n_1(\sigma^2_{I1} + \sigma^2_{C1}) + n_2(\sigma^2_{C2} + \sigma^2_{I2}) + n_3(\sigma^2_{I\phi} + \sigma^2_{I\psi})\right)}{(n_1 + n_2 + n_3)^2}} \,,
    \tag{*\ref{eq:pse_theta_star_S3}} \\
    \eta = \, & n_1(\mu_{C1} - \mu_{I1}) + n_2(\mu_{I2} - \mu_{C2}) + n_3(\mu_{I\psi} - \mu_{I\phi}) \,, \nonumber\\
    \xi = \, &  n_1(\sigma^2_{C1} + \sigma^2_{I1}) + n_2(\sigma^2_{I2} + \sigma^2_{C2}) + n_3(\sigma^2_{I\psi} + \sigma^2_{I\phi}) \,, \textrm{ and} \nonumber\\
    z = \, &  z_{1-\frac{\alpha}{2}} - z_{1 - \pi_{\textrm{min}}}\,. 
    \tag{*\ref{eq:pse_setup23_maths_shorthand}}
\end{align}
We first multiply both sides of Inequality~\eqref{eq:pse_setup23_criterion2_initial} with the fraction $\frac{n_0\sqrt{2}z}{n_1 + n_2 + n_3}$ to get
\begin{align}
    & \frac{n_1 + n_2 + n_3}{n_0}\sqrt{2n_0\sigma^2_{C0} + \xi}
    \frac{n_0\sqrt{2}z}{n_1 + n_2 + n_3}
    >  
    \left(\frac{n_0 + n_1 + n_2 + n_3}{n_0}\sqrt{\xi} - \frac{\eta}{\sqrt{2}z}\right)
    \frac{n_0\sqrt{2}z}{n_1 + n_2 + n_3} \;.
    \label{eq:pse_setup23_criterion2_final_full}
\end{align}
Cancelling terms on both sides of the fractions we have
\begin{align}
    & \sqrt{2n_0\sigma^2_{C0} + \xi} \sqrt{2}z 
    > (n_0 + n_1 + n_2 + n_3) \frac{\sqrt{\xi}\sqrt{2}z}{n_1 + n_2 + n_3} - n_0 \frac{\eta}{n_1 + n_2 + n_3} \;. \tag{\ref{eq:pse_setup23_criterion2_final_full}a}
\end{align}
Again noting the identities for $\Delta_{S3}$ and $\theta^*_{S3}$, i.e.,
\begin{align}
    \Delta_{S3} = \frac{\eta}{n_1 + n_2 + n_3} \,\textrm{ and }\,
    \theta^*_{S3} = \frac{\sqrt{2}\cdot z\cdot\sqrt{\xi}}{n_1 + n_2 + n_3} \,, \nonumber
\end{align}
we can replace the fractions on the RHS to obtain
\begin{align}
    \sqrt{2n_0\sigma^2_{C0} + \xi} \sqrt{2}z 
    > (n_0 + n_1 + n_2 + n_3) \theta^*_{S3} - n_0 \Delta_{S3} \;.
    \tag{\ref{eq:pse_setup23_criterion2_final_full}b}
\end{align}

We then square both sides of Inequality~({\ref{eq:pse_setup23_criterion2_final_full}b})
and move the $2z^2$ term to the RHS:
\begin{align}
    2n_0\sigma^2_{C0} + \xi
    > \frac{\big((n_0 + n_1 + n_2 + n_3) \theta^*_{S3} - n_0 \Delta_{S3}\big)^2}{2z^2} \;.
    \tag{\ref{eq:pse_setup23_criterion2_final_full}c}
\end{align}
Note the squaring still allows the implication to go both ways as both sides of Inequality~({\ref{eq:pse_setup23_criterion2_final_full}c}) are positive.
Based on the identity for $\theta^*_{S3}$, we observe $\xi$ can also be written as
\begin{align}
    \xi = \frac{(n_1 + n_2 + n_3)^2 (\theta^*_{S3})^2}{2z^2} \;.
    \tag{\ref{eq:pse_setup23_criterion2_final_full}d}
\end{align}
Thus, we can group all terms with a $2z^2$ denominator by moving $\xi$ in Inequality~(\ref{eq:pse_setup23_criterion2_final_full}c) to the RHS and substituting Equation~(\ref{eq:pse_setup23_criterion2_final_full}d) into the resultant inequality:
\begin{align}
    2n_0\sigma^2_{C0}
    > \frac{\big((n_0 + n_1 + n_2 + n_3) \theta^*_{S3} - n_0 \Delta_{S3}\big)^2 - \big((n_1 + n_2 + n_3) \theta^*_{S3}\big)^2}{2z^2} \;.
    \tag{\ref{eq:pse_setup23_criterion2_final_full}e}
\end{align}

We finally normalize the inequality to one with unit~$\Delta_{S3}$ and~$\theta^*_{S3}$ to enable effective comparison. 
We divide both sides of Inequality~(\ref{eq:pse_setup23_criterion2_final_full}e) by ${n_0}^2$:
\begin{align}
    \frac{2\sigma^2_{C0}}{n_0}
    > \frac{\left(\frac{n_0 + n_1 + n_2 + n_3}{n_0} \theta^*_{S3} - \Delta_{S3}\right)^2 - \left(\frac{n_1 + n_2 + n_3}{n_0} \theta^*_{S3}\right)^2}{2z^2} \;,
    \tag{\ref{eq:pse_setup23_criterion2_final_full}f}
\end{align}
and split the coefficient of $\theta^*_{S3}$ in the first squared term into an integer~(1) and a fractional ($\frac{n_1 + n_2 + n_3}{n_0}$) part to arrive at
\begin{align}
    \frac{2\sigma^2_{C0}}{n_0} >
    \frac{\left(\theta^*_{S3} - \Delta_{S3} + \frac{n_1 + n_2 + n_3}{n_0} \theta^*_{S3}\right)^2 - \left(\frac{n_1 + n_2 + n_3}{n_0} \theta^*_{S3}\right)^2}{2z^2} \;,
    \nonumber
\end{align}
which is identical to Inequality~\eqref{eq:pse_setup23_criterion2_final}.

\newpage

\section{Calculations Related to Dual Control in Personalisation Strategy Experiment Designs}
\label{sec:miscmath_dual_control}

Similar to the appendix above, this appendix expands the exposition in Section~\ref{sec:pse_expt_comparison_dual_control} by providing the detailed algebraic manipulations of the inequalities involved. In the section, we determine, under the context of designing experiments for personalisation strategies, the sample size required for a dual control experiment setup (aka Setup 4) to emerge superior to a simpler A/B test setup (Setup 3 in this case).

We recap the mathematical arguments presented in the section. Under the evaluation framework introduced in Chapter~\ref{chap:pse}, we can establish superiority of an experiment setup via two evaluation criteria. Setup~4 is superior to Setup~3 if at least one of the following inequalities hold: 
\begin{align}
    \textrm{First criterion: } &  \Delta_{S4} > \Delta_{S3} \textrm{ and } \theta^*_{S4} < \theta^*_{S3} \,, \nonumber \\
    \textrm{Second criterion: } & \Delta_{S4} - \Delta_{S3} > \theta^*_{S4} - \theta^*_{S3} \,, \nonumber
\end{align}
where $\Delta_{S3}$, $\theta^*_{S3}$, $\Delta_{S4}$, and $\theta^*_{S4}$ are that stated in Equations~\eqref{eq:pse_Delta_S3}, \eqref{eq:pse_theta_star_S3}, \eqref{eq:pse_Delta_S4}, and \eqref{eq:pse_theta_star_S4} respectively -- we will restate the equations below where appropriate.

We showed in the main text (Section~\ref{sec:pse_expt_comparison_dual_control}) that $\theta^*_{S4} > \theta^*_{S3}$ always holds, and hence Setup~4 will never be superior to Setup~3 under the first criterion. We thus focus on the second criterion $\Delta_{S4} - \Delta_{S3} > \theta^*_{S4} - \theta^*_{S3}$ and show
\begin{enumerate}
  \item The criterion $\Delta_{S4} - \Delta_{S3} > \theta^*_{S4} - \theta^*_{S3}$ is equivalent to Inequality~\eqref{eq:pse_setup34_criterion2_full}.
\end{enumerate}
By applying various simplifying assumptions on the group response variances (i.e. the $\sigma^2$-terms) and the group sample sizes (i.e. the $n$-terms), we also show:
\begin{enumerate}
  \item[2.] Assuming only the $\sigma^2$-terms are similar in magnitude, we can express the RHS of Inequality~\eqref{eq:pse_setup34_criterion2_full} as Expression~\eqref{eq:pse_setup34_criterion2_sigmaSimplified};
  \item[3.] Assuming only the $n$-terms are similar in magnitude, we can express the LHS of Inequality~\eqref{eq:pse_setup34_criterion2_full} as Expression~\eqref{eq:pse_setup34_criterion2_nSimplified}; and
  \item[4.] Assuming both the $\sigma^2$-terms and the $n$-terms are similar in magnitude, Inequality~\eqref{eq:pse_setup34_criterion2_full} is approximately equivalent to Inequality~\eqref{eq:pse_setup34_criterion2_bothSimplified}.
\end{enumerate}
We will restate all Inequalities and Expressions in full below where appropriate.
Note the appendix only deals with the detailed algebraic manipulation. We refer readers back to Section~\ref{sec:pse_expt_comparison_dual_control} in the main text for the interpretations of and the insights from the resultant inequalities.

\subsection{The master inequality} 

We first show the criterion $\Delta_{S4} - \Delta_{S3} > \theta^*_{S4} - \theta^*_{S3}$, where
\begin{align}
    \Delta_{S3} = \, &   \frac{n_1(\mu_{C1} - \mu_{I1}) + n_2(\mu_{I2} - \mu_{C2}) + n_3(\mu_{I\psi} - \mu_{I_\phi})}{n_1 + n_2 + n_3} \,, 
    \tag{*\ref{eq:pse_Delta_S3}} \\[0.5em]
    \Delta_{S4} = \, &  \frac{n_2(\mu_{I2} - \mu_{C2}) + n_3(\mu_{I\psi} - \mu_{C3})}{n_2 + n_3} - \frac{n_1(\mu_{I1} - \mu_{C1}) + n_3(\mu_{I\phi} - \mu_{C3})}{n_1 + n_3} \,,
    \tag{*\ref{eq:pse_Delta_S4}} \\[0.5em]
    \theta^*_{S3} = \, &  (z_{1-\frac{\alpha}{2}} - z_{1-\pi_{\textrm{min}}}) 
    \sqrt{\frac{2\left(n_1(\sigma^2_{I1} + \sigma^2_{C1}) + n_2(\sigma^2_{C2} + \sigma^2_{I2}) + n_3(\sigma^2_{I\phi} + \sigma^2_{I\psi})\right)}{(n_1 + n_2 + n_3)^2}} \,,
    \tag{*\ref{eq:pse_theta_star_S3}} \\[0.5em]
    \theta^*_{S4} = \, &  2 \cdot (z_{1-\frac{\alpha}{2}} - z_{1-\pi_{\textrm{min}}}) \sqrt{\frac{n_1\left(\sigma^2_{C1} \!+ \sigma^2_{I1}\right) + n_3\left(\sigma^2_{C3} \!+ \sigma^2_{I\phi}\right)}{(n_1 + n_3)^2} +
    \frac{n_2\left(\sigma^2_{C2} \!+ \sigma^2_{I2}\right) + n_3\left(\sigma^2_{C3} \!+ \sigma^2_{I\psi}\right)}{(n_2 + n_3)^2}} \,,
    \tag{*\ref{eq:pse_theta_star_S4}}
\end{align}
is equivalent to
\begin{align}
    &\frac{n_1 \frac{n_2(\mu_{I2}-\mu_{C2}) + n_3(\mu_{I\psi}-\mu_{C3})}{n_2 + n_3} -
           n_2 \frac{n_1(\mu_{I1}-\mu_{C1}) + n_3(\mu_{I\phi}-\mu_{C3})}{n_1 + n_3}}
           {\sqrt{n_1(\sigma^2_{C1} + \sigma^2_{I1}) +
                  n_2(\sigma^2_{C2} + \sigma^2_{I2}) +
                  n_3(\sigma^2_{I\phi} + \sigma^2_{I\psi})}}
    \nonumber\\
    > & \sqrt{2}z 
    \vast[
      \sqrt{
        2 \cdot 
        \frac{
          \begin{array}{l}
            \big(1 + \frac{n_2}{n_1 + n_3}\big)^2 
              \big[n_1(\sigma^2_{C1} + \sigma^2_{I1}) +
              n_3(\sigma^2_{C3} + \sigma^2_{I\phi})\big] + \\[-0.1em]
            \quad \big(1 + \frac{n_1}{n_2 + n_3}\big)^2 
              \big[n_2(\sigma^2_{C2} + \sigma^2_{I2}) +
              n_3(\sigma^2_{C3} + \sigma^2_{I\psi})\big]
            \end{array}
        }
        {n_1(\sigma^2_{C1} + \sigma^2_{I1}) +
         n_2(\sigma^2_{C2} + \sigma^2_{I2}) +
         n_3(\sigma^2_{I\phi} + \sigma^2_{I\psi})}
        }
      - 1
    \vast] \,,
    \tag{*\ref{eq:pse_setup34_criterion2_full}}
\end{align}
where $z = z_{1-\frac{\alpha}{2}} - z_{1 - \pi_{\textrm{min}}}$. There are many terms involved in the inequality, and hence we first simplify the LHS and RHS independently, and combine them in the final step.

We start by writing the LHS (i.e., $\Delta_{S4} - \Delta_{S3}$) in full using  Equations~\eqref{eq:pse_Delta_S3} and~\eqref{eq:pse_Delta_S4}:
\begin{align}
    & \frac{n_2(\mu_{I2} - \mu_{C2}) + n_3(\mu_{I\psi} - \mu_{C3})}{n_2 + n_3} - 
    \frac{n_1(\mu_{I1} - \mu_{C1}) + n_3(\mu_{I\phi} - \mu_{C3})}{n_1 + n_3} - \nonumber\\
    & \frac{n_1(\mu_{C1} - \mu_{I1}) + n_2(\mu_{I2} - \mu_{C2}) + n_3(\mu_{I\psi} - \mu_{I_\phi})}{n_1 + n_2 + n_3} \,.
    \label{eq:pse_setup34_criterion2_full_lhs}
\end{align}
The expression can be rewritten in terms of multiplicative products between the $n$-terms and the (difference between) $\mu$-terms:
\begin{align}
    & n_1(\mu_{I1} - \mu_{C1})\big[-\textstyle\frac{1}{n_1 + n_3} + \textstyle\frac{1}{n_1 + n_2 + n_3}\big] + 
    n_2(\mu_{I2} - \mu_{C2})\big[\textstyle\frac{1}{n_2 + n_3} - \textstyle\frac{1}{n_1 + n_2 + n_3}\big] + \nonumber\\
    & n_3\mu_{I\psi}\big[\textstyle\frac{1}{n_2 + n_3} - \textstyle\frac{1}{n_1 + n_2 + n_3}\big] +
    n_3\mu_{I\phi}\big[-\textstyle\frac{1}{n_1 + n_3} + \textstyle\frac{1}{n_1 + n_2 + n_3}\big] + 
    n_3\mu_{C3}\big[-\textstyle\frac{1}{n_2 + n_3} + \textstyle\frac{1}{n_1 + n_3}\big] \,.
    \tag{\ref{eq:pse_setup34_criterion2_full_lhs}a}
\end{align}
We then extract a $\frac{1}{n_1 + n_2 + n_3}$ term from Expression~(\ref{eq:pse_setup34_criterion2_full_lhs}a):
\begin{align}
    \frac{1}{n_1 + n_2 + n_3}\Big[
    & n_1(\mu_{I1} - \mu_{C1})\big(\!-\!\textstyle\frac{n_1 + n_2 + n_3}{n_1 + n_3} + 1\big) + 
    n_2(\mu_{I2} - \mu_{C2})\big(\textstyle\frac{n_1 + n_2 + n_3}{n_2 + n_3} - 1\big)+ \nonumber\\
    & n_3\mu_{I\psi}\big(\textstyle\frac{n_1 + n_2 + n_3}{n_2 + n_3} - 1\big) +
    n_3\mu_{I\phi}\big(\!-\!\textstyle\frac{n_1 + n_2 + n_3}{n_1 + n_3} + 1\big) + 
    n_3\mu_{C3}\big(\!-\!\textstyle\frac{n_1 + n_2 + n_3}{n_2 + n_3} + \textstyle\frac{n_1 + n_2 + n_3}{n_1 + n_3}\big)
    \Big] \,.
    \tag{\ref{eq:pse_setup34_criterion2_full_lhs}b}
\end{align}
This allows us to perform some cancellation with the RHS, which also has a $\frac{1}{n_1 + n_2 + n_3}$ term, in the final step. Noting
\begin{align}
    \frac{n_1 + n_2 + n_3}{n_1 + n_3} = 1 + \frac{n_2}{n_1 + n_3} \;\;\textrm{and}\;\;
    \frac{n_1 + n_2 + n_3}{n_2 + n_3} = 1 + \frac{n_1}{n_2 + n_3} \,,
    \nonumber
\end{align}
we can write the Expression~(\ref{eq:pse_setup34_criterion2_full_lhs}b) as
\begin{align}
    \frac{1}{n_1 + n_2 + n_3}\Big[
    & n_1(\mu_{I1} - \mu_{C1})\big(-\textstyle\frac{n_2}{n_1 + n_3}\big) +
      n_2(\mu_{I2} - \mu_{C2})\big(\textstyle\frac{n_1}{n_2 + n_3}\big) + \nonumber\\
    & n_3\mu_{I\psi}\big(\textstyle\frac{n_1}{n_2 + n_3}\big) +
      n_3\mu_{I\phi}\big(-\textstyle\frac{n_2}{n_1 + n_3}\big) + n_3\mu_{C3}\big(- 1 - \textstyle\frac{n_1}{n_2 + n_3} + 1 + \frac{n_2}{n_1 + n_3}\big)
    \Big] \,,
    \tag{\ref{eq:pse_setup34_criterion2_full_lhs}c}
\end{align}
and group the $\frac{n_1}{n_2 + n_3}$ and $\frac{n_2}{n_1 + n_3}$ terms to arrive at
\begin{align}
    \frac{1}{n_1 + n_2 + n_3}\bigg[
    & \frac{n_1}{n_2 + n_3} \left(n_2(\mu_{I2}-\mu_{C2}) + n_3(\mu_{I\psi}-\mu_{C3})\right) - \nonumber\\
    & \frac{n_2}{n_1 + n_3} \left(n_1(\mu_{I1}-\mu_{C1}) + n_3(\mu_{I\phi}-\mu_{C3})\right) \bigg] \,.
    \tag{\ref{eq:pse_setup34_criterion2_full_lhs}d}
\end{align}

We also write the RHS (i.e., $\theta^*_{S4} - \theta^*_{S3}$) in full using Equations~\eqref{eq:pse_theta_star_S3} and~\eqref{eq:pse_theta_star_S4}:
\begin{align}
    & 2z\sqrt{
    \frac{n_1(\sigma^2_{C1} + \sigma^2_{I1}) + n_3(\sigma^2_{C3} + \sigma^2_{I\phi})}{(n_1 + n_3)^2} +
    \frac{n_2(\sigma^2_{C2} + \sigma^2_{I2}) + n_3(\sigma^2_{C3} + \sigma^2_{I\psi})}{(n_2 + n_3)^2}} \nonumber\\
    & - \sqrt{2}z
    \sqrt{\frac{n_1(\sigma^2_{I1} + \sigma^2_{C1}) + n_2(\sigma^2_{C2} + \sigma^2_{I2}) + n_3(\sigma^2_{I\phi} + \sigma^2_{I\psi})}{(n_1 + n_2 + n_3)^2}} \,,
    \label{eq:pse_setup34_criterion2_full_rhs}
\end{align}
where $z = z_{1-\frac{\alpha}{2}} - z_{1 - \pi_{\textrm{min}}}$.
We then extract a $\frac{\sqrt{2}z}{n_1 + n_2 + n_3}$ term from Expression~\eqref{eq:pse_setup34_criterion2_full_rhs} to obtain
\begin{align}
    \frac{\sqrt{2}z}{n_1 + n_2 + n_3} \vast[
    & \sqrt{2} \sqrt{
        \begin{array}{l}
          \big(\frac{n_1 + n_2 + n_3}{n_1 + n_3}\big)^2 \left[
              n_1(\sigma^2_{C1} + \sigma^2_{I1}) + n_3(\sigma^2_{C3} + \sigma^2_{I\phi})\right] + \\
          \;\; \big(\frac{n_1 + n_2 + n_3}{n_2 + n_3}\big)^2 \left[
              n_2(\sigma^2_{C2} + \sigma^2_{I2}) + n_3(\sigma^2_{C3} + \sigma^2_{I\psi})\right]
        \end{array}
    } \, - \nonumber\\
    & \sqrt{n_1(\sigma^2_{C1} + \sigma^2_{I1}) +
            n_2(\sigma^2_{C2} + \sigma^2_{I2}) +
            n_3(\sigma^2_{I\phi} + \sigma^2_{I\psi})}
    \,\vast] \,,
    \tag{\ref{eq:pse_setup34_criterion2_full_rhs}a}
\end{align}
where $\frac{n_1 + n_2 + n_3}{n_2 + n_3}$ and $\frac{n_1 + n_2 + n_3}{n_1 + n_3}$ can also be written as $1 + \frac{n_1}{n_2 + n_3}$ and $1 + \frac{n_2}{n_1 + n_3}$, respectively.

We finally combine both sides of the inequality by taking Expressions~(\ref{eq:pse_setup34_criterion2_full_lhs}d) and~(\ref{eq:pse_setup34_criterion2_full_rhs}a):
\begin{align}
    \frac{1}{n_1 + n_2 + n_3}\bigg[
    & \frac{n_1}{n_2 + n_3} \left(n_2(\mu_{I2}-\mu_{C2}) + n_3(\mu_{I\psi}-\mu_{C3})\right) - \nonumber\\
    & \;\;\frac{n_2}{n_1 + n_3} \left(n_1(\mu_{I1}-\mu_{C1}) + n_3(\mu_{I\phi}-\mu_{C3})\right) \bigg]
    \nonumber\\
    > \frac{\sqrt{2}z}{n_1 + n_2 + n_3} \vast[
    & \sqrt{2} \sqrt{
        \begin{array}{l}
          \big(1 + \frac{n_2}{n_1 + n_3}\big)^2 \left[
              n_1(\sigma^2_{C1} + \sigma^2_{I1}) + n_3(\sigma^2_{C3} + \sigma^2_{I\phi}) \right] + \\
          \quad\big(1 + \frac{n_1}{n_2 + n_3}\big)^2 \left[
              n_2(\sigma^2_{C2} + \sigma^2_{I2}) + n_3(\sigma^2_{C3} + \sigma^2_{I\psi}) \right]
        \end{array}
    } \,- \nonumber\\
    & \quad\sqrt{n_1(\sigma^2_{C1} + \sigma^2_{I1}) +
            n_2(\sigma^2_{C2} + \sigma^2_{I2}) +
            n_3(\sigma^2_{I\phi} + \sigma^2_{I\psi})}
    \,\vast] \,.
    \label{eq:pse_setup34_criterion2_full_combined}
\end{align}
Canceling the common $\frac{1}{n_1 + n_2 + n_3}$ terms on both sides, and dividing both sides by\\ ${\sqrt{n_1(\sigma^2_{C1} + \sigma^2_{I1}) + n_2(\sigma^2_{C2} + \sigma^2_{I2}) + n_3(\sigma^2_{I\phi} + \sigma^2_{I\psi})}}$ leads us to
\begin{align}
    &\frac{n_1 \frac{n_2(\mu_{I2}-\mu_{C2}) + n_3(\mu_{I\psi}-\mu_{C3})}{n_2 + n_3} -
           n_2 \frac{n_1(\mu_{I1}-\mu_{C1}) + n_3(\mu_{I\phi}-\mu_{C3})}{n_1 + n_3}}
           {\sqrt{n_1(\sigma^2_{C1} + \sigma^2_{I1}) +
                  n_2(\sigma^2_{C2} + \sigma^2_{I2}) +
                  n_3(\sigma^2_{I\phi} + \sigma^2_{I\psi})}}
    \nonumber\\
    > & \sqrt{2}z 
    \vast[
      \sqrt{
        2 \cdot 
        \frac{
          \begin{array}{l}
            \big(1 + \frac{n_2}{n_1 + n_3}\big)^2 
              \big[n_1(\sigma^2_{C1} + \sigma^2_{I1}) +
              n_3(\sigma^2_{C3} + \sigma^2_{I\phi})\big] + \\[-0.1em]
            \quad \big(1 + \frac{n_1}{n_2 + n_3}\big)^2 
              \big[n_2(\sigma^2_{C2} + \sigma^2_{I2}) +
              n_3(\sigma^2_{C3} + \sigma^2_{I\psi})\big]
            \end{array}
        }
        {n_1(\sigma^2_{C1} + \sigma^2_{I1}) +
         n_2(\sigma^2_{C2} + \sigma^2_{I2}) +
         n_3(\sigma^2_{I\phi} + \sigma^2_{I\psi})}
        }
      - 1
    \vast] \,,
    \nonumber
\end{align}
which is identical to Inequality~\eqref{eq:pse_setup34_criterion2_full}.

\subsection{Simplifying RHS by assuming similar response variances} 
We now simplify the $\sigma^2$-terms in Inequality~\eqref{eq:pse_setup34_criterion2_full} by assuming that they are similar in magnitude, i.e.,
\begin{align}
    \sigma^2_{C1} \approx \sigma^2_{I1} \approx \cdots \approx \sigma^2_{I\psi} \approx \sigma^2_G \,, \nonumber
\end{align}
and show the RHS of the inequality, i.e.,
\begin{align}
    \sqrt{2}z 
    \vast[
      \sqrt{
        2 \cdot 
        \frac{
          \begin{array}{l}
            \big(1 + \frac{n_2}{n_1 + n_3}\big)^2 
              \big[n_1(\sigma^2_{C1} + \sigma^2_{I1}) +
              n_3(\sigma^2_{C3} + \sigma^2_{I\phi})\big] + \\[-0.1em]
            \quad \big(1 + \frac{n_1}{n_2 + n_3}\big)^2 
              \big[n_2(\sigma^2_{C2} + \sigma^2_{I2}) +
              n_3(\sigma^2_{C3} + \sigma^2_{I\psi})\big]
            \end{array}
        }
        {n_1(\sigma^2_{C1} + \sigma^2_{I1}) +
         n_2(\sigma^2_{C2} + \sigma^2_{I2}) +
         n_3(\sigma^2_{I\phi} + \sigma^2_{I\psi})}
        }
      - 1
    \vast] \,,
    \nonumber
\end{align}
is approximately equal to
\begin{align}
   \sqrt{2} z 
    \left[ 
        \sqrt{\frac{n_1 + n_2 + n_3}{n_1 + n_3} + 
        \frac{n_1 + n_2 + n_3}{n_2 + n_3}} - 1 
    \right] \,.
    \tag{*\ref{eq:pse_setup34_criterion2_sigmaSimplified}}
\end{align}
It is safe to apply the simplifying assumption as we know from the evaluation framework specification that there are three classes of parameters in the inequality: the user group sizes~($n$), the mean responses ($\mu$), and the response variances ($\sigma^2$). Among these three classes of parameters, only the user group sizes have the potential to scale in any practical settings, and thus we can effectively treat the means and variances as constants below. 

We begin by substituting $\sigma^2_G$ into Inequality~\eqref{eq:pse_setup34_criterion2_full}:
\begin{align}
    & \frac{n_1 \frac{n_2(\mu_{I2}-\mu_{C2}) + n_3(\mu_{I\psi}-\mu_{C3})}{n_2 + n_3} -
           n_2 \frac{n_1(\mu_{I1}-\mu_{C1}) + n_3(\mu_{I\phi}-\mu_{C3})}{n_1 + n_3}}
           {\sqrt{n_1(2\sigma^2_G) +
                  n_2(2\sigma^2_G) +
                  n_3(2\sigma^2_G)}}
    \nonumber\\
    > & \sqrt{2}z \left[
    \sqrt{
        2 \cdot 
        \frac{\big(1 + \frac{n_2}{n_1 + n_3}\big)^2 
              \big[n_1(2\sigma^2_G) + n_3(2\sigma^2_G)\big] +
              \big(1 + \frac{n_1}{n_2 + n_3}\big)^2 
              \big[n_2(2\sigma^2_G) + n_3(2\sigma^2_G)\big]}
             {n_1(2\sigma^2_G) + n_2(2\sigma^2_G) + n_3(2\sigma^2_G)}}
        - 1
    \right] \,.
    \label{eq:pse_setup34_criterion2_sigmaSimplified_initial}
\end{align}
Moving the common $2\sigma^2_G$ terms out and canceling the common terms in the RHS fraction we have
\begin{align}
    &\frac{n_1 \frac{n_2(\mu_{I2}-\mu_{C2}) + n_3(\mu_{I\psi}-\mu_{C3})}{n_2 + n_3} -
           n_2 \frac{n_1(\mu_{I1}-\mu_{C1}) + n_3(\mu_{I\phi}-\mu_{C3})}{n_1 + n_3}}
           {\sqrt{2\sigma^2_G(n_1 + n_2 + n_3)}} >
    \nonumber\\
    & \sqrt{2}z \left[
    \sqrt{
        2 \cdot 
        \frac{ \big(1 +\frac{n_2}{n_1 + n_3}\big)^2 (n_1 + n_3) +
               \big(1 + \frac{n_1}{n_2 + n_3}\big)^2 (n_2 + n_3)}
            {n_1 + n_2 + n_3}}
        - 1
    \right].
    \tag{\ref{eq:pse_setup34_criterion2_sigmaSimplified_initial}a}
\end{align}
We can already see the LHS of Inequality~(\ref{eq:pse_setup34_criterion2_sigmaSimplified_initial}a) scales along $O(\sqrt{n})$ -- we will demonstrate this result in greater detail in the next subsection.

Focusing on the RHS of the inequality, we express the squared terms as rational fractions and divide each term in the numerator by the denominator to obtain
\begin{align}
    \sqrt{2}z \left[
    \sqrt{
        2  \left[
        \left(\frac{n_1 + n_2 + n_3}{n_1 + n_3}\right)^{2} \frac{n_1 + n_3}{n_1 + n_2 + n_3} +
        \left(\frac{n_1 + n_2 + n_3}{n_2 + n_3}\right)^{2} \frac{n_2 + n_3}{n_1 + n_2 + n_3} \right]}
        - 1 
    \right].
    \label{eq:pse_setup34_criterion2_sigmaSimplified_rhs}
\end{align}
Canceling the common $n_1 + n_2 + n_3$ terms leads to that presented in Expression~\eqref{eq:pse_setup34_criterion2_sigmaSimplified}:
\begin{align}
    \sqrt{2} z \left[ \sqrt{\frac{n_1 + n_2 + n_3}{n_1 + n_3} + \frac{n_1 + n_2 + n_3}{n_2 + n_3}} - 1 \right].
    \nonumber
\end{align}

\subsection{Simplifying LHS by assuming similar sample sizes}
We demonstrate the scaling relation between the LHS of Inequality~\eqref{eq:pse_setup34_criterion2_full} and the number of users in each group by simplifying the $n$-terms (assuming $n_1 \approx n_2 \approx n_3 \approx n$) and showing the LHS of the inequality, i.e.,
\begin{align}
    \frac{n_1 \frac{n_2(\mu_{I2}-\mu_{C2}) + n_3(\mu_{I\psi}-\mu_{C3})}{n_2 + n_3} -
           n_2 \frac{n_1(\mu_{I1}-\mu_{C1}) + n_3(\mu_{I\phi}-\mu_{C3})}{n_1 + n_3}}
           {\sqrt{n_1(\sigma^2_{C1} + \sigma^2_{I1}) +
                  n_2(\sigma^2_{C2} + \sigma^2_{I2}) +
                  n_3(\sigma^2_{I\phi} + \sigma^2_{I\psi})}}
    \nonumber
\end{align}
is approximately equal to
\begin{align}
    \frac{\sqrt{n} \big((\mu_{I2} - \mu_{C2}) - (\mu_{I1} - \mu_{C1}) + \mu_{I\psi} - \mu_{I\phi}\big)}{2\sqrt{\sigma^2_{C1} + \sigma^2_{I1} + \sigma^2_{C2} + \sigma^2_{I2} + \sigma^2_{I\phi} + \sigma^2_{I\psi}}} \,.
    \tag{*\ref{eq:pse_setup34_criterion2_nSimplified}}
\end{align}
Note that unlike the previous subsection, we do not make any simplifying assumptions on  the $\sigma^2$-terms here.
While the relationship (that the LHS of the inequality scales along $O(\sqrt{n})$) is evident by inspecting Inequality~\eqref{eq:pse_setup34_criterion2_full} (or its simplified version as Inequality~(\ref{eq:pse_setup34_criterion2_sigmaSimplified_initial}a)) itself, we believe the simplification allows us to show the relationship more clearly.

We begin by substituting $n$ into Inequality~\eqref{eq:pse_setup34_criterion2_full} to obtain
\begin{align}
    & \frac{n \frac{n(\mu_{I2}-\mu_{C2}) + n(\mu_{I\psi}-\mu_{C3})}{n + n} -
           n \frac{n(\mu_{I1}-\mu_{C1}) + n(\mu_{I\phi}-\mu_{C3})}{n + n}}
           {\sqrt{n(\sigma^2_{C1} + \sigma^2_{I1}) +
                  n(\sigma^2_{C2} + \sigma^2_{I2}) +
                  n(\sigma^2_{I\phi} + \sigma^2_{I\psi})}} 
    \nonumber\\
    > & \sqrt{2}z 
    \vast[
      \sqrt{
        2 \cdot 
        \frac{
            \begin{array}{l}
                \big(1 + \frac{n}{n + n}\big)^2 
                  \big[n(\sigma^2_{C1} + \sigma^2_{I1}) +
                  n(\sigma^2_{C3} + \sigma^2_{I\phi})\big] + \\[-0.1em]
                \quad \big(1 + \frac{n}{n + n}\big)^2 
                  \big[n(\sigma^2_{C2} + \sigma^2_{I2}) +
                  n(\sigma^2_{C3} + \sigma^2_{I\psi})\big]
            \end{array}
        }
        {n(\sigma^2_{C1} + \sigma^2_{I1}) +
            n(\sigma^2_{C2} + \sigma^2_{I2}) +
            n(\sigma^2_{I\phi} + \sigma^2_{I\psi})}
      }
      - 1
    \vast] \,.
    \label{eq:pse_setup34_criterion2_nSimplified_initial}
\end{align}
Moving the common $n$-terms out and canceling them in the fractions where appropriate lead to
\begin{align}
    & \frac{\sqrt{n} \,\frac{1}{2}\left[((\mu_{I2}-\mu_{C2}) + (\mu_{I\psi}-\mu_{C3})) -
           ((\mu_{I1}-\mu_{C1}) + (\mu_{I\phi}-\mu_{C3})) \right]}
           {\sqrt{\sigma^2_{C1} + \sigma^2_{I1} +\sigma^2_{C2} + \sigma^2_{I2} +\sigma^2_{I\phi} + \sigma^2_{I\psi}}}
    \nonumber\\
    > & \sqrt{2}z \left[
    \sqrt{
        2 \cdot 
        \frac{
            \big(1 + \frac{1}{2}\big)^2 
              (\sigma^2_{C1} + \sigma^2_{I1} + \sigma^2_{C3} + \sigma^2_{I\phi}) +
            \big(1 + \frac{1}{2}\big)^2 
              (\sigma^2_{C2} + \sigma^2_{I2} + \sigma^2_{C3} + \sigma^2_{I\psi})
        }
        {\sigma^2_{C1} + \sigma^2_{I1} +
         \sigma^2_{C2} + \sigma^2_{I2} +
         \sigma^2_{I\phi} + \sigma^2_{I\psi}}}
        - 1
    \right] \,,
    \tag{\ref{eq:pse_setup34_criterion2_nSimplified_initial}a}
\end{align}
where the LHS is identical to Expression~\eqref{eq:pse_setup34_criterion2_nSimplified} after canceling the $\mu_{C3}$ terms in the numerator and expressing the half in the numerator as a two in the denominator.

It is clear that there are no $n$-terms left on the RHS of Inequality~(\ref{eq:pse_setup34_criterion2_nSimplified_initial}a), and hence the RHS remains constant as shown in the previous subsection. To set up the inequality to demonstrate the last result -- that the number of users required for a dual control setup to emerge superior is large -- we further simplify the RHS of the inequality by rearranging the terms in the square root:
\begin{align}
    & \frac{\sqrt{n} \left[((\mu_{I2}-\mu_{C2}) + (\mu_{I\psi}-\mu_{C3})) -
           ((\mu_{I1}-\mu_{C1}) + (\mu_{I\phi}-\mu_{C3})) \right]}
           {2\sqrt{\sigma^2_{C1} + \sigma^2_{I1} +\sigma^2_{C2} + \sigma^2_{I2} +\sigma^2_{I\phi} + \sigma^2_{I\psi}}}
    \nonumber\\
    > & \sqrt{2}z \left[
    \sqrt{
        2 \left(\frac{3}{2}\right)^2
        \left(1 + \frac{2\sigma^2_{C3}}
                  {\sigma^2_{C1} + \sigma^2_{I1} + \sigma^2_{C2} +
                   \sigma^2_{I2} +\sigma^2_{I\phi} + \sigma^2_{I\psi}}\right)}
        - 1
    \right] \,.
    \tag{\ref{eq:pse_setup34_criterion2_nSimplified_initial}b}
\end{align}

\subsection{Required number of users for dual control to emerge superior}
We finally show that while Setup 4 could emerge superior to Setup 3 as the number of users increase, the number of users required is large. We do so by assuming both the $\sigma^2$- and $n$-terms are similar in magnitude, i.e. $\sigma^2_{C1} \approx \sigma^2_{I1} \approx \cdots \approx \sigma^2_{I\psi} \approx \sigma^2_G$ and $n_1 \approx n_2 \approx n_3 \approx n$, and show that Inequality~\eqref{eq:pse_setup34_criterion2_full} is approximately equivalent to
\begin{align}
    n > \left(2\sqrt{12}\left(\sqrt{6} - 1\right)z\right)^2 \frac{\sigma^2_G}{\Delta^2} \,,
    \tag{*\ref{eq:pse_setup34_criterion2_bothSimplified}}
\end{align}
where $\Delta = (\mu_{I2} - \mu_{C2}) - (\mu_{I1} - \mu_{C1}) + \mu_{I\psi} - \mu_{I\phi}$ is the actual effect size difference between Setups~4 and~3. Note we are determining when Setup 4 is superior to Setup 3 under the second evaluation criterion -- that \emph{the gain in actual effect} is greater than the loss in sensitivity -- and thus assume $\Delta$ is positive.

The approximate equivalence can be shown by substituting $\sigma^2_G$ into Inequality~(\ref{eq:pse_setup34_criterion2_nSimplified_initial}b), which is Inequality~\eqref{eq:pse_setup34_criterion2_full} with the additional assumption that $n$-terms are similar in magnitude:\footnote{Alternatively, we can substitute $n$ into Inequality~(\ref{eq:pse_setup34_criterion2_sigmaSimplified_initial}a), which is Inequality~\eqref{eq:pse_setup34_criterion2_full} with the additional assumption that the $\sigma^2$-terms are similar in magnitude. Simplifying the resultant inequality would yield the same result.}
\begin{align}
    & \frac{\sqrt{n} \left[((\mu_{I2}-\mu_{C2}) + (\mu_{I\psi}-\mu_{C3})) -
           ((\mu_{I1}-\mu_{C1}) + (\mu_{I\phi}-\mu_{C3})) \right]}
           {2\sqrt{6\sigma^2_G}}
    \nonumber\\
    > & \sqrt{2}z \Bigg[
    \sqrt{
        2 \bigg(\frac{3}{2}\bigg)^2
        \bigg(1 + \frac{2\sigma^2_G}
                  {6\sigma^2_G}\bigg)}
        - 1
    \Bigg] \,.
    \label{eq:pse_setup34_criterion2_bothSimplified_initial}
\end{align}
Noting the expression within the LHS square bracket is equal to $\Delta$, we simplify the expression within the RHS square root, and move every non-$n$ term to the RHS of the inequality to obtain
\begin{align}
    \sqrt{n} > \sqrt{2}{z}\left[\sqrt{6}-1\right]\frac{2\sqrt{6\sigma^2_G}}{\Delta} \,.
    \tag{\ref{eq:pse_setup34_criterion2_bothSimplified_initial}a}
\end{align}
As all quantities in the inequality are positive, we can square both sides and consolidate the coefficients on the RHS to arrive at
\begin{align}
    n > \left(2\sqrt{12}\left(\sqrt{6} - 1\right)z\right)^2 \frac{\sigma^2_G}{\Delta^2} \,,
    \nonumber
\end{align}
which is identical to Inequality~\eqref{eq:pse_setup34_criterion2_bothSimplified}.

\end{document}